\documentclass[11pt,oneside,a4paper]{book}

\usepackage{fullpage}

\usepackage{booktabs}
\usepackage{listings}
\usepackage{lscape}
\usepackage{multicol}
\usepackage{cmll}
\usepackage{amsmath}

\usepackage{varioref}

\renewcommand*{\thepage}{\small\thepage}

\usepackage{pifont}

\usepackage{euler}
\usepackage{titlesec}

\usepackage{parcolumns}

\usepackage{float}
\usepackage[noadjust]{cite}

\usepackage{etex}
\usepackage{color, colortbl}
\usepackage[usenames,dvipsnames]{xcolor}
\usepackage{alltt}

\usepackage[printonlyused]{acronym}
\usepackage[T1]{fontenc}
\usepackage{textcomp}

\usepackage{emptypage}

\definecolor{halfgray}{gray}{0.55}

\usepackage{fancyhdr}
\usepackage{sectsty}

\titleformat{\chapter}[display]
{\normalfont\bfseries\raggedright}{\chaptertitlename \raggedleft\fontsize{45}{0}\selectfont$ \thechapter$}{0ex}{\LARGE}
[\vspace{0.5ex}%
{\titlerule[0.5pt]}]

\usepackage{flafter}   
\usepackage{alltt}
 \definecolor{Gray}{gray}{0.95}

\let\cleardoublepage\clearpage

\usepackage{etoolbox}

\usepackage[T1]{fontenc}
\usepackage{charter}
\usepackage[expert]{mathdesign}

\usepackage[pdftex]{graphicx}
\usepackage{epstopdf}
\usepackage{todonotes}

\usepackage{amsmath}
\usepackage{amssymb}
\usepackage{algorithmic}

\usepackage[titletoc]{appendix}

\usepackage{rotating}

\usepackage[all]{xy}
\usepackage[font=small,format=plain,labelfont=bf,up,textfont=it,up]{caption}
\usepackage{caption}
\usepackage{subcaption}

\usepackage{mathtools}
\usepackage{circuitikz}
\usepackage[pdftex,pdfauthor={M.Singh}, pdftitle={Exploring the fundamental properties of Neutrino from Oscillation Experiments}, pdfkeywords={Neutrino, Oscillation, standar three neutrino-paradigm, DUNE, T2HK, octant, matter effect, CP coverage, flavor-diagonal long-range interaction, coupling vs mass plane, complementarity, synergy between experiments, long-baseline},  colorlinks=true,citecolor=black, urlcolor=black, linkcolor=black]{hyperref}

\usepackage[english]{cleveref}

\usepackage[nottoc]{tocbibind}
\usepackage{setspace}
\usepackage[left=1.5in,right=1in,top=1.3in,bottom=1.3in]{geometry}

\definecolor{dkgreen}{rgb}{0,0.6,0}

\usepackage{etex}
\usepackage{tikz}

\usepackage{xstring}
\usepackage[titles]{tocloft}

\let\oldequation = \equation
\let\endoldequation = \endequation
\AtBeginDocument{\let\oldlabel = \label}
\newcommand{\mynewlabel}[1]{%
  \StrBehind{#1}{eq:}[\Str]
  \myequations{\Str}\oldlabel{#1}}
  \renewenvironment{equation}{%
  \oldequation
  \let\label\mynewlabel
}{\endoldequation}

\newcommand{\listequationsname}{List of equations}
\newlistof{myequations}{equ}{\listequationsname}
\newcommand{\myequations}[1]{%
      \addcontentsline{equ}{myequations}{\protect\numberline{\theequation}#1}}
\usepackage{amsmath}
\usepackage{latexsym}
\usepackage{multirow}
\usepackage{physics}
\usepackage{mathrsfs}
\usepackage{soul}
\usepackage{slashed}
\usepackage{wasysym}

\emergencystretch=\maxdimen
\hyphenchar\font=-1
\newcommand{\beq}{\begin{equation}}
\newcommand{\eeq}{\end{equation}}
\newcommand{\beqa}{\begin{eqnarray}}
\newcommand{\eeqa}{\end{eqnarray}}
\newcommand{\dcp}{\delta_{\mathrm{CP}}}
\newcommand{\Amue}{\mathcal{A}^{\mu e}_{\mathrm{CP}}}
\newcommand{\Amuevac}{[\mathcal{A}^{\mu e}_{\mathrm{CP}}]_{\mathrm{vac}}}
\newcommand{\Amuemat}{[\mathcal{A}^{\mu e}_{\mathrm{CP}}]_{\mathrm{mat}}}

\newcommand{\tzm}{\ensuremath{\theta_{23}}}

\newcommand{\ie}{\textit{i.e.}}

\newcommand{\equ}[1]{Eq.~(\ref{equ:#1})}
\newcommand{\figu}[1]{\fig~\ref{fig:#1}}

\newcommand{\eg}{{\it e.g.}}
\newcommand{\fig}{Fig.}

\newcommand{\Refe}{Ref.}
\newcommand{\Refes}{Refs.}
\begin{document}

\pagenumbering{Alph}
\begin{titlepage}

\centering{
\textsc{\LARGE \textbf{\textcolor{Black}{Exploring the Fundamental Properties of Neutrino from Oscillation Experiments}}}\\

\vspace{2cm}
\ifx\mysynopsis\undefined
\emph{A thesis submitted for the degree of\\}
\else
\emph{Synopsis of the thesis to be submitted to the\\
board of studies in engineering sciences\\
in partial fulfillment of requirements\\
for the degree of}
\fi
{\bf DOCTOR OF PHILOSOPHY}\\
{\bf IN}\\[0.2cm]
{\bf PHYSICS}
 
\vspace{0.7cm}
submitted\\
by\\
\textsc{\large \textbf{{MASOOM SINGH}}}\\

\textsc{{\sc (Regd. No - 02-Physics, 2017-18)}}\\
\textsc{{\sc (DST/INSPIRE/2018/IF180059)}}\\[0.7cm]

\vspace{1cm}
\includegraphics[width=0.26\textwidth]{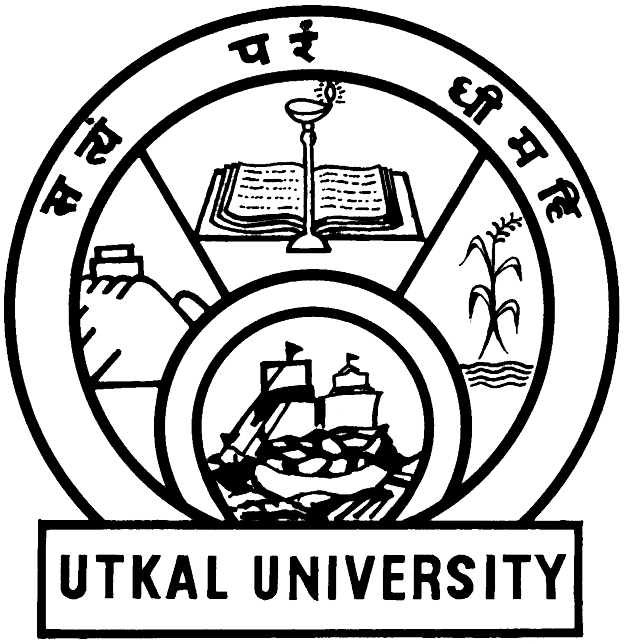}
 
 \singlespacing
 P.G. DEPARTMENT OF PHYSICS\\ [0.3cm]
 UTKAL UNIVERSITY, VANIVIHAR\\ [0.3cm]
 BHUBANESWAR, ODISHA, 751004
 
\vspace{1cm}
{\sc {November $-$ 2023}}
}
\end{titlepage}

\begin{titlepage}

\centering{
\textsc{\LARGE \textbf{\textcolor{Black}{Exploring the Fundamental Properties of Neutrino from Oscillation Experiments}}}\\

\vspace{1.8cm}
\ifx\mysynopsis\undefined
\emph{A thesis submitted for the degree of\\}
\else
\emph{Synopsis of the thesis to be submitted to the\\
board of studies in engineering sciences\\
in partial fulfillment of requirements\\
for the degree of}
\fi
{\bf DOCTOR OF PHILOSOPHY}\\
{\bf IN}\\
{\bf PHYSICS}
 
\vspace{0.7cm}
submitted

\vspace{0.05cm}
by\\
\textsc{\large \textbf{{MASOOM SINGH}}}\\

\textsc{{\sc (Regd. No - 02-Physics, 2017-18)}}\\
\textsc{{\sc (DST/INSPIRE/2018/IF180059)}}\\

\vspace{0.4cm}
\emph{under the guidance of}\\
\textsc{{\sc Guide: Prof. Swapna Mahapatra, Utkal University}}\\
\textsc{{\sc Co-Guide: Prof. Sanjib Kumar Agarwalla, Institute of Physics}}\\[0.7cm]

\vspace{0.5cm}
\includegraphics[width=0.26\textwidth]{gfx/uu_logo.png}
 
 \singlespacing
 P.G. DEPARTMENT OF PHYSICS\\ [0.3cm]
 UTKAL UNIVERSITY, VANIVIHAR\\ [0.3cm]
 BHUBANESWAR, ODISHA, 751004
 
\vspace{1cm}
{\sc {November $-$ 2023}}
}
\end{titlepage}

\pagenumbering{alph}
\clearpage

\thispagestyle{empty}

\chapter*{{DECLARATION}}
\thispagestyle{empty}
I, Mrs. Masoom Singh, hereby declare that the thesis entitled ``\textbf{Exploring the Fundamental Properties of Neutrino from Oscillation Experiments}''\,, submitted to \textbf{Utkal University, Bhubaneswar}, in accordance with academic rules and ethical conduct for the degree of Doctor of Philosophy in Physics, is entirely origial and has been carried out independently by me under the supervision of \textbf{Prof. Swapna Mahapatra}, Utkal University and co-supervision of \textbf{Prof. Sanjib Kumar Agarwalla}, Institute of Physics, Bhubaneswar. I also state that no part of this thesis has been published for the purpose of conferring a degree or diploma at any university or institution. All materials and results not original to this work have been absolutely referred to and cited.


\vspace{5cm}
\begin{flushright}
Masoom Singh\\
No.- 02-Physics 2017-18
\end{flushright}

\begin{titlepage}

\begin{center}
\textsc{\LARGE \textbf{\textcolor{Black}CERTIFICATE}}
\end{center}

\vspace{0.8cm}
This is to certify that the work incorporated in the thesis entitled ``\textbf{Exploring the Fundamental Properties of Neutrino from Oscillation Experiments}'' by \textbf{Mrs. Masoom Singh} (Regd. No.- 02-Physics 2017-18), for the partial fulfilment of the degree of Doctor of Philosophy in Physics being submitted to \textbf{Utkal University, Bhubaneswar, Odisha, India} is a record of her own research work and within the area of registration. The research work reported in this thesis is an original work carried by the candidate and has not been submitted to any university or institution for the award of any degree or diploma.

I am satisfied that the thesis has reached the standards fulfilling the requirement of the regulations relating to the nature of the degree.

\vspace{5cm}
\begin{minipage}{.5\textwidth}
        \textsc{{\sc DATE}}
    \end{minipage}%
    \begin{minipage}{0.5\textwidth}
       \textsc{{\sc Prof. Swapna Mahaptra (Guide)\\
 P.G. Depratment of Physics\\
 Utkal University, Vanivihar\\ Bhubaneswar, Odisha, 751004 }}
    \end{minipage}

\end{titlepage}

\begin{titlepage}

\begin{center}
\textsc{\LARGE \textbf{\textcolor{Black}CERTIFICATE}}
\end{center}

\vspace{0.8cm}
This is to certify that the work incorporated in the thesis entitled ``\textbf{Exploring the Fundamental Properties of Neutrino from Oscillation Experiments}'' by \textbf{Mrs. Masoom Singh} (Regd. No.- 02-Physics 2017-18), for the partial fulfilment of the degree of Doctor of Philosophy in Physics being submitted to \textbf{Utkal University, Bhubaneswar, Odisha, India} is a record of her own research work and within the area of registration. The research work reported in this thesis is an original work carried by the candidate and has not been submitted to any university or institution for the award of any degree or diploma.

I am satisfied that the thesis has reached the standards fulfilling the requirement of the regulations relating to the nature of the degree.

\vspace{5cm}
\begin{minipage}{.4\textwidth}
        \textsc{{\sc DATE}}
    \end{minipage}%
    \begin{minipage}{0.6\textwidth}
       \textsc{{\sc Prof. Sanjib Kumar Agarwalla (Co-Guide)\\
 Institute of Physics, Sachivalaya Marg\\
 Bhubaneswar, Odisha, 751005 }}
    \end{minipage}

\end{titlepage}

\begin{titlepage}

\begin{center}
\textsc{\LARGE \textbf{\textcolor{Black}AREA CERTIFICATE}}
\end{center}

\vspace{0.8cm}
This is to certify that the work incorporated in the thesis entitled ``Exploring the
Fundamental Properties of Neutrino from Oscillation Experiments'' deals with 
a comprehensive study of the mass-mixing parameters of massive neutrinos 
in the next generation of long-baseline experiments. In this thesis, she 
demonstrates the abilities of these future experiments to accurately measure 
the three flavor oscillation parameters, particularly emphasizing the 
determination and significance of the leptonic CP phase, conducting precision 
studies on atmospheric oscillation parameters, exploring the resolution of the 
octant of 2-3 mixing, and investigating flavor-dependent long-range interactions 
of neutrinos with primarily electrons and neutrinos in big celestial bodies.
She conducted this phenomenological research as part of her thesis under 
the able guidance of faculties from the physics department, and submitted her
thesis to the Utkal University, Vani Vihar, Bhubaneswar, adhering to all 
academic rules and ethical conduct for the degree of Doctor of Philosophy 
in Physics. Furthermore, I certify that this thesis work falls within the domain of
Physics and the content of the thesis coincides with the topic selected at the 
time of registration.

\vspace{5cm}
\begin{minipage}{.5\textwidth}
        \textsc{{\sc DATE}}
    \end{minipage}%
    \begin{minipage}{0.5\textwidth}
       \textsc{{\sc Prof. Swapna Mahaptra (Guide)\\
 P.G. Depratment of Physics\\
 Utkal University, Vanivihar\\ Bhubaneswar, Odisha, 751004 }}
    \end{minipage}

\end{titlepage}

\begin{titlepage}

\begin{center}
\textsc{\LARGE \textbf{\textcolor{Black}AREA CERTIFICATE}}
\end{center}

\vspace{0.8cm}
This is to certify that the work incorporated in the thesis entitled ``Exploring the
Fundamental Properties of Neutrino from Oscillation Experiments'' deals with 
a comprehensive study of the mass-mixing parameters of massive neutrinos 
in the next generation of long-baseline experiments. In this thesis, she 
demonstrates the abilities of these future experiments to accurately measure 
the three flavor oscillation parameters, particularly emphasizing the 
determination and significance of the leptonic CP phase, conducting precision 
studies on atmospheric oscillation parameters, exploring the resolution of the 
octant of 2-3 mixing, and investigating flavor-dependent long-range interactions 
of neutrinos with primarily electrons and neutrinos in big celestial bodies.
She conducted this phenomenological research as part of her thesis under 
the able guidance of faculties from the physics department, and submitted her
thesis to the Utkal University, Vani Vihar, Bhubaneswar, adhering to all 
academic rules and ethical conduct for the degree of Doctor of Philosophy 
in Physics. Furthermore, I certify that this thesis work falls within the domain of
Physics and the content of the thesis coincides with the topic selected at the 
time of registration.

\vspace{5cm}
\begin{minipage}{.4\textwidth}
        \textsc{{\sc DATE}}
    \end{minipage}%
    \begin{minipage}{0.6\textwidth}
       \textsc{{\sc Prof. Sanjib Kumar Agarwalla (Co-Guide)\\
 Institute of Physics, Sachivalaya Marg\\
 Bhubaneswar, Odisha, 751005  }}
    \end{minipage}

\end{titlepage}

\clearpage
\chapter*{{Acknowledgments}}
\fancyhead[LO,LE]{{\textbf{\quad\\\hfill \small{Acknowledgments\quad}}{\bf/}}\quad\small\bf\thepage}
It will not be easy to express the depth of my gratitude and sincerity towards everyone who helped me in this significant journey of five years through just few words. Nevertheless, I will try my best. 

Foremost, I extend my heartfelt appreciation and gratitude to my supervisor, Prof. Swapna Mahapatra, and co-supervisor, Prof. Sanjib Kumar Agarwalla, for their exceptional guidance, both academically and personally, throughout this significant period of my life. Prof. Swapna's unwavering support and readiness to assist me in overcoming challenges are truly commendable. Her presence during unavoidable circumstances was invaluable, and I am grateful for the seamless navigation she provided.

I owe a great deal to Prof. Sanjib for his continuous belief in my abilities and the dedicated efforts he invested in propelling me forward on this academic journey. His tireless commitment made this endeavor possible. The countless hours of discussions, the moments of shared tea and fruit-breaks, and the anecdotes from his own life that served as motivation are cherished highlights of this collaborative journey.

Next, I want to extend my sincere gratitude to all my collaborators who have played integral roles in this journey. A special acknowledgment goes to Dr. Soumya C. for generously sharing her knowledge and providing invaluable assistance in simulations during our collaborative endeavors. Her unwavering support during critical times, along with the cherished moments of evening tea breaks and our favorite \textit{vada} time, hold a special place in my memories. I am equally thankful to Dr. Suprabh Prakash for his profound insights and the wealth of knowledge he shared, proving instrumental in various aspects of my work. His responsiveness to queries, even beyond our collaborative period, reflects his dedication. My deepest appreciation goes to Prof. Mauricio Bustamante for his outstanding suggestions and advice. Despite the limited duration of our collaboration, his meticulous approach in discussions, draft editing, and figure edits left a lasting impression on me. I extend my thanks to Alessio Giarnetti and Sudipta Das for their engaging discussions and insightful analyses, which significantly contributed to my work. Ritam Kundu's eccentric questions have been instrumental in enhancing my understanding, and I express my gratitude to him. I also acknowledge Prof. Davide Meloni for numerous crucial suggestions that enriched my work. A heartfelt thanks to Dr. Ashish Narang and Dr. Anil Kumar, even though our collaboration was indirect, for their commendable readiness to provide valuable suggestions. In addition, I appreciate the unwavering support of Sadashiv and Pragyan, particularly for the delightful evening snacks. I extend my thanks to Anuj, Krishna, and Sarmistha for their consistent support whenever needed. I sincerely thank Dr. Sabya Sachi Chatterjee and Dr. Amina Khatun for their insightful suggestions and guidance during the initial stages of this journey.

I am also grateful to Dr. Prafulla Kumar Panda, the head of the physics department, Utkal University. Without his support, my Ph.D. tenure would not have been smooth. I also thank Dr. Shesansu Sekhar Pal, Dr. Pramoda Kumar Samal, Dr. Jagadish Kumar, and Dr. Bhagaban Kisan for their continuous help and support during the coursework tenure and afterwards. I am also thankful to  all the non-teaching staff, especially Rashmi Nani and Loknath Bhaina for their support.

I am equally grateful to Prof. Ajit Mohan Srivastava, Prof. Pankaj Agarwal, Dr. Kirtiman Ghosh, Dr. Debottam Das, Dr. Aruna Kumar Nayak, Dr. Manimala Mitra, Prof. Shamik Banerjee, and Prof. Sudipta Mukherji, the high-energy physics committee members in the Institute of Physics, Bhubaneswar. Their warm welcome and inclusive spirit have consistently made me feel a valued participant in all programs and group activities hosted by the hep-group at IOPB.

I am deeply grateful to SAMKHYA: the High-Performance Computing Facility at the Institute of Physics. The high-performance computing capabilities were instrumental in the success of my simulations during my Ph.D. journey. I would also like to express my sincere gratitude to Makrand Sir and his team for their dedication and assistance whenever needed.

I would like to express my gratitude to Prof. Orlando Peres for his exceptional hospitality, infectious cheerfulness, and the insightful physics discussions that enriched my visit to S$\tilde{a}$o Paulo during the Neutrino school in 2018. Additionally, I am sincerely thankful for the delightful encounters with Prof. Celio Moura, who provided invaluable vegan support during my stay in Brazil (or Brasil, as he fondly referred to it). The time spent with him holds immeasurable value.

Furthermore, I extend my appreciation to Prof. Francesco Vissani, whom I had the privilege of meeting during my stay. The tutorials he conducted and the discussions that ensued were not only encouraging but also exuberant. The vivid memories of his smiling demeanor, even when delving into the most intense topics, remain etched in my mind.

My gratitude extends to the organizers of WHEPP 2019 for graciously allowing my participation, despite my limited contribution to their discussions at that time. I acknowledge the collaborative efforts of Dr. Lakshmi Sandhya Mohan, Dr. Sushant Kumar Raut, Dr. Khushboo Dixit, Rudra Majhi, and Shashank Mishra for the work we initiated during that tenure.

I can't overlook the camaraderie and support of my office-mate, a brief football teacher, and my badminton partner, Dr. Biswajit Das. Alongside Dr. Soumya C., they provided a comforting refuge for me when I needed to recharge. My initial days at IOPB would not have been as joyful without their presence. Gratitude is also due to Dr. Sitender Pratap Kashyap, Dr. Minati Biswal, Dr. Manpreet Kaur, Dr. Tapoja Jha, Dibyendu Rana, and Dr. Vikas Vijaygiri for the shared memories -- birthday cakes, treats, spontaneous trips, and even the planned but unrealized ones. These years would have been notably less vibrant without their companionship. Special thanks to my current office-mates Pujalin, Manish, and Subhadip for upholding the cherished snack tradition in our workplace.

My appreciation also goes to Nilima Priyadarshini and Ankita for their collaboration during the challenging coursework days, which initially appeared formidable. I extend thanks to Ellorika, Jayanti, Kiran, Pritibanya, and Amit for the brief yet valuable time we shared during the coursework.

I take this opportunity to also extend my esteem gratitude towards Prof. S.K.Dash from Regional Institute of Education, N.C.E.R.T. Bhubaneswar and Prof. Mahesh Prakash from Jiwaji University, Gwalior, who nonetheless always supported me to pursue a career in research. It will not be an overstatement to say that I could carry forward in this path only because of them. 

``There is something so beautiful about having long-term friends that have witnessed multiple versions of you''. I cannot forget to mention Sudhansubala and Neeraj who were my `home away from home'. This long journey would have felt much longer if not for you guys. Certainly my generation is a witness of transition from landlines to smart phones. I am grateful from the bottom of my heart for the often group/solo video calls and voice notes from Kriti, Rashmi, Shama, Abhilasha, Shraddha, Rajni, Antima, Anju, Dharmendra, Shalu, Sonu, Srishti and Deepika.    

Finally, I extend my heartfelt gratitude to my parents, sister, and husband for their constant support and encouragement. Their unwavering belief in my pursuits has been a source of inspiration, and I am genuinely fortunate to have such a loving and supportive family.
  
\clearpage

\chapter*{{List of publications}}
\fancyhead[LO,LE]{\vspace*{1mm}{\textbf{\quad\\\hfill \small{List of publications\quad}}{\bf/}}\quad\small\bf\thepage}

\section{Publications in Referred Journal}
\begin{enumerate}
\item {\bf A close look on 2-3 mixing angle with DUNE in light of current neutrino oscillation data},\\Sanjib Kumar Agarwalla, Ritam Kundu, Suprabh Prakash, Masoom Singh\\ \emph{Journal of High Energy Physics, 03 (2022) 206}.
\item {\bf Enhancing Sensitivity to leptonic CP Violation using Complementarity among DUNE, T2HK, and T2HKK},\\Sanjib Kumar Agarwalla, Sudipta Das, Alessio Giarnetti, Davide Meloni, Masoom Singh\\ \emph{ The European Physical Journal C volume 83 (2023) 694}.
\item {\bf Flavor-dependent long-range neutrino interactions in DUNE \& T2HK: alone they constrain, together they discover},\\Masoom Singh, Mauricio Bustamante, Sanjib Kumar Agarwalla \\\emph{Journal of High Energy Physics, 08 (2023) 101} .
\end{enumerate}

\section{Conference Proceedings}
\begin{enumerate}
\setcounter{enumi}{3}
\item {\bf Exploring Earth's Matter Effect in High-Precision Long-Baseline Experiments},\\Masoom Singh, Sanjib Kumar Agarwalla,\\ \emph{PoS EPS-HEP2021 (2022) 191}.
\item {\bf Exploring Matter Effect and Associated Degeneracies at DUNE},\\Masoom Singh, Soumya C., Sanjib Kumar Agarwalla,\\ \emph{PoS NuFact2021 (2022) 069}.
\item {\bf Can Deviation from Maximal $\theta_{23}$ be resolved in DUNE?},\\Masoom Singh, Sanjib Kumar Agarwalla,\\\emph{Phys. Sci. Forum NuFact2022}.
\item {\bf Complementarity between DUNE and T2HK: gateway to improved CP Coverage},\\Masoom Singh, Sudipta Das, Alessio Giarnetti, Sanjib Kumar Agarwalla, Davide Meloni\\\emph{DAE 2022} (under review).
\end{enumerate}

\clearpage

\begin{center}
 \textbf{DEDICATED TO}
 \end{center}

\begin{center}
\emph{Maa, Papa, Di, and Abhi}
\end{center}

\fancyhead[LO,LE]{\vspace*{1mm}{\textbf{\quad\\\hfill\nouppercase \leftmark\quad}{\bf/}}\quad\small\bf\thepage}
\clearpage
\tableofcontents\clearpage

  \makeatletter
  \def\uplabel#1{{\normalfont{\textbf{#1}}\hfill}}
  \renewenvironment{AC@deflist}[1]%
    {\ifAC@nolist%
     \else%
        \raggedright\begin{list}{}%
            {\settowidth{\labelwidth}{\normalfont{\textbf{#1}}\hspace*{3em}}
            \setlength{\leftmargin}{\labelwidth}%
            \addtolength{\leftmargin}{\labelsep}%
            \renewcommand{\makelabel}{\uplabel}}%
      \fi}%
    {\ifAC@nolist%
     \else%
        \end{list}%
     \fi}%
  \makeatother



    \makeatletter   
    \renewcommand\part{%
      \if@openright
        \cleardoublepage
      \else
        \clearpage
      \fi
      \thispagestyle{empty}%
      \if@twocolumn
        \onecolumn
        \@tempswatrue
      \else
        \@tempswafalse
      \fi
      \null\vfil
      \secdef\@part\@spart}
    \makeatother

\pagenumbering{arabic}
\setcounter{page}{0}


\chapter{{Introduction}}
\label{sec:intro}
As the Nature's elusive cosmic voyager, a particle that dances through the universe, unfazed by matter, challenging our very understanding of the fundamental fabric of reality - neutrino, approaches a centennial anniversary, it excites physicists everywhere. In no way would Wolfgang Pauli have imagined that the ``desperate remedy'' he proposed would revolutionize the Standard Model (SM). Despite neutrinos' ubiquitousness, it took nearly 30 years to detect them with precision, while their characteristic features have not yet been fully understood. Neutrinos are intriguing particles within the framework of the SM of particle physics, a comprehensive theoretical framework that successfully describes the electromagnetic, weak, and strong nuclear forces, along with the fundamental particles of the universe. The SM achieved a major breakthrough by unifying electromagnetism and the weak forces into a single electroweak force, described by the electroweak theory, leading to a more cohesive understanding of particle interactions. While the Standard Model has proven immensely successful, it is not a complete theory yet. 

Around the 1920s, researchers observed that the energies of the emitted beta particles and recoil nuclei did not add up to the total energy available from the initial nuclear reaction. It seemed as if some energy was mysteriously missing. In 1930, Pauli attempted to explain this missing energy by proposing a mysterious new particle~\cite{Pauli:1930pc}. Although, he admitted that detecting neutrinos was almost impossible, and as time passed, it was nicknamed the `ghost particle'. It was not until 1953 when Frederick Reines and Clyde Cowan~\cite{Reines:1956rs} got early hints of neutrino detection in their small detector ``\textit{Her Auge}'' (``My Eye'' in German), this ghostly particle got its fame. While the actual saga was commemorated when they further redesigned it to a three-layer stacked 10-ton detector detecting neutrinos sourced from a powerful fission reactor. Finally, in June of 1956, they addressed a telegram  to Pauli that read, `\textit{We are happy to inform you that we have definitively detected neutrinos}.' However, not all discoveries are happy ones; sometimes they are the epitome of even further unanswered basic questions. The discovery of neutrino is one such example. In fact, the neutrinos were so strange that they needed to be re-examined and questioned. This gave rise to a new branch in the ever-evolving field of particle physics, which seeks theories beyond the Standard Model. 

The exploration of particle physics has become a cornerstone of modern physics and has led to a better understanding of the universe. Today, neutrinos are used to study the evolution of the universe, probe the foundations of particle physics, and search for physics beyond the SM. 
The development of multimessenger astronomy with neutrinos detected in a kilometer-deep IceCube~\cite{IceCube:2018fhm} detector in Antarctica has revolutionized this field immensely. In addition to the Sun, radioactive decay, the atmosphere, and supernovas, IceCube has recently validated a nearby galaxy as the origin of neutrinos~\cite{IceCube:2022der}. It is projected that very soon with the IceCube upgrade leading to IceCube-Gen2~\cite{Ishihara:2019aao} along with other telescope experiments~\cite{KM3Net:2016zxf,Avrorin:2014vca}, we will have answers for the unknown sources of huge flux of cosmic rays we receive. Thanks to the excellent angular resolution in ANTARES~\cite{ANTARES:2020srt} for providing constraints on them. Many new-generation instrumentations are in line, including Baikal-GVD~\cite{Baikal-GVD:2021zsq}, KM3NeT-ARCA~\cite{Biagi:2023yoy}, P-ONE~\cite{P-ONE:2020ljt}, and others. Borexino recently reported neutrinos observed from the CNO fusion cycle~\cite{BOREXINO:2020aww} for the first time. This experimental verification paved the way for a crucial understanding of stellar physics. This year also saw the first detection of neutrinos from a new LHC experiment called FASER~\cite{FASER:2023zcr}. Although neutrinos are being abundantly produced in colliders, this is its first time getting detected. Also, SNO+ becomes the first water Cherenkov detector to probe reactor antineutrinos~\cite{SNO:2022qvw}. All previous Cherenkov detectors could not identify reactor electron antineutrinos because of their high detector thresholds (must be lower than $\sim 2.2$ MeV for detection). Just like studying neutrinos from the core of the Sun in real time has led to verification of the Standard Solar Model, the neutrinos from supernova bursts will act as an alarm for the following supernova photons. Once neutrinos from diffuse supernova background get observed, knowing the cosmic evolution in the Universe will not be far~\cite{Beacom:2010kk,Scholberg:2012id}.  Furthermore, the usage of neutrinos is now more than just understanding the basic sciences. The pursuit has spread to applications in nuclear energy and security. Recently, antineutrinos in nuclear non-proliferation have led to a movable neutrino detector, which is perceived as a future tool to detect hidden nuclear reactors~\cite{Subedi:2019shd,Kneale:2022vpw} --- one of the most significant advantages being: 24-hour year-round monitoring. 
 
The usage of neutrinos as a provider yearns for a profound precision in its oscillation parameters. This requires an extensive network of detectors to observe neutrinos from various sources. Advanced computing and analytical tools must be employed to measure the properties of neutrinos with high precision. Several neutrino experiments focus on probing different oscillation parameters depending on the source of the neutrino. Below, we enumerate the present status of these oscillation parameters and the correlations among them.  
 
\section{Present synergies and tensions in oscillation parameters}
\label{sec:present-status}
In this subsection, we briefly visit the epoch-making measurements by various neutrino experiments in the present regime that has paved the way for neutrino to enter the precision era.
\subsection{$\sin^{2}\theta_{12}$ and $\Delta m^{2}_{21}$: The Solar and KamLAND experiment}
\label{sec:solar-param}
\begin{figure}[ht!]
\centering
\includegraphics[width=0.7\linewidth]{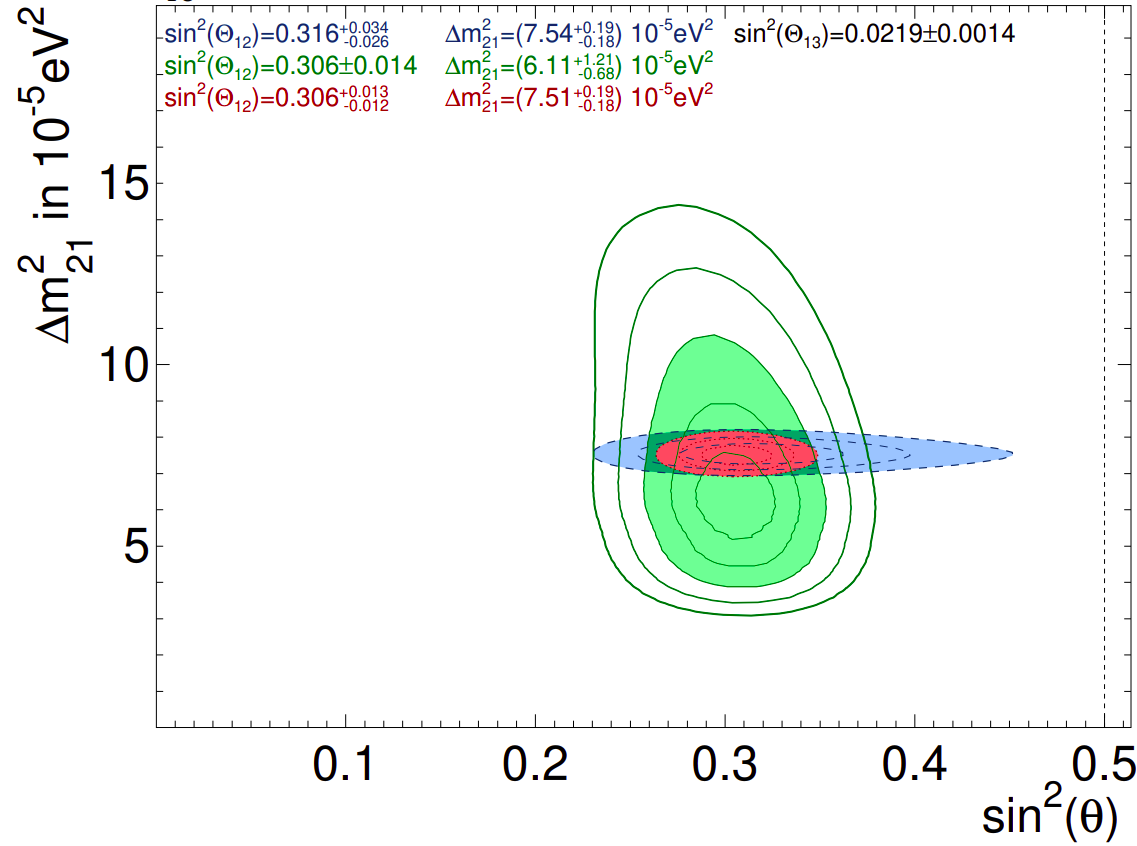}
\caption{Allowed regions in the solar oscillation parameters from Solar (green), KamLAND (blue), and the combined (red) setups. The filled regions correspond to 3$\sigma$, while solid contours are at 1$\sigma$ and 2$\sigma$. Additionally 4$\sigma$ and 5$\sigma$ contours are also shown for Solar.This figure is taken from Ref.~\cite{yasuhiro_nakajima_2020_3959640,SajjadAthar:2021prg}}.
\label{fig:solar-kamland-tension}
\end{figure}

The combined analyses of Solar and KamLAND experiments have dominated the measurement in solar parameters. The last decade witnessed a global tension between the preferred best fit for $\Delta m^{2}_{21}$ by these experiments. The KamLAND preferred relatively higher value. This tension attributes to two crucial disparities,
\begin{itemize}
\item The energy spectrum measured by SNO, SK, and Borexino (Solar experiments) showed no evidence of the low energy upturn as was predicted by the standard LMA-MSW (Large Mixing Angle-Mikheyev Smirnov Wolfenstein)~\cite{Wolfenstein:1977ue,Mikheyev:1985zog} solution for the favored value of $\Delta m^{2}_{21}$ by KamLAND.
\item A non-vanishing day-night asymmetry observed by Super-K~\cite{Super-Kamiokande:2016yck}, thus favoring a higher value of $\Delta m^{2}_{21}$ than KamLAND.
\end{itemize}
KamLAND~\cite{KamLAND:2014gul} is a reactor antineutrino experiment that detects the antineutrinos from reactors located at an average distance of $\sim 140$ km - $215$ km. The effective survival probability of $\bar{\nu}_{e}$ is given by
\begin{eqnarray}
P_{\bar{\nu}_{e} \rightarrow \bar{\nu}_{e}} &=& 1 - 4\abs{U_{e1}}^{2}\abs{U_{e2}}^{2}\sin^{2}\frac{\Delta m_{21}^{2}L}{4E} - 4\abs{U_{e1}}^{2}\abs{U_{e3}}^{2}\sin^{2}\frac{\Delta m_{31}^{2}L}{4E}\nonumber\\
&-& 4\abs{U_{e2}}^{2}\abs{U_{e3}}^{2}\sin^{2}\frac{\Delta m_{32}^{2}L}{4E}\,.
\end{eqnarray}
The typical energy of antineutrinos in KamLAND is of few MeV. So, the contribution from $\Delta m^{2}_{31}$-phase and $\Delta m^{2}_{32}$-phase average out. Therefore, we have~\cite{Esteban:2018azc}
\begin{eqnarray}
P_{\bar{\nu}_{e} \rightarrow \bar{\nu}_{e}} &=&1 - 4\abs{U_{e1}}^{2}\abs{U_{e2}}^{2}\sin^{2}\frac{\Delta m_{21}^{2}L}{4E} - 2\abs{U_{e1}}^{2}\abs{U_{e3}}^{2} - 2\abs{U_{e2}}^{2}\abs{U_{e3}}^{2}\nonumber \\
&=& 1-2\abs{U_{e3}}^{2}(1-\abs{U_{e3}}^{2})-4\abs{U_{e1}}^{2}\abs{U_{e2}}^{2}\sin^{2}\frac{\Delta m^{2}_{21}L}{4E}\nonumber\\
&=& 1-\frac{1}{2}\sin^{2}2\theta_{13}-\sin^{2}2\theta_{12}\cos^{4}\theta_{13}\sin^{2}\frac{\Delta m^{2}_{21}L}{4E}\nonumber\\
&\approx & \sin^{4}\theta_{13} + \cos^{4}\theta_{13}\left(1 - \frac{1}{2}\sin^{2}(2\theta_{12})\sin^{2}\frac{\Delta m^{2}_{21}L}{2E}\right)
\label{eq:Solar-nuebar-survival} 
\end{eqnarray}
Recently, the existing disparity in the measurements of $\Delta m^{2}_{21}$ has reduced to  $\sim 1.1\sigma$ (which was $\sim 2\sigma$ before) after the inclusion of current Super-K data~\cite{yasuhiro_nakajima_2020_3959640} as depicted in Fig.~\ref{fig:solar-kamland-tension}.
Borexino~\cite{BOREXINO:2018ohr} can achieve the most precise measurements of $P_{\bar{\nu}_{e} \rightarrow \bar{\nu}_{e}}$, below 1.5 MeV, where flavor conversion is mostly vacuum-dominated. While at higher energies $\sim 5$ MeV, it is mostly matter-dominated in the Sun; Borexino measurements are in agreement with Super-K~\cite{yasuhiro_nakajima_2020_3959640} and SNO~\cite{SNO:2011hxd}. Therefore, Borexino is the only ongoing experiment that simultaneously probes flavor conversion in neutrino in both vacuum and matter-dominated scenarios. Further, it discards the vacuum-only hypothesis at 98.2\% C.L.~\cite{yasuhiro_nakajima_2020_3959640}.

 \medskip
 \textbf{Day-Night asymmetry}- When the Sun is below the horizon at night, solar neutrinos have to pass through Earth before reaching the detector located on Earth's surface on the other side. Thus, the extensive matter density in the inner layers of Earth leads to the enhancement of $\nu_{e}$ type events through the MSW mechanism at night. While this is not the case during the day, as then, the detector faces the Sun. This asymmetry in the number of events is commonly referred to as the `day/night effect'. It is energy dependent. Currently, from Ref.~\cite{yasuhiro_nakajima_2020_3959640}, Super-K has a 2$\sigma$  preference for non-zero day/night asymmetry.
%
\subsection{$\sin^{2}2\theta_{23}$ and $\Delta m^{2}_{31}$: Long-baseline accelerator, atmospheric, and reactor experiments} 
\label{sec:th23-Deltam31}
The atmospheric mixing angle ($\theta_{23}$) and mass squared splitting ($\Delta m^{2}_{32}$) are mostly constrained by atmospheric and long-baseline (LBL) experiments as shown later in Table~\ref{tab:sources-and-constraints}. The expression of $P_{\nu_{\mu}\rightarrow \nu_{\mu}}$ in a simplified manner can be given as below (refer to Eq.~\ref{eq:3flavor-numu-disapp-with-matter} for a detailed discussion later) 
\begin{eqnarray}
P_{\nu_{\mu}\rightarrow \nu_{\mu}} &=& 1 - 4 \abs{U_{\mu 3}}^2(1-\abs{U_{\mu 3}}^{2})\sin^{2}\frac{\Delta m^{2}_{32}L}{4E}\nonumber\\
&=& 1 - \sin^{2}2\theta_{\mu\mu}\sin^{2}\frac{\delta m^{2}_{\mu\mu}L}{4E}\,,\\
\text{where,}~
\sin^{2}\theta^{\text{eff}}_{\mu\mu} &=& \cos^{2}\theta_{13}\sin^{2}\theta_{23\,,}\nonumber\\
\text{and}~
\delta m^{2}_{\mu\mu} &=& \sin^{2}\theta_{12}\Delta m^{2}_{31} + \cos^{2}\theta_{12}\Delta m^{2}_{32} + \cos\delta_{\mathrm{CP}}\sin\theta_{13}\tan\theta_{23}\Delta m^{2}_{21}\nonumber
\end{eqnarray}
%

In the current scenario, while T2K and NO$\nu$A favor the maximal mixing (MM) solution of $\theta_{23}$ ($\sin^{2}\theta_{23}=0.5$), it is disfavored by MINOS. From the global oscillation fit, we observe that the addition of LBL accelerator and MBL reactor favors NMO over IMO with $\sim 1.3\sigma$ ~\cite{Huber:2003pm}. Further, the addition of atmospheric $\nu$ data especially from Super-K (Phase IV), induces more preference in favor of NMO ($\sim 2.5 \sigma$), and favors $\theta_{23}< \pi/4$ in NMO at $\sim 1.6\sigma$. However, the recent Phase V update takes $\theta_{23}$ close to maximal mixing (this has not yet been considered in the global analysis).
\subsection{$\sin^{2}2\theta_{13}$: Reactor experiments}
\label{sec:reactor-param}
There are many ongoing medium-baseline reactor experiments like Daya Bay~\cite{DayaBay:2012fng}, RENO~\cite{RENO:2012mkc}, and Double Chooz~\cite{DoubleChooz:2011ymz}. They measure the electron antineutrino coming from nearby reactors ($\sim 1$ km), thus analyzing $P_{\bar{\nu}_{e}\rightarrow \bar{\nu}_{e}}$ survival probability. Eq.~\ref{eq:Solar-nuebar-survival} can be further simplified to
\begin{eqnarray}
P_{\bar{\nu}_{e}\rightarrow \bar{\nu}_{e}} &=& 1 - \cos^{4}\theta_{13}\sin^{2}2\theta_{12}\sin^{2}\Delta m^{2}_{21}\nonumber\\
 &-& \sin^{2}2\theta_{13}(\cos^{2}\theta_{12}\sin^{2}\Delta m^{2}_{31} + \sin^{2}\theta_{12}\sin^{2}\Delta m^{2}_{32})\nonumber \\
 &=& 1 - \cos^{4}\theta_{13}\sin^{2}2\theta_{12}\sin^{2}\Delta m^{2}_{21} - \sin^{2}2\theta_{13}\sin^{2}\Delta m^{2}_{ee}\,,
\end{eqnarray} 
where, $\Delta m^{2}_{ee} = \cos^{2}\theta_{12}\Delta m^{2}_{31} + \sin^{2}\theta_{12}\Delta m^{2}_{32}$. Here, $\Delta m^{2}_{ee}$ does not depend on the choice of mass ordering and uncertainty in the measurement of $\theta_{12}$~\cite{DayaBay:2016ggj}. Depending upon the baseline $L$, the dominant parameters vary. If $L\sim$ 1 - 2 km, then $P_{\bar{\nu}_{e}\rightarrow \bar{\nu}_{e}}$ gives information on $\sin^{2}\theta_{13}$, as oscillations induced over this baseline are due to $\Delta m^{2}_{31}$ (as considered here). However, when $L \geq 50$ km, then $P_{\bar{\nu}_{e}\rightarrow \bar{\nu}_{e}}$ gives information on $\sin^{2}\theta_{12}$, as oscillations induced over this baseline are due to $\Delta m^{2}_{21}$ (as discussed in Sec.~\ref{sec:solar-param}).
The current precision from Daya Bay stands undisputable at about 2.8\%. The best precision in Daya Bay surmises to the most powerful nuclear-power complex that provides very intense antineutrino flux. Also, it has water-proof automated calibration units, which are fully submerged in water without any chimney. So, it did not experience any background coming from external radioactivity or the Michel electrons from decays of stopped muons~\cite{Cao:2016vwh}. Moreover, the usage of Gadolinium has enhanced the achievable precision even more further.
While experiments from other sources are also sensitive to $\theta_{13}$, none of them can challenge the Daya Bay precision. The constraints from Solar and KamLAND do not get affected due to the feeble ongoing tension. Their measurements are consistent with the reactor experiments. $\theta_{13}$ can also be determined by LBL accelerator experiments through studying $\nu_{\mu}\rightarrow\nu_{e}$ (refer to Eq.~\ref{eq:3flavor-nue-app-with-matter} later). However, as this channel broadly depends on the CP phase which has the most uncertainty among each oscillation parameter, the precision with which LBL accelerators can determine $\theta_{13}$ is very less. 
\subsection{$\delta_{\mathrm{CP}}$: LBL accelerators}
The presently allowed region in normal mass ordering (NMO) is comparatively larger that include CP-conserving case of $\delta_{\mathrm{CP}} = \pi$ at $\sim 2\sigma$, while adding atmospheric data shifts the preference to $\delta_{\mathrm{CP}} =3 \pi/2$, disfavoring $\delta_{\mathrm{CP}} = \pi$ at 1.6$\sigma$. In contrast, the inverted mass ordering (IMO) shows an overall inclination for $\delta_{\mathrm{CP}} = 3\pi/2$. This difference comes from the correlation of CP phase with $\theta_{23}$~\cite{Minakata:2013eoa,Coloma:2014kca}. However, there is an ongoing tension between the two LBL experiments: T2K and NO$\nu$A, wherein T2K favors  $\delta_{\mathrm{CP}} = 3\pi/2$ for NMO and NO$\nu$A has a preference for $\delta_{\mathrm{CP}} = \pi/2$. 
%
\subsection{Current oscillation parameters}
\begin{figure}[htb!]
\includegraphics[width=1.0\linewidth]{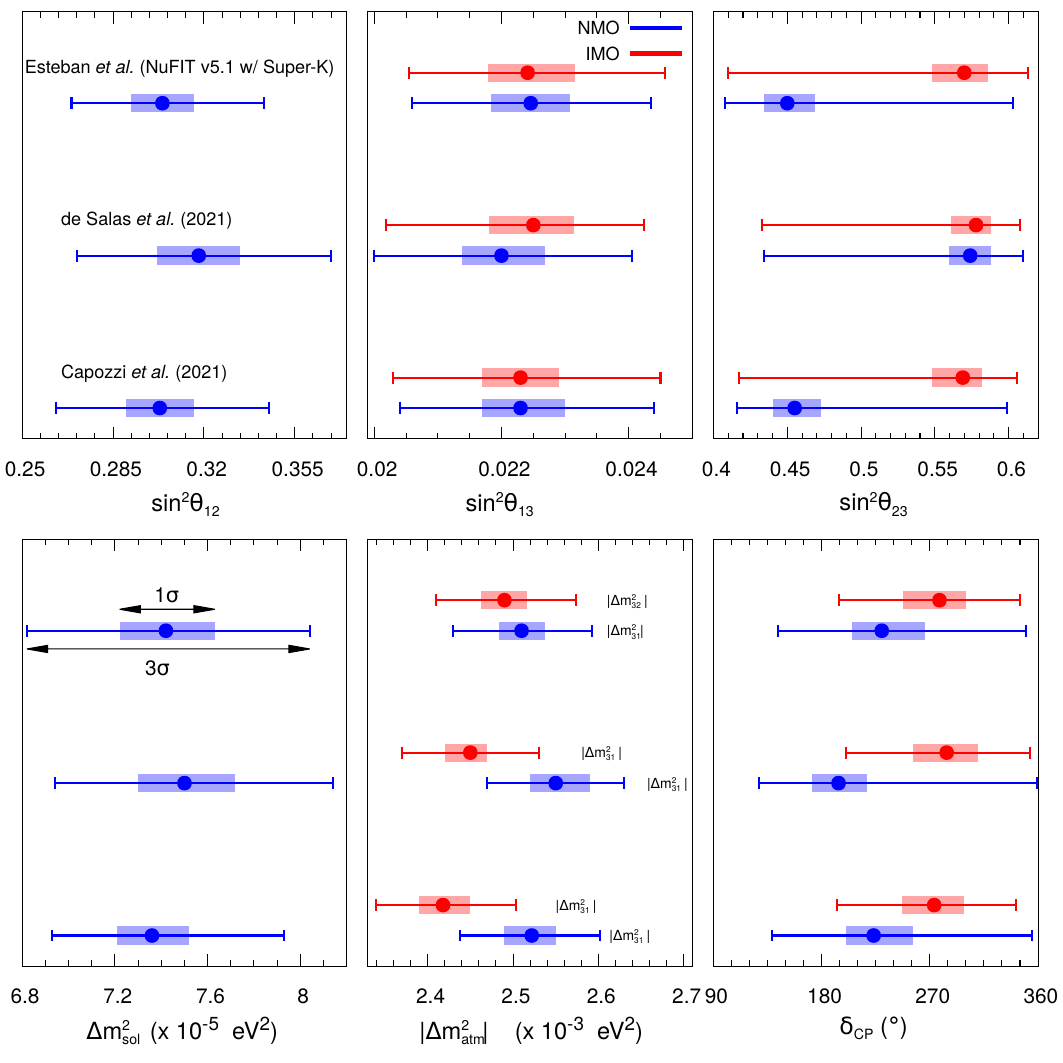}
\caption{Current 1$\sigma$ and 3$\sigma$ allowed ranges following the global oscillation data~\cite{Esteban:2020cvm,deSalas:2020pgw,Capozzi:2021fjo} for both NMO and IMO. Refer Table~\ref{tab:best-fit} for details.}
\label{fig:current-allowed-osc-params}
\end{figure} 
Following the above discussion on constraints from different experiments and the global oscillation data in Ref.~\cite{Esteban:2020cvm,deSalas:2020pgw,Capozzi:2021fjo}\,, we reproduce Fig.~\ref{fig:current-allowed-osc-params}. For reference, we quote the corresponding best fit and allowed ranges at 1$\sigma$ and 3$\sigma$ in the Table~\ref{tab:best-fit} from Ref.~\cite{Capozzi:2021fjo}, assuming NMO and IMO.
%
\begin{table}[htb!]
		\label{tableprecision}

		\centering
		\resizebox{\columnwidth}{!}{%
			\begin{tabular}{|c|c|c|c|c|}
				\hline 
				\textbf{Parameter} & \textbf{Best fit} & \textbf{1$\sigma$ range} & \textbf{3$\sigma$ range} \\
				\hline 
				$\Delta m^2_{21}/10^{-5}$ $\mathrm{eV^{2}}$ & 7.36 & 7.21 - 7.52 & 6.93 - 7.93 \\
				\hline
				$\sin^{2}\theta_{12}/10^{-1}$ & 3.03 & 2.90 - 3.16  & 2.63 - 3.45 \\
				\hline
				$\sin^{2}\theta_{13}/10^{-2}$ & 2.23 (2.23) & 2.17 - 2.30  & 2.04 - 2.44 \\
				\hline
				$\sin^2\theta_{23}/10^{-1}$ & 4.55 (5.7) & 4.40 - 4.73 (5.48 - 5.82)  & 4.16 - 5.99 (4.17 - 6.06) \\
				\hline
				$\Delta m^2_{31}/10^{-3}$ $\mathrm{eV^2}$ & 2.522 (2.418) & 2.490 - 2.545 (2.394 - 2.448) & 2.436 - 2.605 (2.341 - 2.501) \\
				\hline
				$\delta_{\text{CP}}$/$^\circ$ & 223 (274)  & 200 - 256 (247 - 299)  & 139 - 355 (193 - 342) \\
				\hline \hline
			\end{tabular}
			}
			\caption{\footnotesize{Values of oscillation parameters from Ref.~\cite{Capozzi:2021fjo} assuming NMO. Bracketed values correspond to the same in IMO. Refer Fig.~\ref{fig:current-allowed-osc-params} for more details.}}
			\label{tab:best-fit}
	\end{table}

This era has enabled the study of neutrinos with unprecedented precision and accuracy. However, the current oscillation data cannot decide whether the atmospheric mass splitting ($\Delta m^{2}_{31}$) is positive ($\Delta m^{2}_{31} > 0$) or negative ($\Delta m^{2}_{31} < 0$). The first possibility gives rise to the neutrino mass pattern: $m_{3}>m_{2}>m_{1}$, known as normal mass ordering (NMO) and for the second possibility, we have $m_{2}>m_{1}>m_{3}$, labeled as inverted mass ordering (IMO). Knowing the ordering of the neutrino masses is of prime importance, because it dictates the structure of the neutrino mass matrix, and hence could give vital clues towards the underlying theory of neutrino masses and mixing. Information on the neutrino mass ordering could have other far-reaching phenomenological consequences. For instance, if it turns out that $\Delta m^{2}_{31} < 0$, and yet no neutrino-less double beta-decay is observed even in the very far future experiments, that would be a strong hint that neutrinos are not \textit{Majorana particles}.

Determining the value of $\delta_{\mathrm{CP}}$ to explore the possibility of CP violation in the leptonic sector (to question if $\delta_{\mathrm{CP}}$ differs from both $0^{\circ}$ and $180^{\circ}$) has emerged as the top priority in the field of neutrino oscillation physics. Substantial CP violation in the lepton sector may point toward the possibility that neutrinos are involved in generating the observed matter-antimatter asymmetry in the universe dynamically. The present generation long-baseline experiments: T2K and NO$\nu$A hint towards a non-zero leptonic CP phase, however it is only through next-generation long-baseline experiments that these hints could turn into discovery. If CP violation is observed, it would open up a new era of neutrino physics. Further, it can shed light on the nature of the neutrino mass ordering and provide a deeper understanding of the origin of matter in the universe.

Another recent and crucial development related to neutrino oscillation parameters is the hint of non-maximal $\theta_{23}$ from the global fits of world neutrino data, which indicates two nearly degenerate solutions for 2-3 mixing angle: one $< 45^{\circ}$, termed as lower octant (LO), and other $>45^{\circ}$, denoted as higher octant (HO). Accurate measurement of $\theta_{23}$ and the resolution of its correct octant (if it turns out to be non-maximal) are also very important issues that need to be addressed in the current and next-generation neutrino oscillation experiments. This information will provide crucial inputs to the theories of neutrino masses and mixings. This is essential to determine the nature of neutrinos, whether they are Dirac or Majorana particles, and to probe the underlying flavor symmetry of the neutrino sector. Moreover, $\theta_{23}$ is a key element of the leptonic mixing matrix and is necessary for understanding leptonic CP violation. 

However, before we address these generic properties of neutrino, one must ascertain the ability of these experiments to exclude the vacuum hypothesis or in other words to establish the Earth matter effect with significant precision. This can be done by performing careful measurements over a broad range of energies and distances. Moreover, the comparison of the data obtained with theoretical predictions is essential to accurately determine the strength of the Earth matter effect. Finally, the neutrino oscillation parameters can be determined with greater accuracy. This information is then used to improve predictions and refine the existing neutrino models. This helps to further our understanding of the fundamental properties of neutrinos and their interactions with matter. 

Discovering a new fundamental interaction would be striking evidence of physics beyond the Standard Model. Yet, because new interactions are likely feeble, they are difficult to detect. Neutrinos have immense potential to reveal new physics~\cite{Coloma:2022dng}. Their capability to look for new neutrino-matter interactions is good because, in the Standard Model, neutrinos interact only weakly. Therefore, the presence of an additional neutrino interaction may be more easily spotted, even if it is feeble. Neutrinos also offer a unique way to explore new physics because they can traverse large distances and interact with matter in ways that other particles cannot. This makes them ideal probes to search for evidence of new physics beyond the Standard Model. One such interaction is flavor-dependent long-range neutrino interactions mediated by ultra-light mediators~\cite{He:1990pn,Foot:1990uf}. 

In this thesis, we try to answer these fundamental properties of neutrinos using the upcoming long-baseline oscillation experiments.

 \section{Motivation behind this thesis}
 \label{sec:motivation}
Neutrino physics has progressed dramatically after its first discovery in the year 1956 at the Savannah River~\cite{Reines:1956rs}, propelled by the most surprising discovery that neutrinos have mass, thereby providing us with an absolute evidence for extending physics beyond the Standard Model. The year 2011 came with the discovery of non-zero 1-3 mixing angle~\cite{DoubleChooz:2011ymz,DayaBay:2012fng,RENO:2012mkc} that represents a crucial milestone in establishing the three-flavor oscillation picture of neutrinos which by far explains most of the oscillation data. It opened up exciting prospects for the current and future neutrino oscillation experiments to address the remaining fundamental unknowns. Neutrino physics is now poised to move into the precision regime, where the thrust has now drifted to establishing a detailed understanding of the framework of neutrino mass matrix. Certainly, precise reconstruction of this mass matrix will shed immense light on the underlying new physics which will elaborate on neutrino mass and mixing. A number of high-precision neutrino oscillation experiments are
currently taking data and improving our knowledge about the oscillation parameters day-by-day. Several ambitious next-generation oscillation experiments are driven by the mission of deciphering the fundamental properties of these elusive particles, and their peculiar interactions.

A study of $\nu_{\mu} (\bar{\nu}_{\mu}) \rightarrow \nu_{e} (\bar{\nu}_{e})$ and $\nu_{\mu} (\bar{\nu}_{\mu}) \rightarrow \nu_{\mu} (\bar{\nu}_{\mu})$ oscillations accessible to the long-baseline super-beam experiments is a convenient way to probe three-flavor effects, including the sub-leading ones. The present generation of ongoing long-baseline experiments includes: Tokai-to-Kamioka (T2K) in Japan, bearing a baseline of 295 km, and the NuMI Off-axis Neutrino Appearance (NO$\nu$A) in the United States with a baseline of 810 km. In recent years, they have performed magnificently well, giving hints for the non-zero value of the Dirac CP phase~\cite{T2K:2023smv}. Together, they along with other atmospheric experiments, have achieved a strong indication towards non-maximal of $\theta_{23}$, thus leading to the issue of determining the correct octant of 2-3 mixing angle. The question of correct mass ordering still remains unanswered in the current realm of global neutrino oscillation data, very weakly favoring normal mass ordering~\cite{deSalas:2020pgw, Esteban:2020cvm,Capozzi:2021fjo}. These three are the big questions in the field of neutrino oscillation physics. In this thesis, I broadly traverse exploring these big questions. Furthermore, the interaction of neutrinos with Earth's matter induces significant alterations in their propagation, exerting a profound impact on oscillation phenomena. Hence, a precise measurement of Earth's Matter effect is indispensable for accurately interpreting data from upcoming high-precision long-baseline experiments aimed at unraveling profound questions. I delve into this aspect further. The continual questioning of the existence of only four fundamental forces in Nature has led to stringent constraints on the potential existence of a fifth fundamental force~\cite{Lee:1955vk, Okun:1995dn, Williams:1995nq, Dolgov:1999gk}. There is no better candidate to explore this new force than the neutrino itself, whose distinctive features have already challenged the foundations of the existing Standard Model. This thesis also provides insights in this direction. 

The upgrade of T2K to Tokai-to-Hyper-Kamiokande (T2HK)~\cite{Hyper-Kamiokande:2018ofw} in Japan and Deep Underground Neutrino Experiment (DUNE)~\cite{DUNE:2021cuw} are the two most eagerly awaited high-precision, next-generation long-baseline neutrino experiments of the upcoming decade (2030s), expected to display immensely rich physics programs, both within the standard three-neutrino paradigm and beyond. In this thesis, we analyze the proficiency of these upcoming long-baseline experiments. Our focus is on leveraging the strengths of both experiments to unravel the mysteries within neutrino oscillation physics and extend our exploration into realms that transcend existing boundaries. 
 \section{Structure of this thesis}
 \label{sec:structure}
We begin by revisiting the extensive literature on neutrinos, retracing the theoretical foundations that mark the inception of this field. In Chapter~\ref{sec:intro}, we analyze the present synergies and tensions in the ongoing neutrino experiments, surmising the present status of oscillation parameters. We also summarize the current issues in three-neutrino oscillation and beyond. In Chapter~\ref{sec:sec2}, we delve into the quantum-mechanical formulations that elucidate the phenomenon of neutrino oscillations. Starting with the two-flavor scenario, we systematically extend our discussion to encompass three flavors in both vacuum and matter. Additionally, we provide detailed insights into various conceivable neutrino sources. In Chapter~\ref{ch:physics-richness}, we shed light on the critical characteristic features inherent in long-baseline experiments. Furthermore, we offer a condensed summary detailing the current progress of completed, ongoing, and anticipated long-baseline experiments. Chapter~\ref{sec:ch4} discusses our first work, wherein we study the efficiency of the upcoming experiment, DUNE in establishing the Earth matter effect. We also compute the achievable precision on the Earth matter density using DUNE and elaborate on various degeneracies that we come across in this study. In Chapter~\ref{sec:ch5}, the spotlight is on the complementarity between the two forthcoming extensive long-baseline experiments: DUNE and T2HK, particularly when combined (DUNE + T2HK) to establish the Earth matter effect. Building upon our earlier analysis, this chapter explores how the synergy between DUNE and T2HK can effectively resolve inherent degeneracies in standalone experiments. The result is a substantial enhancement in the precision of measurements concerning Earth matter density. Moving forward, Chapter~\ref{sec:Ch6} delves into the capacity of DUNE to determine the accurate octant of the atmospheric mixing angle. Additionally, we conduct computations to estimate its expected precision concerning the atmospheric mixing angle and mass-squared differences. Subsequently, Chapter~\ref{sec:ch7} tackles the intricacies of leptonic CP violation within the upcoming long-baseline experiments, examining both standalone and combined scenarios. The chapter emphasizes the advantages derived from the complementary features of DUNE and T2HK, particularly within the combined DUNE + T2HK configuration. In Chapter~\ref{sec:ch8}, we discuss the possibility of existence of a new kind of interaction: the flavor-dependent neutrino-matter interactions. We analyze the capability of DUNE and T2HK in predicting the constraints on these new long-range interactions. We also analyze the complementary features of DUNE + T2HK in discovering these new interactions. In the concluding Chapter~\ref{sec9:conclusion}, we offer a comprehensive summary and outline prospective directions for future research. 
 \chapter{{Neutrino and its Interaction}}
 \label{sec:sec2}
\section{Neutrino oscillations in vacuum}
\label{sec:nu_vacuum}
Neutrinos get produced and detected in their weak flavor eigenstates: $\ket{\nu_{\alpha}}$ where $\alpha = e, \mu, \tau$. In principle, one can estimate their masses by kinematically studying the corresponding interactions through which neutrinos are produced. However, when the uncertainty in this measurement surpasses the difference in the energies of the mass eigenstates: $\ket{\nu_{k}}$ where $k=1,2,3$, then the flavor eigenstates have to be determined as a linear superposition of the three mass eigenstates which in neutrino mode can be expressed as
\begin{equation}
\ket{\nu_{\alpha}} = \sum_{k} U^*_{\alpha k}\ket{\nu_{k}}.
\label{eq:nut-propagation}
\end{equation} 
While the same in antineutrino mode can be written as 
\begin{equation}
\ket{\bar{\nu}_{\alpha}} = \sum_{k} U_{\alpha i}\ket{\bar{\nu}_{k}},
\label{eq:antinut-propagation}
\end{equation}
wherein $U$ is a unitary Pontecorvo-Maki-Nakagawa-Sakata matrix. This superposition of the three flavor or mass eigenstates of neutrino forms the basis of neutrino oscillations. Before proceeding further in our derivation of the standard neutrino oscillation framework, below, we enumerate the assumptions of this model.
\begin{itemize}
\item Neutrinos produced or detected via weak interaction processes are fully described by their flavor eigenstates.
\item All flavor neutrinos have a definite momentum $\vec{p}$, so we consider an equal momentum assumption for all massive neutrino components. This ensures that neutrinos travel in wavepackets, and, wavepackets associated with each mass eigenstate overlap with other, corroborating coherent nature. 
\item The time of propagation, $t$ is same as the distance traveled, $L$ by the neutrino beam.
\end{itemize}
Following the unitarity of $U$ in equations~\ref{eq:nut-propagation} and~\ref{eq:antinut-propagation}, flavor states must be orthonormal, thus
\begin{equation}
\bra{\nu_{\alpha}}\ket{\nu_{\beta}} = \delta_{\alpha\beta}\,.
\end{equation}
The neutrino mass eigenstates can be written in the form of massive neutrino states, $\ket{\nu_{k}}$ as,
\begin{equation}
\mathscr{H}\ket{\nu_{k}} = E_{k}\ket{\nu_{k}}\,,
\end{equation}
with energy eigenvalues $E_{k} = \sqrt{\abs{\vec{p}}^{2} + m^{2}_{k}}$\,. Following  the time-dependent Schr\"{o}dinger equation, the mass eigenstates evolve to
\begin{equation}
i\hbar\frac{\partial}{\partial t}\ket{\nu_{k} (t)} = \mathscr{H}\ket{\nu_{k}(t)}\,,
\label{eq:ch2a}
\end{equation}
implying evolution of massive neutrino states as plane waves
\begin{equation}
\ket{\nu_{k}(t)} = e^{-iE_{k}t}\ket{\nu_{k}}\,.
\label{eq:ch2b}
\end{equation}
From the above set of equations (equations~\ref{eq:ch2a} -~\ref{eq:ch2b}), the evolution of flavor eigenstate $\ket{\nu_{\alpha}(t)}$ starting at $t=0$ is given by
\begin{equation}
\ket{\nu_{\alpha}(t)} = \sum_{k}U^{*}_{\alpha k}e^{-i E_{k}t}\ket{\nu_{k}}\,,
\label{eq:time-dependent-evolution}
\end{equation}
following the convention, $\ket{\nu_{\alpha}(t=0)} = \ket{\nu_{\alpha}}$.  We know
\begin{equation}
\sum_{\alpha}U^{*}_{\alpha k}U_{\alpha j} = \delta_{jk}\,.
\end{equation}
To express mass eigenstates in terms of the flavor states, we can write
\begin{equation}
\ket{\nu_{k}} = \sum_{\alpha}U_{\alpha k}\ket{\nu_{\alpha}}\,.
\end{equation}
Using the above, we can further write
\begin{equation}
\ket{\nu_{\alpha}(t)} = \sum_{\beta = e, \mu, \tau}\left(\sum_{k}U^{*}_{\alpha k}U_{\beta k}e^{-iE_{k}t}\right)\ket{\nu_{\beta}}\,.
\end{equation}
So to determine the amplitude of neutrino flavor transition from flavor $\alpha$ to $\beta$ at time $t$, we compute,
\begin{align}
A_{\nu_{\alpha}\rightarrow \nu_{\beta}} (t) \ = & \ \bra{\nu_{\beta}}\ket{\nu_{\alpha}(t)} = \sum_{k}U^{*}_{\alpha k}U_{\beta k}e^{-i E_{k} t}
\end{align}
and thus the probability of this flavor transition is given by, 
\begin{align}
P_{\nu_{\alpha}\rightarrow \nu_{\beta}} (t) \ = & \ \abs{A_{\nu_{\alpha}\rightarrow \nu_{\beta}} (t)}^2\\
= & \ \sum_{k}U^{*}_{\alpha k}U_{\beta k}e^{iE_{k} t}\sum_{j}U_{\alpha j}U^{*}_{\beta j}e^{-iE_{j} t}\,. 
\end{align}
The speed with which neutrinos travel is ultra-relativistic. So, following the relation $E_{k} = \sqrt{\abs{\vec{p}}^{2} + m^{2}_{k}}$\,, and keeping in mind the equal momentum assumption,
\begin{eqnarray}
E_{k} & \approx & E + \frac{m_{k^2}}{2E}\nonumber\\
\Rightarrow E_{k} - E_{j} & \approx & \frac{\Delta m^{2}_{kj}}{2E}\,,
\end{eqnarray}
where we define $\Delta m^{2}_{kj} = m^{2}_{k} - m^{2}_{j}$, termed as the mass splitting or mass-squared difference. Further, $E = \abs{\vec{p}}$\,, is a valid approximation as long as we neglect the minute mass contribution. In a defined neutrino experiment, the measured quantity is $L$, which denotes the distance traveled by neutrino instead of the propagation time $t$. We know that under natural units system $t = L$, thus the above equation can be re-written as a function of energy,
\begin{align*}
P_{\nu_{\alpha}\rightarrow \nu_{\beta}} \ = & \ \sum_{k}U^{*}_{\alpha k}U_{\beta k}e^{\frac{-im^{2}_{k} L}{2E}}\sum_{j}U_{\alpha j}U^{*}_{\beta j}e^{\frac{im^{2}_{j} L}{2E}} \\
= & \ \sum_{k}\abs{U_{\alpha k}}^2\abs{U_{\beta k}}^2 + \sum_{j>k}\left(U_{\alpha k}U^{*}_{\beta k}U^{*}_{\alpha j}U_{\beta j}e^{\frac{-i\Delta_{jk}L}{2E}} + \overline{U_{\alpha k}U^{*}_{\beta k}U^{*}_{\alpha j}U_{\beta j}e^{\frac{-i\Delta_{jk}L}{2E}}}\right)\\
= & \ \abs{\sum_{k}U^{*}_{\alpha k}U_{\beta k}}^{2} - \sum_{j>k}\left(U_{\alpha k}U^{*}_{\beta k}U^{*}_{\alpha j}U_{\beta j} + \overline{U_{\alpha k}U^{*}_{\beta k}U^{*}_{\alpha j}U_{\beta j}}\right)\\
+ & \ \sum_{j>k}\left(U_{\alpha k}U^{*}_{\beta k}U^{*}_{\alpha j}U_{\beta j}e^{\frac{-i\Delta_{jk}L}{2E}} + \overline{U_{\alpha k}U^{*}_{\beta k}U^{*}_{\alpha j}U_{\beta j}e^{\frac{-i\Delta_{jk}L}{2E}}} \right)
\end{align*} 
where $\Delta_{kj} = m^{2}_{k}-m^{2}_{j}$. Using the unitarity relation on $U$, we get $\sum_{k}U^{*}_{\alpha k}U_{\beta k} =  \delta_{\alpha\beta}\,,$ So,
\begin{align*}
 P_{\nu_{\alpha}\rightarrow \nu_{\beta}} \ = & \ \delta_{\alpha\beta} - \sum_{j>k}\left[\left(U_{\alpha k}U^{*}_{\beta k}U^{*}_{\alpha j}U_{\beta j}\right)\left(1 - e^{-i\frac{\Delta_{jk}L}{2E}}\right)+ \overline{\left(U_{\alpha k}U^{*}_{\beta k}U^{*}_{\alpha j}U_{\beta j}\right)\left(1 - e^{-i\frac{\Delta_{jk}L}{2E}}\right)}\right]\\
= & \ \delta_{\alpha\beta} - \sum_{j>k}2 \mathscr{R}\left[\left(U_{\alpha k}U^{*}_{\beta k}U^{*}_{\alpha j}U_{\beta j}\right)\left(1-e^{-i\frac{\Delta_{jk}L}{2E}}\right)\right]\\
= & \ \delta_{\alpha\beta} - 2\sum_{j>k} \mathscr{R}\left[\left(U_{\alpha k}U^{*}_{\beta k}U^{*}_{\alpha j}U_{\beta j}\right)\mathscr{R}\left(1-e^{-i\frac{\Delta_{jk}L}{2E}}\right) - \mathscr{I}\left(U_{\alpha k}U^{*}_{\beta k}U^{*}_{\alpha j}U_{\beta j}\right)\mathscr{R}\left(1-e^{-i\frac{\Delta_{jk}L}{2E}}\right)\right]\\
= & \ \delta_{\alpha\beta} - 2 \sum_{j>k} \mathscr{R}\left(U_{\alpha k}U^{*}_{\beta k}U^{*}_{\alpha j}U_{\beta j}\right)2\sin^{2}\frac{\Delta_{jk}L}{2E}  + 2 \sum_{j>k} \mathscr{I}\left(U_{\alpha k}U^{*}_{\beta k}U^{*}_{\alpha j}U_{\beta j}\right)\sin\frac{\Delta_{jk}L}{2E}
\end{align*} 
\begin{equation}
 P_{\nu_{\alpha}\rightarrow \nu_{\beta}} \ = \ \delta_{\alpha\beta} - 4 \sum_{j>k} \mathscr{R}\left(U_{\alpha k}U^{*}_{\beta k}U^{*}_{\alpha j}U_{\beta j}\right)\sin^{2}\frac{\Delta_{jk}L}{2E}  + 2 \sum_{j>k} \mathscr{I}\left(U_{\alpha k}U^{*}_{\beta k}U^{*}_{\alpha j}U_{\beta j}\right)\sin\frac{\Delta_{jk}L}{2E}\,.
\end{equation}
\subsection{Two-flavor regime}
\label{sec:two_flavor}
Let us discuss the most easy and idealistic scenario of having only two flavors involved in the neutrino mixing. Then following the definition of rotation matrices in Euclidean space, we can also define a $2 \times 2$ unitary mixing matrix as
\begin{center}
$ U = \begin{bmatrix}
\cos \theta & -\sin \theta\\
\sin \theta & \cos \theta
\end{bmatrix}$\,,  where $0\leq\theta\leq \pi/2$
\end{center}
Substituting this in Eq.~\ref{eq:nut-propagation}\,, we get
\begin{align}
\begin{bmatrix}
\nu_{e} \\ \nu_{\mu} 
\end{bmatrix} \ = & \ \begin{bmatrix}
\cos \theta & -\sin \theta\\
\sin \theta & \cos \theta
\end{bmatrix}\begin{bmatrix}
\nu_{1} \\ \nu_{2} 
\end{bmatrix}
\end{align}
After time $t$, the corresponding flavor eigenstates will supposedly evolve as
\begin{align}
\begin{bmatrix}
\nu_{e} (t) \\ \nu_{\mu} (t) 
\end{bmatrix} \ = & \ \begin{bmatrix}
\cos \theta & -\sin \theta\\
\sin \theta & \cos \theta
\end{bmatrix}\begin{bmatrix}
\nu_{1} (0) \\ \nu_{2} (0) 
\end{bmatrix}\,,
\end{align}
assuming that the unitary mixing matrix is time-independent. Then following Eq.~\ref{eq:ch2b}\,, we get
\begin{align}
\begin{bmatrix}
\nu_{e} (t) \\ \nu_{\mu} (t) 
\end{bmatrix} \ = & \ \begin{bmatrix}
e^{-\frac{m^{2}_{1}L}{2E}}\cos^{2}\theta + e^{-\frac{m^{2}_{2}L}{2E}}\sin^{2}\theta & (e^{-i\frac{m^{2}_{1}L}{2E}}+e^{-i\frac{m^{2}_{2}L}{2E}})\cos\theta\sin\theta\\
(e^{-i\frac{m^{2}_{1}L}{2E}}+e^{-i\frac{m^{2}_{2}L}{2E}})\cos\theta\sin\theta & e^{-\frac{m^{2}_{1}L}{2E}}\sin^{2}\theta + e^{-\frac{m^{2}_{2}L}{2E}}\cos^{2}\theta
\end{bmatrix}\begin{bmatrix}
\nu_{e} \\ \nu_{\mu}  
\end{bmatrix}\,.
\label{eq:flavor-mass-explicit-U-2flavor}
\end{align}
The above equation explicitly conveys that none of the flavor eigenstates purely contain only one mass eigenstate. In each flavor eigenstate, there is some admixture from both the mass eigenstates: $m_{1}$ and $m_{2}$. Using the above equation, we can infer that the probability of probing $\nu_{e}$ after time $t$ as $\nu_{e}$ is given by
\begin{align}
P(\nu_{e}\rightarrow \nu_{e}) \ = & \ \abs{e^{-\frac{m^{2}_{1}L}{2E}}\cos^{2}\theta + e^{-\frac{m^{2}_{2}L}{2E}}\sin^{2}\theta}^{2}\\
= & \ 1 - \sin^{2}2\theta\sin^{2}\frac{(m^{2}_{2}-m^{2}_{1}L)}{4E}
\label{eq:nu-e-survival-2flavor}
\end{align}
This probability of detecting the same flavor $\nu_{\alpha}$ after propagation is termed as the \textit{survival} or \textit{disappearance probability}. Following this discussion, the probability of $\nu_{e}$ flavor getting detected as $\nu_{\mu}$ flavor after time $t$ is given by
\begin{align}
P(\nu_{e}\rightarrow \nu_{\mu}) \ = & \ \abs{-e^{-i\frac{m^{2}_{1}L}{2E}} + e^{-i\frac{m^{2}_{2}L}{2E}}\cos\theta\sin\theta}^{2}\\
= & \ \sin^{2}2\theta\sin^{2}\frac{(m^{2}_{2}-m^{2}_{1})L}{4E}\,.
\label{eq:nu-e-transition-2flavor}
\end{align}
This probability of detecting some other flavor $\nu_{\beta}$ after propagation of $\nu_{\alpha}$ for distance $L$ with energy $E$ is termed as the \textit{transition probability} of $\nu_{\alpha}$ or \textit{appearance  probability }of $\nu_{\beta}$. Eq.~\ref{eq:nu-e-transition-2flavor} shows that when $\nu_{e}$ has traveled for a distance $L_{osc}/2 = 2\pi E/\Delta m^{2}$\,, the probability for detecting it as $\nu_{\mu}$ is maximal. This is true for any flavor. Also, after propagating for a full oscillation length $L_{osc}$\,, the system is back in the initial state.

It must be noted from Eqs.~\ref{eq:nu-e-transition-2flavor} and~\ref{eq:nu-e-survival-2flavor} that these oscillation probabilities are neither sensitive to the sign of $\Delta m^{2}$ nor to the octant of $\theta$, i.e. it cannot tell if $\theta\in \left]0,\pi/4\right[$ (lower octant) or if $\theta\in \left]\pi/4,\pi/2\right[$ (higher octant). Moreover, the frequency term that drives the oscillation is dependent on the difference between the mass squares and not individual mass eigenstates. Further, it does not depend independently on the propagation length ($L$) or the neutrino energy ($E$), instead it is influenced by the ratio $L/E$. Since, we will be needing this term to understand most of the oscillation phenomena, it will be convenient to estimate it in the conventional units by reinstating the usage of $\hbar$ and $c$. Thus, the phase becomes
\begin{equation}
\label{eq:phase-normal-units}
\mathscr{\phi} = 1.27\frac{\Delta m_{jk}^{2} ( \text{eV}^2)\, L(\text{km})}{E (\text{GeV})}\,.
\end{equation}
\subsection{Three-flavor regime}
\label{sec:three-flavor-vac}
For three flavors of neutrino, we subsequently need to define a $3 \times 3$ unitary matrix which will depict the mixing matrix, $U$. We know that for an $n \times n$ complex unitary matrix, total number of mixing angles, Dirac CP-phase, and Majorana phases are given by $\frac{N(N-1)}{2},\, \frac{(N-1)(N-2)}{2},\, \text{ and } (N-1)$, respectively. So, for a $3 \times 3$ matrix, we have 3 mixing angles ($\theta_{12},\theta_{13},\theta_{23}$), 1 Dirac CP phase ($\delta_{\mathrm{CP}}$), and 2 Majorana phases (say $\vartheta_{1}, \vartheta_{2}$). So, just like before, we can use the $3 \times 3$ rotation matrices. The question of which order to follow while writing the three rotation matrices is defined by the ease of extraction of information. Conventionally we follow the following parameterization given by Pontecorve-Maki-Nakagawa-Sakata (PMNS)~\cite{Pontecorvo:1957cp,Pontecorvo:1957vz,Maki:1962mu}
\begin{align*}
U' = & \ U^{\text{PMNS}}\cdot U^{\text{Maj}}\\
= & \ U_{23} (\theta_{23})\cdot U_{13}(\theta_{13},\delta_{\mathrm{CP}})\cdot U_{12}(\theta_{12})\cdot U^{\text{Maj}}\\
= & \ \begin{bmatrix}
1 & 0 & 0\\
0 & c_{23} & s_{23}\\
0 & -s_{23} & c_{23}
\end{bmatrix}\begin{bmatrix}
c_{13} & 0 & s_{13}e^{-i\delta_{\mathrm{CP}}}\\
0 & 1 & 0\\
-s_{13}e^{i\delta_{\mathrm{CP}}} & 0 & c_{13}
\end{bmatrix}\begin{bmatrix}
c_{12} & s_{12} & 0\\
-s_{12} & c_{12} & 0\\
0 & 0 & 1
\end{bmatrix}\begin{bmatrix}
e^{i\vartheta_{1}/2} & 0 & 0\\
0 & e^{i\vartheta_{2}/2} & 0\\
0 & 0 & 1
\end{bmatrix}\\
= & \ \begin{bmatrix}
c_{12}c_{13} & s_{12}c_{13} & s_{13}e^{-i\delta_{\mathrm{CP}}}\\
-s_{12}c_{23}-c_{12}s_{13}s_{23}e^{i\delta_{\mathrm{CP}}} & c_{12}c_{23} - s_{12}s_{13}s_{23}e^{i\delta_{\mathrm{CP}}} & c_{13}s_{23}\\
s_{12}s_{23} - c_{12}s_{13}c_{23}e^{i\delta_{\mathrm{CP}}} & -c_{12}s_{23}-s_{12}s_{13}c_{23}e^{i\delta_{\mathrm{CP}}} & c_{13}c_{23}
\end{bmatrix}\begin{bmatrix}
e^{i\vartheta_{1}/2} & 0 & 0\\
0 & e^{i\vartheta_{2}/2} & 0\\
0 & 0 & 1
\end{bmatrix}\,,
\end{align*}
where $s_{ij}=\sin\theta_{ij}$ and $c_{ij}=\cos\theta_{ij}$. Since, Majorana phases do not have any role in the standard three-neutrino oscillations, we will be dropping these phases henceforth. The choice of parameterization of mixing angles while determining $U$ has its own benefits. From the $U^{\text{PMNS}}$, we have

\begin{enumerate}
\item with regards to solar mixing angle ($\theta_{12}$)
\begin{itemize}
\item  $\tan\theta_{12} = \frac{U_{12}}{U_{11}}$
\item $\cos^{2}\theta_{12} = \frac{\abs{U_{11}}^2}{1-\abs{U_{13}}^2}$
\item $\sin^{2}\theta_{12} = \frac{\abs{U_{12}}^2}{1-\abs{U_{13}}^2}$
\end{itemize}
\item with regards to atmospheric mixing angle ($\theta_{23}$)
\begin{itemize}
\item  $\tan\theta_{23} = \frac{U_{23}}{U_{33}}$
\item $\cos^{2}\theta_{23} = \frac{\abs{U_{33}}^2}{1-\abs{U_{13}}^2}$
\item $\sin^{2}\theta_{23} = \frac{\abs{U_{2 3}}^2}{1-\abs{U_{13}}^2}$
\end{itemize}
\end{enumerate}
Hence, the choice of above parameterization. One can certainly choose parameterization of other choices, but then these simple ratios will change keeping the probabilities same.

%
\begin{figure}[t!]
\includegraphics[width=0.5\linewidth]{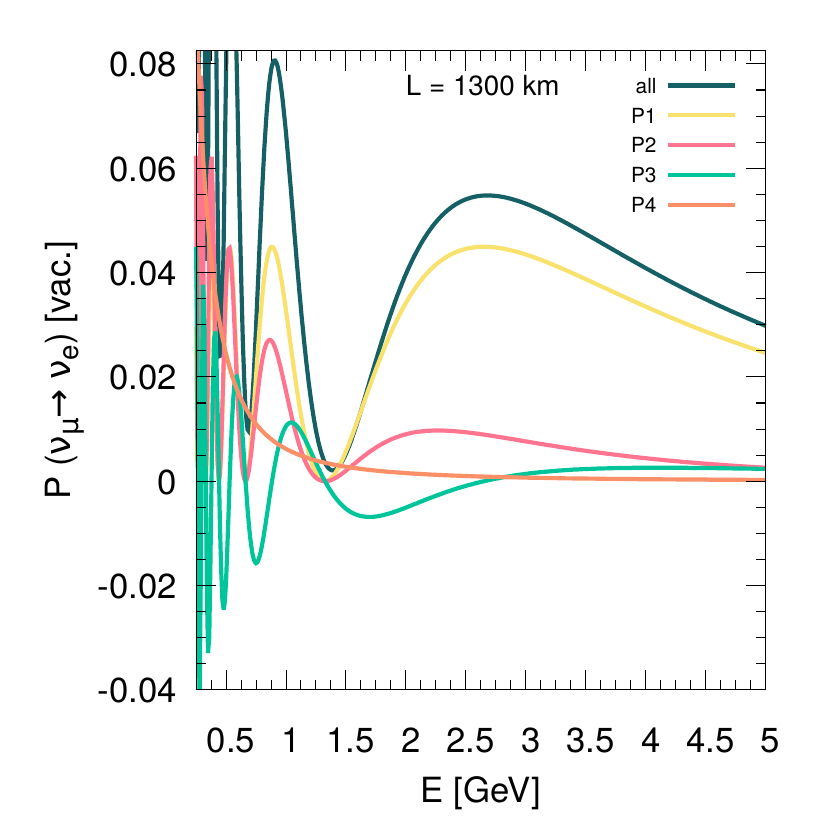}
\includegraphics[width=0.5\linewidth]{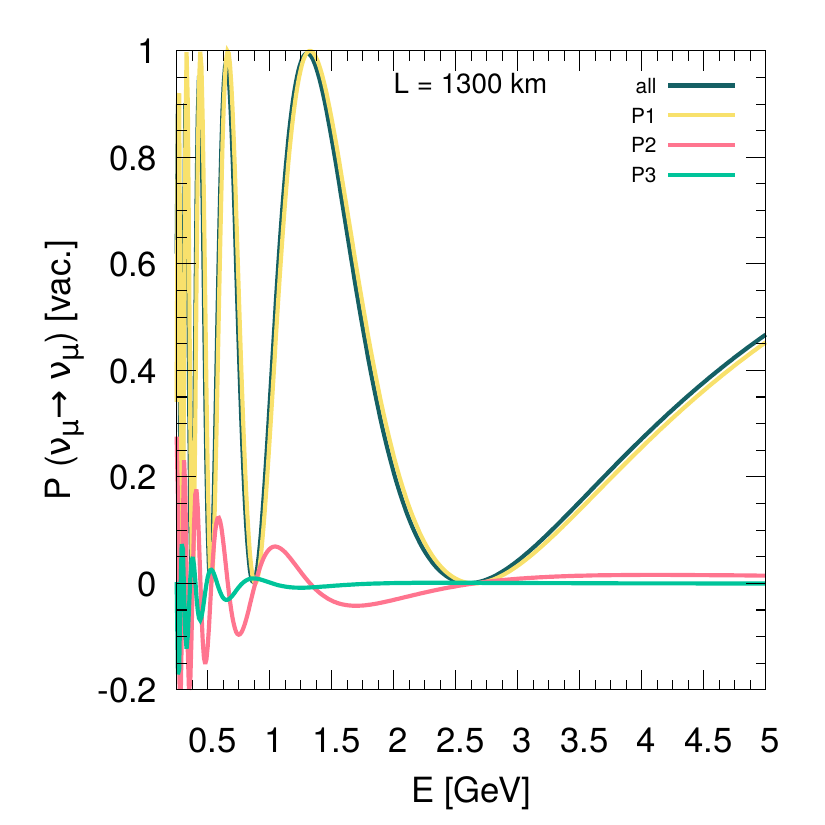}
\caption{Refer to Eq.~\ref{eq:nue-app-3flavor-vacuum} and Eq.~\ref{eq:numu-disapp-3flavor-vacuum} for the corresponding labels in left and right panel, respectively. Curves are drawn for illustrative purpose assuming $\delta_{\mathrm{CP}} = 45^{\circ}$, NMO and benchmark values from Ref.~\cite{Capozzi:2021fjo}.}
\label{fig:separate-prob-terms-vacuum}
\end{figure}
%
The two most important channels that we will be analyzing throughout this thesis are the appearance of $\nu_{e}$ in $\nu_{\mu}$ and the disappearance of $\nu_{\mu}$ as $\nu_{\mu}$ in both neutrino and antineutrino modes. Thus, following Ref.~\cite{Akhmedov:2004ny}\,, wherein we consider only upto second order in $\alpha$ and when $\cos \theta_{13}\sim 1$, we obtain these two expressions as
\begin{eqnarray}
P^{\mathrm{vac}}_{\nu_{\mu} \rightarrow \nu_{e}} &\simeq& \underbrace{\sin^{2}{2\theta_{13}}\sin^{2}\theta_{23}\sin^{2}\Delta}_{\text{P1}}   \nonumber \\ &+& \underbrace{(\alpha\Delta)\sin2\theta_{13}\sin2\theta_{12}\sin2\theta_{23}\sin\Delta\sin\Delta \sin\delta_{\mathrm{CP}}}_{\text{P2}}\nonumber \\
&+& \underbrace{(\alpha\Delta)\sin2\theta_{13}\sin2\theta_{12}\sin2\theta_{23}\sin\Delta\cos\Delta \cos\delta_{\mathrm{CP}}}_{\text{P3}}\nonumber \\ &+&\underbrace{\sin^{2}2\theta_{12}\cos^{2}\theta_{23}\sin^{2}(\alpha\Delta)}_{\text{P4}}\,,
\label{eq:nue-app-3flavor-vacuum}
\end{eqnarray}
\begin{eqnarray}
P^{\mathrm{vac}}_{\nu_{\mu} \rightarrow \nu_{\mu}} &\simeq& \underbrace{1 - \sin^{2}2\theta_{23}\sin^{2}\Delta + 4\sin^{2}\theta_{13}\sin^{2}\theta_{23}\cos 2\theta_{23}\sin^{2}\Delta}_{\text{P1}}\nonumber\\
&+& \underbrace{(\alpha\Delta)\sin 2\Delta[\cos^{2}\theta_{12}\sin^{2}2\theta_{23}
- 2\sin\theta_{13}\sin 2\theta_{12}\sin^{2}\theta_{23}\sin 2 \theta_{23}\cos\delta_{\mathrm{CP}}]}_{\text{P2}}\nonumber\\
&-& \underbrace{\alpha^{2}\Delta^{2}\left[\sin^{2}2\theta_{12}\cos^{2}\theta_{23} + \cos^{2}\theta_{12}\sin^{2}2\theta_{23}(\cos 2\Delta - \sin^{2}\theta_{12}) \right]}_\text{P3}\,,
\label{eq:numu-disapp-3flavor-vacuum}
\end{eqnarray}
where $\alpha = \frac{\Delta m^{2}_{21}}{\Delta m^{2}_{31}}$ and $\Delta =  \frac{\Delta m^{2}_{31} L}{4E}$. In the left panel of Fig.~\ref{fig:separate-prob-terms-vacuum}\,, we observe that maximum contribution in the $P^{\mathrm{vac}}_{\nu_{\mu} \rightarrow \nu_{e}}$ comes from first term (P1) in Eq.~\ref{eq:nue-app-3flavor-vacuum}. This is because the first term does not have any $\alpha$ suppression unlike P2, P3, and P4. While, the magnitude of P1 is dominated by $\sin^{2}\theta_{23}$, the $\sin^{2}2\theta_{13}$ is relatively small. Therefore, the first term is often referred as the \textit{atmospheric} term, giving insights on the atmospheric parameters ($\theta_{23}\,, \Delta m^{2}_{31}$). The P2 term is $\alpha$ suppressed, while it also has a $\sin\delta_{\mathrm{CP}}$ term. This makes this term as \textit{CP-violating} term, since this term can differentiate between the change of sign of intrinsic or real CP phase. Contrastingly, P3 has a $\cos\delta_{\mathrm{CP}}$, which acts as \textit{CP-conserving} term, since this term does not give any idea about the change of sign in CP phase. Finally, the P4 gives information on the solar parameters ($\theta_{12}\,,\Delta m^{2}_{21}$), thus often called as \textit{solar} term. However, it is suppressed by an order of O(10$^{-3}$) when considered for the first oscillation maximum. For the second oscillation maximum, it is suppressed by the order of O(10$^{-2}$). Therefore, probing the appearance of $\nu_{e}$ in $\nu_{\mu}$ beam is immensely rich in physics potentials. For a further elaborate discussion, refer to Chapter~\ref{sec:ch7}.

Further, the right panel in Fig.~\ref{fig:separate-prob-terms-vacuum} depicts P1 as the leading term for similar reasons as discussed previously. P2 and P3 are $\alpha$ suppressed. Further, even in P1, the second term corresponding to $\sin^{2}2\theta_{23}\sin^{2}\Delta$ is the most contributing of all, because the third term ($4\sin^{2}\theta_{13}\sin^{2}\theta_{23}\cos 2\theta_{23}\sin^{2}\Delta$) has $\sin^{2}\theta_{13}$ in its magitude which again suppresess. The disappearance channel does not have any $\sin\delta_{\mathrm{CP}}$, it only has dependence on $\cos \delta_{\mathrm{CP}}$, which is again $\alpha$ suppressed. It must be noted that all these discussion considers a vacuum-like scenario. We discuss more in Sec.~\ref{sec:nu-matter}, that depicts a more realistic scenario with the inclusion of matter.  
 Following the discussion of action of discrete symmetries on oscillation probabilities in Ref.~\cite{Akhmedov:2001kd}, we have
 \begin{align}
 P_{\nu_{\alpha} \rightarrow \nu_{\beta}} \ \xrightarrow{~CP} & \ P_{\bar{\nu}_{\alpha} \rightarrow \bar{\nu}_{\beta}}\,,\\
\xrightarrow{~~T~} & \ P_{\nu_{\beta} \rightarrow \nu_{\alpha}}\,,\\ 
\xrightarrow{CPT} & \ P_{\bar{\nu}_{\beta} \rightarrow \bar{\nu}_{\alpha}}\,.
\label{eq:CPT-osc-prob}
\end{align}  
Thus, if we assume that CPT conservation holds then $P_{\nu_{\alpha} \rightarrow \nu_{\beta}}  = P_{\bar{\nu}_{\beta} \rightarrow \bar{\nu}_{\alpha}}$, which in turn implies that the CP and T asymmetries are equal in neutrino oscillations. So CP asymmetry is given by,
\begin{equation}
\mathcal{A}_{\alpha\beta} \equiv \frac{P_{\nu_{\alpha} \rightarrow \nu_{\beta}} - P_{\bar{\nu}_{\alpha} \rightarrow \bar{\nu}_{\beta}}}{P_{\nu_{\alpha} \rightarrow \nu_{\beta}} + P_{\bar{\nu}_{\alpha} \rightarrow \bar{\nu}_{\beta}}} = \frac{P_{\nu_{\alpha} \rightarrow \nu_{\beta}} - P_{\nu_{\beta} \rightarrow \nu_{\alpha}}}{P_{\nu_{\alpha} \rightarrow \nu_{\beta}} + P_{\nu_{\beta} \rightarrow \nu_{\alpha}}}
\label{eq:cp-asymmetry}
\end{equation}
Also, from CPT conservation, we have $ P_{\nu_{\alpha} \rightarrow \nu_{\alpha}} =  P_{\bar{\nu}_{\alpha} \rightarrow \bar{\nu}_{\alpha}}$\,, which implies that genuine CP violation does not exist in the disappearance channel (more on this later in Chapter~\ref{sec:ch7}), which can be verified from Eq.~\ref{eq:numu-disapp-3flavor-vacuum}. For convenience of studying CP violation, one can define $\Delta P_{\alpha\beta} \equiv  P_{\nu_{\alpha} \rightarrow \nu_{\beta}} -  P_{\bar{\nu}_{\alpha} \rightarrow \bar{\nu}_{\beta}}$, which can be further simplified into
\begin{equation}
\Delta P_{\alpha\beta} = \pm 16J\sin\frac{\Delta m^{2}_{21}L}{4E}\sin\frac{\Delta m^{2}_{31}L}{4E}\sin\frac{\Delta m^{2}_{32}L}{4E}\,,~~ \text{where } J \equiv \text{ Im }[U_{e1}U^{*}_{\mu 1}U^{*}_{e2}U_{\mu 2}]\,,
\label{eq:DeltaP-Jarlskog}
\end{equation}
following Ref.~\cite{Bilenky:1980cx,Barger:1980jm}. The $+$ sign is used when $(\alpha, \beta, \gamma)$ follows $(e,\mu,\tau)$ or its even permutation, while the sign is reversed if it traces an odd permutation. The quantity $J$ is thus invariant under the change of basis and is coined as the \textit{Jarlskog invariant} which under the standard parameterization is given by
\begin{equation}
J = \frac{1}{8}\cos\theta_{13}\sin 2\theta_{12}\sin 2 \theta_{13} \sin 2 \theta_{23}\sin\delta_{\mathrm{CP}}\,.
\label{eq:Jarlskog-invariant}
\end{equation} 
Therefore from Eqs.~\ref{eq:Jarlskog-invariant} and ~\ref{eq:DeltaP-Jarlskog}, we can infer the necessary conditions for CP violation as
\begin{itemize}
\item non-zero and  $\neq 90^{\circ}$ for any mixing angle
\item non-zero value of mass-splittings
\item non conserving values of $\delta_{\mathrm{CP}}$\,, i.e $\delta_{\mathrm{CP}} \neq 0^{\circ}$ and $\pm 180^{\circ}$ 
\end{itemize}
Further the unitarity of the PMNS matrix and Eq.~\ref{eq:DeltaP-Jarlskog} implies that the CP-violating term is independent of the channel through which it is probed (except for a change in sign). So
\begin{eqnarray}
\text{ Im }[U_{e1}U^{*}_{\mu 1}U^{*}_{e2}U_{\mu 2}] &=& -\text{ Im }[U_{e1}U^{*}_{\tau 1}U^{*}_{e2}U_{\tau 2}] = \text{ Im }[U_{\mu 1}U^{*}_{\tau 1}U^{*}_{\mu 2}U_{\tau 2}]\,,\\
&\Rightarrow & \Delta P_{e\mu} = \Delta P_{e\tau} = \Delta P_{\mu\tau}
\end{eqnarray}
\section{Neutrino oscillations in matter}
\label{sec:nu-matter}
Time-dependent Schr\"{o}dinger equation can be used to describe the neutrino propagation in matter as well,
\begin{equation}
i\frac{d}{dt}\ket{v(t)} = \mathscr{H}\ket{v(t)}
\label{eq:evolution-eq}
\end{equation}
where the Hamiltonian $\mathscr{H}$ is a sum of free kinetic energy ($\mathscr{H}_{0}$) part that determines neutrino propagation in vacuum and an additional interaction Hamiltonian $\mathscr{H}_{I}$ which arises due to the interaction of neutrino with the medium\cite{Wolfenstein:1977ue},
\begin{equation}
\mathscr{H} = \mathscr{H}_{0} + \mathscr{H}_{I}
\label{eq:Hamiltonian}
\end{equation}
Now if we express Eq.~\ref{eq:evolution-eq} in the flavor eigenbasis we get
\begin{equation}
i\frac{d}{dt}\nu_{\alpha}(t) = \sum_{\beta}\mathscr{H}_{\alpha\beta}\nu_{\beta}(t)\,,
\label{eq:evolution-flavor-basis}
\end{equation}
where $\alpha, \beta = e,\mu,\tau$. Also, $\mathscr{H}_{\alpha\beta} \equiv \bra{\nu_{\alpha}}\mathscr{H}\ket{\nu_{\beta}}$ are the matrix elements in the flavor basis, $\bra{\nu_{\alpha}}\ket{\nu(t)}$ is the amplitude of determining flavor eigenstate $\ket{\nu_{\alpha}}$ at time $t$. Further, the massive neutrino eigenstates, $\nu_{k}$ bearing momentum $\vec{p}$ can be expressed as one of the eigenstates of the vacuum Hamiltonian as
\begin{equation}
\mathscr{H}_{0}\ket{\nu_{k}} = E_{k}\ket{\nu_{k}}\,,
\end{equation} 
where $E_{k} = \lvert\sqrt{\vec{p}^2 + m^{2}_{k}}\rvert$\,.
and the eigenstates of interaction Hamiltonian in flavor basis is given as
\begin{equation}
\mathscr{H}_{I}\ket{\nu_{\alpha}} = V_{\alpha}\ket{\nu_{\alpha}}\,,
\end{equation}
The matter potential $V_{\alpha}$ is a result of the coherent forward scatterings of neutrino with the electrons and nucleons in the medium, leaving the neutrino momentum unchanged~\cite{Wolfenstein:1977ue}. The electron neutrinos can directly interact with electrons in the medium mediated by the SM $W^{\pm}$ boson via the charged current (CC) interaction. While the dearth of muons and taus in ordinary matter restricts this CC current interaction to only $\nu_{e}$, all the flavors can interact by mediating through the SM $Z$ boson via neutral current (NC) interaction identically. The contribution of CC and NC interaction leads to a diagonal matter potential in the flavor basis~\cite{Wolfenstein:1977ue,Barger:1980tf}. Summarizing this, we have,
\begin{eqnarray}
V_{\text{CC},\alpha} &=& \begin{cases}
\sqrt{2}G_{F}N_{e}~~~\alpha = e\\
0\qquad\qquad~\alpha = \mu,\tau
\end{cases}\\
V_{\text{NC},\alpha} &=& -\frac{G_{F}}{\sqrt{2}}N_{n}\qquad \alpha = e,\mu,\tau\,,
\end{eqnarray}
where, $G_{F} (\text{Fermi coupling constant}) = 1.166 \times 10^{-5}$ GeV$^{-2}$, $N_{e}\,, N_{n}$ are the electron and neutron number\footnote{$V_{\text{NC}}$ depends only on $N_{n}$ as the contributions due to electron density ($N_{e}$) and proton density ($N_{p}$) cancel out due to the assumption that the medium under consideration is neutral. We also assume that the medium is isoscalar, $N_{p} = N_{n}$\,, and unpolarized made of non-relativistic particles.} densities in the medium, respectively. The above CC potential can be refurbished in terms of line-averaged constant Earth matter density $(\rho_{\mathrm{avg}})$ as
\begin{equation}
V_{\text{CC}} \simeq 7.6 \times \left(\frac{N_{e}}{N_{p} + N_{n}}\right) \left(\frac{\rho_{\mathrm{avg}}}{10^{14} \text{ g/cm}^{3}} \right)~\text{eV}\,.
\label{eq:VCC-rho-avg1}
\end{equation}
For an isoscalar and electrically neutral medium, we can simplify it further 
\begin{equation}
V_{\text{CC}} \simeq 7.6 \times 0.5 \left(\frac{\rho_{\mathrm{avg}}}{10^{14} \text{ g/cm}^{3}} \right)~\text{eV}\,.
\label{eq:VCC-rho-avg2}
\end{equation}
Following the charge conjugation, the sign changes when one switches to antineutrino mode. So
\begin{equation}
V_{\alpha}(\bar{\nu}) = -V_{\alpha}(\nu)\,.
\end{equation}
Since the NC contribution is universal, the matter interaction Hamiltonian $(\mathscr{H}_{I})$ reduces to
\begin{equation}
\mathscr{H} = \frac{1}{2E}(UM^{2}U^{\dagger} + \mathcal{A})\,,
\end{equation} 
The propagation eigenstates here are not the mass eigenstates ($\ket{\nu_{k}}$) discussed in Sec~\ref{sec:nu_vacuum}, but the eigenstates of new matter interaction Hamiltonian $\ket{\nu_{k}^{m}}$. Similarly, the unitary mixing matrix in matter is $U_{m}$ which diagonalizes $\mathscr{H}$ as 
\begin{eqnarray}
\mathscr{H}\ = \ U_{m}\begin{bmatrix}
E^{m}_{1} &  & 0\\
0 & E^{m}_{2} & 0\\
0 & 0 & E^{m}_{3}\end{bmatrix}U^{\dagger}_{m}\,,\qquad
\begin{bmatrix}
\ket{\nu_{e}}\\
\ket{\nu_{\mu}}\\
\ket{\nu_{\tau}}
\end{bmatrix} \ = \ U^{\dagger}_{m}\begin{bmatrix}
\ket{\nu_{1}^{m}}\\
\ket{\nu_{2}^{m}}\\
\ket{\nu_{3}^{m}}
\end{bmatrix}\,,
\label{eq:evolution-matrix-matter}
\end{eqnarray}
Where $E^{m}_{k}$ are the energy levels in matter. Similarly generalizing Eq.~\ref{eq:time-dependent-evolution} in vacuum scenario to matter we get
\begin{equation}
\nu_{\alpha}(t) = \sum_{k}(U_{m})_{\beta k}\nu^{m}_{k}(t)\,.
\end{equation}
For convenience, we will initiate with discussing two flavor and then subsequently extending it to three flavor paradigm.
\subsection{Two-flavor scenario}
\label{sec:nu-matter-2flavor}
In two flavor scenario, following Eqs.~\ref{eq:flavor-mass-explicit-U-2flavor} and~\ref{eq:evolution-matrix-matter}, we have in the flavor eigen basis (say $\ket{\nu_{e}}, \ket{\nu_{\mu}}$), the matter Hamiltonian is
\begin{equation}
\mathscr{H}_{I}^{2-\text{flavor}} = \begin{bmatrix}
\frac{\Delta m^{2}}{2E} \cos 2\theta \pm \sqrt{2}G_{F}N_{e} & \frac{\Delta m^{2}}{4E}\sin 2\theta\\
\frac{\Delta m^{2}}{4E} \sin 2\theta & 0
\end{bmatrix}
\end{equation}
wherein $\Delta m^{2}$ and $\theta$ are the vacuum oscillation mass-squared splitting and mixing angle, respectively. Also, $+$ (--) corresponds to neutrino (antineutrino) mode. This is further diagonalized by its corresponding evolved mixing angle in matter~\cite{Wolfenstein:1977ue}
\begin{equation}
\mathscr{H}_{I}^{2-\text{flavor}} = U_{m}\begin{bmatrix}
E_{1}^{m} & 0\\
0 & E_{2}^{m}
\end{bmatrix}U^{\dagger}_{m}\,,
\end{equation}
where $U_{m} = \begin{bmatrix}
\cos\theta_{m} & \sin\theta_{m}\\
-\sin \theta_{m} & \cos\theta_{m}
\end{bmatrix}$\,, So
\begin{eqnarray}
E^{m}_{2} - E^{m}_{1} &=& \frac{\Delta m^{2}}{2E}\sqrt{\left(1 \mp \frac{N_{e}}{N_{\text{res}}} \right)^{2} \cos 2\theta + \sin^2 2\theta}\,,\\
\sin 2\theta_{m} &=& \frac{\sin 2\theta}{\sqrt{\left(1 \mp \frac{N_{e}}{N_{\text{res}}}\right)^2 \cos^{2}2\theta + \sin^{2}2 \theta}}\,,\\
\cos 2 \theta_{m} &=& \frac{\left(1 \mp \frac{N_{e}}{N_{\text{res}}}\right)\cos 2 \theta}{\left(1 \mp \frac{N_{e}}{N_{\text{res}}}\right)^{2} \cos^{2}2\theta + \sin^{2} 2\theta}\,,
\end{eqnarray}
where the sign, $-$ corresponds to neutrino and $+$ corresponds to antineutrino mode. Also,
\begin{equation}
N_{\text{res}} = \frac{\Delta m^{2}\cos 2\theta}{2\sqrt{2}G_{F}E}
\label{eq:MSW-resonance}
\end{equation} 
is the resonance density, wherein the mixing angle in matter, $\theta_{m}$ is maximal when $N_{e} = N_{\text{res}}$, irrespective of the value of vacuum mixing angle ($\theta$)~\cite{Mikheyev:1985zog,Mikheev:1986wj}. So the resonance condition in neutrino mode is achieved when 
\begin{equation}
\Delta m^2 \cos 2 \theta > 0\,.
\end{equation} 
This is the well-established \textit{Mikhev-Smirnov-Wolfenstein} or commonly referred as \textit{MSW resonance}~\cite{Wolfenstein:1977ue,Barger:1980tf}. Analogicaly for antineutrinos this condition reverses, i.e. $\Delta m^2 \cos 2 \theta < 0$ and so the resonance occurs when $N_{e} = -N_{\text{res}}$.
Thus we finally get the oscillation probabilty as
\begin{equation}
P_{\nu_{e}\rightarrow \nu_{\mu}} = \sin^{2}2\theta_{m}\sin^{2}\left(\frac{\Delta m^{2}L}{4E}\sqrt{\sin^{2}2\theta + \left(\cos 2\theta - \frac{2EV}{\Delta m^{2}}\right)^2} \right)\,.
\end{equation}
It can be easily verified that in the limit $V \rightarrow 0$, the above equation reduces to Eq.~\ref{eq:nue-app-3flavor-vacuum} in vacuum.
\subsection{Three-flavor scenario}
\label{sec:three-flavor-mat}
Following the above discussion, one can extend it to three flavor oscillations in matter as
\begin{equation}
i\frac{d}{dt}\begin{bmatrix}
\ket{\nu_{e}}\\
\ket{\nu_{\mu}}\\
\ket{\nu_{\tau}}
\end{bmatrix} = \left(\frac{1}{2E}U\begin{bmatrix}
m^2_{1} & 0 & 0\\
0 & m^{2}_{2} & 0\\
0 & 0 & m^{2}_{3}
\end{bmatrix}U^{\dagger} + 
\begin{bmatrix}
V_{CC} & 0 & 0\\
0 & 0 & 0\\
0 & 0 & 0
\end{bmatrix}\right)
\begin{bmatrix}
\ket{\nu_{e}} \\
\ket{\nu_{\mu}}\\
\ket{\nu_{\tau}}
\end{bmatrix}\,.
\end{equation}
As long as $\rho_{\mathrm{avg}}$ in Eq.~\ref{eq:VCC-rho-avg1} is constant, we can use the above equation to determine various appearance and disappearance channels.
%
\begin{figure}[t!]
\includegraphics[width=0.5\linewidth]{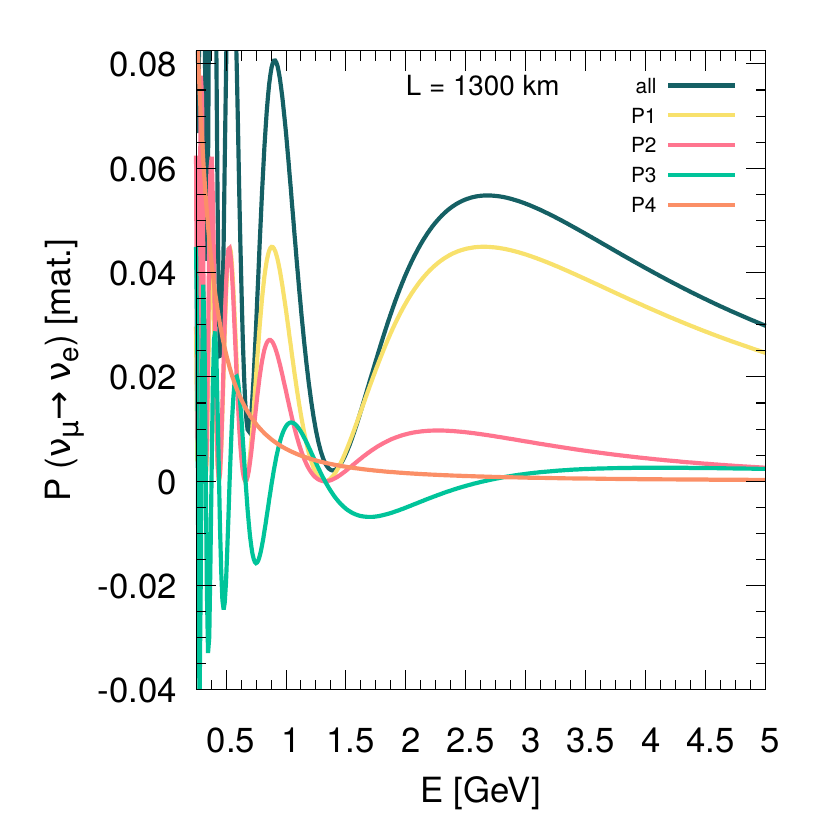}
\includegraphics[width=0.5\linewidth]{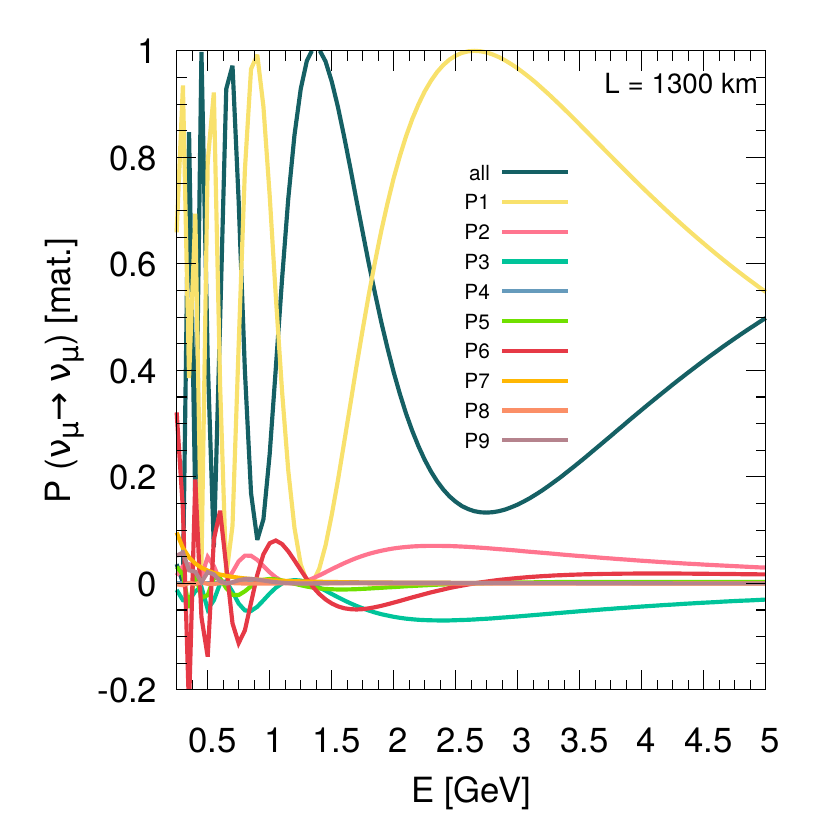}
\caption{Refer to Eq.~\ref{eq:3flavor-nue-app-with-matter} and Eq.~\ref{eq:3flavor-numu-disapp-with-matter} for the corresponding labels in left and right panel, respectively. Curves are drawn for illustrative purpose assuming $\delta_{\mathrm{CP}} = 45^{\circ}$, NMO and benchmark values from Ref.~\cite{Capozzi:2021fjo}.}
\label{fig:separate-prob-terms-matter}
\end{figure}
%

We give here expressions for appearance of $\nu_{\mu}$ and disappearance of $\nu_{\mu}$ following Ref.~\cite{Akhmedov:2004ny} in matter, wherein we consider only upto second order in $\alpha$ and when $\cos \theta_{13}\sim 1$
\begin{eqnarray}
P_{\nu_{\mu}\rightarrow \nu_{e}} &\simeq& \underbrace{\sin^{2}\theta_{23}\sin^{2}2\theta_{13}\frac{\sin^{2}A\Delta}{A^2}}_{\text{P1}}\nonumber\\
&+& \underbrace{\alpha\sin 2\theta_{13}\sin 2\theta_{23}\sin 2\theta_{12}\sin \delta_{\mathrm{CP}}\sin\Delta\frac{\sin(\hat{A}\Delta)}{\hat{A}}\frac{\sin[1-\hat{A}]\Delta}{(1-\hat{A})\Delta}}_{\text{P2}}\nonumber\\
&+& \underbrace{\alpha\sin 2\theta_{13}\sin 2\theta_{23}\sin 2\theta_{12}\cos \delta_{\mathrm{CP}}\cos\Delta\frac{\sin(A\Delta)}{A}\frac{\sin(A-1)\Delta}{(A-1)}}_{\text{P3}}\nonumber\\
&+& \underbrace{\alpha^2\cos^2 \theta_{23}\sin^{2}2\theta_{12}\frac{\sin^{2}((A - 1)\Delta)}{(A-1)^2}}_{\text{P4}}\,,
\label{eq:3flavor-nue-app-with-matter} 
\end{eqnarray}
\begin{eqnarray}
P_{\nu_{\mu}\rightarrow\nu_{\mu}}&\simeq& 1 - \underbrace{\sin^{2}2\theta_{23}\sin^{2}\Delta}_{\text{P1}} - \underbrace{4\sin^{2}\theta_{13}\sin^{2}\theta_{23}\frac{\sin^{2}(A-1)\Delta}{(A-1)^{2}}}_{\text{P2}}\nonumber\\
&-& \underbrace{\frac{2}{A-1}\sin^{2}\theta_{13}\sin^{2}2\theta_{23}\left(\sin\Delta\cos A \Delta\frac{\sin (A-1)\Delta}{A-1} -\frac{A\Delta}{2}\sin 2\Delta\right)}_{\text{P3}}\nonumber\\
&+& \underbrace{\frac{1}{A-1}\alpha\sin\theta_{13}\sin 2\theta_{12}\sin 4 \theta_{23}\cos\delta_{\mathrm{CP}}\sin\Delta\left(A\sin\Delta - \frac{\sin A\Delta}{A}\cos(A-1)\Delta\right)}_{\text{P4}}\nonumber\\
 &-& \underbrace{2 \alpha\sin\theta_{13}\sin 2\theta_{12}\sin 2\theta_{23}\cos\delta_{\mathrm{CP}}\cos\Delta\frac{\sin A\Delta}{A}\frac{\sin (A-1)\Delta}{(A-1)}}_{\text{P5}}\nonumber\\
 &+& \underbrace{\alpha\Delta\cos^{2}\theta_{12}\sin^{2}2\theta_{23}\sin 2\Delta}_{\text{P6}}\nonumber\\
&-& \underbrace{\alpha^2\sin^{2}2\theta_{12}\cos^{2}\theta_{23}\frac{\sin^{2}A\Delta}{A^{2}}}_{\text{P7}} - \underbrace{(\alpha\Delta)^{2}\cos\theta_{12}^{4}\sin^{2}2\theta_{23}\cos 2 \Delta}_{\text{P8}}\nonumber\\
&+& \underbrace{\frac{1}{2A}\alpha^{2}\sin^{2}2\theta_{12}\sin^{2}2\theta_{23}\left(\sin\Delta\frac{\sin A \Delta}{A}\cos(A-1)\Delta-\frac{\Delta}{2}\sin 2\Delta\right)}_{\text{P9}}\,,
\label{eq:3flavor-numu-disapp-with-matter}
\end{eqnarray}
where $\Delta = \frac{\Delta m^{2}_{31}L}{4E}$, $A = \frac{2\sqrt{2}G_{F}N_{e}E}{\Delta m^{2}_{31}}$. As discussed in Sec.~\ref{sec:three-flavor-vac}\,, the nature of the terms contributing as a whole in $P_{\nu_{\mu}\rightarrow \nu_{e}}$ in presence of matter is similar to vacuum-like scenario. In the right panel, we assume additional higher order terms to depict their contribution in $P_{\nu_{\mu}\rightarrow \nu_{\mu}}$ channel. The interesting point to note is in the appearance channel, \textit{leading term $\propto \sin^{2}\theta_{23}$}, while in the disappearance, \textit{leading term $\propto \sin^{2}2\theta_{23}$}. The effect of this is studied later in Chapter~\ref{sec:Ch6}. Following the charge conjugation in antineutrino mode
\begin{itemize}
\item $A \rightarrow -A$
\item $\delta_{\mathrm{CP}} \rightarrow -\delta_{\mathrm{CP}}$
\end{itemize}
This change in sign due to both CP phase and matter potential when one goes from neutrino mode of detection to antineutrino, leads to a confusion between genuine (intrinsic) CP violation (refer to Sec.~\ref{sec:three-flavor-vac}) and fake (extrinsic) CP violation. This is further elaborated and exploited in great detail in Chapter~\ref{sec:ch7}.  
 Next few sections are dedicated to explaining the current scenario in various neutrino sources. Furthermore, we also elaborate on neutrino interactions beyond the above-mentioned standard three-flavor scenario.

 \section{Physics of neutrino sources}
 \label{sec:neutrino_sources}
 The remarkable presence of neutrinos everywhere is next to none. We, humans, have further added by building high-power reactors that source intense electron antineutrinos ($\sim 2 \times 10^{20}~\bar{\nu}_{e}$ per GW thermal power). Largely, Earth also falls into this category emitting electron antineutrinos from the `natural' radioactive beta decay. Next, the core of the Sun is one of the proactive sources of electron neutrinos ($\sim 7 \times 10^{10}~\nu_{e}/\text{cm}^{2}/\text{sec}$). Then, the neutrinos from supernova bursts are one of a kind. Further, the unstoppable neutrinos from the upper atmosphere via interaction with the cosmic rays is another crucial source. In recent years, other neutrinos that originate from astrophysical sources, such as gamma-ray bursts, have also been added to this research area. Still expected but not yet detected, the cosmogenic neutrinos through cosmic neutrino background comes next in line. Finally, the accelerator neutrinos that are artificially produced are the current workhorses for the yet unknowns in neutrino physics. In this section, we will briefly study the prospects of all of these above-mentioned sources of neutrinos. We also give Table~\ref{tab:sources-and-constraints} summarizing various sources of neutrinos and different oscillation parameters they constrain in the 3-neutrino paradigm.
  \subsection{Reactor antineutrinos}
 \label{sec:reactor}
 It was with reactor antineutrinos as a source that neutrinos got discovered in 1956. Almost 70 years down the lane, and yet we do not completely understand their absolute flux and spectral shape. There are four major isotopes: ${}^{235}$U, ${}^{239}$Pu, ${}^{238}$U, and ${}^{241}$Pu, that act as a source for producing more than 99\% $\bar{\nu}_{e}$ following the radioactive beta decay in them. Each fission leads to $\sim 200$ MeV~\cite{Kopeikin:2004cn,Ma:2012bm} while producing six $\bar{\nu}_{e}$, each with an energy of about 10 MeV. The usual inverse beta decay (IBD) followed in the reactors is given by
 \begin{equation}
 \bar{\nu}_{e} + p \rightarrow e^{+} + n \,,
 \label{eq:IBD}
 \end{equation}
where the threshold energy for parent antineutrinos is $\sim 1.8$ MeV. After the reaction, positron takes away most of the parent antineutrino energy, annihilating immediately and producing a prompt signal. The neutron is then scattered away, further getting captured by a H (or Gd, Li, Cd --- if scintillators are doped with either of these) nucleus, producing a delayed signal at $\sim 2.2$ MeV (or 8 MeV). This prompt, followed by a delayed signal (time of delay varies depending upon the capturing nucleus), confirms occurrence of an event in reactors. The initial vouching for reactor neutrino experiments was done through Ref.~\cite{Bemporad:2001qy}. Out of eight reactors proposed, three got accepted: Double Chooz, Daya Bay, and RENO. It was in the year 2012, that the three-neutrino paradigm got certified with the discovery of non-zero value of $\theta_{13}$ by Daya Bay~\cite{DayaBay:2012fng} and RENO~\cite{RENO:2012mkc}, following the earlier upper bounds from T2K~\cite{Catanesi:2013fxa}, MINOS~\cite{MINOS:2011amj}, and Double Chooz~\cite{DoubleChooz:2019qbj}. The discovery of non-zero $\theta_{13}$ opened up the prospects for leptonic CP violation in neutrino sector. 

The present reactor antineutrinos face two major anomalies:

\medskip
\textbf{Absolute Reactor antineutrino flux Anomaly}.--- It is Huber-Mueller (HM) model that predicts theoretically the expected reactor antineutrino flux for all the four major isotopes~\cite{Mueller:2011nm}. Comparing this with the flux observed experimentally from Daya Bay~\cite{DayaBay:2018heb}, Double Chooz~\cite{DoubleChooz:2019qbj}, and RENO~\cite{RENO:2020dxd}, currently a deficit of about 6\% exists~\cite{Mueller:2011nm} than what is predicted. This 6\% deficit is the reactor antineutrino anomaly (RAA)~\cite{Huber:2016fkt}. Following Ref.~\cite{Kopeikin:2021ugh}, the community is currently suggesting to re-evaluate the predicted flux. 

\medskip
\textbf{Shape Anomaly.}--- The anomaly in the spectral shape of reactor antineutrinos around 5 MeV, or colloquially referred as `5 MeV excess' was first observed in 2014~\cite{DoubleChooz:2014kuw}. Recently, Daya Bay reported an excess of 3.2$\sigma$~\cite{DayaBay:2019yxq} and RENO 4$\sigma$~\cite{RENO:2020dxd} evidence of correlation between the 5 MeV excess and the flux anomaly. Other very small-baseline reactors like NEOS also observed excess~\cite{Kim:2021sfe}. While PROSPECT~\cite{PROSPECT:2020sxr}, STEREO~\cite{STEREO:2020hup} also have reported along the same line. The upcoming JUNO-TAO, which is expected to start taking data by the end of 2023, has a main goal to determine the reactor antineutrino spectrum with a high precision and excellent energy resolution of $< 2\%$~\cite{JUNO:2020ijm}, thus explaining the 5 MeV excess and reactor antineutrino flux anomalies.

However, a recent work in Ref.~\cite{Giunti:2021kab} shows a modified calculation of the HM model which are consistent with the predictions of the Estienne-Fallot summation model~\cite{Estienne:2019ujo} and the Kurchatov Institute conversion model~\cite{Kopeikin:2021ugh}, thereby almost leading to the elimination of the RAA. 
\subsection{Solar neutrinos}
\label{sec:solar}
It was the solar neutrino anomaly (deficit in observed neutrinos than predicted from the standard solar model: SSM) that gave the first hint of neutrino oscillations~\cite{Bahcall:2002ng}. Detectors based on the principles of liquid-scintillator and water-Cherenkov provide the best scenario for analyzing the precision spectroscopy of solar neutrinos. In the solar core, the fusion of protons to helium gives rise to solar neutrinos following
\begin{equation}
4p + 2e^{-} \rightarrow {}^{4}\text{He} + 2\nu_{e} + 26.73 ~\text{MeV}\,.
\end{equation}
The above follows the two main chain reactions inside the Sun:

\medskip
\textbf{pp chain.}--- This is the primary reaction with a branching ratio of $\sim 86\%$ following,
\begin{equation}
p + p \rightarrow (H^{2})^{+} + e^{+} + \nu_{e}\,,
\end{equation}
having the flux of neutrinos in the order of $10^{10} \text{ s}^{-1} \text{ cm}^{-2}$ bearing continuous energy spectrum with 420 keV endpoint. The above reaction also has probabilities of producing a mono-energetic ${}^{7}$Be (0.384 MeV and 0.862 MeV), \textit{pep} neutrino (1.44 MeV), and ${}^{8}$B with low flux ($10^{6} \text{ s}^{-1} \text{ cm}^{-2}$) and a spectrum ending around 15 MeV. 

\medskip
\textbf{CNO cycle.}--- Carbon-Nitrogen-Oxygen cycle is dominant for stars being 1.3 times heavier than the Sun. Probing neutrinos from the CNO cycle is relatively tough as it is degenerate with ${}^{210}$Bi present in the liquid scintillator as contamination and of \textit{pep} solar neutrinos. The reactions follow as:
\begin{eqnarray}
(N^{13})^{+} &\rightarrow & C^{13} + e^{+} + \nu_{e}\,, \nonumber \\
(O^{15})^{+} &\rightarrow & N^{15} + e^{+} + \nu_{e}\,, \nonumber \\
(F^{17})^{+} &\rightarrow & O^{17} + e^{+} + \nu_{e}\,, 
\end{eqnarray}
where neutrinos bear energy between (0 - 2.7) MeV.

\medskip
The most precise measurements of the \textit{pp}, $^{7}$Be, and \textit{pep} neutrinos are provided by Borexino~\cite{BOREXINO:2018ohr}. While the high-precision measurements of $^{8}$B exist from SNO~\cite{SNO:2011hxd} and Super-K~\cite{yasuhiro_nakajima_2020_3959640,Super-Kamiokande:2016yck}. As the flux count for \textit{hep} neutrinos is very low, we only have upper limits in the literature from SNO~\cite{SNO:2011hxd}, which is still three times greater than the predicted SSM flux. Till now, neutrinos are the only direct evidence of nuclear fusion being the powerhouse in the Sun . The issue of \textit{metallicity}, in which the abundance of elements heavier than Hydrogen has been observed less than predicted, is still an ongoing issue~\cite{Vinyoles:2016djt}. Recent observation of the CNO solar neutrinos by Borexino is expected to help in this direction~\cite{Ranucci:2022xvk}. Further, the upcoming JUNO, THEIA, and DARWIN are a few other projects that can help resolve this issue in the near future.
\subsection{Atmospheric neutrinos}
\label{sec:atmospheric}
A high abundance of cosmic protons and other heavier nuclei in the atmosphere on interacting with atoms from the Earth's atmosphere leads to energetically charged pions, kaons, and other heavier mesons that further decays to produce huge flux of neutrinos~\cite{Hamilton:1943dne,Reines:1965qk} following,
\begin{eqnarray}
\pi^{+},K^{+} &\rightarrow & \nu_{\mu}\mu^{+} \rightarrow \nu_{\mu}e^{+}\nu_{e}\bar{\nu}_{\mu}\,,\nonumber \\
\pi^{-},K^{-} &\rightarrow & \bar{\nu}_{\mu}\mu^{-} \rightarrow  \bar{\nu}_{\mu}e^{-}\bar{\nu}_{e}\bar{\nu}_{\mu}\,. 
\end{eqnarray}
The very first experiments to observe atmospheric neutrinos were implemented in the Kolar Gold Field mines in India~\cite{Krishnaswamy:1971be}. The atmospheric neutrino anomaly was pioneered by Super-K in 1998 that gave initial hints for neutrino oscillations~\cite{Super-Kamiokande:1998kpq}. The flux peaks around $\sim$ GeV for a typical atmospheric neutrino spectrum. Existing uncertainties in the measurements of pion-to-kaon ratio, composition, and spectrum of primary cosmic ray, together contributes to the uncertainties in the prediction of atmospheric neutrino flux. Conventionally, studies exist over a huge range of energy, covering between a few GeV to hundreds of TeV for studying the atmospheric neutrino flux~\cite{Super-Kamiokande:2015qek}. Decay of heavier mesons like charm produces \textit{prompt} neutrinos. This source of neutrinos has been predominantly explored by Super-Kamiokande (Super-K)~\cite{Super-Kamiokande:2015qek}, ANTARES~\cite{ANTARES:2021cwc}, and IceCube~\cite{IceCube:2019cia}. 

The striking feature of atmospheric neutrinos is the massive range of energy and baselines that they cover before hitting the detector. The flux peaks for zenith angle $90^{\circ}$, and is least near the horizon. Also, due to the atmospheric neutrino oscillation, tau neutrinos appear, which have recently been observed by both Super-K~\cite{Super-Kamiokande:2017edb} and IceCube-DeepCore~\cite{IceCube:2019dqi}. The upcoming new generation atmospheric neutrino detectors like ORCA, IceCube-Upgrade, and Hyper-Kamiokand, will certainly be crucial additions.
\subsection{Accelerator neutrinos}
\label{sec:accelerator}
These artificial neutrino sources were explored as early as the 1960's~\cite{Schwartz:1960hg}. They achieve the beam of neutrinos through either of the following three categories- 

\medskip
\textbf{Pion Decay in Flight.}--- Typically, there is a proton beam which is aimed at a target, responsible for producing charged pions and kaons. These are collimated towards a decay region wherein they decay to neutrinos and corresponding charged lepton~\cite{Dore:2018ldz}. The majority of pion decays are dominated by two-body decays, wherein 
\begin{eqnarray}
\pi^{+} &\rightarrow & \nu_{\mu} + \mu^{+}\nonumber\,,\\
\pi^{-} &\rightarrow & \bar{\nu}_{\mu} + \mu^{-}\,.
\end{eqnarray}
Kaon and muon decays also produce contamination of electron neutrinos. The two ongoing accelerators are: T2K and NO$\nu$A that receive intense beamline from NuMI at Fermilab~\cite{MINOS:2014rjg} and J-PARC~\cite{T2K:2020txr}, respectively. There is also TeV II~\cite{NuTeV:2003kth} working at high energy ($\sim 70$ GeV). An on-axis beam gives the maximum integrated flux over a broad spectrum of energy, while an off-axis beam exploits the decays transverse to the direction of motion~\cite{McDonald:2001mc} (refer to Sec.~\ref{sec:wide-band} for further details). Depending on the proton beam, energy of neutrinos can vary from few MeV to hundreds of GeV.

\medskip
\textbf{Muon Decay at Rest ($\mu-$DAR).}--- Here, neutrino beams are generated by stopping pions and kaons. This has been used in the LSND, KARMEN, and at SNS, using pions at low energy that further generate muon neutrinos at 30 MeV. Recently, the MiniBooNE detector exhibited kaon-DAR~\cite{MiniBooNE:2018dus}.

\medskip
\textbf{Novel neutrino beams.}--- Following three-body decay as
\begin{equation}
\mu^{-} \rightarrow \bar{\nu}_{e}+e^{-}+\nu_{\mu}\,,
\end{equation}
an intense and highly precise neutrino beam can be produced using the muon storage rings where a monochromatic muon beam decays. This principle works behind the neutrino-factories~\cite{DeRujula:1998umv}. Although, it is still expected in future as the principle of cooled muon beams need to be feasible. One of its kind is $\nu$STORM~\cite{nuSTORM:2013cqr,Ahdida:2020whw}. 

Further, there have been recent proposals on radioactive ion storage rings being used as electron neutrino `beta-beams'~\cite{Benedikt:2011za}. 
\subsection{Supernova neutrinos}
\label{sec:supernova}
Stars, much more massive than the Sun ($M_{\odot}$), about eight times more, are unable to handle the gravitational force, thus leading to gravitational collapse forming a neutron star or a black hole. Whist the cores of stars with mass $\geq 10M_{\odot}$ can no longer ignite nuclear fusion and thus no thermodynamic equilibrium is reached that could support the outer envelope. As the surrounding matter falls inward due to gravity, the core temperature rises, while dissociating iron into $\alpha$ particles and nucleons. This leads to a massive electron capture, producing electron neutrinos: about $10^{58}$ neutrinos during the burst. This magnificent display of neutrino burst was last observed in the year 1987 A~\cite{Kamiokande-II:1987idp}. Although the number of neutrino interactions detected was quite small, it confirmed our understanding of core collapse. We comprehend that this explosion occurs due to delayed mechanisms, which are primarily neutrino-driven~\cite{Bethe:1985sox}. Owing to its complexity, stellar core collapses are expected  very few times in a century. We expect the next burst to bring a wealth of information and understandings still left to comprehend in this direction~\cite{Tamborra:2020cul}. Just like before, the burst in 1987 A put stringent limits on exotic physics~\cite{Schramm:1990pf}, the same could be expected in future. The Supernova Early Warning System~\cite{Antonioli:2004zb,Scholberg:2008fa} (SNEWS) is network stretched over neutrino detectors world-wide, which will provide warning to astronomers everywhere about the neutrino burst, followed by the first electromagnet shower hours later. 

Although in our Galaxy, we are stuck with only two-three supernovae per century, other galaxies in the Universe have supernova explosions going on every second. The flux from these past explosions is termed as Diffuse Supernova Background (DSNB). Many work exists in literature where they try to measure DSNB flux~\cite{Moller:2018kpn}.     
\subsection{Astrophysical neutrinos}
\label{sec:astrophysical}
High-energy neutrinos ($E_{\nu} \geq 10^{15}$ eV) pass through dense astrophysical environments unadulterated, which is totally opaque to electromagnetic radiations. These are the neutrinos produced from the interaction of cosmic rays with matter or radiation. The neutrino flavors produced in the ratio $\nu_{e}:\nu_{\mu}:\nu_{\tau} = 1:2:0$ (where neutrinos and antineutrinos are indistinguishable), following the flavor mixing described by averaged probability, $P_{\alpha\beta} = \sum_{i}\abs{U_{\alpha i}}^{2}\abs{U_{\beta i}}^{2}$, we expect the flavor ratio at the point of detection to be 1:1:1~\cite{Farzan:2008eg}. Since neutrinos do not get affected by magnetic fields, detection of these very high-energy neutrinos will thus point towards cosmic-ray acceleration. Ultra high-energy (UHE) neutrinos are believed to be produced by the interaction of high-energy protons that leads to charged pions via $p\gamma$ or $p\bar{p}$ interactions.

The first detection of high-energy neutrinos of extraterrestrial origin was done in the year 2013 by the IceCube~\cite{IceCube:2013low} located in the South Pole. ANTARES has been taking data from 2006, located in the Nothern Hemisphere. However, the sparsity of these events can be understood from the fact that IceCube, with a decade-long years of observation, could detect only two tau neutrinos~\cite{IceCube:2020fpi} and one electron antineutrino at the Glashow resonance ($\sim 6.3$ PeV)~\cite{Glashow:1960zz}. Future experiments that are either proposed or are under construction include KM3NeT-ARCA, Baikal GVD, and P-ONE~\cite{P-ONE:2020ljt}. 

\subsection{Geo-antineutrinos}
\label{sec:geoneutrinos}
The natural radioactivity, predominantly of ${}^{40}$K and a minor contribution from ${}^{232}$Th and ${}^{238}$U in the Earth, produces geoneutrinos, which are in principle antineutrinos~\cite{Fiorentini:2007te}. The followed models~\cite{Mantovani:2003yd,YuHuang:2013} believe that geoneutrino fluxes are largest over continents while being smallest over ocean basis. This observation is a result of higher ratio of Th to U in a continental crust than in the mantle, which might be a result of ongoing biological activities~\cite{Sleep:2013vwz}. 

Geoneutrino fluxes from Th and U have been performed by KamLAND~\cite{Bellini:2021sow} and Borexino~\cite{Borexino:2019gps}. The lithophilic nature of K, U, and Th restricts the origin of geoneutrino fluxes only from crust and mantle. SNO+ is expected to also have sensitivity towards geoneutrinos~\cite{Chen:2005zza}. They estimate an interaction rate of about 20 events per year. Further, JUNO is another upcoming experiment. With about 25 times the size of SNO+, it is estimated that JUNO will have better sensitivity. The main challenges for the future geoneutrino experiments include a combination of uncertainty in the measurements of K flux, developing sensitivity to directional analysis of geoneutrinos, thereby leading to better reconstruction of geological models.

\subsection{Relic neutrinos}
\label{relic}
The Standard Cosmology Model depicts that following the hot Big Bang, within 1 second neutrinos got decoupled, these were colloquially termed as the Big Bang neutrinos. This happened when the rate of weak interaction process became slower than the rate of expansion of the Universe. These neutrinos coupling out formed the present day Cosmic Neutrino Background (C$\nu$B) or the relic neutrinos. Since neutrinos are believed to have finite mass-squared differences observed through flavor oscillations, it is believed that atleast two massive states of neutrinos have transitioned from relativistic to non-relativistic energies. However, with the present generation of experiments, we do not have any evidence of C$\nu$B, with an expected average number density of 336 particles per cm$^{3}$. Although, relic neutrinos pervade space, their temperature $T^{0}_{\nu}$ is miutely small ($\sim$1.9 K. Many ideas have been suggested along the line~\cite{Weinberg:1962zza}. However, to have an understanding of how inconspicuous the detection of relic neutrinos is one can find from the literature that KATRIN, the world-leading neutrino mass experiment has the meagre capture rate of relic neutrinos in the order of $10^{-6}$ per year. Recently, PTOLEMY has some ongoing suggestions for detection as suggested in the following references~\cite{Lisanti:2014pqa,Akhmedov:2019oxm}. 
%
\begin{table}[htb!]
\begin{tabular}{|c|c|c|}
\hline
Osc. parameters & Strongest constraints from & Other constrainig experiments\\
\hline
$\Delta m^{2}_{21}$ & KamLAND & other Solar experiments\\
$\theta_{12}$ & Solar & KamLAND\\
$|\Delta m^{2}_{31}|$ & LBL & Atm. + Reactor\\ 
$\theta_{23}$ & Atm. + LBL & -\\
$\theta_{13}$ & Reactor & (Solar+KamLAND) \&  (Atm.+LBL)  \\
$\delta_{\mathrm{CP}}$ & LBL & Atm.\\
mass ordering & (LBL+Reactor) \& Atm. & Cosmology \& 0$\nu\beta\beta$\\
\hline
\end{tabular}
\caption{The dominant and sub-dominant contribution to each oscillation parameter from various experiments designed to use different neutrino sources.}
\label{tab:sources-and-constraints}
\end{table} 

\section{Beyond 3-$\nu$ paradigm}
Apart from the standard 3-$\nu$ paradigm, there are experimental suggestions, though not conclusive, hinting towards exotic physics beyond the standard three-neutrino framework. We discuss these briefly below. 

\subsection{Non Standard neutrino Interactions (NSIs)}
\label{sec:NSI}
The existence of non-zero neutrino mass and mixing points towards physics beyond the minimal Standard Model. As such extended models often come with additional energy scales they usually introduce additional terms to the standard neutrino interaction Lagrangian. While few new neutrino physics scenarios are well motivated, others remain speculative. In any non-standard interaction (NSI), an additional vector-like interaction is introduced for the left-handed SM neutrinos with other fermions say $f$, following~\cite{Wolfenstein:1977ue}
\begin{equation}
-\mathscr{L} = 2\sqrt{2}G_{F}\varepsilon^{f}_{\alpha\beta}[\bar{\nu}_{\alpha}\gamma_{\mu}P_{L}\nu_{\beta}][\bar{f}\gamma^{\mu}f]\,,
\end{equation}
where the new interaction's strength is determined through dimensionless (complex) $\varepsilon$ parameters. $f$ can induce additional coherent forward scattering of neutrinos, generically arising by exchange of some mediator state; either heavy enough ($M_{Z'}\geq 10^{-12}$ eV) for the contact information to be crucial or too light~\cite{Coloma:2020gfv} ($M_{Z'}\leq 10^{-12}$ eV) so that the contact information is no longer valid and the interaction becomes long-range. In both the scenarios, the interaction modifies neutrino oscillation phenomena that can be studied in long-baseline experiments~\cite{Ohlsson:2012kf}. The effect of latter scenario is discussed in Chapter~\ref{sec:ch8}. In principle, NSI could affect starting from the production, propagation to detection processes. It is very difficult to disentangle the effects of NSI from the standard three-flavor scenarios with the current oscillation experiments. One of the major objectives of future neutrino program is to identify and put strong constraints on these interactions~\cite{Coloma:2020gfv}. 
\subsection{Unitarity Violation}
\label{sec:unitarity}
The $3 \times 3$ PMNS matrix is unitary for: a) three Dirac neutrinos that generate mass following the scheme of their corresponding charged counterparts, and b) three left-handed Majorana neutrinos with no additional fermionic states responsible for mass generation. Nevertheless, participation of additional fermionic states will typically lead to unitarity violation. Presently, there are various mass mixing models~\cite{Akhmedov:1995vm}, that allow this scenario of non-unitary neutrino mixing. Leaving certain special flavor structures, it is expected that deviations from unitarity should scale as the powers of the ratio of the electroweak scale with the new physics scale. Under see-saw models, the $3\times 3$ PMNS matrix becomes a non-unitary sub-matrix of a larger mixing matrix.~\cite{Escrihuela:2015wra}.

 The upcoming experiment JUNO expects to achieve very high-precision measurements on the unitarity tests of the first row of the PMNS matrix~\cite{Ellis:2020hus}. Further, loosely constrained $U_{\alpha\tau}$ elements expect stronger bounds from tau related atmospheric neutrino studies~\cite{IceCube-PINGU:2014okk,KM3Net:2016zxf}, long-baseline studies~\cite{DeGouvea:2019kea}, and SHiP experiment~\cite{SHiP:2015vad}. There are extensive studies which suggest that the precision measurements in long-baseline experiments can significantly waver due to unitarity violation~\cite{Escrihuela:2016ube}.
\subsection{Sterile neutrinos}
\label{sec:sterile}
Since the last three decades, we have witnessed the existence of short-baseline flavor transformation anomalies. The LSND~\cite{LSND:1995lje} observed an excess of $\bar{\nu}_{e}$ events following the suspected  oscillation $\bar{\nu}_{\mu}\rightarrow \bar{\nu}_{e}$ at 3.3$\sigma$. They explain this excess by introducing the presence of oscillation corresponding to the mass-squared difference, $\Delta m^{2}_{L} \sim 1 \text{ eV}^{2}$. As this mass splitting is pretty different from the usual solar and atmospheric mass-squared differences, it suggested existence of a fourth neutrino mass eigenstate; $sterile$ neutrino. This anomaly was further favored by MiniBooNE~\cite{MiniBooNE:2012maf}. Similar anomaly in the solar sector remains, where SAGE and GALLEX~\cite{SAGE:1994ctc,GALLEX:1998kcz} observed lower event rates than expected, pointing towards $\nu_{e}$ disappearance corresponding to the oscillations with $\Delta m^{2} \geq 1 \text{ eV}^{2}$; coined as \textit{Gallium anomaly}. Further, the mismatching in reactor neutrino flux as described in Sec.~\ref{sec:reactor} led to \textit{reactor anomaly}, being somewhat consistent with a sterile neutrino proposition~\cite{Tamborra:2020cul}. However, recent measurements from reactor experiments rather point towards issue in the calculation of reactor antineutrino flux~\cite{Schirber:2023aei}. References~\cite{Abazajian:2012ys,Drewes:2017zyw} summarizes the physics studies of sterile neutrinos. While, recent measurements from STEREO reject the hypothesis of light sterile neutrino excluding the best-fit points of Neutrino-4 and NEOS-RENO at 3.3$\sigma$ and 2.8$\sigma$, respectively, other experiments like PROSPECT and DANSS weakly favor STEREO.
\subsection{Neutrino decay}
\label{sec:neutrino-decay}
In principle, massive neutrinos can decay to lighter neutrinos radiatively~\cite{Pal:1981rm} following
\begin{equation}
\nu_{k} \rightarrow \nu_{j} + \gamma\,,
\end{equation}
with extremely long half-life. Nevertheless, presence of new physics, can enhance this decay mode, given the decay product belongs to beyond the Standard Model particles; a sterile neutrino or a Majoron. When global lepton number symmetry breaks, it gives rise to a Goldstone boson, Majoron. Decays like this will have significant effect on both oscillation experiments~\cite{Fogli:1999qt} and physics of astrophysical neutrinos~\cite{Bustamante:2016ciw,DeGouvea:2019kea}. 
\subsection{Neutrino Decoherence}
As discussed in Sec.~\ref{sec:two_flavor} and further, that neutrino oscillations consider quantum-mechanical treatment, wherein we assume that different wave packets of different mass eigenstates share equal momenta and thus overlap. However, once these wave packets do not overlap, coherence is lost. There have been many studies on analyzing the effect of this decoherence, obtaining limits on decoherence related parameters~\cite{Fogli:2003th,Stuttard:2020qfv} (to cite a few).
\subsection{CPT violation and lorentz invariance}
\label{sec:LIV}
One of the exotic studies that can be performed with neutrinos is discovering and constraining the possibility of violation of the fundamental basis: relativity and quantum field theory~\cite{Diaz:2016xpw}. It must be noted that CPT violation leads to Lorentz invariance violation (LIV), but not vice-versa. CPT can be violated by breaking causality, keeping Lorentz invariant true. CPT violation refers to $P_{\nu_{\alpha} \rightarrow \nu_{\beta}} \neq P_{\bar{\nu}_{\alpha} \rightarrow \bar{\nu}_{\beta}}$\,. However, this effect can also be confused with the fake CP generated due to matter effects (discussed elaborately in Chapter~\ref{sec:Ch6}). In literature, we have CPT violation origin described in the context of LIV~\cite{Diaz:2009qk,Agarwalla:2023wft}, Extra dimensions~\cite{Arkani-Hamed:1998wuz}, and NSI~\cite{Diaz:2015dxa}. 
\chapter{Long-baseline Neutrino Oscillation Experiments}
\label{ch:physics-richness}
It was as early as 1960 when Melvin Schwartz for the first time suggested a realistic scheme of a neutrino beam for the study of neutrino interactions~\cite{Schwartz:1960hg}. He showed how the forward production of pions and kaons at high-energy proton accelerator would subsequently decay into a collimated beam of neutrinos. The first commissioned AGS accelerator was at Brookhaven in 1962, thus marking the inception of the first artificial neutrino beam~\cite{Shiltsev:2019rfl}. This accelerator was intended to analyze the existence of two different types of neutrinos. Remarkably, it also measured a number of quasielastic neutrinos. In 1988, Melvin Schwartz, Leon M. Lederman, and Jack Steinberger got awarded with the Nobel Prize for ``\textit{the neutrino beam method and the demonstration of the doublet structure of the leptons through the discovery of the muon neutrino}''. In 1961, Simon van der Meer proposed the concept of `horn' magnet. The method of determining the flux of neutrinos as a function of protons on target (P.O.T.), began in the year 1965 at BNL. 1970s brought revolutionary changes in particle physics with three powerful beams: a 350 - 400 GeV proton accelerator at Fermilab in 1972, a 70 GeV proton accelerator at Serpukhov, and a 300 GeV Super Proton Synchrotron at CERN. 
The electroweak theory demanded the existence of the \textit{neutral current interaction}, following $\nu_{l} + N \rightarrow \nu_{l} + p + n$. The search for this interaction around 1973 became one of the first crucial results with neutrino beams in favor of the electroweak theory. Around the same time there were interests in studying the structure of nucleon through neutrinos in the SLAC's e - p experiment. As the V-A theory formulated that neutrinos scatter quarks and antiquarks differently. This led to \textit{deep inelastic charged current scattering} of $\nu_{\mu}$ or $\bar{\nu}_{\mu}$ on the nucleon, following $\nu_{\mu} (\bar{\nu}_{\mu}) + N \rightarrow \mu^{-} (\mu^+)$. In the following sections, we describe the characteristic features of neutrino beam that essentially set long-baseline experiments apart.
\section{Characteristic features }
\subsection{Wide-band neutrino beam}
\label{sec:wide-band}
Historically, this has been the most common type of neutrino beam in use at proton accelerators. The essential principle driving this is the focusing of pions and kaons over a wide range of momenta, which needs a maximal neutrino flux. The intensity, energy spectrum, and composition of a wide-band beam depend on the flux of muons monitored with counters that are located at varying depths and radial positions in the dump after the decay tunnel. Finally, the characteristics of neutrino beam are concluded by measuring the muon flux and combining them with models devised for the production of kaons and pions in the target. \textit{A wide-band beam is specifically designed for maximizing intensity, thus reducing statistical errors}. Historically, significant contribution of this beam type is seen while studying the interaction $\nu_{\mu}+e \rightarrow \nu_{\mu}+e$. The scattering of neutrino on electrons giving rise to a single electron in the product was very difficult to detect (as the cross-section of the process is $\sim 2000$ times smaller than neutrino-nucleon scattering. However, E734 in 1980 collected hundreds of events in a liquid scintillator calorimeter, thus leading to high precision measurements of electroweak theory~\cite{Ahrens:1990fp}. 
\subsection{Narrow-band neutrino beam}
\label{sec:narrow-band}
An alternate solution to the above was first tested in 1974 at Fermilab, a dichromatic beam~\cite{Edwards:1976ij}. In this beam, a 300 GeV protons hit unaligned target material. The mesons of only selected momentum were focussed further on the decay channel. The decay of pions and kaons produces two separated broad distibutions of neutrinos. In 1977, CERN initiated a similar beam named the narrow-band beam, as the parent mesons were selected in a narrow energy interval~\cite{Nobel:2018}. \textit{At the peak, the intensity of neutrinos in a wide-band beam is about two orders greater than narrow-band beam,} however, becoming comparable at high-energy edge. Nevertheless, because of the bending of mesons in narrow-band beam, it has much less contamination of wrong-sign neutrinos, and of neutrinos from neutral kaons. The selective meson momenta and bending of beam induces high-precision in flux measurements. Thus \textit{narrow-band beam predicts intensity, composition, and energy spectrum of the neutrino beam with much better accuracy}~\cite{NuTeV:1998aks}. Precision measurements of total cross-section and electroweak mixing angle $\sin^{2}\theta_{W}$ are some of the essential features of narrow-band beam.
\subsection{On and Off-axis beam}
\label{sec:off-axis}
\begin{figure}
\centering
\includegraphics[width=0.5\linewidth]{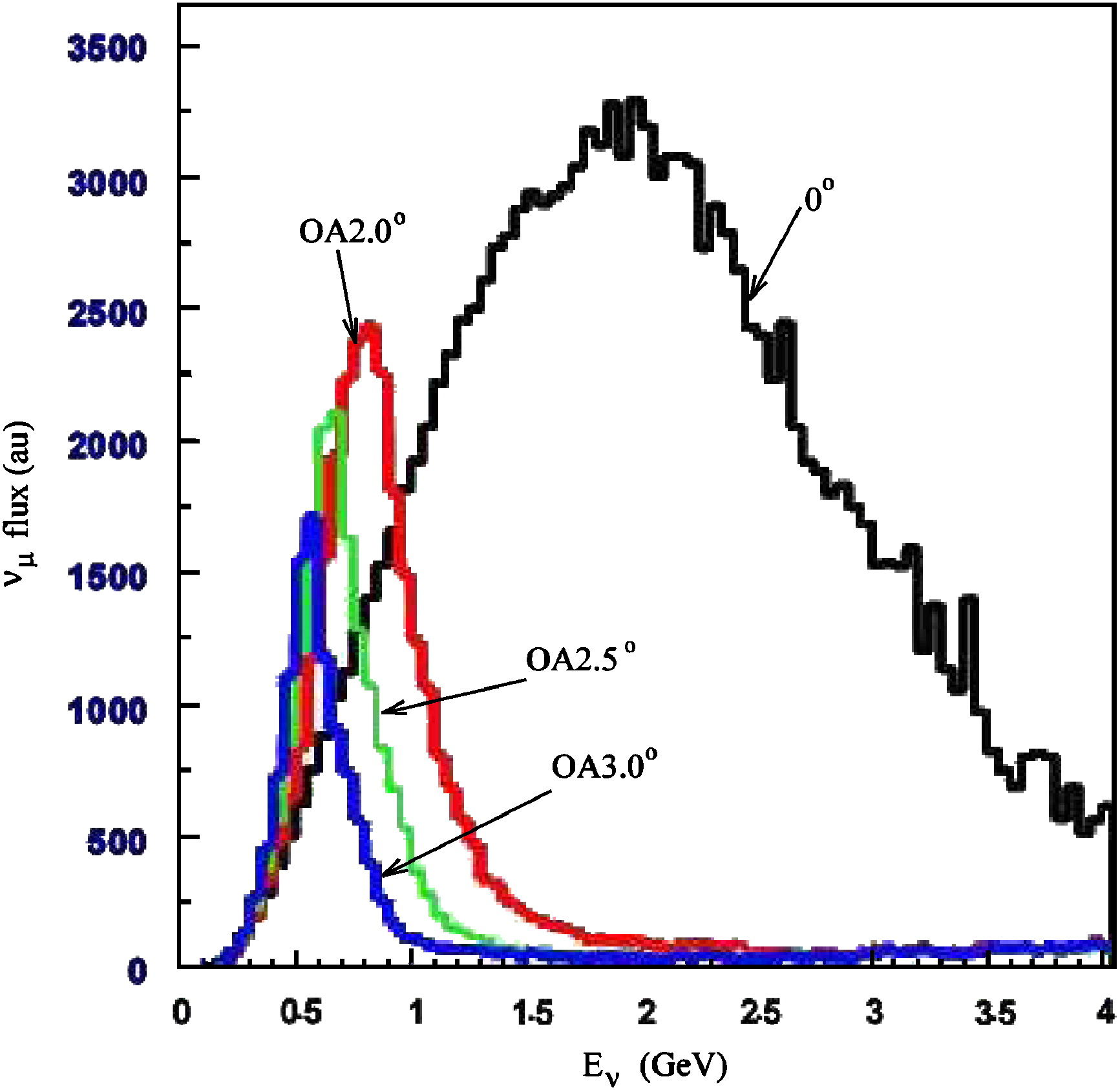}
\caption{Spectrum of neutrino energy for an illustrative on-axis and off-axis angle for the T2K experiment~\cite{Kudenko:2008ia}. The figure is taken from Ref.~\cite{Dore:2018ldz}.}
\label{fig:off-axis}
\end{figure}
It was proposed in 1995 at Brookhaven to set the target material, secondary beam optics, and the decay tunnel, a few degrees off the detector such that the detector becomes off-axis to the incoming neutrino beam. This setup came widely to be recognised as the off-axis beam. Instinctively, one would expect that the integrated neutrino flux will be maximum when the detector is along the axis of the beam or commonly known as on-axis beam. Here, the neutrino energy is directly proportional to the energy carried by parent meson. So in an on-axis beam the broad spectrum perceived by the detector is the reflection of the broad spectrum of decaying parent pions itself. However, if the detector is instead placed a few degrees off the beam axis, then the maximum neutrino energy becomes a function of the chosen off-axis angle. Moreover, despite that the expected neutrino yield from any pion will be smaller at an off-axis angle $\theta$ than on-axis, contribution of almost all pions even in a broad energy setup is maximally allocated around a narrow energy interval. Therefore, the \textit{maximum peak in the neutrino energy flux is larger in an off-axis beam than on-axis at the same energy.} This is illustrated in Fig.~\ref{fig:off-axis} using the simulated data for T2K. By choosing an off-axis angle in a long-baseline experiment, we can tune the experiment for a given $L/E$. The ongoing long-baseline experiments: T2K~\cite{T2K:2011qtm} and NO$\nu$A~\cite{NOvA:2007rmc} are both located off-axis, bearing an angle of 2.4$^{\circ}$ and 0.8$^\circ$, respectively. This helps them to tune the maximum flux of neutrinos to their corresponding first oscillation maxima $E$. Therefore, an \textit{off-axis beam provides an intense flux of neutrinos in a narrow spectrum of energy tunable to a desired $L/E$. }However, an \textit{on-axis beam provides an access to both first and second oscillation maxima, helping in better reconstruction of neutrino energy in every energy bin.} Further, as the high-energy tail in the off-axis is very less, the neutral current and $\tau$ backgrounds are efficiently reduced.

\subsection{CP trajectory diagram}
\label{sec:bi-probability}
\begin{figure}
\centering
\includegraphics[width=\linewidth]{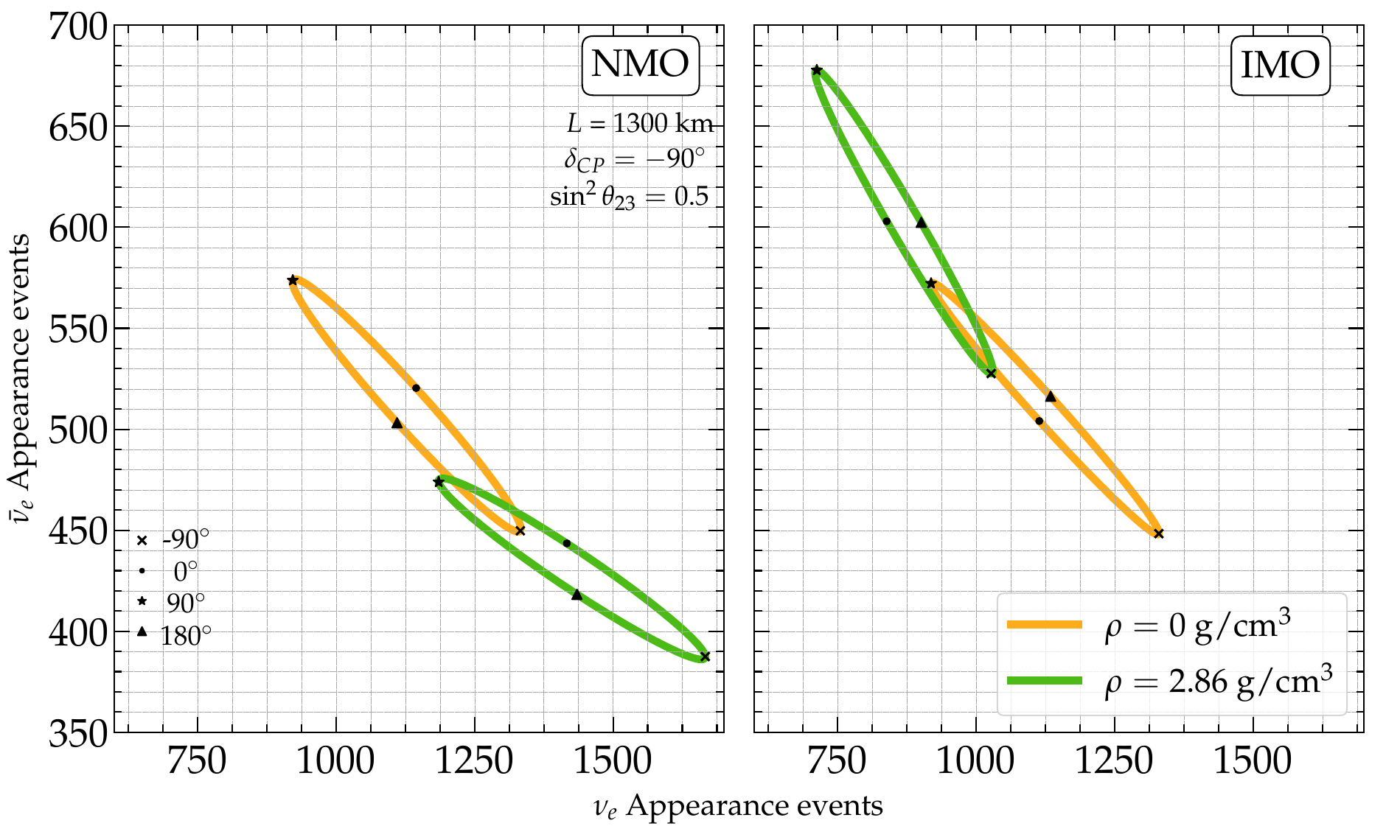}
\caption{Bi-events computed assuming NMO and IMO in the left and right panel using DUNE~\cite{DUNE:2020jqi}, $\sin^{2}\theta_{23} = 0.5$, and other benchmark parameters from Ref.~\cite{Capozzi:2021fjo}.}
\label{fig:bi-events-intro}
\end{figure}

CP trajectory diagram~\cite{Minakata:2001qm,Machida:2003ft} or colloquially renowned as the bi-probability is a crucial pictorial representation of three major effects: intrinsic or genuine CP-violation phase, $\delta_{\mathrm{CP}}$ induced due to the presence of $\sin \delta$ term, a CP-conserving effect due to the presence of $\cos\delta$, and the matter effect. CP-trajectory forms an ellipse and this shape has been previously justified~\cite{Arafune:1996bt,Minakata:1997td}. This can be further extended to the plane containing total event rates instead. This is termed as bi-events. Few crucial characteristic features of bi-events that will be exploited later in Chapter~\ref{sec:ch4} are highlighted here.
\begin{itemize}
\item The trajectory is closed because it is essential that the probability must be a periodic function of $\delta$, which is a phase variable.
\item For a vacuum scenario, the major (or minor) axis always makes an angle of $\pi/4$.
\item In presence of matter, while CP trajectories in vacuum move downward-right for NMO case ($\Delta m^{2}_{31}>0$), it moves upward-left under IMO ($\Delta m^{2}_{31}<0$). 
\item Furthermore, comparing both panels in Fig.~\ref{fig:bi-events-intro}, it can be concluded that the degree of nonoverlapping of trajectories in neutrino and antineutrino modes provides almost the measurement of matter effect. 
\item Also, changing the value of matter potential, neither affects the major axis (that gives measurement of coefficent of $\sin\delta$) nor the minor axis (that gives measurement of coefficient of $\sin\delta$).
\item The chirality of trajectory, i.e. whether the trajectory unfolds in a clockwise or counter-clockwise direction on varying $\delta_{\mathrm{CP}} \in [0, 2\pi]$ is decided by the sign of $\Delta m^{2}_{31}$.
\item In a broader perspective, the trajectory in both the cases follows same nature, but is perceived differently. To understand this, let us begin at $\delta_{\mathrm{CP}} = 0^{\circ}$ in both NMO and IMO scenarios, and start traveling towards $\delta_{\mathrm{CP}}=90^{\circ}$. In both, the probability in neutrino mode decreases, while increasing in antineutrino mode, but as $\delta_{\mathrm{CP}} = 0^{\circ}$  is located near (far) from the origin in NMO (IMO), they unfold in opposite directions.
\end{itemize}
\section{Previous, ongoing, and upcoming Experiments:}
\label{sec:experiments-overview}

\subsection{MINOS/MINOS+}
\label{subsec:MINOS}
MINOS is one of the first generation long-baseline experiments that initiated data taking from February 2005 in phases till April 2012~\cite{MINOS:2014rjg}. This then got upgraded to MINOS+ that started from 2013 and ran until June 2016~\cite{MINOS:2020llm}. It consisted of a near detector located 1.04 km away from the NuMI at Fermilab and a 4.2 kt far detector located 735 km apart in Minnesota. Both detectors are steel-scintillators, wherein the magnetic field leads to the bending of charged muon depending upon its nature of charge. This differentiated between neutrinos and antineutrinos.

MINOS used a 3 GeV proton energy beam that hit a water-cooled graphite, producing mesons, which were then focussed by magnetic horns depending upon their charge. MINOS was able to accumulate 1.07 $\times\, 10^{21}$ protons-on-target (P.O.T.) in $\nu_{\mu}$ and 0.33 $\times\, 10^{21}$ P.O.T. in $\bar{\nu}_{\mu}$ modes, respectively. While the NuMI beam also got upgraded on going from MINOS to MINOS+, receiving a 6 GeV proton energy beam, leading to an exposure of 0.58 $\times\, 10^{21}$ P.O.T. in $\nu_{\mu}$ mode in the first two years of run. The remaining analysis is still ongoing~\cite{Evans:2017brt}. The flux peaks around 1.5 GeV.
\subsection{T2K}
\label{subsec:T2k}
T2K is one of the first superbeam long-baseline experiments consisting of near detectors: on-axis INGRID and 2.5$^{\circ}$ off-axis ND280, both around 280 m away from the neutrino production at Japan Proton Accelerator Research Complex (J-PARC). It has a 50 kt water Cherenkov far detector, Super-Kamiokande, 295 km apart and about 1 km underground in the Mozumi Mine in Hida's Kamioka~\cite{Yamada:2006fz}. The near detector comprises of two fine-grained detectors: plastic scintillators FGD1 and FGD2. The near detector is mainly responsible for an efficient reconstruction of muons from neutrino interactions. Currently T2K is upgrading ND280 to improve this reconstruction of muons isotropically. They also have plans to further lower the energy threshold from 450 MeV/c~ to 300 MeV/c~\cite{Chakrani:2023goj}, improving the neutron resolution to (15 - 30)\%~\cite{Munteanu:2019llq}.

T2K uses a 2.5$^{\circ}$ off-axis J-PARC neutrino beam, produced in Toaki, Ibiraki, alternating between neutrino and antineutrino modes since 2014. The 515 kW beam of 30 GeV protons is fired onto a monolithic graphite target, producing charged mesons that further decay in flight to neutrinos. The resulting neutrino flux is narrow-band, with a peak energy of 0.6 GeV. The polarity of horns can either focus on positive or negative charged particles, thus consequently the far detector receives either neutrino or antineutrino-rich beam. Starting from 2010, T2K has accumulated a total of 3.6 $\times\, 10^{21}$ protons-on-target (P.O.T.) until 2020, wherein there are (1.97 and 1.63)  $\times\, 10^{21}$ P.O.T. in antineutrino  and neutrino modes, respectively at the far detector~\cite{T2K:2023smv}. We consider a total projected exposure of $7.8 \times 10^{21}$ P.O.T. with a beam power of 750 kW, equally divided in neutrino and antineutrino modes following Ref.~\cite{T2K:2014xyt}. We assume uncorrelated 5\% and 10\% systematic errors on signal and background events, respectively, for both the appearance and disappearance channels.
\subsection{NO$\nu$A}
\label{subsec:NOvA}
NO$\nu$A consists of a near detector 1 km away, and an off-axis (0.8$^{\circ}$) far detector, 100 m underground, 810 km apart at Minnesota from the source of NuMI beam in Fermilab. It has a 14 kt liquid scintillator far detector~\cite{NOvA:2004blv}. The liquid scintillator used in NO$\nu$A is cheap and easier to assemble than the solid scintillator used in MINOS. 

NO$\nu$A uses a 700 kW beam bearing 120 GeV proton energy, striking a carbon target, producing mesons. The magnetic horns help to choose between positively or negatively charged pions, thereby leading to neutrino or antineutrino-rich beams peaking around 2 GeV. The neutrino flux is a narrow-band ranging between a few GeV. Starting from 2014 until May 2021, NO$\nu$A has collected 1.7 $\times\, 10^{21}$ P.O.T. in the neutrino and 1.27 $\times\, 10^{21}$ P.O.T. in the antineutrino mode~\cite{Shanahan:2021jlp}. NO$\nu$A expects to double their exposure in both neutrino and antineutrino modes by 2026. we consider a full projected exposure of NO$\nu$A, $3.6 \times 10^{21}$ P.O.T. equally divided in neutrino and antineutrino modes following Ref.~\cite{NOvA:2007rmc}. We assume an uncorrelated 5\% and 10\% systematic errors on signal and background events, respectively, for both the appearance and disappearance channels. 
\subsection{DUNE}
\label{subsec:DUNE}
DUNE will consist of a near detector, about 600~m downstream of the neutrino production point on the Fermilab site, and a far detector, 1285~km away and about 1.5~km underground, in the Sanford Underground Research Facility in South Dakota~\cite{DUNE:2018tke}.  The near detector will monitor and characterize the neutrino beam (though it has physics capabilities itself, too~\cite{DUNE:2021tad, DUNE:2022aul}).  We focus on the far detector since it offers prime sensitivity to neutrino oscillations.  It is a state-of-the-art liquid-argon time projection chamber with a net volume of 40~kt; to generate our results, we consider single-phase detection only~\cite{DUNE:2020jqi}.  Neutrino detection is via charged-current neutrino-argon interaction.  Detector deployment will be phased~\cite{DUNE:2021mtg}, but in our simulations we consider only the final, total detector volume.  

DUNE will use the Long Baseline Neutrino Facility (LBNF) neutrino beam produced at Fermilab.  There, the Main Injector of the LBNF fires a 1.2-MW beam of protons of 120~GeV onto a graphite target, producing charged mesons that decay in flight to neutrinos. The resulting neutrino flux is wide-band, ranges from a few hundreds of MeV to a few tens of GeV, and is expected to peak at 2.5~GeV, with most neutrinos in the 1--5~GeV range.  By changing the polarity of the focusing horns~\cite{DUNE:2018hrq, DUNE:2018mlo}, the experiment can run in neutrino or antineutrino mode.  Following the DUNE Technical Design Report~\cite{DUNE:2020jqi}, we adopt a run time of 5 years in neutrino mode and 5 years in antineutrino mode.  This amounts to $1.1 \times 10^{21}$ protons-on-target per year and a net exposure of 480 kt~MW~year.  
To produce our results, we use the DUNE simulation configuration from~\cite{DUNE:2021cuw}.
\subsection{T2HK}
\label{subsec:T2HK}
T2HK will consist of near detectors, about 280~m downstream from the neutrino production point at the J-PARC, and a far detector, 295~km away and about 1.7~km underground, in the Tochibora mines of Japan, 8~km from Super-Kamiokande~\cite{Hyper-Kamiokande:2022smq}.  Like in DUNE, the near detectors will monitor and characterize the neutrino beam, and we focus on the far detector.  It will be a tank filled with purified water, with a net volume of 187~kt, whose internal wall is lined with photomultipliers (PMTs).  Neutrino detection is via quasielastic charged-current scattering (QECC),  $\nu_l + n \rightarrow p + l^-$ and $\bar{\nu}_l + p \rightarrow n + l^+$ ($l = e, \mu, \tau$), and via charged-current deep inelastic scattering (DIS), $\nu_l + N \rightarrow l^- + X$ and $\bar{\nu}_l + N \rightarrow l^+ + X$ ($l = e, \mu$), where $X$ represents final-state hadrons ($\nu_\tau$ DIS is suppressed due to the large tauon mass).  Electrons emit gamma rays by bremsstrahlung and $e^+e^-$ annihilation, which register as a fuzzy ring on the PMTs.  Muons emit Cherenkov light, which registers as a sharply defined ring.

Like its predecessor, T2K (Tokai-to-Kamioka)~\cite{T2K:2023smv}, T2HK will use the 2.5$^{\circ}$-off-axis J-PARC neutrino beam~\cite{McDonald:2001mc}.  To produce it, J-PARC fires a 1.3-MW beam of protons of 30~GeV onto a graphite target.  The resulting neutrino flux is narrow-band, ranges from a few~MeV to a few~GeV, and is expected to peak at 600~MeV, with most neutrinos in the 100--3000~MeV range.  As in DUNE, by changing the polarity of the focusing horns, T2HK can run in neutrino or antineutrino mode~\cite{T2K:2012bge}.  Following~\cite{Hyper-Kamiokande:2016srs}, we adopt a run time of 2.5 years in neutrino mode and 7.5 years in antineutrino mode, in accordance with the default 1:3 ratio planned for them.  This amounts to $2.7 \times 10^{22}$ protons-on-target per year and a net exposure of 2431 kt~MW~year.  To produce our results, we match the binned event spectra that we generate under standard oscillations with those of~\cite{Hyper-Kamiokande:2016srs}.

T2HKK setup is a combination of two far detectors: one in Japan called as the Japanese Detector (JD) and the other in Korea called as the Korean Detector (KD). This proposed second detector KD receives the same flux as JD with a baseline of $L$ = 1100 km, and is expected to work at the second oscillation maximum. KD is also proposed to host a 187 kt water Cherenkov detector, which will accumulate huge statistics, with expected signal normalization uncertainties of 5\% in the appearance channel and 3.5\% in the disappearance. Further, the sources of backgrounds and their contribution to normalization uncertainties have been taken from Ref.~\cite{Hyper-Kamiokande:2016srs}. T2HKK is also expected to have a run time distribution similar to T2HK distributed in the ratio of 1:3.

\begin{table}[htb!]
\resizebox{\columnwidth}{!}{%
    \centering
    \begin{tabular}{|c|c|c|}
    \hline \hline
       Characteristics  & DUNE & T2HK / second det. in Korea (KD)    \\
       \hline \hline
       Baseline (km) & 1285 & 295 (1100)\\
       \hline
       $\rho_{\mathrm{avg}}$ (g/cm$^{3}$) & 2.848 & 2.7 (2.8)\\
       \hline
       Beam & LBNF~\cite{DUNE:2020lwj} & J-PARC~\cite{Hyper-Kamiokande:2018ofw}\\
       \hline
       Beam Type & wide-band, on-axis & narrow-band, 2.5$^{\circ}$ off-axis\\
       \hline
       Beam Power & 1.2 MW & 1.3 MW\\
       \hline
       Proton Energy & 120 GeV & 30 GeV\\
       \hline
       P.O.T./year & 1.1 $\times$ 10$^{21}$ & 2.7 $\times$ 10$^{22}$\\
       \hline
       Flux peaks at (GeV) & 2.5  & 0.6  \\
       \hline
       1$^{\mathrm{st}}$ ( 2$^{\mathrm{nd}}$) oscillation maxima & \multirow{2}{*}{2.6 (0.87) }& \multirow{2}{*}{0.6 (0.2) / 1.8 (0.6)}  \\
       for appearance channel (GeV) & &\\
       \hline
       Detector mass (kt) & 40, LArTPC & 187 each, water Cherenkov\\
       \hline
       Runtime ($\nu + \bar{\nu}$) yrs & 5 + 5 & 2.5 + 7.5\\
       \hline
       Exposure (kt$\cdot$MW$\cdot$yrs)  & 480 & 2431\\
       \hline
       Signal Norm. Error (App.) & 2\% & 5\%\\
       \hline
       Signal Norm. Error (Disapp.) & 5\% & 3.5\%\\
       \hline
       Binned-events & \multirow{2}{*}{~\cite{DUNE:2021cuw} }& \multirow{2}{*}{~\cite{Hyper-Kamiokande:2016srs}} \\
       matched with & &\\
       \hline \hline
    \end{tabular}}
    \caption{Essential experimental features of various long-baseline experiments considered in our analysis. \textbf{Charateristic features represent contrasting features in DUNE and T2HK.}}
    \label{table:contrasting-features-dune-t2hk}
\end{table}

\section{Statistical analysis}
\label{sec:stats}
In all our analysis, we make use of the publicly available General Long Baseline Experiment Simulator (GLoBES) \cite{Huber:2004ka,Huber:2007ji}. Herein, to compare true and test event spectra in each of the analyses, we use the Poissonian $\chi^2$ function that follows \Refes~\cite{Baker:1983tu, Cowan:2010js, Blennow:2013oma}. So for each experiment $e$, and for each detection channel $c = \{{\rm app}~\nu,~{\rm app}~\bar{\nu},~{\rm disapp}~\nu,~{\rm disapp}~\bar{\nu}\}$, this is
\begin{eqnarray}
 \chi_{e,c}^{2}
 ( \boldsymbol{\theta}, o)
 =
 &&
 \underset{\left\{\xi_{s}, \{\xi_{b, c, k}\}\right\}}{\mathrm{min}} 
 \left\{
 2\sum^{N_e}_{i=1}
 \left[
 N_{e, c, i}^{{\rm test}}
 (\boldsymbol{\theta}, o, \xi_s, \{\xi_{b,c,k}\})
 \right. \right.
 \nonumber \\
 &&
 \left. \left.
 -
 N_{e, c, i}^{{\rm true}}
 \left( 
 1
 +
 \ln
 \frac{N_{e, c, i}^{{\rm test}}
 (\boldsymbol{\theta}, o, \xi_s, 
 \{\xi_{b, c, k}\})}
 {N_{e, c, i}^{{\rm true}}}
 \right)
 \right]
 +
 \xi^{2}_{s}
 +
 \sum_k \xi^{2}_{b, c, k} 
 \right\} \;,
 \label{eq:chi2-in-all}
\end{eqnarray}
where  $N_{e, c, i}^{{\rm true}}$ and $N_{e, c, i}^{{\rm test}}$ are the true and test event rates in the $i$-th bin of $E_{\rm rec}$, $N_e$ is the number of bins of $E_{\rm rec}$, $\boldsymbol{\theta}$ are the test values of the most relevant mixing parameters (more on this later which depends upon the type of analysis pursued), $o = \{ \text{NMO, IMO} \}$ is the test mass ordering, and $\xi_{s}$ and $\xi_{b, c, k}$ are, respectively, pull terms for the systematic uncertainties on the signal and the $k$-th background contribution to detection channels $c$.  The pull terms have the same values in neutrino and antineutrino mode, and are uncorrelated with one another.  The true number of events is
\begin{equation}
 N_{e, c, i}^{{\rm true}}
 = 
 N_{e, c, i}^{s, {\rm true}}
 + 
 N_{e, c, i}^{b, {\rm true}} \;,
\end{equation}
where $N_{e, c, i}^{s, {\rm true}}$ and $N_{e, c, i}^{b, {\rm true}}$ are, respectively, the number of signal ($s$) and background ($b$) events, summed over all channels, computed using the true values of the mixing parameters and mass ordering.

The test number of events is
\begin{equation}
 \label{equ:num_test}
 N_{e, c, i}^{{\rm test}}
 (\boldsymbol{\theta}, o, \xi_s, \left\{\xi_{b,c,k}\right\})
 =
 N^s_{e,c,i}( \boldsymbol{\theta}, o)
 (1+\pi_{e,c}^s\xi_s)
 +
 \sum_k
 N^{b}_{e,c,k,i}(\boldsymbol{\theta}, o)
 \left(
 1+\pi_{e,c,k}^b\xi_{b,c,k}
 \right) \;,
\end{equation}
where $\pi_{e,c}^s$ and $\pi_{e,c,k}^b$ are normalization errors on the signal and background rates (more on this later). Subsequent computations are analysis-specific, and as such, they will be referenced and elaborated further as needed in the forthcoming chapters. 
\chapter{{Earth Matter Effect and Related Degeneracies in DUNE}}
\label{sec:ch4}
 Neutrino physics is driven by the six major fundamental parameters: two independent mass-squared differences 
$ (\Delta m^2_{31} = m^2_3 - m^2_1)\, , (\Delta m^2_{21} = m^2_2 - m^2_1)$, three mixing angles - solar, reactor, and atmospheric $(\theta_{12},\,\theta_{13},\,\theta_{23})$, and a Dirac CP phase $(\delta_{\mathrm{CP}})$~\cite{ParticleDataGroup:2012pjm}. Among the two unknown mass-squared differences, one exploits the solar regime $(\Delta m^2_{21})$, while the other explains atmospheric domain $(\Delta m^2_{31})$. It is well established that, with the inclusion of variation in inelastic properties and the dense matter in the interior of Earth, there is generation of anisotropy and an ineleastic dispersion~\cite{Arafune:1997hd,Zaglauer:1988gz}. This affects the propagation of neutrino in presence of matter, in the form of a new potential, the so-called matter potential or Mikheyev-Smirnov-Wolfenstein (MSW) potential~\cite{Wolfenstein:1977ue,Barger:1980tf} (refer to Sec.~\ref{sec:nu-matter-2flavor}). The observation of MSW resonant matter in Sun~\cite{Mikheev:1986wj} played a crucial role in deciding the sign of $\Delta m^2_{21}$; being positive, $m_{2}$ mass eigenstate is heavier than $m_{1}$ mass eigenstate. However, the existing global fit data~\cite{Esteban:2020cvm,Capozzi:2021fjo,deSalas:2020pgw} is not sufficient enough to explore the sign of the atmospheric mass splitting, leading to two possible prospects. First, when the atmospheric mass-splitting ($\Delta m^{2}_{31}$) is positive ($\Delta m^2_{31} > 0$) or NMO and the second scenario is when $\Delta m^2_{31}$ is negative ($\Delta m^2_{31} < 0$) or IMO (as explained in Chapter~\ref{sec:intro}). Discovery of finite value of $\theta_{13}$ - the smallest mixing angle, connects the two major sectors of three-flavor neutrino paradigm (solar and atmospheric). Just like dense matter in Sun, played a paramount figure in determining the sign of $\Delta m^2_{21}$, determining the sign of $\Delta m^2_{31}$ needs adequate measurement of the interference between the CP phase and matter effect, thus necessitates a sufficiently long baseline~\cite{Minakata:2001qm}. Undoubtedly, the Earth matter effect plays an indispensable part in the precision estimation of the defined 3-$\nu$ oscillation framework. Matter potential brings significant enhancement (decrement) in the probability case of neutrino (antineutrino), when it changes from vacuum scenario to a medium with finite matter density, for a certain energy and baseline. Therefore, for a given experiment, we need to understand the role of matter effect, before we plan to extract information on oscillation parameters. For an instance, probing Dirac CP asymmetry phase becomes quite messy because of the production of extrinsic CP phase, in addition to the genuine or intrinsic CP asymmetry in oscillation physics. This extrinsic or fake CP asymmetry is induced by matter~\cite{Gonzalez-Garcia:2001snt}. This happens as in presence of matter, $P_{\nu_{\mu}\rightarrow \nu_{e}} - P_{\bar{\nu}_{\mu}\rightarrow \bar{\nu}_{e}} \neq 0$, even if we reduce the Dirac CP phase ($\delta_{\mathrm{CP}}) = 0^{\circ}$. The prevailing framework~\cite{Esteban:2020cvm,Capozzi:2021fjo,deSalas:2020pgw}, suggests that the reactor mixing angle ($\theta_{13}$) and the solar regime ($\theta_{12}$ and $\Delta m^{2}_{21}$) are already into the precision era. Likewise, as the precision on the mass-squared differences and mixing angles, continues to ameliorate, similary the uncertainty in the measurement of CP violation is most certainly going to improve in the upcoming years, validating the existence of Earth matter effect comes next. 
Since we have entered the precision era in neutrino physics, we can expect that the data from numerous ongoing and next generation high-precision experiments will be able to certify the value of $\rho_{\mathrm{avg}}$, which otherwise we consider as an input.
 
We consider the simplest model describing Earth matter potential$-$ the Preliminary Reference Earth Model (PREM)~\cite{Dziewonski:1981xy}. It is frequency dependent, being transversely isotropic for upper mantle. In this simplified version of PREM~\cite{Super-Kamiokande:2017yvm}, the inner core stretches for about 1220 km, with $\rho_{\mathrm{avg}}$ = 13.0 g/cm$^3$. While, the outer core expands for nearly 2260 km ($\rho_{\mathrm{avg}}$ = 11.3 g/cm$^3$). The mantle of Earth has been assigned an extent of 2221 km bearing 5 g/cm$^3$ as the Earth matter density. The sweeping expanse of upper most surface of Earth - the Earth crust, is taken to be 5801 km ($\rho_{\mathrm{avg}}$ = 3.3 g/cm$^3$). Among all the layers, Earth crust and upper mantle houses all the long-baseline experiments. The matter effect in long-baseline neutrino experiments have been explored recently in Ref.~\cite{Bharti:2020gnu}\footnote{There have been also studies on the effects of different matter density profiles in neutrino oscillations for T2HK or Japanese detector and T2HKK or the Korean detector in Ref~\cite{King:2020ydu} and that in the case of solar neutrino has been studied in~\cite{Bakhti:2020tcj}.}. The commencement of long-baseline experiment - DUNE, has been awaited by physicists all over the world. The neutrino is expected to start its journey from Fermi National Accelerator Laboratory (FNAL) in Batavia, covering a total distance of about 1300 km, it will reach the Sanford Underground Research Laboratory in Lead. Owing to its long baseline, the process of extraction of information on obtaining CP violation, mass ordering, and octant of $\theta_{23}$ (refer to Chapter~\ref{sec:intro}), is dependent on our assumption of matter profile. Therefore, it becomes imperative to question will the data collected by DUNE detector really feel the realm of Earth matter effect through which neutrino travels? If yes, then how precisely can it determine the matter potential? By matter potential, we shall be discussing here the line-averaged constant Earth matter density ($\rho_{\mathrm{avg}}$) which is a sufficient profile type to characterize the Earth matter along the DUNE baseline as verified in Ref.~\cite{Ioannisian:2017dkx,Bakhti:2020tcj}. In this chapter, we mainly address two major questions: first, how efficient is DUNE in ruling out the vacuum hypothesis solutions, and second, how precisely can DUNE measure the line-averaged constant Earth matter density ($\rho_{\mathrm{avg}}$). For quantifying the later, we introduce a free generic parameter $\mathrm{\beta}_{\mathrm{SF}}$, where SF stands for scaling factor. Using this scaling factor we scale the standard layers of Earth density defined by the PREM profile~\cite{Dziewonski:1981xy} as follows
\begin{equation}
\mathrm{\rho}_{\mathrm{avg}}\rightarrow \mathrm{\beta}_{SF} \cdot \mathrm{\rho}_{\mathrm{avg}}\, ,
\label{scalingfactor}
\end{equation}

throughout the baseline of DUNE. Therefore, if our hypothesis of data on Earth density layers is exactly as predicted by the PREM profile then generically $\beta_{\mathrm{SF}}$ must equal to 1. For instance, if this scaling factor tends to 2, then this simply points that we scale up each layer of Earth density as prescribed in the PREM profile by a factor of 2. Conservatively, $\beta_{\mathrm{SF}} = 0$ signifies vacuum hypothesis. By analyzing the variation in scaling factor $(\mathrm{\beta}_{SF})$, we come across other degeneracies between $\rho_{\mathrm{avg}}$ and uncertain oscillation parameters: $\delta_{\mathrm{CP}}$ and $\sin^{2}\theta_{23}$~\cite{Singh:2021kov}. Using scaling factor have been previously studied in the context of atmospheric~\cite{Super-Kamiokande:2017yvm,Nizam:2019dex}, solar~\cite{Maltoni:2015kca}, and long-baseline~\cite{Kelly:2018kmb} experiments. Also, a number of analytical studies exist in the literature, citing a few~\cite{Akhmedov:2004ny,Minakata:2015gra,Parke:2019jyu}. In the present chapter, we follow a very simple analytical approach for quantifying the difference between the standard three-flavor neutrino paradigm in finite matter density, as defined by PREM profile ($\mathrm{\beta}_{\mathrm{SF}} \sim$ 1) and the vacuum oscillation hypothesis ($\mathrm{\beta_{SF}} \sim$ 0). We expect that in future as our understanding on three-flavor neutrino oscillation framework improves, the data itself will be able to reveal about $\rho_{\mathrm{avg}}$ with high significance. The main aim of this chapter is to unravel several interesting issues along this direction in the context of the upcoming DUNE facility.  

This chapter is catalogued as follows. In Sec.~\ref{probability}, we give a simple analytical approach for  differentiating vacuum hypothesis from the finite matter density scenario at the probablity level. In Sec.~\ref{setup}, we point out the characteristic features of DUNE which we consider in our simulations. We also discuss here, our statistical approach briefly. In Sec.~\ref{bi-event-plot}, we scrutinize the impact of possible degeneracies between $\rho_{\mathrm{avg}}-\mathrm{\delta_{\mathrm{CP}}} - \mathrm{\theta_{23}}$ using bi-events. Sec.~\ref{results} discusses our results: a) with how much significance DUNE can establish Earth matter effect, b) the precision measurement in the density parameter ($\rho_{\mathrm{avg}}$), c) role of exposure in precisely determining $\rho_{\mathrm{avg}}$, and d) type of new degeneracies relevant in our discussion. We culminate our key findings and outline concluding remarks in Sec.~\ref{work1:summary-conclusions}.
%
\section{Matter versus vacuum neutrino oscillation probabilities}
\label{probability}
Nothing could be more generic than initiating our discussion with the effect on oscillation probability. The long-baseline experiment we are discussing, mostly stresses on the transition probability in both neutrino $(P_{\nu_{\mu}\rightarrow \nu_{e}})$ and antineutrino $(P_{\bar{\nu}_{\mu}\rightarrow \bar{\nu}_{e}})$ modes. These channels not only provide us with ample opportunities in our discussion of Earth matter effect, but also addresses almost all of the present major issues in neutrino physics as discussed in Sec~\ref{sec:three-flavor-vac} and~\ref{sec:three-flavor-mat}. Therefore, in this section, we present a simplistic analytical calculation for differentiating matter and vacuum neutrino oscillation in the context of long-baseline experiments.

For all the simulations in this chapter, we make use of subsequent values of defined six oscillation parameters until mentioned otherwise:~$\sin^2\theta_{23}=0.5$, $\sin^22\theta_{13}=0.085$, $\sin^2\theta_{12}= 0.307$, $\delta_{\mathrm{CP}}= [-180^{\circ}:180^{\circ}]$, $\Delta m^2_{31}=2.5\, (-2.4)\times 10^{-3}\,\mathrm{eV}^2$ for NMO (IMO), and $\Delta m^2_{21}=7.4 \times 10^{-5}\,\mathrm{eV}^2$. These parameters are very well in agreement with the current global-fit ranges of oscillation parameters~\cite{Esteban:2020cvm,Capozzi:2021fjo,deSalas:2020pgw}.
 It should be noted that while going from NMO to IMO, we make use of the relation~\cite{deGouvea:2005hk}
\begin{eqnarray}
\Delta m^{2}_{31}\,(\mathrm{IMO})= 2 \,\Delta m^{2}_{21}\cos^{2}\theta_{12} - \Delta m^{2}_{31}\, (\mathrm{NMO}) \,.
\label{eq:imo-relation-with-nmo}
\end{eqnarray}
As the exact expression of neutrino oscillation probability in a full three-flavor framework is quite complicated, in this work, we consider uptill second order terms of $\sin\theta_{13}\,\sim 0.149$ 
and $\alpha\,(\equiv \Delta m^2_{21}/\Delta m^2_{31}) \, \sim 0.03$, which themselves are very small~\cite{Mikheyev:1985zog,Wolfenstein:1977ue}
\begin{eqnarray}
P^{\mathrm{mat}}_{\nu_{\mu} \rightarrow \nu_{e}} &\simeq& {\underbrace{\sin^{2}{2\theta_{13}}\sin^{2}\theta_{23}\frac{\sin^{2}[(1-\hat{A})\Delta]}{(1-\hat{A})^{2}}}_{\mathrm{leading}}}  +  {\underbrace{ \alpha^{2} \sin^{2}2\theta_{12}\cos^{2}\theta_{23}\frac{\sin^{2}\hat{A}\Delta}{\hat{A^{2}}}}_{\mathrm{least}}} \nonumber \\ &+& {\underbrace{\alpha\cos\theta_{13}\sin2\theta_{13}\sin2\theta_{23}\sin2\theta_{12}\frac{\sin[(1-\hat{A})\Delta]}{1-\hat{A}}\frac{\sin(\hat{A}\Delta)}{\hat{A}}\cos(\Delta+\delta_{\mathrm{CP}})}_{\mathrm{sub-leading}}} \,. 
\label{eq:2}
\end{eqnarray}
In the above equation, $\Delta \equiv \frac{\Delta m^{2}_{31}L}{4E}$, $\hat{A} \equiv \frac{A}{\Delta m^2_{31}}$, here $A = 2 \sqrt{2} G_F N_e E$ is the Wolfenstein matter potential. Note that $\hat{A}$ is a dimensionless quantity and it can be numerically calculated for a given baseline ($L$) in an experiment with neutrino energy ($E$) and the line-averaged constant Earth matter density ($\rho_{\mathrm{avg}}$) as
\begin{equation}
\label{eq:3}
 \hat{A} = \left(\frac{7.6 \times 10^{-5}\, (\mathrm{eV^2})}{\Delta m^2_{31}}\right)\times \left(\frac{\rho_{\mathrm{avg}}}{\mathrm{g/cm^3}}\right) \times \left(\frac{\mathrm{E}}{\mathrm{GeV}}\right).
\end{equation} 
The Eq.~\ref{eq:2} reduces to 
vacuum neutrino oscillation probability, in the limit $\hat{A}\to 0$ as given by~\cite{Akhmedov:2004ny}
\begin{eqnarray}
P^{\mathrm{vac}}_{\nu_{\mu} \rightarrow \nu_{e}} &\simeq& {\underbrace{\sin^{2}{2\theta_{13}}\sin^{2}\theta_{23}\sin^{2}\Delta}_{\mathrm{leading}}} + {\underbrace{\sin^{2}2\theta_{12}\cos^{2}\theta_{23}\sin^{2}(\alpha\Delta)}_{\mathrm {least}}} \nonumber \\ &+& {\underbrace{(\alpha\Delta)\sin2\theta_{13}\cos\theta_{13}\sin2\theta_{12}\sin2\theta_{23}\sin\Delta\cos(\Delta+\delta_{\mathrm{CP}})}_{\mathrm{sub-leading}}} \,.
\label{eq:1}
\end{eqnarray}
In both the Eq.~\ref{eq:2} and \ref{eq:1}, the terminology `leading' refers to the most contributing term as it is independent of $\alpha$. While the `sub-leading' term here is $\alpha$ suppressed, however it would have been the main contributor, if we were determining the $\delta_{\mathrm{CP}}$. Proceeding further to the `least' term, it is $\alpha^{2}$ suppressed, thus its contribution is minimal. Therefore, we focus only on the leading term which gets substantially modified on going from vacuum to matter. The approximated difference between the leading terms in matter and vacuum oscillation probability is given by
\begin{eqnarray}
\Delta P &=& (P^{\mathrm{mat}}_{\nu_{\mu} \rightarrow \nu_{e}})_{\mathrm{leading}} - \, (P^{\mathrm{vac}}_{\nu_{\mu} \rightarrow \nu_{e}})_{\mathrm{leading}} \nonumber \\
&=& \sin^{2}\theta_{23}\sin^{2}{2\theta_{13}}\left[ \frac{\sin^{2}[(1-\hat{A})\Delta]}{(1-\hat{A})^{2}} -\sin^{2}\Delta\right] \,.
\label{dp1}
\end{eqnarray}
Now, using this definition, we estimate $\Delta P$ for various ongoing and upcoming long-baseline experiments as shown in Table \ref{work1:table:1}. DUNE being a wide-band beam, it has access to both first and second oscillation maxima ~\cite{DUNE:2021cuw}, while NuMI Off-Axis $\nu_{e}$ Appearance (NO$\nu$A)~\cite{Ayres:2002ws} receives typically an off-axis (0.8$^{\circ}$), narrow-band beam having most of the flux accomodated around the first oscillation maximum. Bearing a baseline of 295 km and receiving a narrow-band and an off-axis (2.5$^{\circ}$) beam, Tokai to Hyper-Kamiokande (T2HK) ~\cite{ Hyper-KamiokandeProto-:2015xww, Hyper-Kamiokande:2018ofw} plans to have its first detector in Japan (JD). It will be observing only the first oscillation maximum, while it also has future plans of setting a second narrow-band beam detector in Korea (KD) bearing baseline of 1100 km ~\cite{Hyper-Kamiokande:2016srs}, and being sensitive to only second oscillation maxima. The European Spallation Source neutrino Super Beam (ESS$\nu$SB)~\cite{ESSnuSB:2013dql} has been designed to work at the second oscillation maximum which is very good to probe CP-violation. From Table~\ref{work1:table:1}, we also observe that the value of $\hat{A}<1$ for DUNE and NO$\nu$A at first oscillation maximum, while $\hat{A}<<1$ at DUNE's second oscillation maxima (see sixth column). In all other experiments $\hat{A}<<1$, thus we can safely simplify Eq.~\ref{dp1} further by making use of the binomial expansion of $(1-\hat{A})^{-2}$. We only consider uptill second order in $\hat{A}$,

\begin{eqnarray}
\label{eq:4}
\Delta P = \dfrac{1}{2}\sin^{2}\mathrm{\theta_{23}}\sin^{2}{2\mathrm{\theta_{13}}}\left[(3\hat{A}^{2}+2\hat{A}-1)+\cos\left[(2n+1)\pi\hat{A}\right](3\hat{A}^{2}+2\hat{A}+1)\right].
\end{eqnarray}
From the above expression, we observe that the matter potential comes in the oscillation maxima with a cosine term. The above equation makes it easier to calculate the difference between transition probability of neutrino wandering in matter with that in vacuum for any oscillation maxima. It should be stressed that $\Delta P$ is somewhat dependent on the parameters: $\theta_{23}$ and $\theta_{13}$. In the standard 3$\nu$ oscillation-paradigm with matter, the mixing angle $\theta_{13}^{m}$ changes\footnote{`m' refers to the presence of finite matter density, so this superscript differentiates reactor mixing angle in finite matter density with vacuum like scenario. \label{c}} significantly with energy as discussed in~\cite{Chatterjee:2015gta}, which is indeed very important. As $\Delta P$ is proportional to $\sin^2\theta_{23}$, the relative difference between the matter and vacuum solutions are quite higher for values 
of $\theta_{23}$ in HO than that in LO. The last two columns in Table~\ref{work1:table:1} depicts approximate analytical $\Delta P$.

\begin{table}[htb!]
\centering
\resizebox{\columnwidth}{!}{%
\begin{tabular}{|c|c|c|c|c|c|c|c|c|}
\hline\hline 
\multirow{2}{*}{Facility} &  \multirow{2}{*}{$L$ (km)} & \multicolumn{2}{c|}{$E$\,(GeV) } & \multicolumn{2}{|c|}{$\hat{A}=2\sqrt{2}G_FN_e E/\Delta m_{31}^2$} & 
\multicolumn{2}{c|}{$\Delta P = P^{\mathrm{mat}}_{\nu_{\mu} \rightarrow \nu_{e}} -  P^{\mathrm{vac}}_{\nu_{\mu} \rightarrow \nu_{e}}$} \\
\cline{3-8}
& & \small 1$^{\mathrm{st}}$osc. max. & \small 2$^{\mathrm{nd}}$osc. max. & \small 1$^{\mathrm{st}}$osc. max. & \small 2$^{\mathrm{nd}}$osc. max. & \small 1$^{\mathrm{st}}$osc. max. & \small 2$^{\mathrm{nd}}$osc. max.\\
\hline
\hline
DUNE & 1300    &  2.63 & 0.88     &   0.2284 & 0.0761    & 0.0176 & 0.0011 \\ 
NO$\nu$A    & 810     & 1.64 & -      &  0.1389 & -     & 0.0116 & - \\ 
T2K/JD     & 295     &  0.60 & - &  0.0506 & -     &  0.0043 & - \\ 
KD           & 1100    &  - & 0.74  & - &0.0629     & - & 0.0017  \\
ESS$\nu$SB         &  540    &  - &  0.36 & - & 0.0309  & -  &  0.0018\\ 
\hline\hline 
\end{tabular}
}
\caption{ Charateristic features like $L$ and $E$ of various ongoing and upcoming long-baseline experiments are used to compute $\hat{A}$ at the two oscillation maxima. We also depict $\Delta P$ following Eq.~\ref{eq:4} assuming NMO. For benchmark values of oscillation parameters refer text.}
\label{work1:table:1}
\end{table}
\begin{figure}[htb!]
\includegraphics[width=0.48\textwidth,height=0.5\textwidth]{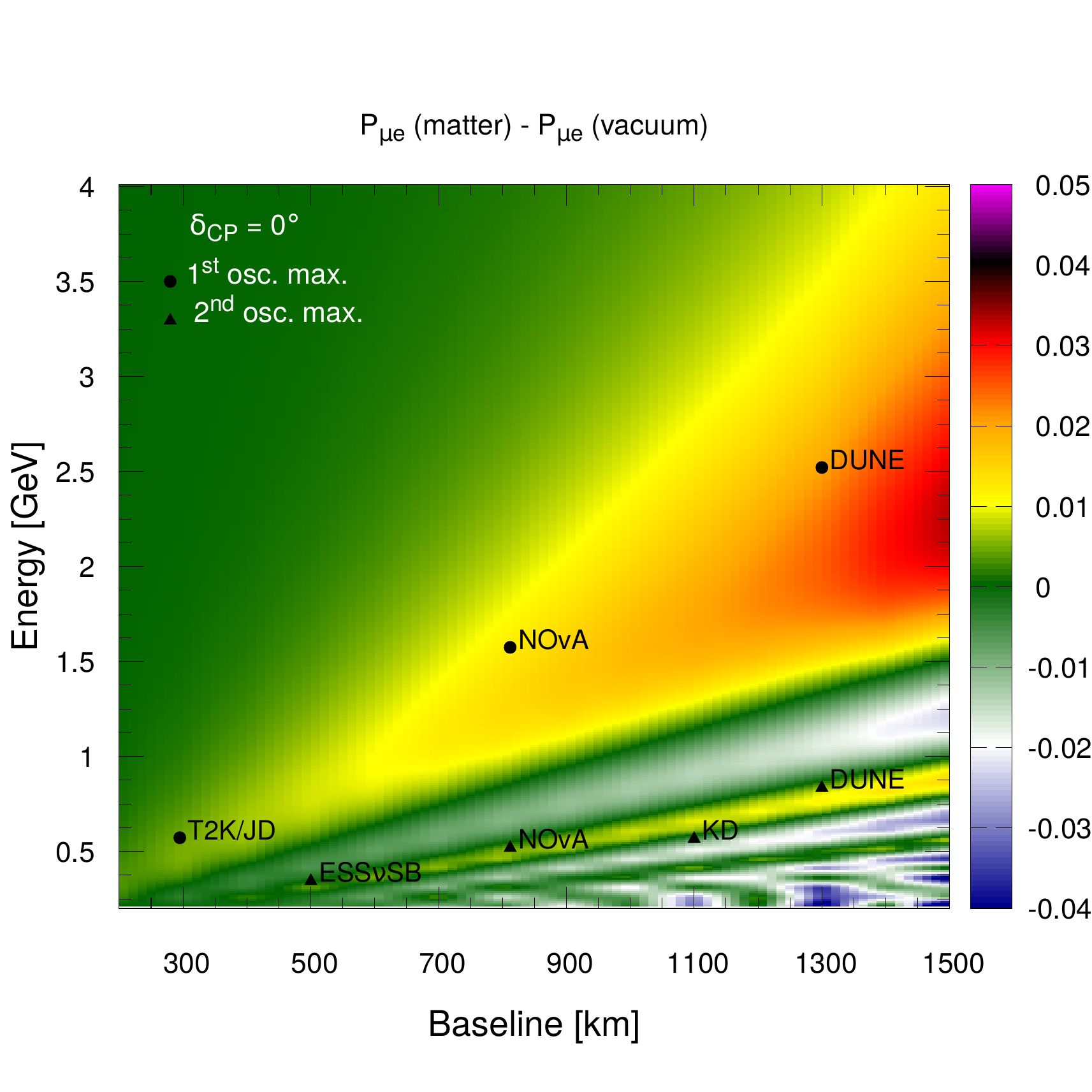} 
\hspace{0.1cm}
\includegraphics[width=0.48\textwidth,height=0.5\textwidth]{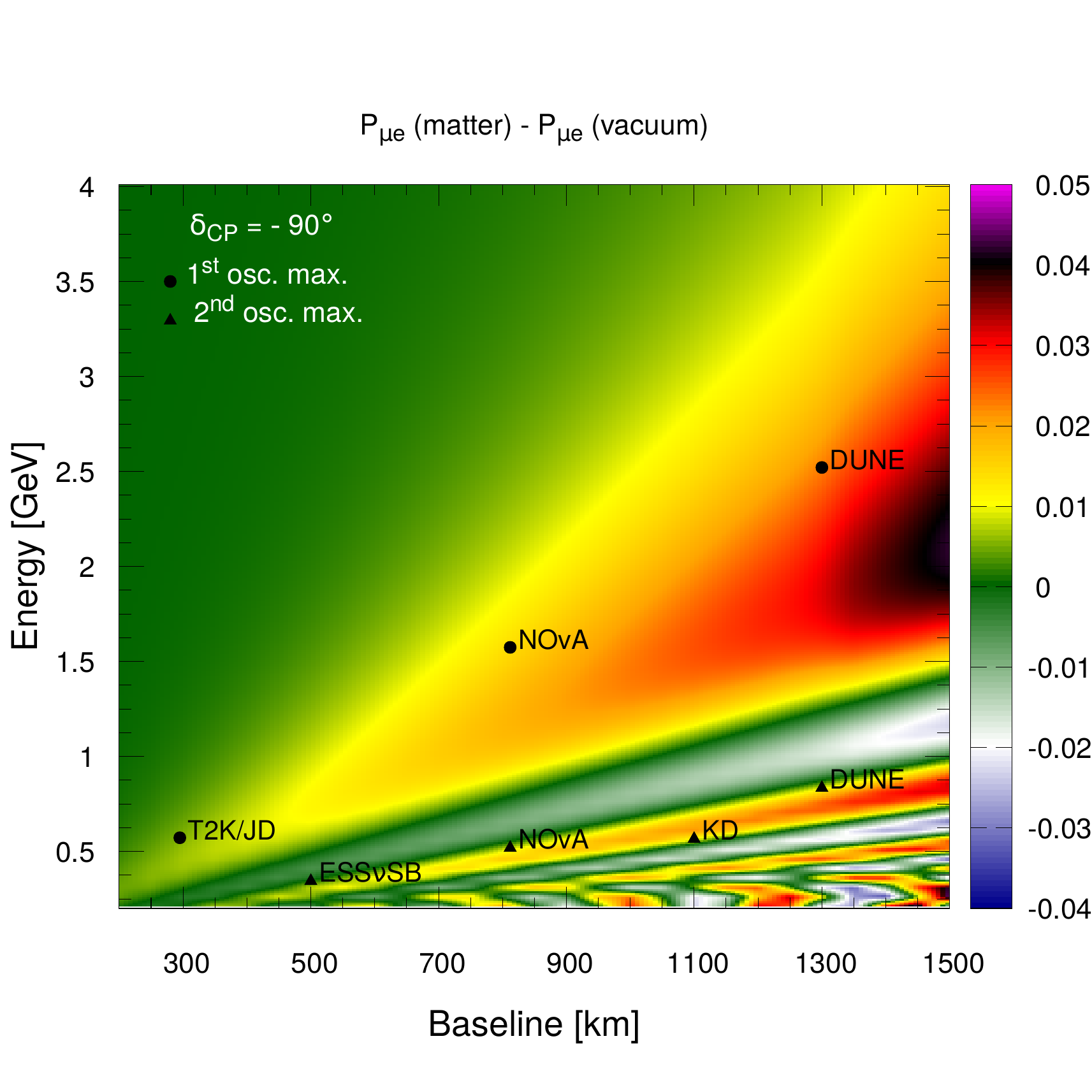}
\caption{Difference between $P_{\nu_{\mu}\rightarrow \nu_{e}}$ in matter with $P_{\nu_{\mu}\rightarrow \nu_{e}}$ in vacuum is shown as a function of baseline, $L$ and neutrino energies, $E$, assuming NMO. We consider the benchmark values as mentioned before in the text, and $\rho_{\mathrm{avg}}$ = 2.86 g/cm$^3$ for DUNE and 2.79 g/cm$^3$ for all other experiments. The left (right) panel depicts $\delta_{\mathrm{CP}} = 0^{\circ} (-90^{\circ})$.}
\label{work1:fig:1}
\end{figure}  

In Fig.~\ref{work1:fig:1}, we show the difference between transition probability in matter with vacuum estimating it numerically using the full three-flavor neutrino paradigm using the possible baselines and energies of present (T2K and NO$\nu$A) and future (T2HK, DUNE and ESS$\nu$SB) long-baseline experiments. In order to obtain this, we numerically simulate $\Delta P$ for different baselines and neutrino energies with our benchmark values in the full 3-$\nu$ scenario. While incorporating matter effect, we consider $\rho_{\mathrm{avg}}=2.86$ g/cm$^3$ for DUNE and $\rho_{\mathrm{avg}}=2.79$ g/cm$^3$ for all other experiments, corresponding to each baseline. In the figure, the degree of color represents the distribution of $\Delta P$ in $L-E$ plane. On comparing the numerically simulated $\Delta P$ in Fig.~\ref{work1:fig:1} with our analytically approximated $\Delta P$ in the last two columns of Table~\ref{work1:table:1}, we see that the first oscillation maximum in DUNE is orange in color which is close to 0.017 as depicted analytically, while NO$\nu$A is more towards yellow side, matching almost to the analyically estimated value of 0.011. So we observe that both the numerically simulated $\Delta P$ by considering the full 3-$\nu$ paradigm and the analytically approximated $\Delta P$, matches quite well. We can also observe from the Fig.~\ref{work1:fig:1}, that the maximum difference (violet-colored region) which is around $\Delta P \sim$ 0.05 for $\delta_{\mathrm{CP}} = 0^{\circ}$, baseline 1500 km and beam energy 0.11 GeV changes to a range of baselines $[1150:1500]$ km bearing energies between $[1.8:2.3]$ GeV for $\delta_{\mathrm{CP}} = -90^{\circ}$. However, the minimum difference (blue-colored region) value is - 0.04, which mostly spans the lowest energy range in both $\delta_{\mathrm{CP}} = 0^{\circ}$ and $-90^{\circ}$. The positive value of $\Delta P$, which is signifying matter oscillation dominating over vacuum, is much greater than negative value of $\Delta P$, where vacuum oscillation dominates. The vacuum oscillation dominance is observed only at lower energy. This confirms the general notion that the Earth matter effect enhances the neutrino oscillation probability with increase in energy. This nature is distinctly visible more for CP-violating choices ( $\delta_{\mathrm{CP}} = -90^{\circ}$), than CP-conserving values of $\delta_{\mathrm{CP}} ( \sim 0^{\circ})$. However, it is the complementarity of matter potential and energy not far from oscillation maxima which brings most enhanced neutrino oscillation as can be easily understood from the Fig.~\ref{work1:fig:1}. Elaborating further, in Fig.~\ref{work1:fig:1}, T2K and T2HK exhibits $\Delta P$ of about 0.004 at their first oscillation maximum, while NO$\nu$A generates $\Delta P$ of around 0.011 trailing behind DUNE, which shows $\Delta P$ as 0.017 at their first oscillation maximum. As expected, the second oscillation maximum exhibits way lower value of $\Delta P$ for the prior experiments, though they are in the consistent range in Fig.~\ref{work1:fig:1}. Moreover, DUNE exhibits maximum $\Delta P$, owing to its longest baseline, therefore we shall be only focussing on DUNE only in this chapter.
\begin{figure}[htb!]
\centering
\includegraphics[width=0.49\textwidth,height=0.45\textwidth]{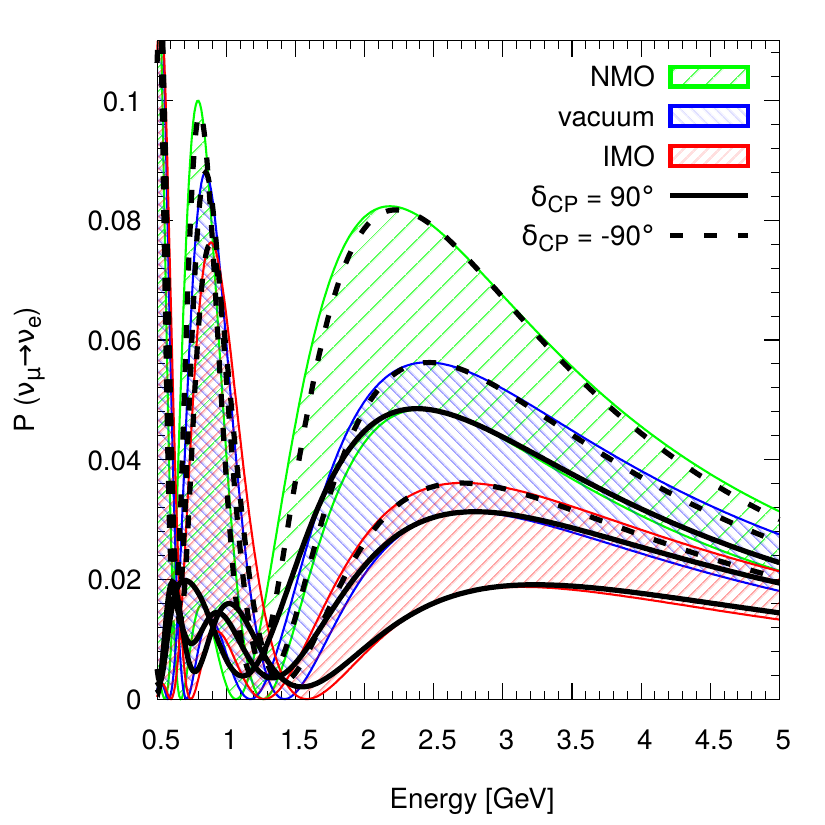}
\includegraphics[width=0.49\textwidth,height=0.45\textwidth]{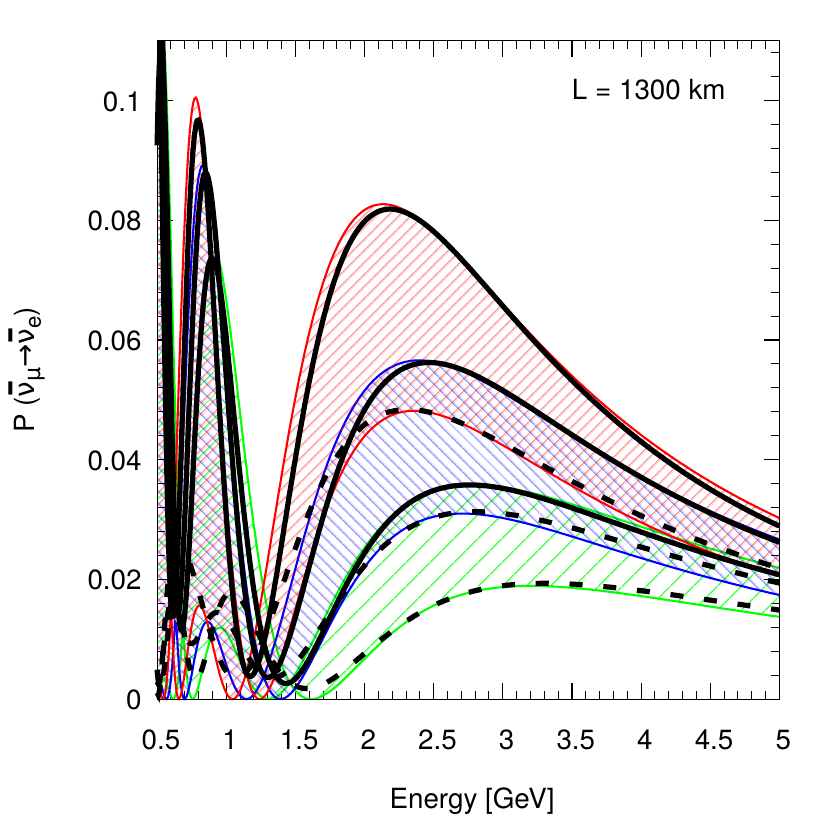}
\vskip0.1cm
\includegraphics[width=0.49\textwidth,height=0.45\textwidth]{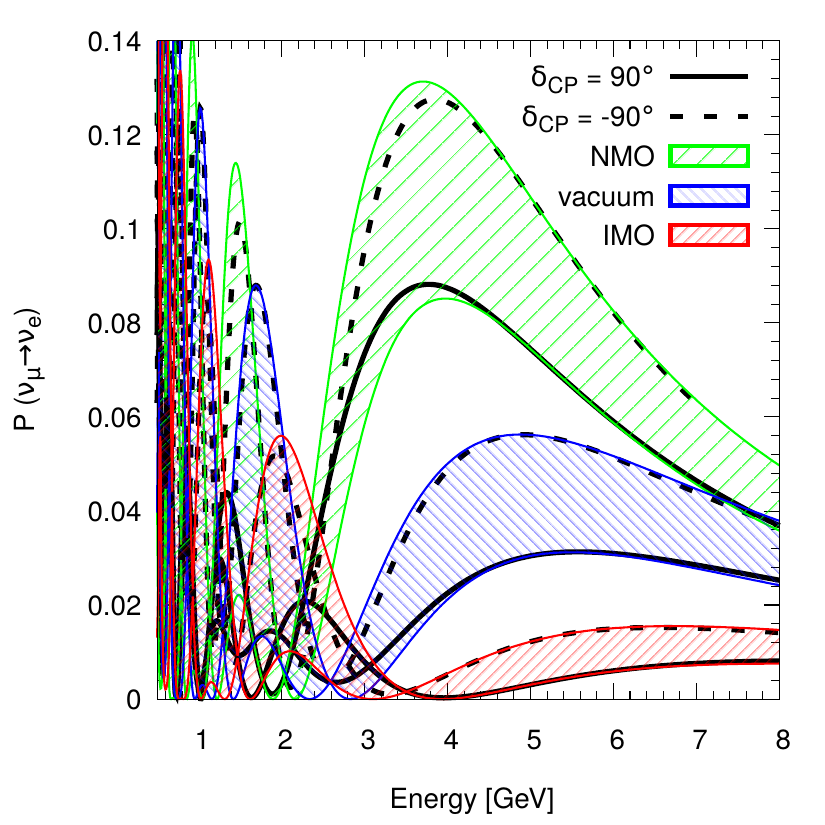}
\includegraphics[width=0.49\textwidth,height=0.45\textwidth]{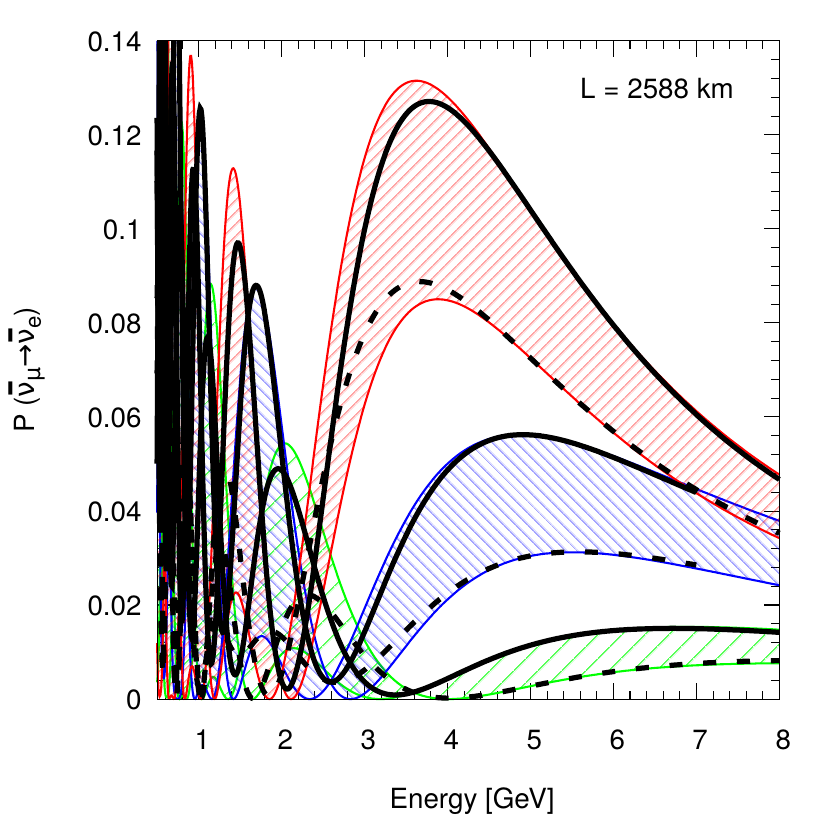}
\caption{ Transition neutrino (antineutrino) probability as a function of energy is depicted in the left (right) panels. The top (bottom) panels correspond to oscillation probability for baseline 1300 km (2588 km). Bands correspond to the uncertainty in $\delta_{\mathrm{CP}}$ over the range of $\delta_{\mathrm{CP}} \in [-180^{\circ}:180^{\circ}]$ in three scenarios: matter with $\mathrm{\rho}_{\mathrm{\mathrm{avg}}}\, = 2.86$ g/cm$^3$ assuming NMO and IMO, and in vacuum ($\mathrm{\rho}_{\mathrm{avg}}$ = 0 g/cm$^3$). Note that top and bottom panels have different x and y ranges.}
\label{work1:fig:2}
 \end{figure}

Next, to elaborate further on the contribution from neutrino (antineutrino) mode, we exhibit transition probability in left (right) panel of  Fig.~\ref{work1:fig:2}. The bands in the figure is obtained by varying $\delta_{\mathrm{CP}}$ in the range $[-180^{\circ}:180^{\circ}]$. The top panel in this figure represents significant overlapping between matter and vacuum bands, thus depicting how difficult it would be to differentiate matter and vacuum oscillation, in either of the mass orderings. This overlapping region will act as an unfavorable domain in resolving the matter solutions from vacuum. Overlapping region refers to certain discrete values of $\delta_{\mathrm{CP}}$  in matter exhibiting similar probabilities with $\delta_{\mathrm{CP}}$  in vacuum. This gives us a preview that eventhough, DUNE is expected to be very efficient in segregating the mixing of NMO and IMO solutions because of the large matter effect, it does not expel out the solutions of vacuum oscillation from matter significantly. 

Further, we observe that the probability with matter for $\delta_{\mathrm{CP}}= -90^{\circ}~(90^{\circ})$ with NMO (IMO) can be easily segregated from the vacuum solutions from the top left panel of the same figure. This will thus act as the favorable region for establishing matter effect or for vacuum exclusion in DUNE. While the reverse is true in the antineutrino mode (upper right panel in Fig.~\ref{work1:fig:2}). However, the solid (dashed) line for $\delta_{\mathrm{CP}}= 90^{\circ}~(-90^{\circ})$ in neutrino channel assuming NMO (IMO) is totally overlapping with vacuum band. This will act as the unfavorable zone for establishing matter effect in DUNE. Similar, but in reverse, observations can be concluded from the $\bar{\nu}$ mode (see top right panel of the Fig.~\ref{work1:fig:2}). Nevertheless, it can be pointed out that at the probability level, overlapping of matter and vacuum solutions is more in NMO (IMO) than in IMO (NMO) for $\nu$ ($\bar{\nu}$) mode.

There is a possibility to disentangle matter from vacuum band if we consider wide range of energies and almost double the baseline, as depicted in the bottom panels of Fig.~\ref{work1:fig:2}. One such proposed experiment with baseline of 2588 km is Protvino to ORCA (P2O)~\cite{Akindinov:2019flp}, which also culminates the important features of a bimagic baseline experiment~\cite{Dighe:2010js,Joglekar:2011zz}. As we observe, with this baseline and consequential matter effect, the overlapping between bands with finite matter density in both NMO and IMO with vacuum are lifted in both neutrino and antineutrino modes. We can expect this experiment to perform better in establishing matter effect than we could observe in DUNE.  

In order to certify our probability-level discussion further, we proceed with the event-level analysis in the following section.
\section{Simulation details}
\label{setup}

In this work, we consider all the spectral information for a 150 kt$\cdot$MW$\cdot$years of exposure, in each $\nu$ and $\bar{\nu}$ modes, thus summing to a net 300 kt$\cdot$MW$\cdot$years of exposure. This exposure considers a proton beam power of about 1.07 MW bearing energy of about 80 GeV, following the previous proposed setup in \cite{DUNE:2021cuw} which is expected to produce 1.47$\times 10^{21}$ protons on target per year. We use the upgraded flux files from the previous version as mentioned in the collaboration paper~\cite{DUNE:2016ymp} for an total run time of 7 years. We divide this run time equally in both the $\nu$ (3.5 years) and $\bar{\nu}$ modes (3.5 years). Reconstructed energy range is taken from 0.1 GeV - 20 GeV unless mentioned otherwise. For this chapter, we incorporate identical systematics for both the modes. We have matched the binned event spectra for the previous version of files from DUNE collaboration, follwing Ref.~\cite{DUNE:2016rla}.\\
For all our simulations, General Long Baseline Experiment Simulator (GLoBES) \cite{Huber:2004ka,Huber:2007ji} is the tool. Throughout the work, full three-flavor $\nu$ oscillation phenomena has been considered.

 \subsection{Statistical analysis}
 \label{stat}
Following the discussion in Sec.~\ref{sec:stats}, we define $\boldsymbol{\theta} = \{\theta_{23},~\delta_{\mathrm{CP}},~ \Delta m_{31}^2\}$  which gives the subset of oscillation parameters over which we minimize in the fit. The corresponding range of minimization taken are $\delta_{\mathrm{CP}} \in [-180^{o}:180^{o}]$, $\sin^2\theta_{23} \in [0.4:0.6]$, and $\Delta m_{31}^2 \in [0.36:0.64]$, respectively. 
 
 The intrinsic electron neutrino or antineutrino contamination in the beam, backgrounds due to presence of tau neutrino and antineutrino appearance, Neutral Current (NC) events, and finally the mis-identified charged muon events as charged electron events are being considered as backgrounds in the appearance channel simulation. Whereas, the major backgrounds in the disappearance channel is coming from NC events and  tau neutrino and antineutrino appearance. It shoud be noted that the normalization uncertainties in both the modes ($\nu$ and $\bar{\nu}$) for appearance and disappearance channels are independent of each other. This estimates to around 5$\%$ in disappearance, while around 2$\%$ in appearance in each mode, respectiely. Further, the correlated background normalization uncertainties from  distinctive sources varies from 5$\%$ to 20$\%$.

\section{Identifying new degeneracies using bi-events}
\label{bi-event-plot}

In this section, we exhibit degeneracies existing between the oscillation parameters: $\delta_{\mathrm{CP}}$ and $\theta_{23}$ with $\rho_{\mathrm{avg}}$. We study the bi-events in two extreme scenarios: first in vacuum and second with finite matter density. We also extend our discussion to explain the role of these new degeneracies while establishing matter effect in DUNE. For an elaborate discussion on importance of bi-events in general, refer to Sec.~\ref{sec:bi-probability}.

\begin{figure}[htb!]
\centering
\includegraphics[width=0.8\textwidth,height=0.75\textwidth]{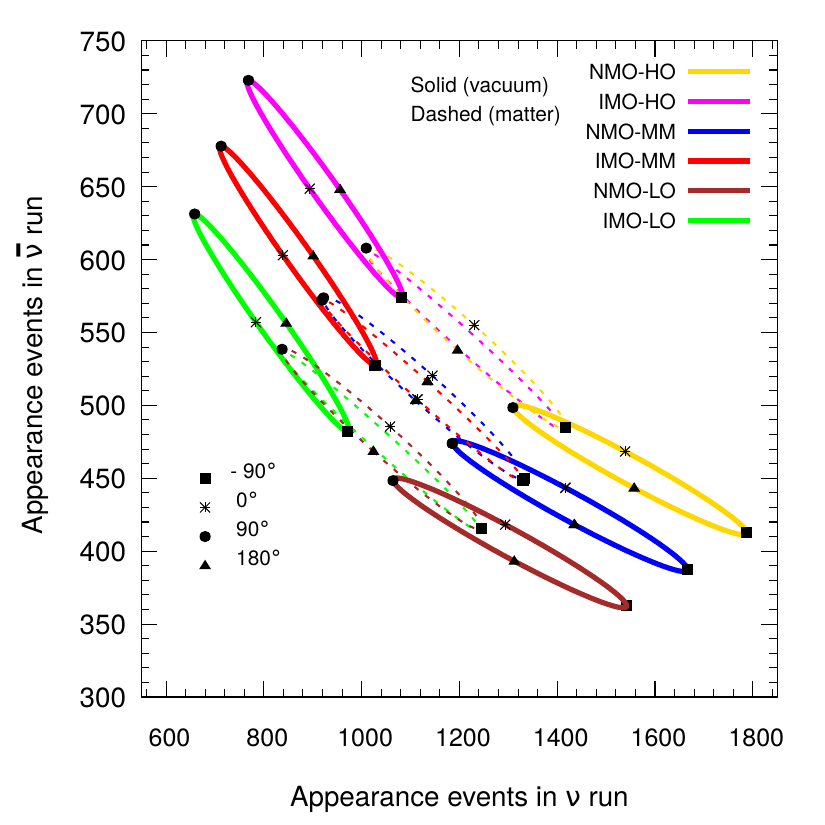}
\caption{Bi-events in the canvas of appearance neutrino and antineutrino modes, for DUNE, assuming 300 kt$\cdot$MW$\cdot$years of exposure (splitting equally between $\nu$ and $\bar{\nu}$). The ellipses are obtained by varying $\delta_{\mathrm{CP}}$ in the whole range. The three sets are generated with three probable choices of $\sin^2\theta_{23}$: 0.44, 0.5, and 0.56, depicting illustrative choices in LO, MM, and HO, respectively in both NMO and IMO.}
\label{work1:fig:3}
\end{figure}

The bi-event curves for DUNE is given in Fig.~\ref{work1:fig:3}. 
The ellipses in the figure are obtained by calculating appearance 
event rates in both neutrino and antineutrino modes for each value 
of $\delta_{\mathrm{CP}}$ in the range [-180$^{\circ}$ : 180$^{\circ}$]. The bi-event curves also represent pronounced overlapping of matter and vacuum event rates in both NMO and IMO, similar to what we observed in Fig.~\ref{work1:fig:2}. Perceiving more, the overlapping region of vacuum and matter in NMO is relatively more than that in IMO. Apart from this, following informations can be inferred from the figure:
\begin{itemize}
\item The bi-events depicting separate mass ordering, each in presence of matter are safely distinct, while the vacuum bi-events show considerable overlapping with each of them. This hints that if somehow DUNE does not receive the expected Earth matter density, then the sensitivity to differentiate between the two mass ordering will get affected.

\item Overlapping of bi-events in presence of matter with the same in vacuum renders that, some combination of [$\delta_{\mathrm{CP}},\rho_{\mathrm{avg}}$] in matter is degenerate with certain combination in vacuum, in \textit{both neutrino and antineutrino events}. This hints towards a mild $\rho_{\mathrm{avg}}-\delta_{\mathrm{CP}}$ degeneracy. 

\item Similarly, there is a significant overlapping between the bi-events in one octant of $\sin^{2}\theta_{23}$-vacuum with other octants of $\sin^{2}\theta_{23}$-matter. Thereby hinting towards a probable $\rho_{\mathrm{avg}}-\theta_{23}$ degeneracy. However, the degeneracy observed is only in either of the modes: neutrino or antineutrino at a time. Therefore, combining the neutrino and antineutrino statistics might help in breaking the $\rho_{\mathrm{avg}}-\theta_{23}$ degeneracy, which was not feasible in the previous scenario.  

\item The bi-events exhibiting hints of a probable $\rho_{\mathrm{avg}}-\delta_{\mathrm{CP}}$ degeneracy, illustratively, in neutrino mode [(0 g/cm$^3$, -151$^{\circ}$), (2.86 g/cm$^3$, 60$^{\circ}$)] and in antineutrino mode [(0 g/cm$^3$, -64$^{\circ}$), (2.86 g/cm$^3$, 24$^{\circ}$)], one can expect that the sensitivity to rule out vacuum solutions might fall in danger. However, incorporating spectral information might play a crucial role in such cases. 

\item 
It should be also noted from the figure that  the event rates which correspond to  $\delta_{\rm{CP}}=-90^\circ $ for NMO and 90$^\circ$ in IMO in matter is lying far away from the vacuum solution irrespective of value of $\sin^2\theta_{23}$. Whereas, the event rates corresponding to  $\delta_{\rm{CP}}=90^\circ$ for NMO and $-90^\circ$ for IMO, in matter is lying very close to the vacuum solutions. This hints that the sensitivity to exclude vacuum solutions, will be higher in the Lower Half Plane (Upper Half Plane) of $\delta_{\mathrm{CP}}$ region \textit{i.e.,} $-180^\circ<\delta_{\rm{CP}}<0^\circ$ ($0^\circ<\delta_{\rm{CP}}<180^\circ$) for NMO (IMO). Moreover, we notice that the neutrino and antineutrino modes have their own favorable and unfavorable regions in $\delta_{\mathrm{CP}}$, thus complementing each other as a whole, in improving the exclusion sensitivity.

\item Further, illustratively we checked that decreasing the value of matter density parameter ($\rho_{\mathrm{avg}}$), shifts the bi-events with finite matter density towards the vacuum case. This we term as the inward flow. Analogously, increasing the $\rho_{\mathrm{avg}}$, shifts the bi-events with finite matter away from the vacuum solutions. This for reference we coin as the outward flow. This nature of shifting is observed to be true in all the octant of $\sin^{2}\theta_{23}$: LO, MM, and HO. This phenomena of inward and outward flow performs an essential role while discussing the precision studies of   $\rho_{\mathrm{avg}}$. This is furthur ellaborated in Sec.~\ref{sec5.3}.
\end{itemize}

\section{Results}
\label{results}

In this segment, we address: a) the efficiency of DUNE in establishing matter effect by ruling out vacuum hypothesis, b) the ability of DUNE in precisely measuring the $\rho_{\mathrm{avg}}$ and further quantifying it by comparing with other standard measurements on $\rho_{\mathrm{avg}}$ in the literature, c) studying the variation in precision measurements of $\rho_{\mathrm{avg}}$ as a function of exposure, and finally d) summarizing with some noticeable new degeneracies among the oscillation parameters involved in this study.

\subsection{Sensitivity of DUNE towards the Earth matter effect}

To determine the sensitivity of DUNE in establishing the matter effect by excluding the vacuum solutions, we follow the following definition of statistical significance 
\begin{equation}
\Delta \chi^2_{\mathrm{ME}}= \underset{\boldsymbol{\theta}}{\mathrm{min}}\, 
\{ \chi^2(\rho_{\mathrm{avg}}^{\mathrm{true}} \neq 0) - \chi^2(\rho_{\mathrm{avg}}^{\mathrm{test}} = 0) \}\,,
\label{eq:chi2-establishig-matter-effect} 
\end{equation}
where, $\boldsymbol{\theta} = \{\theta_{23},~\delta_{\mathrm{CP}},~ \Delta m_{31}^2\}$ 
is the set of parameters on which minimization is executed in the fit (refer to Sec.~\ref{stat}). We follow the following ranges for minimization, $\sin^2\theta_{23} \in [0.4:0.6]$, $\delta_{\mathrm{CP}} \in [-180^{\circ}:180^{\circ}]$, and $\Delta m_{31}^2 \in [0.36:0.64] \times 10^{-3}$ eV$^{2}$. A fit to the ``observed data'' is executed to obtain $ \chi^2(\rho_{\mathrm{avg}} \neq 0)$ and $\chi^2(\rho_{\mathrm{avg}} = 0)$ by supposing standard 3-$\nu$  oscillation framework in matter and vacuum like scenarios, respectively. Suppressing statistical fluctuations, we expect $ \chi^2~(\rho_{\mathrm{avg}} \neq 0) \approx 0$. We also represent the statistical significance in terms of n$\sigma$, where $n\equiv \sqrt{\Delta \chi^2_{\mathrm{ME}}}$. This has also been discussed to provide median sensitivity of hypothesis by following the frequentist approach in Ref.~\cite{Blennow:2013oma}. 

\begin{figure}[htb!]
\centering
\includegraphics[width=0.495\textwidth]{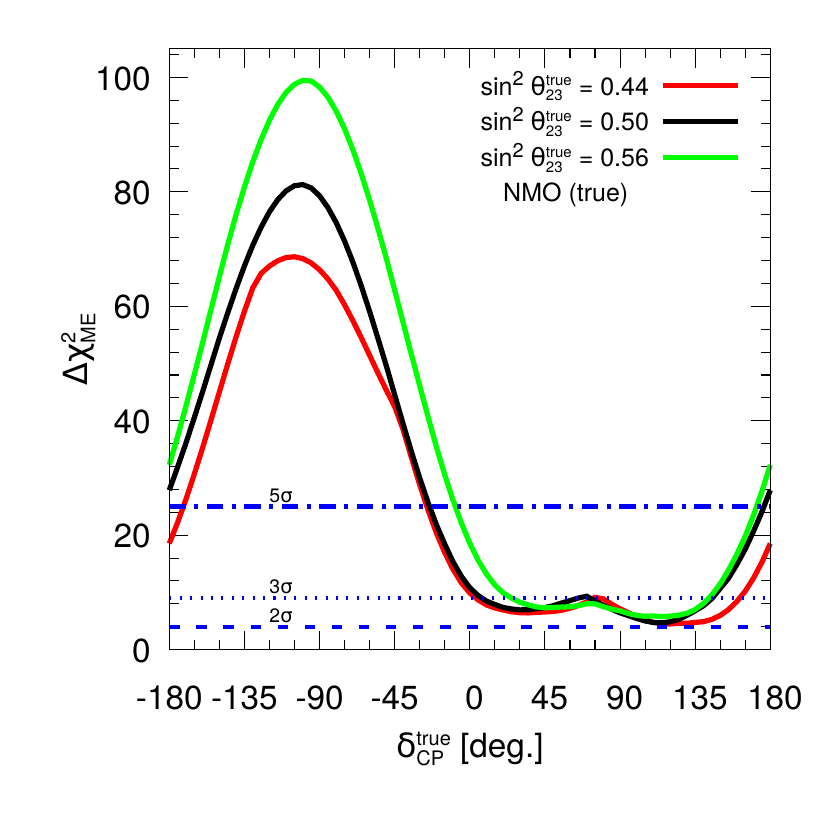} 
\includegraphics[width=0.495\textwidth]{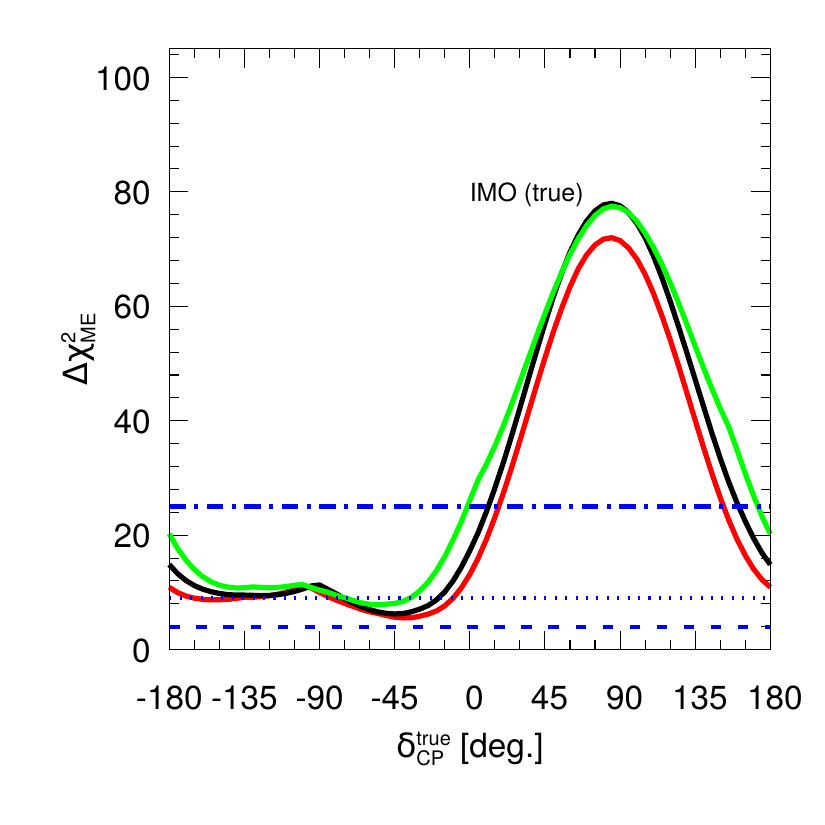}
\caption{Potential of DUNE to establish matter effect ($\Delta \chi^{2}_{\mathrm{ME}}$) as a function of true values of $\delta_{\mathrm{CP}}$. This is depicted for three probable choices of 
$\sin^2\theta_{23}$ assuming both true NMO and IMO following Eq.~\ref{eq:chi2-establishig-matter-effect}. We follow the benchmark values and allowed regions as mentioned in Sec.~\ref{probability}. \textbf{Irrespective of the values of $\delta_{\mathrm{CP}}, \sin^{2}\theta_{23}$, and sign of $\Delta m^{2}_{31}$\,, DUNE can establish Earth matter effect with at least $2\sigma$.}}
\label{work1:fig:4}
\end{figure}

  In the literature, we often see similar discussion around the sensitivity for segregating different mass ordering in DUNE~\cite{DUNE:2015lol}, where the statistical signficance (value of $\Delta\chi^2$) is much higher than the sensitivity involved in this discussion. A comparative study between the ability of DUNE to establish Earth matter effect ($\Delta \chi^{2}_{\mathrm{ME}}$) and its efficacy in ruling out wrong mass ordering, denoted by $\Delta \chi^{2}_{\mathrm{MO}}$\,, has been discussed in Table~\ref{work1:table:2}. The table clearly quantifies the two sensitivities. We observe that owing to large matter effect, the sensitivity of DUNE in establishing the correct mass oredering by excluding the wrong mass ordering solutions is way higher than its corresponding sensitvity in establishing matter effect by ruling out vacuum solutions. This difference is more vivid for the unfavorable choices in $\dcp$. The numbers are illustrative and can be matched from Fig.~\ref{work1:fig:4}. 

\begin{table}[htb!]
\centering
\begin{tabular}{|c|c|c|}
\hline\hline
True Parameters & $\Delta \chi^{2}_{\mathrm{ME}}$ & $\Delta \chi^{2}_{\mathrm{MO}}$\\
\hline
NMO, $\delta_{\mathrm{CP}}$ = -90$^{\circ}$ & 79.3 & 386  \\
\hline
NMO, $\delta_{\mathrm{CP}}$ = 90$^{\circ}$ & 6.4 & 54.4 \\
\hline
IMO, $\delta_{\mathrm{CP}}$ = -90$^{\circ}$ & 11.3 & 46.7 \\
\hline
IMO, $\delta_{\mathrm{CP}}$ = 90$^{\circ}$ & 77.6 & 316 \\
\hline\hline
\end{tabular}
\caption{Comparison study to exhibit sensitivity of DUNE towards Earth matter effect and in establishing the correct mass ordering for four illustrative choices of $\delta_{\mathrm{CP}}$. This assumes maximal mixing in $\sin^{2}\theta_{23}$ (0.5) and NMO. \textbf{Extracting correct mass ordering is comparatively much more feasible than establishing matter effect using DUNE.}
}
\label{work1:table:2}
\end{table}

\begin{table}[htb!]
\centering
\begin{tabular}{|c|c|c|c|c|c|}
\hline\hline
& \multicolumn{5}{c|}{$\Delta \chi^{2}_{\mathrm{ME}}$} \\
\cline{2-6}
$\delta_{\mathrm{CP}}^{\mathrm{true}}$ & Fixed Parameter & \multicolumn{4}{c|}{Minimizing Over} \\
\cline{3-6}
&  & $\delta_{\mathrm{CP}}^{\mathrm{test}}$ & $\theta_{23}^{\mathrm{test}}$ & $(\Delta m^2_{31})^{\mathrm{test}}$ & All \\
\hline\hline
$0^{\circ}$  & 86.4 & 13.4 & 72.7 & 86.4 & 10.8 \\ \hline
$90^{\circ}$  & 103.9 & 6.8 & 84.2 & 103.9 & 6.4 \\
\hline
$-90^{\circ}$  & 90.6 & 90.3 & 79.8 & 90.5 & 79.3 \\
\hline
$180^{\circ}$  & 109.8 & 31.4 & 94.6 & 109.7 & 27.9 \\
\hline\hline
\end{tabular} 
\caption{ Effect of minimization of different oscillation parameters on the test-statistic for sensitivity towards $\rho_{\mathrm{avg}}$. The test-statistic is computed using Eq.~\ref{eq:chi2-establishig-matter-effect}, assuming $\rho_{\mathrm{avg}} = 2.86$ g/cm$^3$. In this table, we report the values of the test-statistic calculated for four illustartive choices of $\delta_{\mathrm{CP}}$. Our main results are obtained by minimizing over $\delta_{\mathrm{CP}}$, $\theta_{23}$, $\Delta m_{31}^2$ and the wrong mass-ordering following Eq.~\ref{eq:imo-relation-with-nmo} in the fit. Results for partial minimization are shown only for the purpose of singling out the effect of different parameters. The values of the parameters that are not minimized over are fixed to the benchmark values as disussed in Sec.~\ref{probability}.}
\label{work1:table:3}
\end{table}

Table~\ref{work1:table:3} summarizes the effect of minimization over each oscillation parameter: 
$\delta_{\mathrm{CP}}$, $\sin^2\theta_{23}$, and $\Delta m_{31}^2$ in both mass ordering, on the senstivity in $\Delta \chi^{2}_{\mathrm{ME}}$. From the table, we observe that the effect of $\delta_{\mathrm{CP}}$ minimization is much more prominent than other oscillation parameters. This effect is much more crucial for the true values of $\delta_{\mathrm{CP}}$ belonging to the 
unfavorable region \textit{ i.e.,} $90^{\circ}$ (-90$^{\circ}$) in true NMO (IMO), while, minimization over values of $\delta_{\mathrm{CP}}$ in favorable regions \textit{ i.e.,} $-90^{\circ}$($90^{\circ}$) in true NMO (IMO) does not effect $\Delta \chi^{2}_{\rm{ME}}$ sensitivity much. In addition to $\delta_{\mathrm{CP}}$, we also notice that minimization over uncertainty in $\theta_{23}$ also plays an essential role in the study of $\Delta \chi^{2}_{\rm{ME}}$ in DUNE. Letting the uncertain parameters free in the fit leads to decrease in the sensitivity as expected. 

In Fig.~\ref{work1:fig:4}, we depict the sensitivity of DUNE in establishing the matter effect, following Eq.~\ref{eq:chi2-establishig-matter-effect} as a function of the CP phase, $\delta_{\mathrm{CP}}$ in data. Assuming NMO, DUNE can establish matter effect at 5$\sigma$ for a CP coverage of $\sim 46\%$ that mostly comes from the favorable region \textit{i.e.,} $\delta_{\mathrm{CP}} \in [-180^{\circ}, 0^{\circ}]$ (Lower Half Plane). Analogously, in IMO, a 5$\sigma$ sensitivity in establishing matter effect is achievable with a CP coverage of $\sim 43\%$, wherein  $\delta_{\mathrm{CP}} \in [0^{\circ}, 1800^{\circ}]$ (Upper Half Plane). This follows from the discussion in Sec.~\ref{bi-event-plot}. Moreover, the unexpected sensitivity ($\Delta \chi^2_{\mathrm{ME}} > 0$) to establish matter effect in the unfavorable region is enhanced by the correlation between the neutrino and antineutrino modes along with the incorporation of the spectral informations. Therefore, irrespective of the mass ordering, $\sin^{2}\theta_{23}$ and CP phase, DUNE is able to rule out vacuum hypothesis solutions above 2$\sigma$. Also, it is expected that with the increase in statistics, the sensitivity to establish matter effect will further improve, which we notice by transitioning from $\sin^{2}\theta_{23} = 0.44$ to 0.5 and further 0.55 in NMO. This happens as the appearance statistics in neutrino mode gets enhanced with increase in $\sin^{2}\theta_{23}$. However, in the case of IMO, appearance statistics in antineutrino mode gets enhanced, while reducing in neutrino mode. Therefore, we do not observe a clear increment in the sensitivity to establish $\rho_{\mathrm{avg}}$ in IMO, with increase in $\sin^{2}\theta_{23}$ as it depends mostly on neutrino statistics.

\subsection{Measuring Earth matter density}

In the previous result section, we find that DUNE can establish the Earth matter effect with a significance above 2$\sigma$ irrespective of the true values of the 
$\delta_{\mathrm{CP}}$,  mass ordering, and octant of $\sin^{2}\theta_{23}$. Therefore, now it becomes imperative to question that how well we can measure the line-averaged constant Earth matter density parameter ($\rho_{\mathrm{avg}}$). 
 
\subsubsection{Achievable precision on $\rho_{\mathrm{avg}}$ using DUNE}
\label{sec:work1:precision}

\begin{figure}[htb!]
\centering
\includegraphics[width=0.495\textwidth]{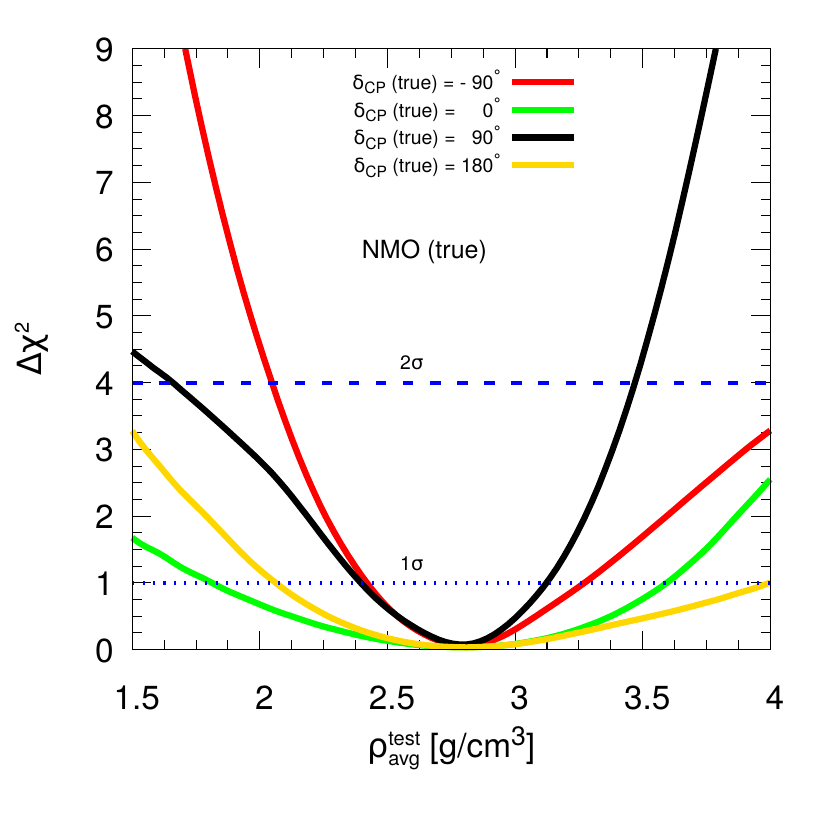} 
\includegraphics[width=0.495\textwidth]{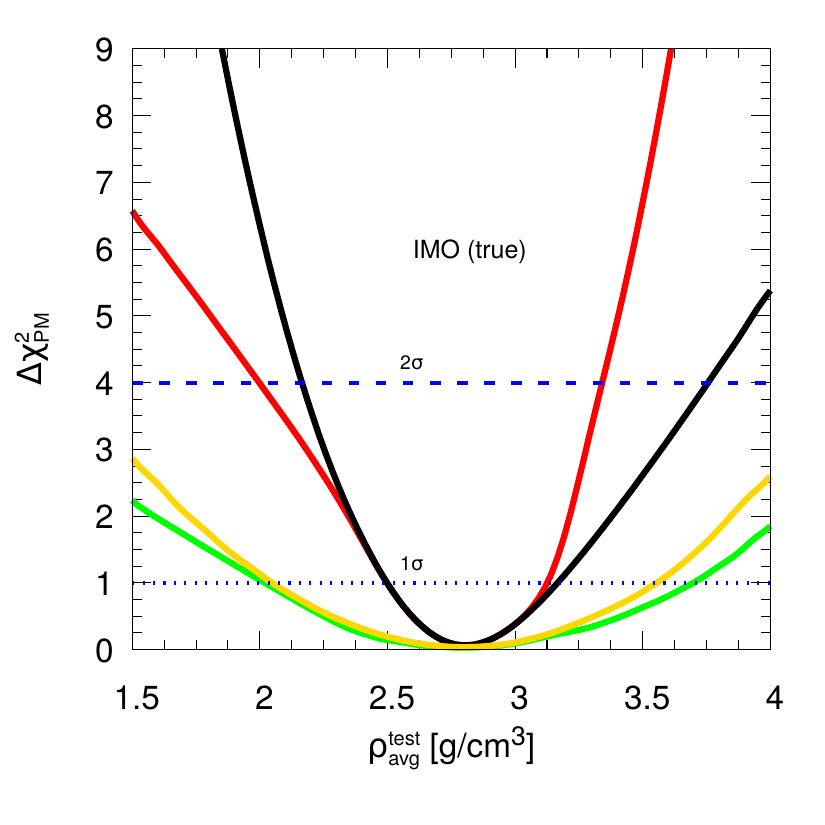}
\caption{Precision in the measurement of  $\rho_{\mathrm{avg}}$ for four illustrative true values of $\delta_{\mathrm{CP}}$. We follow Eq.~\ref{eq:chi2-achievable-precision-rho-avg}\,, and the benchmark values along with the allowed ranges for minimization from Sec.~\ref{probability}.  We assume $\sin^2\theta_{23}$ (true) = 0.5 and NMO (IMO) in the left (right) panel. \textbf{Precision in $\rho_{\mathrm{avg}}$ will be highest if CP is maximally violated.}}
\label{work1:fig:5}
\end{figure}

Fig \ref{work1:fig:5} depicts the attainable precision by DUNE in measuring $\rho_{\mathrm{avg}}$ for four illustrative values of $\delta_{\mathrm{CP}}$ in data, assuming both NMO and IMO (one at a time). We define
\begin{equation}
\Delta \chi^2_{\mathrm{PM}}(\rho_{\mathrm{avg}}) = \underset{\boldsymbol{\theta}}{\mathrm{min}}\{\chi^2(\rho_{\mathrm{avg}}) -\chi^2_0\} \,,
\label{eq:chi2-achievable-precision-rho-avg}
\end{equation}
where $\chi^2(\rho_{\mathrm{avg}})$ is computed by performing a fit 
to the ``observed data'' by assuming $\rho_{\mathrm{avg}} =2.86$ g/cm$^3$. $\chi^2_0$ is the minimum value of $\chi^2(\rho_{\mathrm{avg}})$\,, 
wherein $\rho_{\mathrm{avg}} \in [1.5:4]$ g/cm$^3$. Here, $\boldsymbol{\theta}$ = $\{\delta_{\mathrm{CP}}, \,\sin^2\theta_{23}, \, \Delta m_{31}^2\}$ and also the wrong mass ordering, is the set of oscillation parameters over which we perform minimization in the fit over the ranges as mentioned in Sec.~\ref{probability}. For quantifying the precision further, we define a parameter $p({\rho_{\mathrm{avg}}})$~\cite{Fogli:2012ua} such that it depicts the relative 1$\sigma$ precision in measurement of $\rho_{\mathrm{avg}}$, as defined in the following equation.
\begin{equation}
p({\rho_{\mathrm{avg}}}) = \left(\dfrac{\rho_{\mathrm{avg}}^{\mathrm{test}}(\mathrm{max})-\rho_{\mathrm{avg}}^{\mathrm{test}}(\mathrm{min})}{2\times\rho_{\mathrm{avg}}^{\mathrm{true}}}\right)\times 100\% \, , 
\label{eq:work1:precision}
\end{equation}
where we define the maximum and minimum allowed values of $\rho_{\mathrm{avg}}$ as $\rho_{\mathrm{avg}}^{\mathrm{test}}(\mathrm{max})$ and $\rho_{\mathrm{avg}}^{\mathrm{test}}(\mathrm{min})$ at 1$\sigma$, respectively. Also, $\rho_{\mathrm{avg}}^{\mathrm{true}}$ is the true choice 
of line-averaged constant Earth matter density for DUNE 
which is 2.86 g/cm$^3$.
\begin{table}[htb!]
\centering
\begin{tabular}{|c|c|c|c|c|}\hline
$\delta_{\mathrm{CP}}^{\mathrm{true}}$ & -90$^{\circ}$ & 90$^{\circ}$ & 0$^{\circ}$ & 180$^{\circ}$ \\ \hline
$p(\rho_{\mathrm{avg}})$ in \% & 15 (11) & 13 (12) &31 (30) & 34 (26) \\ \hline
\end{tabular}
\caption{Considering four illustrative true choices of $\delta_{\mathrm{CP}}$ NMO, and $\sin^2\theta_{23}$ (true) = 0.5, we tabulate relative 1$\sigma$ precision in determination of $\rho_{\mathrm{avg}}$ following Eq.~\ref{eq:chi2-achievable-precision-rho-avg}. IMO is assumed to be true for the bracketed values. Values are extracted from Fig.~\ref{work1:fig:5}.}
\label{work1:table:4}
\end{table}

\begin{figure}[htb!]
\centering
\includegraphics[width=\textwidth,height=\textwidth]{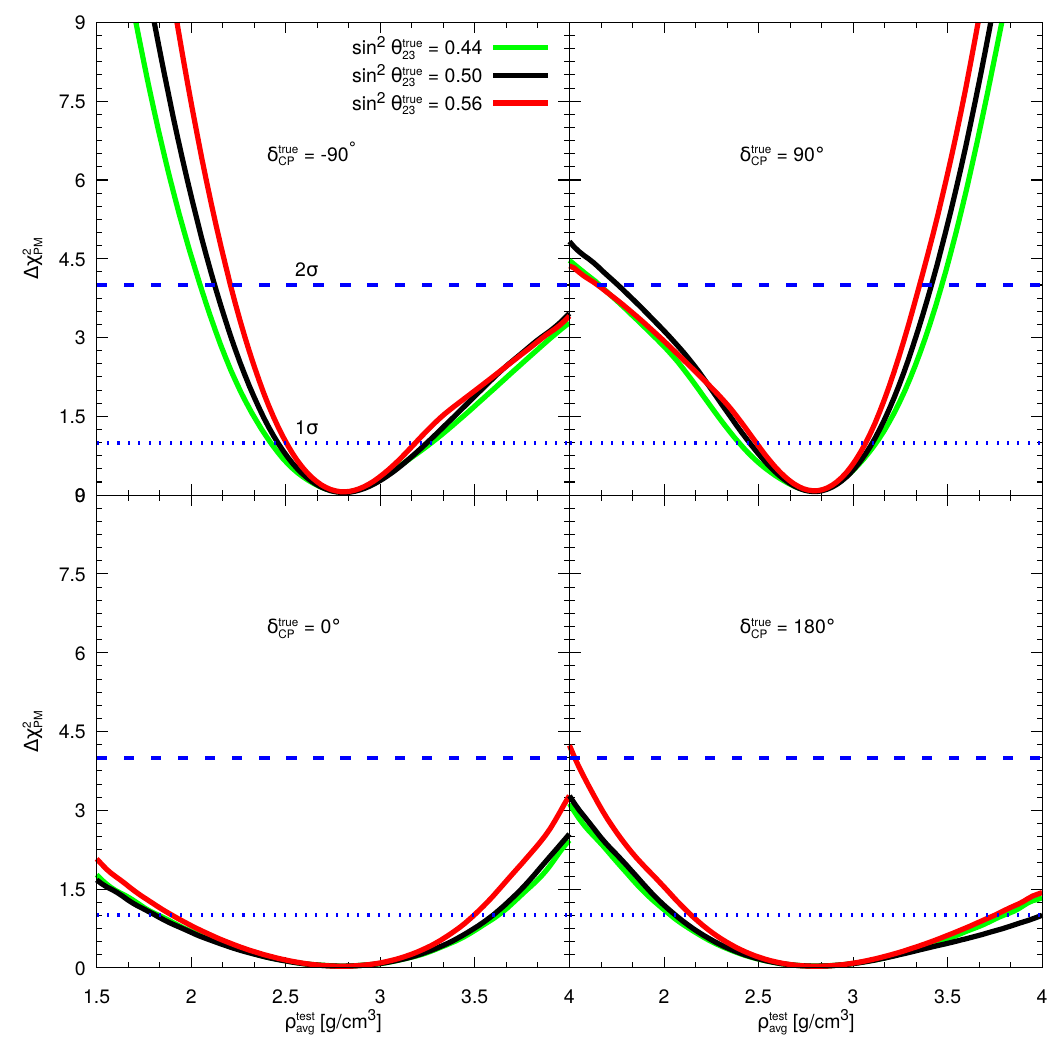}
\caption{Precision in the measurement of  $\rho_{\mathrm{avg}}$ for three illustrative values of $\sin^{2}\theta_{23}$ corresponding to each octant, while every panel represents an illustrative choice of $\delta_{\mathrm{CP}}$. We follow Eq.~\ref{eq:chi2-achievable-precision-rho-avg}\,, and the benchmark values along with the allowed ranges for minimization from Sec.~\ref{probability}.  We assume NMO. \textbf{Dependence of octant of $\sin^{2}\theta_{23}$ on precision measurements in $\rho_{\mathrm{avg}}$ is subtle.}}
\label{work1:fig:6}
\end{figure}

The relative 1$\sigma$ precision in the measurement of $\rho_{\mathrm{avg}}$ is tabulated in Table~\ref{work1:table:4} following Eq.~\ref{eq:work1:precision} and Fig.~\ref{work1:fig:5}. The values in the bracket correspond to true IMO. The numbers indicate that the achievable relative 1$\sigma$ precision in $\rho_{\mathrm{avg}}$ is better for CP-violating values of $\delta_{\mathrm{CP}}$ \textit{i.e.,} $-90^{\circ}$ and $90^{\circ}$  than CP-conserving values of $\delta_{\mathrm{CP}}$ \textit{i.e.,} $0^{\circ}$ and $180^{\circ}$. This is because maximal CP violation has less correlation with $\rho_{\mathrm{avg}}$ as we will see in the later sections, where we study the degeneracies. Further, for completeness, we also show the expected precision with four illustrative choices of CP phase for other values of octant of $\sin^{2}\theta_{23}$ in Fig.~\ref{work1:fig:6}. We expect the same nature in IMO as well. Fig.~\ref{work1:fig:6} reciprocates the takeaway from Fig.~\ref{work1:fig:5}. However, meagre dependence be, the HO choice has consistently better precision over other octants of $\sin^{2}\theta_{23}$ for all illustrative choices of CP phase, as explained in the previous section.

\subsubsection{Comparison with other experiments}

\begin{figure}[htb!]
\centering
\includegraphics[width=0.8\textwidth,height=0.75\textwidth]{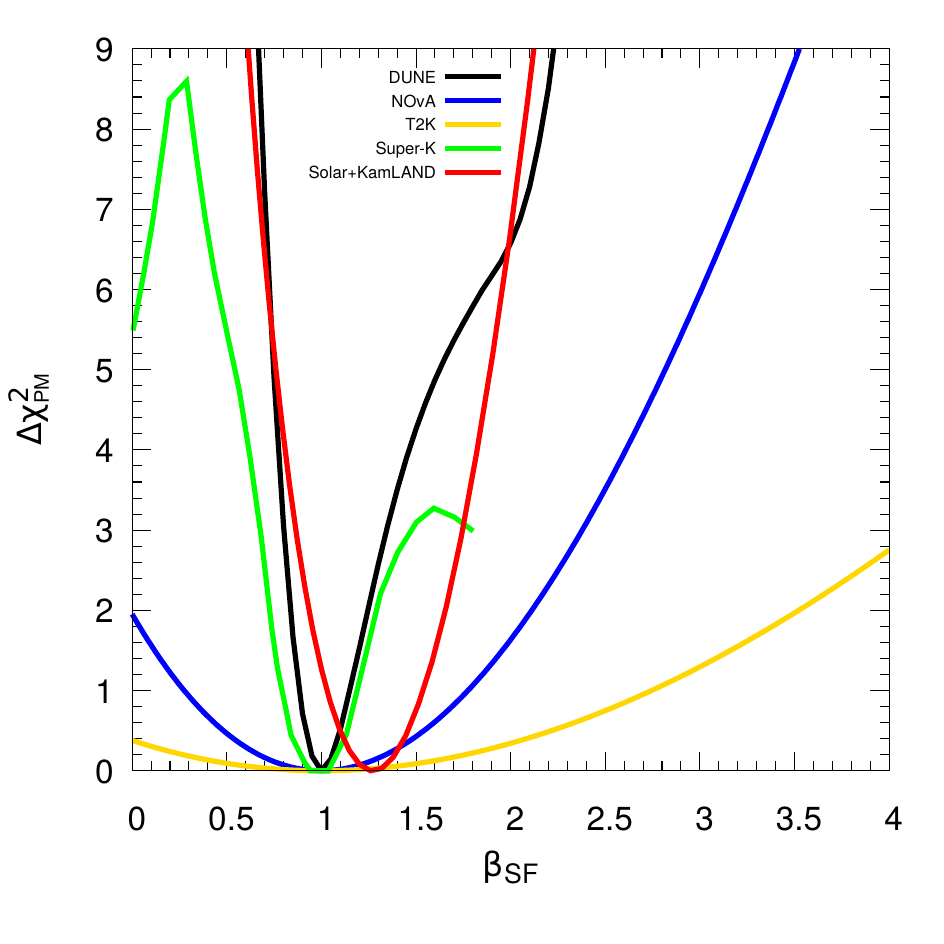}
\caption{Precision in the measurement of $\rho_{\mathrm{avg}}$ depicted as a function of a scaling parameter $\beta_{\mathrm{SF}}$ following Eq.~\ref{scalingfactor}. Existing curves for Solar + KamLAND are taken from Ref.~\cite{Maltoni:2015kca} and Super-Kamiokande from Ref.~\cite{Abe:2017aap,Fukuda:1998mi}. For T2K and NO$\nu$A, we follow benchmark values of oscillation parameters with $\rho_{\mathrm{avg}}= 2.79$ g/cm$^3$ and $\delta_{\mathrm{CP}} = -90^{\circ}$. \textbf{Given the considered benchmark values, DUNE can most precisely measure $\rho_{\mathrm{avg}}$ around the true value.}}
\label{work1:fig:7}
\end{figure}
Next, we exhibit the precision in the  measurement of $\rho_{\mathrm{avg}}$ in terms of scaling parameter $\beta_{\mathrm{SF}}$ (as defined previously in Eq.~\ref{scalingfactor}). Using this parameter we try to compare precision measurement of $\rho_{\mathrm{avg}}$ in DUNE with similar studies performed in the literature such as in Solar \cite{Maltoni:2015kca}, reactor, and atmospheric  \cite{Super-Kamiokande:2017yvm,Super-Kamiokande:1998kpq}. To obtain the $\Delta \chi^2_{\mathrm{PM}}$, we use our benchmark values as true values and compare it with test values by varying $\beta_{\mathrm{SF}}$ in the range [0 : 4]. We perform minimization over $\sin^2\theta_{23}$, $\delta_{\mathrm{CP}}$, and $\Delta m_{31}^2$ along with wrong mass ordering in the fit. For T2K, we assume an equally divided exposure in both $\nu$ and $\bar{\nu}$ modes, thus summing to a net P.O.T. of $7.8\times 10^{21}$. For the simulation of T2K, we closely follow the details as given in~\cite{T2K:2014xyt,Agarwalla:2016mrc}. While for NO$\nu$A, we assume a full projected exposure of $3.6\times 10^{21}$ P.O.T. which is shared equally in both neutrino and antineutrino modes. To simulate NO$\nu$A, we use the experimental features as given in~\cite{Agarwalla:2012bv,Patterson:2012zs,Agarwalla:2013ju}. In order to have a direct comparison, we also show precision measurement of $\rho_{\mathrm{avg}}$ for Super-Kamiokande and Solar + KamLAND experiments in the same figure. In \cite{Maltoni:2015kca}, the authors had a similar approach of scaling matter potential and performing a fit on the Solar and KamLAND combined data. Thus, as explained before in the discussion of Eq.~\ref{scalingfactor}, if the prediction on Earth matter density matches the standard PREM profile, then $\beta_{\mathrm{SF}}$ must equal 1. However, as visible in Fig.4 of Ref.~\cite{Maltoni:2015kca}, the minima in their study cuts the scaling factor at 1.7 instead of 1. This indicates that their data prefers more matter than it has been predetermined by the PREM profile. In their work, they explain this discrepancy due to the lingering tensions in the measurement of $\Delta m^{2}_{21}$ between Solar and KamLAND experiments and the insufficient data for addressing the predicted upturn in $^{8}$B spectrum (by LMA MSW mechanism) by the Super-K data. However, recently after Super-K announced its Phase (IV)~\cite{Super-Kamiokande:2016yck}, in which their data is consistent with the predicted upturn by the MSW resonance curve, reduced day-night asymmetry, and further reduction in the tension between Solar and KamLAND sensitivity of $\Delta m^{2}_{21}$~\cite{NuFIT,Esteban:2020cvm,Capozzi:2021fjo,deSalas:2020pgw}, have combinedly produced a positive effect on existing studies~\cite{Maltoni:2015kca}. This improvement can be seen in the present study, as depicted in Fig~\ref{work1:fig:7} (see red curve ). In the current combined precision fit\footnote{This curve has been obtained through private communication with the authors.} of Solar and KamLAND (red curve) in Fig~\ref{work1:fig:7}, the minima cuts-off scaling factor at 1.2. From the figure, it is observed that the data in considered long-baseline experiments (DUNE, NO$\nu$A, and T2K) and Super-Kamiokande, prefer Earth density identical to PREM profile, unlike in Solar + KamLAND. Moreover, we can infer from Fig~\ref{work1:fig:7} that considering our benchmark values, precision in quantifying matter density is better in DUNE than any other experiments.

\subsection{Effect of exposure}

This subsection exhibits performance of DUNE in establishing $\rho_{\mathrm{avg}}$ as a function of exposure. 
\begin{figure}[htb!]
\centering
\includegraphics[width=0.8\textwidth,height=0.75\textwidth]{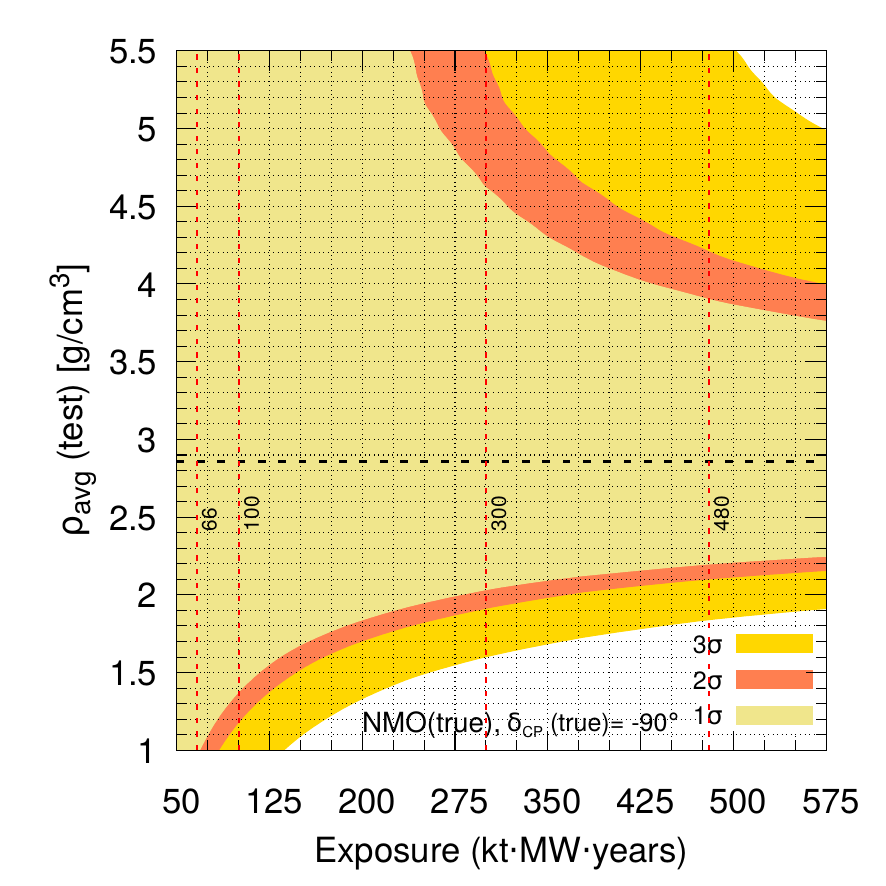}
\caption{Efficiency of DUNE to precisely measure $\rho_{\mathrm{avg}}$ as a function of exposure (kt$\cdot$MW$\cdot$years). We generate data by assuming true NMO, MM choice of $\theta_{23}$, and $\delta_{\mathrm{CP}}$ (true) = -90$^{\circ}$. Minimization have been performed over $\theta_{23}$, $\delta_{\mathrm{CP}},$ $\Delta m^{2}_{31}$ (NMO and IMO). While considering the exposure, we assume equally divided run-time in both $\nu$ and $\bar{\nu}$ modes. \textbf{Precision in $\rho_{\mathrm{avg}}$ is highly statistically dominated.} }
\label{work1:fig:8}
\end{figure} 
The orchestration and deployment of DUNE detector is planned to be phased by gradually increasing its fiducial mass and beam power~\cite{DUNE:2020jqi,DUNE:2021mtg}. Nevertheless, this incremental approach is quite advisable considering the huge challenges that they will be facing while deploying such enormous liquid argon detector. Currently the collaboration proposes~\cite{DUNE:2020jqi} to double their commencing beam power of 1.2 MW once they finish 6 years of operation. Also, they suggest a moduled set-up of far detectors (FD). In their first year, they plan to house two FD bearing a net fiducial mass of 20 kt, to which they recommend on adding one more of 10 kt in its second year and another in its fourth year of deployment. 

Through Fig.~\ref{work1:fig:8} we portray how the efficiency of DUNE changes in determining $\rho_{\mathrm{avg}}$ as we go on increasing its exposure expressed in kt$\cdot$MW$\cdot$years. The outcomes are drawn at 1$\sigma$, 2$\sigma$, and 3$\sigma$ with true NMO, MM choice of $\theta_{23}$\,, and $\delta_{\mathrm{CP}}$ (true) = -90$^{\circ}$. Uncertainties in atmospheric parameters ($\theta_{23}$ and $\Delta m^{2}_{31}$), wrong mass ordering (IMO) and CP phase have been examined by minimizing over their uncertain ranges as discussed earlier in Sec.~\ref{probability}. We notice from the Fig.~\ref{work1:fig:8} that precision in measuring $\rho_{\mathrm{avg}}$ barely reaches 2$\sigma$ from both the higher and lower pobable choices of $\rho_{\mathrm{avg}}$ (test) with our benchmark exposure of 300 kt$\cdot$MW$\cdot$years. This nature can be compared and understood with the previously discussed Fig.~\ref{work1:fig:5} (red curves drawn for $\delta_{\mathrm{CP}}$ (true) = -90$^{\circ}$). The C.L. with which DUNE can precise $\rho_{\mathrm{avg}}$ is not quite good if we consider the hypothesis of $\rho_{\mathrm{avg}}$ with higher range in our theory. Hence with the current planning of moduled set-up, increasing statistics significantly helps in precisely determining $\rho_{\mathrm{avg}}$, increasing from 45.80\% to 29.7\%  at 1$\sigma$ (the numbers are quoted following the definition of $p({\rho_{\mathrm{avg}}})$ from Eq.~\ref{eq:work1:precision}) on increasing exposure from 300 kt$\cdot$MW$\cdot$years to 480 kt$\cdot$MW$\cdot$years. In the following subsection, we discuss the degeneracies among oscillation parameters which plays crucial role in determining $\rho_{\mathrm{avg}}$.

\subsection{Degeneracies in $(\rho_{\mathrm{avg}}^{\mathrm{test}} - \delta_{\mathrm{CP}}^{\mathrm{test}})$ and $(\rho_{\mathrm{avg}}^{\mathrm{test}} - \sin^2\theta_{23}^{\mathrm{test}})$ planes}
\label{sec5.3}

In this subsection, we show the allowed regions in the test-statistic of $\rho_{\mathrm{avg}}-\delta_{\mathrm{CP}}$ and $\rho_{\mathrm{avg}}-\theta_{23}$ planes, study correlations among them, and reflect upon their consequential impact on studying the sensitivity towards $\rho_{\mathrm{avg}}$ ($\Delta \chi^2_{\mathrm{ME}}$ ).

\begin{figure}[htb!]
\centering
\includegraphics[width=\textwidth,height=\textwidth]{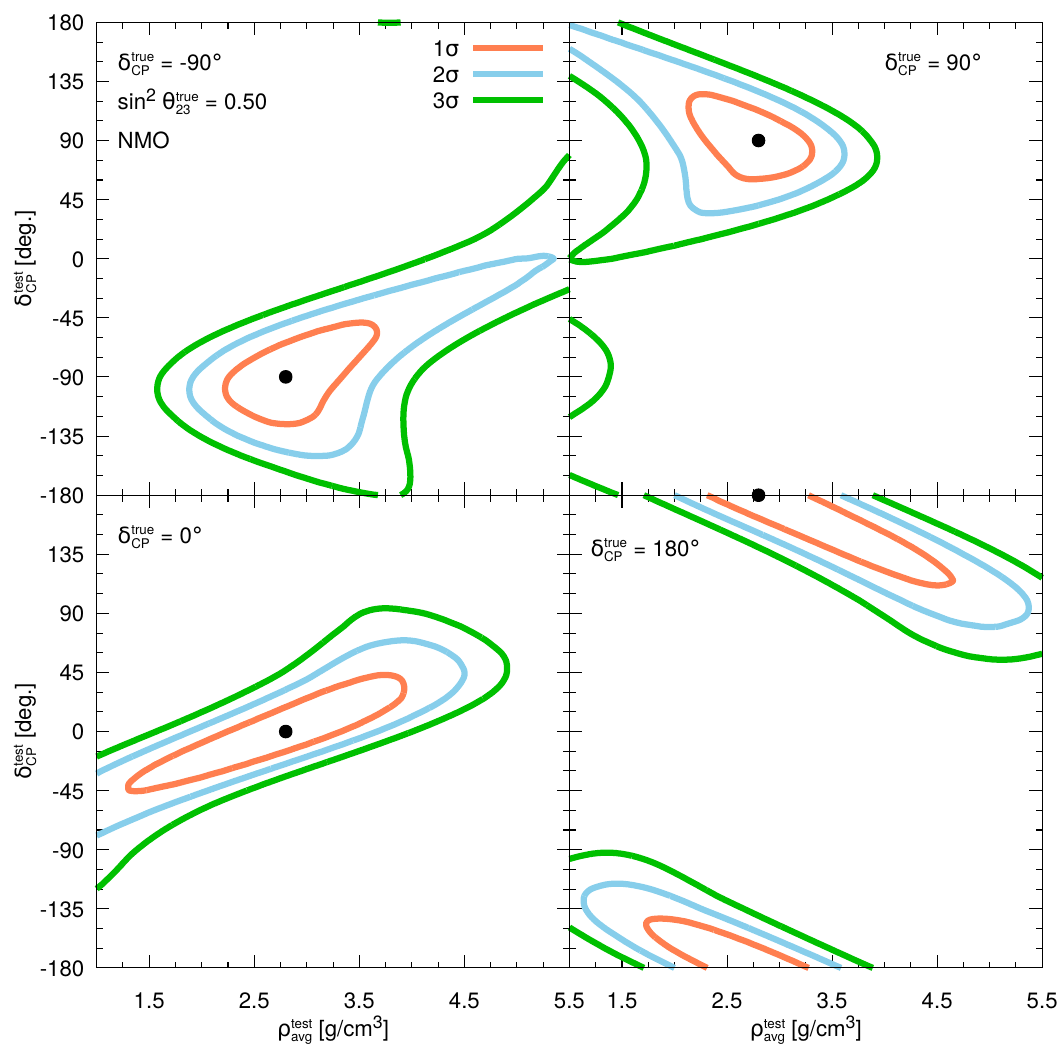}
\caption{Allowed regions of matter density, $\rho_{\mathrm{avg}}$ and CP-violating phase, $\delta_{\mathrm{CP}}$. The illustrative true values of $\delta_{\mathrm{CP}}$ is same as Fig.~\ref{work1:fig:5} and the atmospheric mixing angle is illustratively chosen to be maximal mixing (0.5). The test-statistic is minimized over $\sin^2\theta_{23}$ and $\Delta m_{31}^2$, and the mass ordering; see Eq.~\ref{eq:imo-relation-with-nmo}. See Section~\ref{sec5.3} for details. \textbf{Uncertainty in the values of $\delta_{\mathrm{CP}}$ critically affects the allowed region in $(\rho_{\mathrm{avg}}^{\mathrm{test}} - \delta_{\mathrm{CP}}^{\mathrm{test}})$ plane.} }
\label{work1:fig:9}
\end{figure} 
\begin{figure}[htb!]
\centering
\includegraphics[width=\textwidth,height=\textwidth]{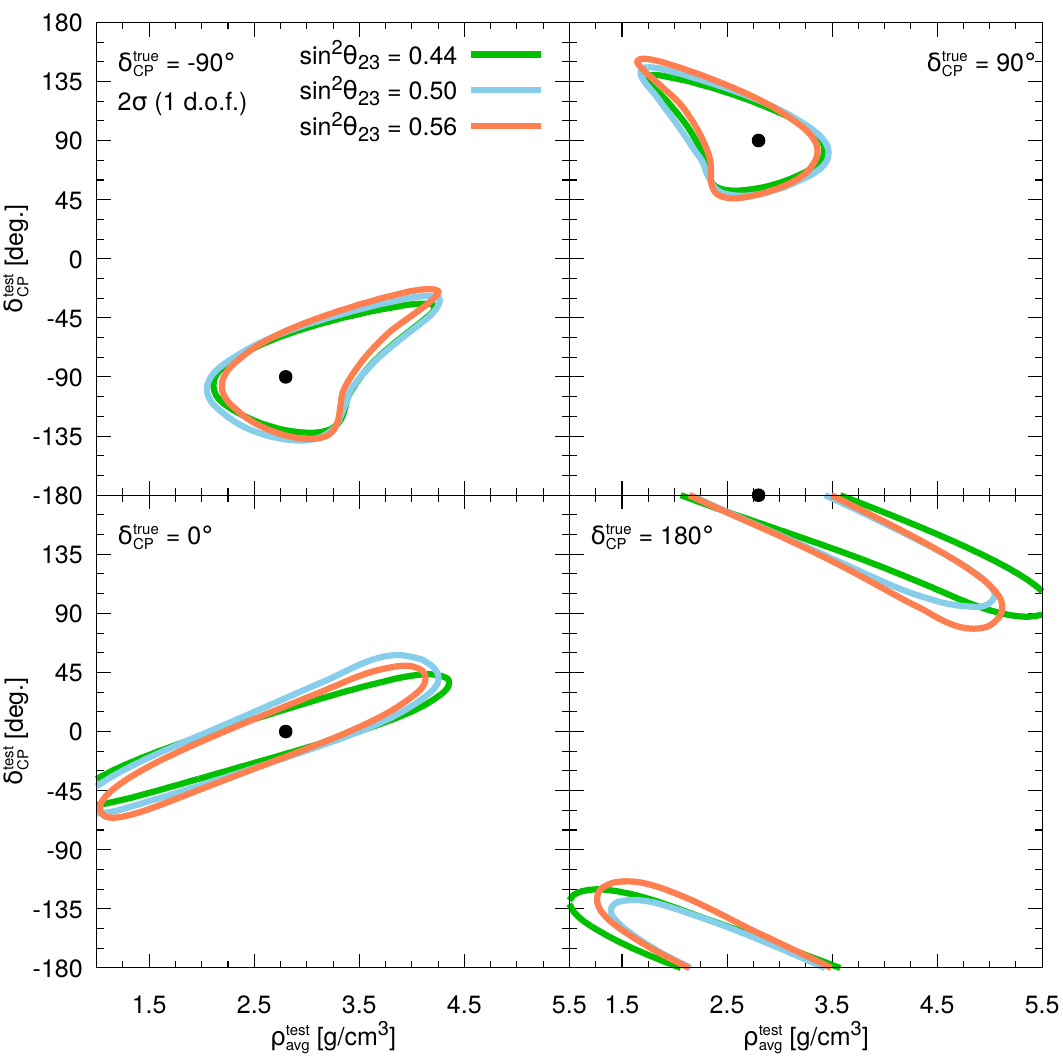}
\caption{Supplement to Fig.~\ref{work1:fig:9}, it depicts the allowed regions of matter density, $\rho_{\mathrm{avg}}$ and CP-violating phase, $\delta_{\mathrm{CP}}$ for other illustrative choices of $\sin^{2}\theta_{23}$. The illustrative true values of $\delta_{\mathrm{CP}}$ is same as Fig.~\ref{work1:fig:9}. The test-statistic is minimized over $\sin^2\theta_{23}$ and $\Delta m_{31}^2$, and the mass ordering; see Eq.~\ref{eq:imo-relation-with-nmo}. See Section~\ref{sec5.3} for details. \textbf{For a given CP phase, uncertainty in the values of $\sin^{2}\theta_{23}$ does not have any essential impact on the allowed region in $(\rho_{\mathrm{avg}}^{\mathrm{test}} - \delta_{\mathrm{CP}}^{\mathrm{test}})$ plane.}}
\label{work1:fig:10}
\end{figure}
In Fig.~\ref{work1:fig:9}, we exhibit the allowed region in $\rho_{\mathrm{avg}}^{\mathrm{test}}$ and $\delta_{\mathrm{CP}}^{\mathrm{test}}$ for four different illustrative choices of $\delta_{\mathrm{CP}}$ and defined true value of $\rho_{\mathrm{avg}}=2.86$ g/cm$^3$. In order to obtain the allowed region, we vary test values of $\delta_{\mathrm{CP}}$ and $\rho_{\mathrm{avg}}$ in the range $[-180^{\circ}:180^{\circ}]$ and $[1:5.5]$, respectively, while performing minimization over $\sin^2\theta_{23}$ and $\Delta m_{31}^2$. We generate data by assuming true values of NMO and $\sin^2\theta_{23}=0.5$. From the  figure, it can be seen that, irrespective of the values of $\delta_{\mathrm{CP}}$\,, there exists a circumscribed region for test value of $\delta_{\mathrm{CP}}$ with respect to $\rho_{\mathrm{avg}}$. To understand the nature of allowed region in $\rho_{\mathrm{avg}}-\delta_{\mathrm{CP}}$ plane, we make use of the bi-events spectra (refer to Fig.~\ref{work1:fig:3}) where we have discussed the inward and outward flows depending upon the variation in the value of $\rho_{\mathrm{avg}}$. For instance, considering the case for $\delta_{\mathrm{CP}}=-90^{\circ}$ with NMO and vacuum (refer to square label on dashed-brown colored ellipse in Fig.~\ref{work1:fig:3}). On traversing from vacuum to matter case, the ellipse shifts far away from overlapped region (outward flow) due to an enhancement in oscillation probability which results in increased $\Delta\chi^2$. In contrast, as explained earlier in Sec.~\ref{bi-event-plot}, following the inward flow, we expect a deterioration in the oscillation probability and hence resulting in lowering of $\Delta\chi^2$. Further, from the Fig.~\ref{work1:fig:9}, we also observe that the nature of allowed region is exactly opposite for $\delta_{\mathrm{CP}}= 90^{\circ}$, assuming true NMO. Therefore, as we go towards higher values of $\rho_{\mathrm{avg}}$, the uncertainty in $\delta_{\mathrm{CP}}$ increases (decreases) for true value of $\delta_{\mathrm{CP}} = -90^{\circ}~ (90^{\circ})$, thereby making the $(\rho_{\mathrm{avg}}-\delta_{\mathrm{CP}})$ plane asymmetrical. We also observe that the 1$\sigma$ allowed range of $\rho_{\mathrm{avg}}$ for true values of $\delta_{\mathrm{CP}}=0^{\circ}$ and $180^{\circ}$ is more than that of $\delta_{\mathrm{CP}}=-90^{\circ}$ and $90^{\circ}$. This indicates that the uncertainty in average matter density can be more precisely measured if in Nature we have maximal CP-violating values (as seen from Fig.~\ref{work1:fig:4} as well). For completeness, we also depict the nature of isocontours for other values of $\sin^{2}\theta_{23}$ in Fig.~\ref{work1:fig:10}. Just like precision studies in Fig.~\ref{work1:fig:6}, here also we observe subtle dependence of allowed regions around the true value on different octant of $\sin^2\theta_{23}$.

\begin{figure}[htb!]
\centering
\includegraphics[width=\textwidth,height=0.4\textwidth]{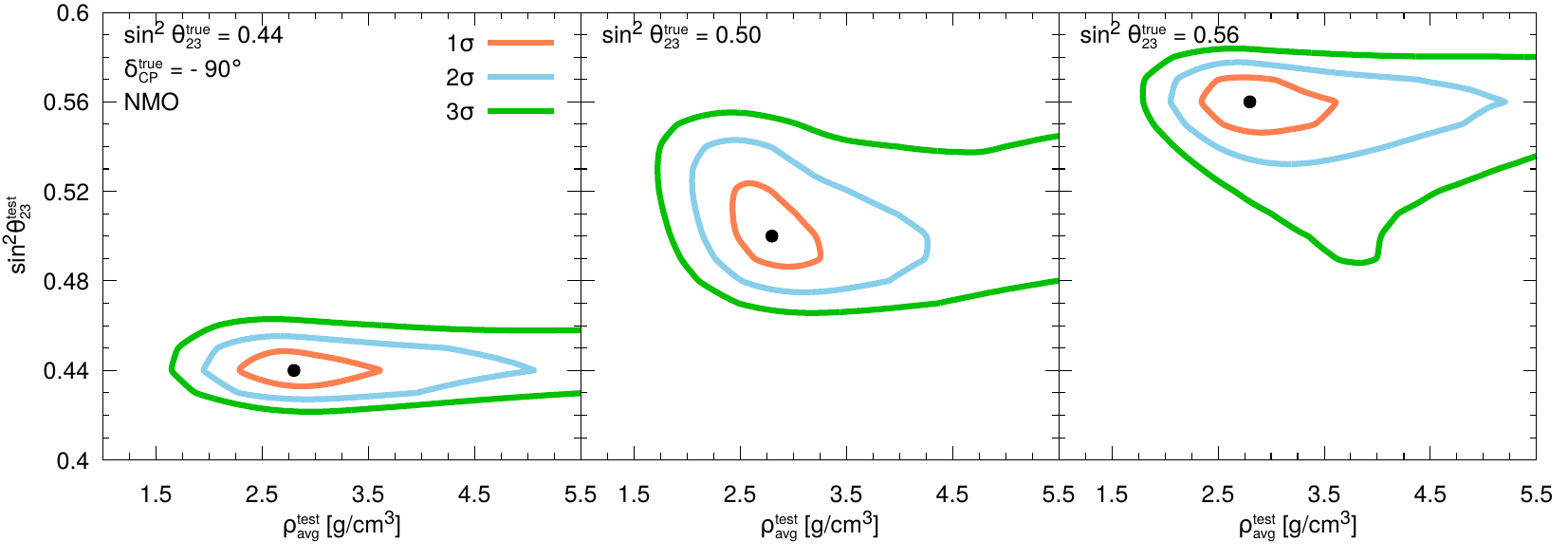}
\caption{Allowed regions of matter density, $\rho_{\mathrm{avg}}$ and atmospheric mixing angle, $\sin^2\theta_{23}$. The true values of $\sin^{2}\theta_{23}$ is same as Fig.~\ref{work1:fig:4} and the CP phase is illustratively chosen to be maximally violated ($-90^{\circ}$). The test-statistic is minimized over $\delta_{\mathrm{CP}}$, $\Delta m_{31}^2$, and the mass ordering; see Eq.~\ref{eq:imo-relation-with-nmo}. See Section~\ref{sec5.3} for details. \textbf{Uncertainty in the octant of $\sin^{2}\theta_{23}$ does not affect the allowed values in $\rho_{\mathrm{avg}}^{\mathrm{test}}$\,, however given the present uncertainty, LO is most precisely measured in $(\rho_{\mathrm{avg}}^{\mathrm{test}} - \sin^2\theta_{23}^{\mathrm{test}})$ plane. }
}
\label{work1:fig:11}
\end{figure}

\begin{figure}[htb!]
\centering
\includegraphics[width=\textwidth,height=0.4\textwidth]{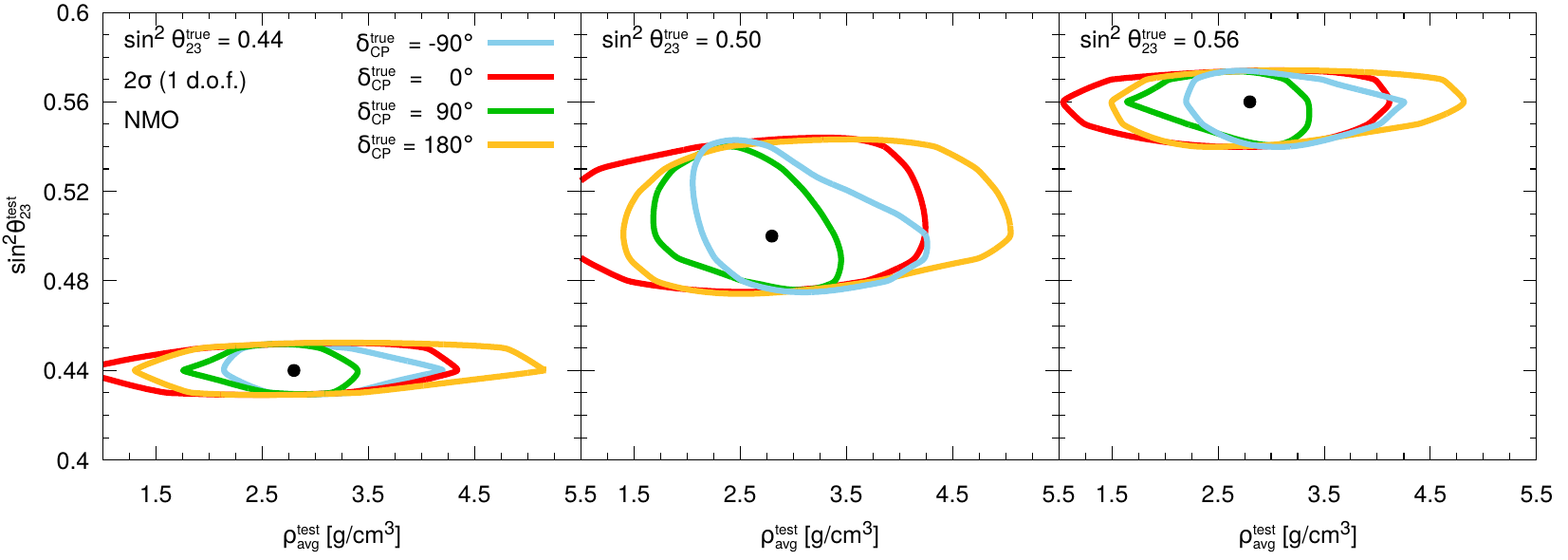}
\caption{Supplement to Fig.~\ref{work1:fig:11}, it depicts the allowed regions of matter density, $\rho_{\mathrm{avg}}$ and atmospheric mixing angle, $\sin^2\theta_{23}$ for other illustrative choices of $\delta_{\mathrm{CP}}$. The illustrative true values of $\sin^2\theta_{23}$ is same as Fig.~\ref{work1:fig:11}. The test-statistic is minimized over $\delta_{\mathrm{CP}}$, $\Delta m_{31}^2$, and the mass ordering; see Eq.~\ref{eq:imo-relation-with-nmo}. See Section~\ref{sec5.3} for details. \textbf{For a given value of $\sin^{2}\theta_{23}$\,, maximally CP-violating phases are able to provide stringent allowed regions in $(\rho_{\mathrm{avg}}^{\mathrm{test}} - \sin^2\theta_{23}^{\mathrm{test}})$ plane in comparison to CP-conserving values.}}
\label{work1:fig:12}
\end{figure} 

Next, in Fig.~\ref{work1:fig:11} we exemplify the allowed test-statistic in the $\rho_{\mathrm{avg}}-\sin^2\theta_{23}$ plane to exhibit the allowed regions for three probable choices of $\sin^2\theta_{23}$, assuming true $\rho_{\mathrm{avg}} = 2.86$ g/cm$^3$.  While obtaining the $\Delta \chi^2$, we perform minimization over $\delta_{\mathrm{CP}}$, $\Delta m_{31}^2$, and the wrong mass ordering in the fit by assuming NMO in the data. For generating the contours, we scan over $\sin^2\theta_{23}$ and $\rho_{\mathrm{avg}}$ in the range $[0.4:0.6]$ and $[1:5.5]$ g/cm$^3$ respectively. From the figure, we observe that the spread of contour along $\rho_{\mathrm{avg}}^{\mathrm{test}}$ (x-axis), follows almost similar pattern for all the three scenarios of octant of $\sin^{2}\theta_{23}$: LO, MM, and HO, while the spread along $\sin^{2}\theta_{23}^{\mathrm{test}}$ (y-axis) is least for LO. The allowed region at 3$\sigma$ C.L. for true choice of HO, has an extended tail towards maximal choice of $\theta_{23}$. This concludes that with true choice of HO, possibility of MM solution still remais allowed at 3$\sigma$ confidence level. Being directly proportional, higher the value of $\rho_{\mathrm{avg}}$ higher will be $A=2\sqrt{2}G_{F}N_{\mathrm{e}}$, which subsequently magnifies transition channel (in neutrino mode) and thus results in large $\Delta\chi^2$. We note from the figure that uncertainty in $\sin^{2}\theta_{23}$ around the best-fit value is less when compared to uncertainty in $\rho_{\mathrm{avg}}$. To explicitly understand how different choices of $\delta_{\mathrm{CP}}$ in data might affect this degeneracy in $(\rho_{\mathrm{avg}}-\theta_{23})$ plane, we explore the same for other values of $\delta_{\mathrm{CP}}$ in Fig.~\ref{work1:fig:12}. We observe that the allowed values for $\rho^{\rm{test}}_{\mathrm{avg}}$ (along x-axis) are more for the true CP-conserving values. This is in accordance with the discussion for Fig.~\ref{work1:fig:5}, which explains that, DUNE has relatively less ability in precision measurements of $\rho_{\mathrm{avg}}$ when CP-conserving choices are considered as true solutions. This confirms that true choices of $\delta_{\mathrm{CP}}$ affects the measurement in $\rho_{\mathrm{avg}}$ signficantly. 

\section{Summary}
\label{work1:summary-conclusions}

The planned Deep Underground Neutrino Experiment 
is the most awaited flagship mega-science project in the future
neutrino roadmap which will measure the mass-mixing parameters 
with unmatched precision and establish the standard three-flavor
oscillation picture on a strong footing. Apart from probing a 
landscape of various beyond the Standard Model (BSM)
scenarios~\cite{DUNE:2020fgq}, the primary science goals of 
DUNE are to determine the pattern of neutrino masses 
(normal or inverted), mixing angles, and CP violation in the 
lepton sector at high confidence level~\cite{DUNE:2020ypp,DUNE:2020jqi}. The Earth matter effect felt by neutrinos and antineutrinos while 
travelling a distance of 1300 km from Fermilab to the Homestake Mine in South Dakota through the Earth's crust plays an important role  
in these measurements. Therefore, it makes complete sense to ask
how much sensitivity the DUNE has towards the Earth matter effect. In this thesis, we address this pressing issue for the first time and
investigate in detail the potential of DUNE to establish the matter 
oscillation by excluding the vacuum oscillation. We find that this performance indicator largely depends on the true choices of $\delta_{\mathrm{CP}}$ and $\theta_{23}$ with which the prospective data are generated. Our analyses reveal that with the optmized neutrino beam design and using an exposure of 300 kt$\cdot$MW$\cdot$years, DUNE can corroborate the existence of Earth's matter effect at 2$\sigma$ C.L. irrespective of the true choices of mass ordering, $\delta_{\mathrm{CP}}$, and $\theta_{23}$. Besides, DUNE is cabable to exclude the vacuum oscillation hypothesis at 3$\sigma$ (5$\sigma$) significance for 64\% (46\%) choices of true $\delta_{\mathrm{CP}}$ assuming true NMO and $\sin^2\theta_{23}$ (true) = 0.5. If the true mass ordering turns out to be inverted in Nature and $\sin^2\theta_{23}$ (true) = 0.5, then, DUNE can reject the vacuum oscillation at 3$\sigma$ (5$\sigma$) significance for 82\% (43$\%$) choices of true $\delta_{\mathrm{CP}}$. On the other hand, DUNE can exclude the wrong mass ordering in the
presence of Earth matter at more than 6.5$\sigma$ significance irrespective of the true choices of mass ordering, $\delta_{\mathrm{CP}}$ and assuming $\sin^2\theta_{23}$ (true) = 0.5. It suggests that 
the task of establishing the matter oscillation by excluding the vacuum oscillation is always difficult as compared to measuring the neutrino mass ordering assuming matter oscillation.

As we will achieve better precision on the mass-squared differences and mixing angles, and the uncertainty in the measurement of the CP phase, $\delta_{\mathrm{CP}}$, is going to be reduced in the near future, the prospective DUNE data may reveal important information on the Earth matter density that the neutrino beam will travel through from Fermilab to Lead, South Dakota. We find that for maximal CP-violating choices of true $\delta_{\mathrm{CP}}$,
DUNE can measure the line-averaged constant Earth matter density ($\rho_{\mathrm{avg}}$) with a relative 1$\sigma$ precision of 10\% to 15\% depending on the choice of true neutrino mass ordering. The same for CP-conserving values of true $\delta_{\mathrm{CP}}$, is around 25\% to 30\%. We further observe that the achievable 
precision on $\rho_{\mathrm{avg}}$ at DUNE depends strongly on the true choice of $\delta_{\mathrm{CP}}$, but it also has a mild dependence on the true choice of $\sin^2\theta_{23}$. We find that if $\delta_{\mathrm{CP}}$ (true) turns out to be around $-90^{\circ}$ or $90^{\circ}$, the precision with which DUNE can measure $\rho_{\mathrm{avg}}$ is better than the the precision which can be achieved using the atmospheric data from Super-Kamiokande, combined data from Solar and KamLand, and from the full exposure of T2K and NO$\nu$A.

We also identify new degeneracies in ($\rho_{\mathrm{avg}}^{\mathrm{test}} - \delta_{\mathrm{CP}}^{\mathrm{test}}$) and ($\rho_{\mathrm{avg}}^{\mathrm{test}} - \sin^2\theta_{23}^{\mathrm{test}}$) planes for different true values of $\delta_{\mathrm{CP}}$ and $\sin^2\theta_{23}$. We notice that the uncertainty in $\delta_{\mathrm{CP}}$ affects the measurement of $\rho_{\mathrm{avg}}$ more than that of $\theta_{23}$. Needless to mention that a detailed study of these parameter degeneracies are indispensable to have an accurate assessment of the DUNE data.
\chapter{{Exploring Earth's Matter Effect using the Synergy between DUNE and T2HK}}
\label{sec:ch5}

In the preceding chapter, we delved into a comprehensive exploration of the intricate interplay between uncertainty in oscillation parameters, specifically focusing on the critical variables of the CP phase and atmospheric mixing angle, using DUNE. This detailed analysis illuminated the profound impact these uncertainties can have on the precision of measurements in the line-averaged constant Earth matter density. By establishing a nuanced understanding of how these oscillation parameters contribute to uncertainties, we laid the groundwork for a more informed and accurate interpretation of measurements related to Earth's matter effect in subsequent analyses. This insight serves as a crucial foundation for refining our grasp of neutrino properties and advancing the precision of measurements in the dynamic realm of neutrino physics. 

Building on our prior investigations, we discerned that degeneracies within the $\rho_{\mathrm{avg}}-\delta_{\mathrm{CP}}$ and $\rho_{\mathrm{avg}}-\theta_{23}$ planes impose limitations on achieving enhanced precision in $\rho_{\mathrm{avg}}$. Notably, the uncertainty associated with the CP phase significantly diminishes the confidence level with which we can ascertain $\rho_{\mathrm{avg}}$ in unfavorable zones (such as $\delta_{\mathrm{CP}} = 90^{\circ}$ in the Normal Mass Ordering (NMO) and $-90^{\circ}$ in the Inverted Mass Ordering (IMO), respectively).

This underscores the imperative for a strategic interplay among long-baseline experiments. One experiment, operating in a region with a substantial matter potential, can complement another that excels in precise measurements of the intrinsic $\delta_{\mathrm{CP}}$. This complementary approach has been succinctly demonstrated in Ref.~\cite{Singh:2021kov}, showcasing the symbiotic relationship between different experimental setups.

In our endeavor to establish $\rho_{\mathrm{avg}}$, we capitalize on the complementarity between two highly anticipated high-precision long-baseline experiments: DUNE and Tokai to Hyper-Kamiokande (T2HK/JD)~\cite{Hyper-Kamiokande:2018ofw}. This collaborative utilization of experiments with diverse strengths enhances our capability to accurately determine $\rho_{\mathrm{avg}}$, showcasing the significance of complementarity in advancing our understanding of neutrino properties. DUNE and T2HK are strategically designed experiments with distinct beam characteristics, each contributing uniquely to our exploration of neutrino properties. DUNE, located at Fermilab in Illinois, is set to receive a high-intensity, on-axis, wide-band beam covering both the first and second oscillation maxima. In contrast, T2HK, proposed to operate with an off-axis (2.5$^{\circ}$), narrow-band beam, focuses primarily on the first oscillation maximum. The different beam profiles of these experiments offer complementary advantages.

In DUNE, where the neutrino energy ($E$) is fixed at the first oscillation maximum, an exploration of varying $L/E$ (baseline with neutrino energy) peaks becomes possible. This exploration is intricately linked to the value of the CP phase. By studying the envelope of differing $L/E$ ratios in DUNE, we gain valuable insights, particularly into the impact of the CP phase.

On the other hand, T2HK, with its relatively shorter baseline and statistical advantage at the first oscillation maximum, is well-suited to establish intrinsic CP violation and precisely probe the CP phase. The complementary nature of these experiments becomes apparent in the variations of $L/E$ ratios, providing a unique perspective on the interplay between baseline, neutrino energy, and the CP phase.

Throughout this chapter, we demonstrate how the collaboration between these two experiments, each characterized by its wide-band and narrow-band beams, respectively, helps mitigate the impact of degeneracies within the $(\rho_{\mathrm{avg}}-\delta_{\mathrm{CP}})$ and $(\rho_{\mathrm{avg}}-\theta_{23})$ planes. This collaborative approach enhances the precision of measurements in $\rho_{\mathrm{avg}}$ and underscores the significance of complementarity in advancing our understanding of neutrino oscillations. 

Following the discussion in the previous chapter, herein also we consider a comprehensive independent parameter: $\beta_{\mathrm{SF}}$. SF in the subscript stands for a scaling factor. This parameter helps in traversing through the two major likelihoods that we are stressing on: state with finite line-averaged constant Earth matter density ($\beta_{\mathrm{SF}}$ = 1) and the other being equivalent to the vacuum state ($\beta_{\mathrm{SF}}$ = 0). This is attained by defining,
\begin{equation}
\rho_{\mathrm{avg}} \rightarrow \beta_{\mathrm{SF}} \cdot \rho_{\mathrm{avg}}\, .
\end{equation}

The different sections in this chapter is organized as follows. We initiate our discussion with Sec.~\ref{sec:1} by addressing the issue at probability level. Then with Sec.~\ref{sec:2} we extend our understanding with bi-events. Then in Sec.~\ref{sec:establish-rho} we communicate our findings, unraveling the C.L. with which the possible choices of combined DUNE + T2HK can address the issue of exploring Earth matter effect. Also, in Sec.~\ref{sec:prec-rho} we compute the precision that complementarity between DUNE and T2HK can attain, while Sec.~\ref{sec:work2-allowed-region} discusses tackling the inkling degeneracies in $(\rho_{\mathrm{avg}} - \delta_{\mathrm{CP}})$ and $(\rho_{\mathrm{avg}} - \theta_{23})$ planes.

\section{Bi-events in T2HK}
\label{work2:sec2}

In this section, we exhibit the bi-events in the canvas of neutrino-antineutrino appearance events for differentiating between the finite matter density and vacuum like scenario. For simulating T2HK events, we use the details mentioned in Sec.~\ref{subsec:T2HK} The values in Table~\ref{work2:table:one} are used for all the simulations unless mentioned otherwise. These values are at par with the current global best-fit values as given in Ref.~\cite{Esteban:2018azc,Capozzi:2021fjo,deSalas:2020pgw}. Further, wile discussing our results in IMO, we follow the modified expression for atmospheric mass-squared splitting ($\Delta m^{2}_{31}$) from Eq.~\ref{eq:imo-relation-with-nmo}.
\begin{table}[ht!]
 \begin{center}
\begin{tabular}{|c|c|c|}
\hline \hline
\textbf{Parameter} & \textbf{True Value}& \textbf{Range of Minimization} \\
\hline \hline
$\sin^{2}\theta_{12}$ & $0.307$& -\\
$\sin^{2}\theta_{13}$ & $0.085$& -\\
$\sin^{2}\theta_{23}$ & $0.5$& [0.4 , 0.6] \\
$\delta_{\mathrm{CP}}$ & [-$180^{\circ},180^{\circ}]$& [-$180^{\circ},180^{\circ}]$\\
$\Delta m^{2}_{21}$ & 7.4 $\times$ $10^{-5}\mathrm{eV}^{2}$& -\\
$\Delta m^{2}_{31}$ & 2.5 $\times$ $10^{-3}\mathrm{eV}^{2}$(NMO)& [2.36 , 2.64]\\
& -2.4 $\times$ $10^{-3}\mathrm{eV}^{2}$(IMO)& [-2.64 , -2.36]\\
$\rho_{\mathrm{avg}}$ & 2.8 g/cm$^{3}$ (T2HK) & -\\
& 2.86 g/cm$^{3}$ (DUNE) & -\\
\hline \hline
 \end{tabular}
 \caption{Benchmark values of oscillation parameters and their respective uncertain ranges that have been manoeuvred in the present chapter unless mentioned otherwise.}
\label{work2:table:one}
\end{center}
\end{table}

\begin{figure}[htb!]
\includegraphics[width=0.49\linewidth]{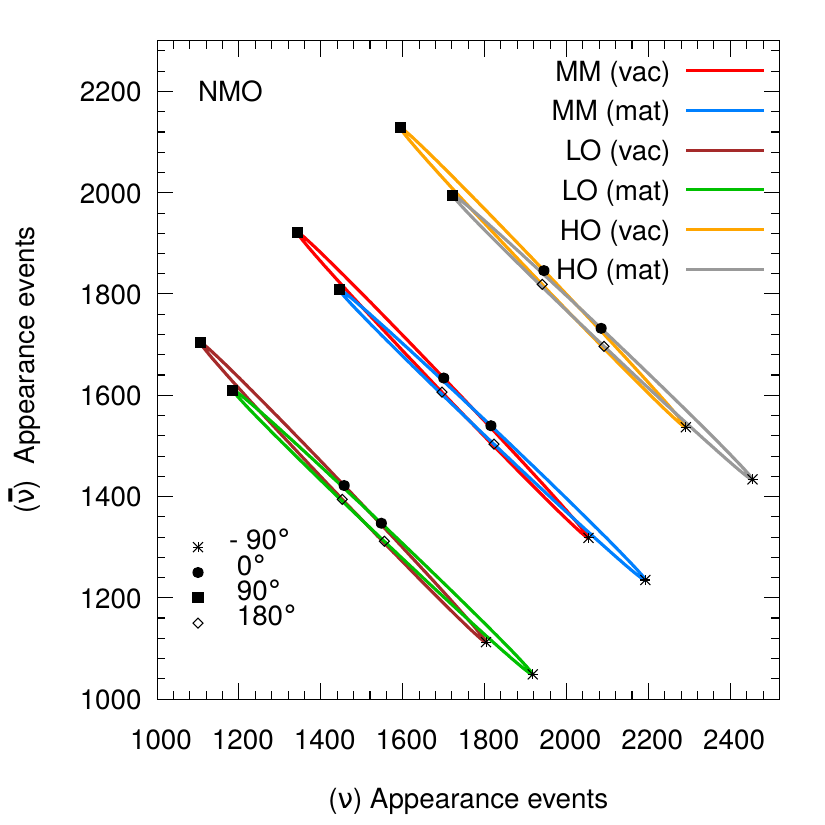}
\includegraphics[width=0.49\linewidth]{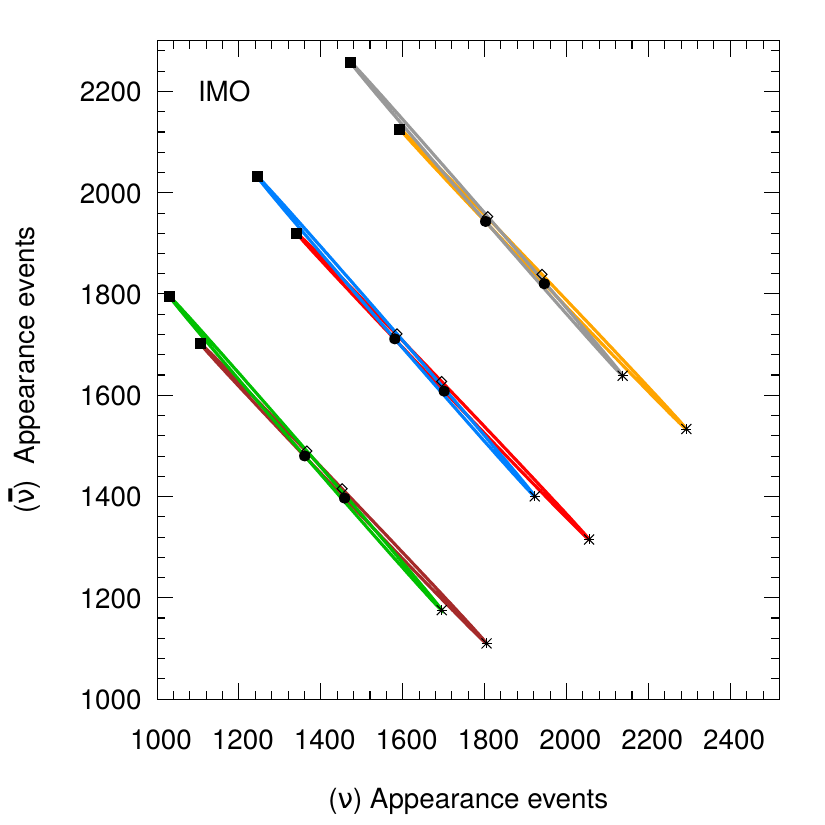}
\caption{Bi-events in the canvas of appearance neutrino and antineutrino modes, for T2HK, assuming 2431 kt$\cdot$MW$\cdot$years of exposure (splitting total run time of 10 years in the ratio of 1:3 between $\nu$ and $\bar{\nu}$) (refer to Table~\ref{table:contrasting-features-dune-t2hk}). The ellipses are obtained by varying $\delta_{\mathrm{CP}}$ in the whole range. The three sets are generated with three probable choices of $\sin^2\theta_{23}$: 0.44, 0.5, and 0.56, depicting illustrative choices in LO, MM, and HO, respectively in both NMO and IMO. }
\label{work2:fig:1}
\end{figure}

In Fig.~\ref{work2:fig:1}, bi-events are given for both the extremum: with finite matter density ($\rho_{\mathrm{avg}} = 2.8$ g/cm$^{3}$) and vacuum-like scenario ($\rho_{\mathrm{avg}} = 0$ g/cm$^{3}$) for T2HK, corresponding to three different choices of $\sin^{2}\theta_{23}$ in both NMO and IMO. As expected, the maximum number of events in $\nu$ ($\bar{\nu}$) mode is observed in its favorable region, i.e. NMO (IMO) and $\delta_{\mathrm{CP}} = -90^{\circ}$ (90$^{\circ}$) with finite matter density. Also, following the expression in Eq.~\ref{eq:2}, $P_{\nu_{\mu}\rightarrow \nu_{e}} \propto \sin^{2}\theta_{23}$ in the leading term. Therefore, we expect the number of appearance events to increase as we choose the $\sin^{2}\theta_{23}$ towards HO. Furthermore under both NMO and IMO assumption, we observe that the CP-conserving choices ($\delta_{\mathrm{CP}} = 0^{\circ},\,180^{\circ}$) have degenerate events corresponding to both vacuum-like and finite matter density scenarios, while, CP-violating choices ($\delta_{\mathrm{CP}} = \pm 90^{\circ}$) in finite matter density does not have overlapping with the vacuum-like bi-events. This is because, the most enhancement and reduction due to the presence of matter potential occurs around CP-violating values in both NMO and IMO. For an instance, presence of matter in NMO, $\nu$ mode, gets enhanced (reduced) when CP phase is $-90^{\circ} (90^{\circ})$ in the presence of finite matter density, while the nature is exactly opposite in antineutrino mode. Similarly, the nature reverses when we change the asssumption to IMO. Moreover, the angle of inclination of each bi-event with both the axes implies that correlation between neutrino and antineutrino modes will play a crucial role in establishing the matter effect.

\section{Sensitivity to establish $\rho_{\mathrm{avg}}$}
\label{sec:establish-rho}
%
\begin{figure}[htb!]
\centering
\includegraphics[width=\linewidth]{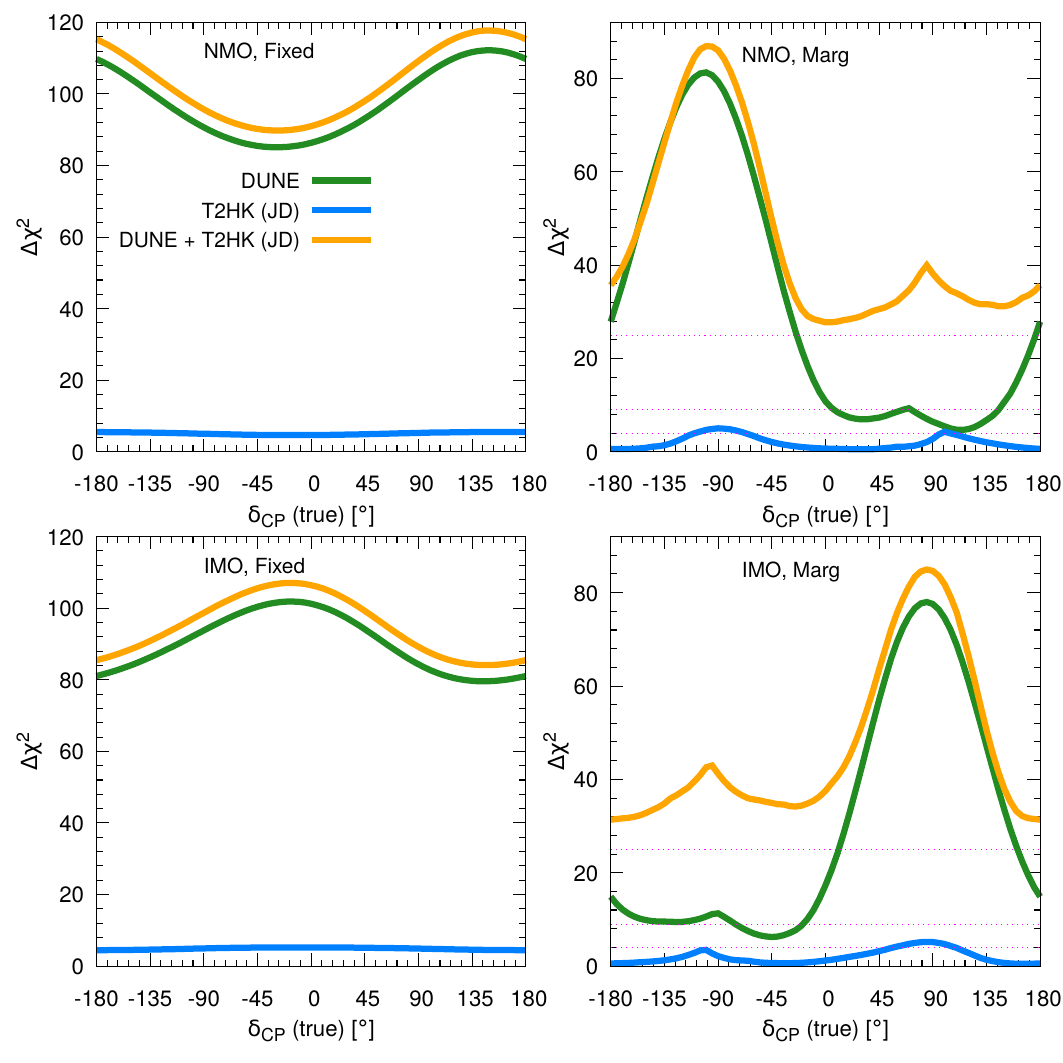}
\caption{The left (right) panels of figure shows the ability of DUNE, T2HK, and DUNE + T2HK to establish $\rho_{\mathrm{avg}}$ under fixed parameter (minimized) scenarios (refer text for further explanation), assuming true choices as: $\delta_{\mathrm{CP}}$ = -90$^{\circ}$, and NMO (IMO) in top (bottom) panels. Please note that the y-ranges are different between the left and right panels. Refer to Table~\ref{work2:table:one} for other benchmark values. \textbf{Complementarity in DUNE + T2HK is the only solution to achieve discovery-level significance in the sensitivity of $\rho_{\mathrm{avg}}$\,, irrespective of mass ordering, atmospheric mixing angle, and CP phase.}}
\label{work2:fig:2}
\end{figure}

\begin{table}[t!]
\centering
\begin{tabular}{|c|c|c|c|c|c|}
\hline
& \multicolumn{5}{c|}{$\Delta \chi^{2}_{\mathrm{ME}}$} \\
\cline{2-6}
$\delta_{\mathrm{CP}}^{\mathrm{true}}$ & Fixed & \multicolumn{4}{c|}{Marginalizing Over} \\
\cline{3-6}
& Parameter  & $\delta_{\mathrm{CP}}^{\mathrm{test}}$ & $\sin^{2}\theta_{23}^{\mathrm{test}}$ & $(\Delta m^2_{31})^{\mathrm{test}}$ & All \\
\hline\hline
$-90^{\circ}$  & 96 & 95.4 & 89 & 93.1 & 86 \\ \hline
$0^{\circ}$  & 91 & 31.4 & 81 & 91 & 27.7 \\
\hline
$90^{\circ}$  & 109 & 43.7 & 92.4 & 107 & 38.1 \\
\hline
$180^{\circ}$  & 115 & 45.6 & 104.2 & 115 & 35.6\\
\hline\hline
\end{tabular} 
\caption{In this table we provide $\Delta \chi^{2}_{\mathrm{ME}}$ for four illustrative choices of true $\delta_{\rm{CP}}$ in DUNE + T2HK. Under `All' we minimize over $\delta_{\mathrm{CP}}$, $\theta_{23}$,
and $\Delta m_{31}^2$, while under other columns we compute $\Delta \chi^{2}_{\mathrm{ME}}$ by partially minimizing for the purpose of singling out the effect of different parameters. When performing partial minimization, remaining parameters are fixed to the benchmark vaues in Table~\ref{work2:table:one}.}
\label{work2:table:2}
\end{table}
%
Fig.~\ref{work2:fig:2} displays the sensitivity with which we can establish the line-averaged constant Earth matter density ($\rho_{\mathrm{avg}}$) in DUNE, T2HK, and the combined DUNE + T2HK setup. Following the definition of $\chi^2$ in Sec.~\ref{sec:stats}, we define statistical significance for establishing $\rho_{\mathrm{avg}}$ as:
 \begin{equation}
\Delta \chi^2_{\rm DM}= \underset{\boldsymbol{\theta}}{\mathrm{min}}\left\{ \chi^2\left(\rho_{\mathrm{avg}}^{\mathrm{test}}= 0\right) - \chi^2\left(\rho_{\mathrm{avg}}^{\mathrm{true}} \neq 0 \right) \right\},  
\end{equation}
where the set $\boldsymbol{\theta} = {\delta_{\mathrm{CP}}}, \sin^{2}\theta_{23}, |\Delta m^{2}_{31}|, \pm \Delta m^{2}_{31}$ represents the uncertain parameters on which we minimize. In the fit we consider, vacuum hypothesis, while generate the data with finite matter density, keeping all other parameters fixed, in the fixed-paramter scenario. While in the minimized case, we vary $\vec{\lambda}$ in the defined uncertain range from Table~\ref{work2:table:one}. As expected if we know the values of each parameter in Nature, then indeed the significance will be higher in comparison with the minimized scenario. Thus, the higher value of $\Delta \chi^2_{\rm DM}$ in fixed-parameter  scenario for each setup. Further, following the discussion around Fig.~\ref{work2:fig:1}\,, when there is no uncertainty in oscillation parameter, the sensitivity to establish $\rho_{\mathrm{avg}}$ is least around $\delta_{\mathrm{CP}}$ = 0$^{\circ}$ (180$^{\circ}$) under NMO (IMO) assumption. Therefore, if we already all the oscillation parameters are known then, the significance to establish  $\rho_{\mathrm{avg}}$ will be very high, irrespective of $\delta_{\mathrm{CP}}$ under both NMO and IMO. But with the inclusion of uncertainty through minimization, we observe the sensitivity of DUNE deteriorating immensely around the unfavorable combination. However, the sensitivity in T2HK around CP-violating values remains highest. This is because in T2HK, the baseline is relatively much lesser than DUNE, which leads to less matter effect, thereby less contamination from extrinsic CP phase. This ensures better precision in $\delta_{\mathrm{CP}}$. Hence, the complementarity between DUNE and T2HK achieves sensitivity to establish matter effect by rejecting vacuum hypothesis with a greater than 5$\sigma$ for all values of CP pahse under both NMO and IMO. 

Table~\ref{work2:table:2}\,, gives an insight into the effect of oscillation parameter in detail. For favorable combination ($\delta_{\rm CP} = -90^{\circ}$)\,, the effect of uncertainty in CP phase is the least as this is fixed by both DUNE and T2HK, with sensitivity getting affected mostly (which in itself is very small) by the uncertainty in $\sin^{2}\theta_{23}$. Under the least favorable scenario ($\delta_{\rm CP} = 90^{\circ}$)\,, the uncertainty in CP phase affects predominantly the sensitivity in $\Delta \chi^{2}_{\mathrm{ME}}$\,, this is because although DUNE does not perform well, the excellent precision of T2HK helps. While the CP-conserving scenarios are worst hit, wherein the sensitivity to $\rho_{\mathrm{avg}}$ predominantly comes from T2HK.

To surmise, lack of significant matter effect in T2HK helps to establish better precision in $\delta_{\mathrm{CP}}$ and huge detector statistics ensures improved measurements in $\sin^{2}\theta_{23}$\,, both of which are relatively less  in DUNE. In contrast, DUNE provides the sufficient matter effect necessary for establishing the $\rho_{\mathrm{avg}}$ at high significance, but uncertainty in oscillation parameters deteriorates it. Therefore, combination is necessary for establishing $\rho_{\mathrm{avg}}$ over the entire landscape of CP phase.

\section{Precision in $\rho_{\mathrm{avg}}$}
\label{sec:prec-rho}
\begin{figure}[htb!]
\centering
\includegraphics[width=\linewidth]{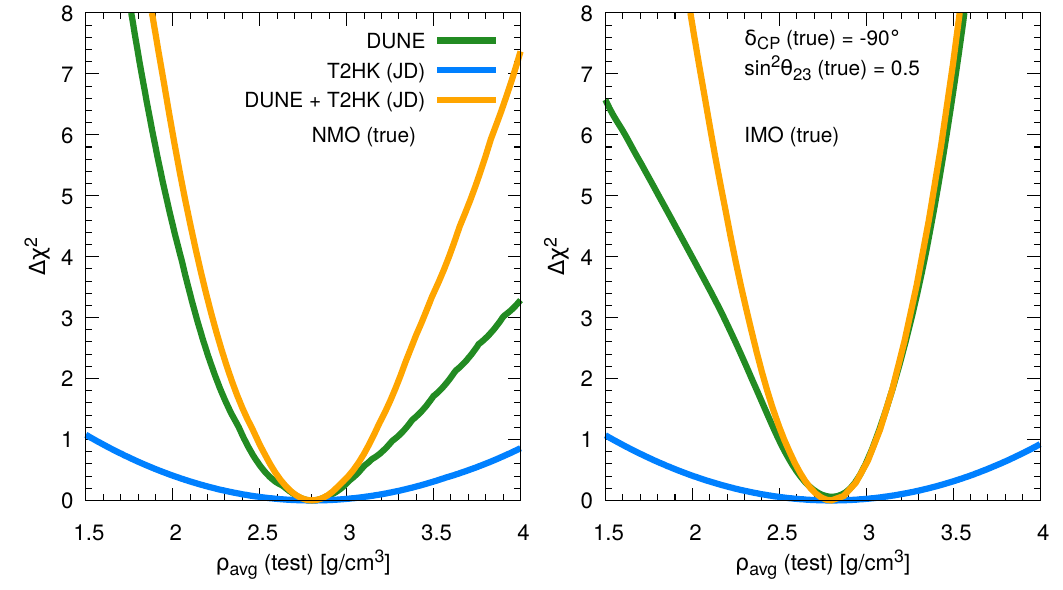} 
\caption{Precision in the measurement of  $\rho_{\mathrm{avg}}$. We follow Eq.~\ref{eq:chi2-achievable-precision-rho-avg}\,, and the benchmark values along with the allowed ranges for minimization from Table~\ref{work2:table:one}.  We assume $\sin^2\theta_{23}$ (true) = 0.5 and NMO (IMO) in the left (right) panel. \textbf{Complemenatrity between DUNE + T2HK provides degeneracy-free precision in $\rho_{\mathrm{avg}}$\,, thus improving sensitivity even for the unfavorable ranges for standalone DUNE or T2HK.}}
\label{work2:fig:3}
\end{figure}


Following the discussion in Sec.~\ref{sec:work1:precision}\,, we show in Fig.~\ref{work2:fig:3}\,, 
the affect of the combination of DUNE + T2HK in pecision measurements of $\rho_{\mathrm{avg}}$. We generate data assuming the benchmark value of $\rho_{\mathrm{avg}}$ mentioned in Table~\ref{work2:table:one}. $\chi^2_0$ is the minimum value of $\chi^2(\rho_{\mathrm{avg}})$\,, 
wherein $\rho_{\mathrm{avg}} \in [1.5:4]$ g/cm$^3$. Here, $\boldsymbol{\theta}$ = $\{\delta_{\mathrm{CP}}, \,\sin^2\theta_{23}, \, \Delta m_{31}^2\}$ and also the wrong mass ordering, is the set of oscillation parameters over which we perform minimization in the fit over the ranges as mentioned in Sec.~\ref{probability}. For quantifying the precision further, we use definition of $\Delta \chi^2_{\mathrm{PM}}$ from Eq.~\ref{eq:chi2-achievable-precision-rho-avg}, we find that the relative 1$\sigma$ precision in measurements of $\rho_{\mathrm{avg}}$ that DUNE + T2HK achieves assuming NMO is 11\%. This further improves on considering IMO to about 9\%. This precision is the best-case scenario since maximum achievable precision favors maximal CP violation. 

For the sake of completion, we also checked the achievable precision for other illustrative choices of $\delta_{\mathrm{CP}}$, which followed the nature of fig.~\ref{work1:fig:5}. The precision in  $\rho_{\mathrm{avg}}$ is predominantly dependent on the uncertainty in $\delta_{\mathrm{CP}}$ and $\sin^{2}\theta_{23}$. The narrow-band beam, positioned off-axis in T2HK, helps to receive maximum neutrino flux in comparison to the on-axis, wide-band beam in DUNE. Furthermore, the relatively less matter effect does not contaminate the intrinsic CP measurements. This leads to a much precise measurement of $\delta_{\mathrm{CP}}$ in T2HK than DUNE. The huge statistics in T2HK is because of the large detector (refer to Sec.~\ref{subsec:T2HK}), which also improves in measurements of $\sin^{2}\theta_{23}$ along with DUNE. However, the correct mass ordering is fixed by DUNE owing to large baseline. The interplay between DUNE and T2HK thus helps to achieve better precision in  $\rho_{\mathrm{avg}}$ even for CP-conserving values. Also, the complementarity between DUNE and T2HK immensely reduces the dependency of precision measurements in  $\rho_{\mathrm{avg}}$ on the uncertainties of $\delta_{\mathrm{CP}}$ and $\sin^{2}\theta_{23}$, leading to better achievable precision. This can be further testified by studying the allowed regions in the these planes, as anayzed in the following sections.

\section{Allowed regions in $(\rho_{\mathrm{avg}}^{\mathrm{test}} - \delta_{\mathrm{CP}}^{\mathrm{test}})$ and $(\rho_{\mathrm{avg}}^{\mathrm{test}} - \sin^2\theta_{23}^{\mathrm{test}})$ planes}
\label{sec:work2-allowed-region}
\begin{figure}[htb!]
\centering
\includegraphics[width=\textwidth,height=\textwidth]{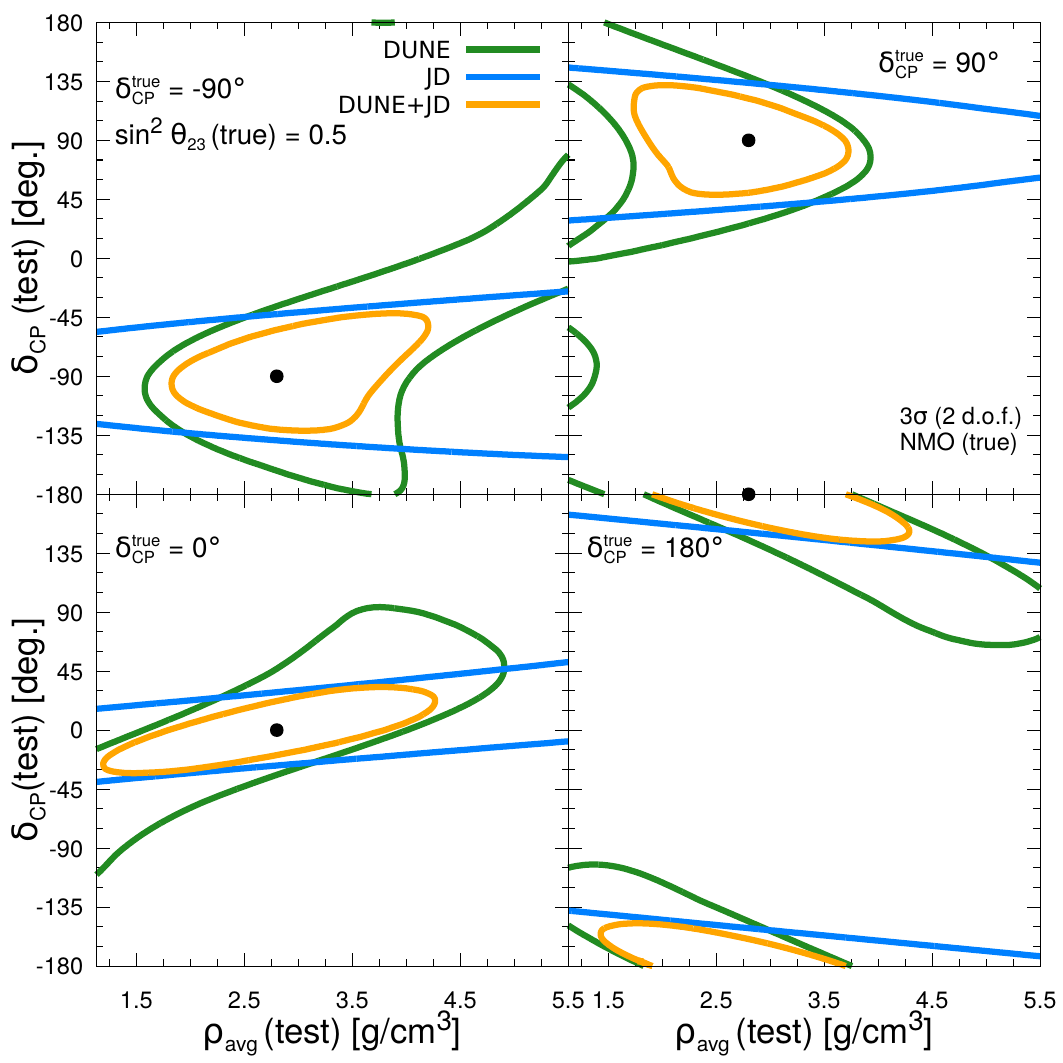}
\caption{Allowed regions of matter density, $\rho_{\mathrm{avg}}$ and CP-violating phase, $\delta_{\mathrm{CP}}$. The illustrative true values of $\delta_{\mathrm{CP}}$ are taken as $\pm 90^{\circ}, 0^{\circ}, 180^{\circ}$ and the atmospheric mixing angle is illustratively chosen to be maximal mixing (0.5). The test-statistic is minimized over $\sin^2\theta_{23}$, $\Delta m_{31}^2$, and the mass ordering; see Eq.~\ref{eq:imo-relation-with-nmo}. See Section~\ref{sec5.3} and~\ref{sec:work2-allowed-region} for details. \textbf{Complementarity in DUNE + T2HK helps to achieve a highly stringent allowed region in $(\rho_{\mathrm{avg}} (\mathrm{test}) - \delta_{\mathrm{CP}} (\mathrm{test}))$ plane, otherwise unattainable by standalone experiments. } }
\label{work2:fig:5}
\end{figure} 

\begin{figure}[htb!]
\centering
\includegraphics[width=\textwidth,height=0.4\textwidth]{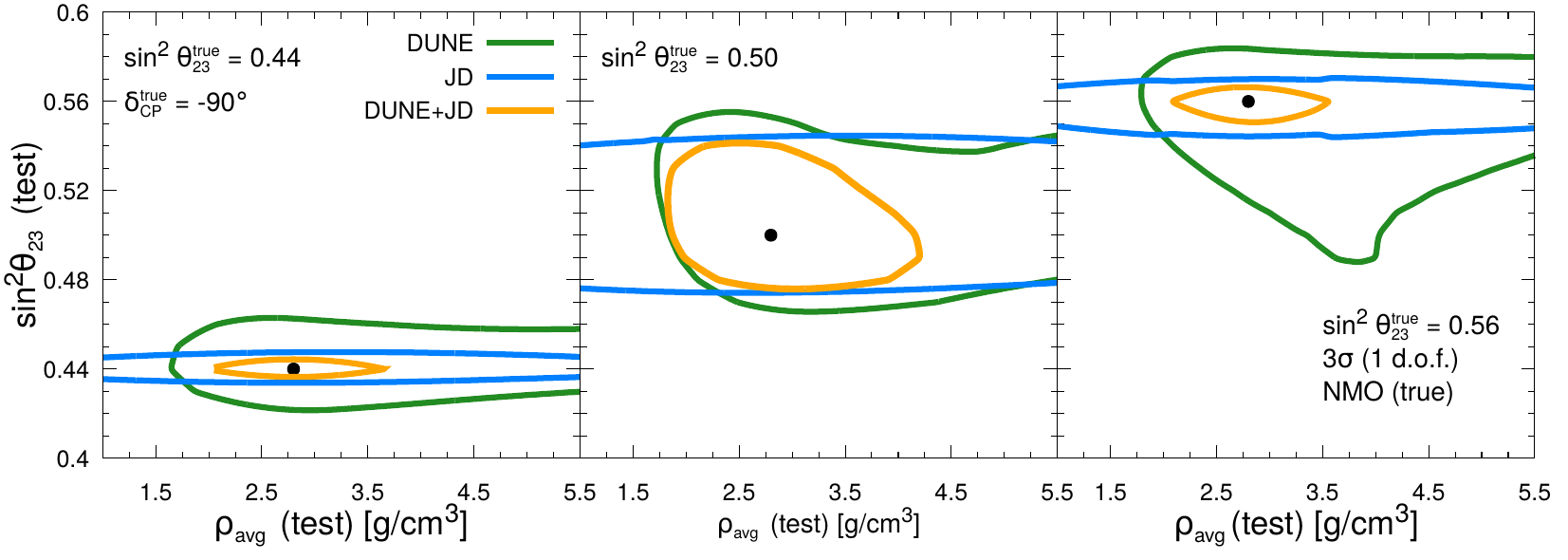}
\caption{Allowed regions of matter density, $\rho_{\mathrm{avg}}$ and atmospheric mixing angle, $\sin^2\theta_{23}$. The true values of $\sin^{2}\theta_{23}$ is same as Fig.~\ref{work1:fig:4} and the CP phase is illustratively chosen to be maximally violated ($-90^{\circ}$). The test-statistic is minimized over $\delta_{\mathrm{CP}}$, $\Delta m_{31}^2$, and the mass ordering; see Eq.~\ref{eq:imo-relation-with-nmo}. See Section~\ref{sec5.3} and~\ref{sec:work2-allowed-region} for details. \textbf{DUNE + T2HK when combined, reduces the allowed region by the standalone experiments by manifold}, while individual DUNE and T2HK are insensitive to change in the octant of $\sin^{2}\theta_{23}$, combined exhibits most stringent contour in LO, followed by HO, and then MM. See text in Sec.~\ref{sec:work2-allowed-region} for details.}
\label{work2:fig:6}
\end{figure}
In this section, we show the allowed regions in test $\rho_{\mathrm{avg}}-\delta_{\mathrm{CP}}$ and $\rho_{\mathrm{avg}}-\theta_{23}$ planes, study correlations among them, and reflect upon its consequential effect on studying the sensitivity for establishing matter effect ($\Delta \chi^2_{\mathrm{ME}}$ ) in standalone DUNE, T2HK, and the combined DUNE + T2HK experiments. 

In order to obtain the allowed region in Fig.~\ref{work2:fig:5}, we scan the test-statistic for the values of $\delta_{\mathrm{CP}}$ and $\rho_{\mathrm{avg}}$ in the range $[-180^{\circ}:180^{\circ}]$ and $[1:5.5]$, respectively. We perform minimization over $\sin^2\theta_{23}$ and $\Delta m_{31}^2$ in the fit. We assume true values of NMO and MM solutions of $\sin^2\theta_{23}$ in the fit. We observe that owing to less baseline, the sensitivity of T2HK along the x-axis, i.e. $\rho_{\mathrm{avg}}$ is very weak, pertaining to allowed region extending throughout the scanned canvas. However, the allowed region by T2HK in $\delta_{\mathrm{CP}}$ axis is highly stringent. Moreover, the almost parallel axes (both major and minor) of the allowed regions in T2HK signifies that there is very less correlation between $\rho_{\mathrm{avg}}$ and $\delta_{\mathrm{CP}}$. However, DUNE exhibits a diagonal behavior, thus exhibiting high correlation as explained elaborately in Sec.~\ref{sec5.3}. Large matter effect in DUNE, leads to better allowed regions for $\rho_{\mathrm{avg}}$ than T2HK. Therefore, the interplay between DUNE and T2HK helps in achieving such a small allowed uncertainty around the true  values of $\rho_{\mathrm{avg}}$ and $\delta_{\mathrm{CP}}$. Again, following the discussion in Sec.~\ref{sec5.3}\,, maximally violated CP phase achieves better precision than CP-conserving values in DUNE + T2HK as well.

Next, in Fig.~\ref{work2:fig:6}, we exhibit the allowed test-statistic in the $\rho_{\mathrm{avg}}-\sin^2\theta_{23}$ plane to display the allowed regions for three probable choices of $\sin^2\theta_{23}$, assuming true $\rho_{\mathrm{avg}}$ as mentioned in Table~\ref{work2:table:one}.  While obtaining the $\Delta \chi^2$, we perform minimization over $\delta_{\mathrm{CP}}$, $\Delta m_{31}^2$, and the wrong mass ordering in the fit by assuming NMO in the data. For generating the contours, we scan over $\sin^2\theta_{23}$ and $\rho_{\mathrm{avg}}$ in the range $[0.4:0.6]$ and $[1:5.5]$ g/cm$^3$ respectively. Likewise in Fig.~\ref{work2:fig:5}, we observe that due to large matter effect in DUNE, it faces contamination from fake CP asymmetry, while on the other hand, the large matter effect also facilitates fixing the mass ordering. However, T2HK in contrast, precisely measures $\sin^{2}\theta_{23}$, because of huge disappearance statistics, as visible from the relatively lesser stretch of contours along the y-axis. This leads to a fruitful complementarity between DUNE and T2HK, thus smaller allowed regions in $(\rho_{\mathrm{avg}}^{\mathrm{test}} - \sin^2\theta_{23}^{\mathrm{test}})$ plane. Relatively, most stringent regions are obtained if data is generated with $\sin^2\theta_{23}$ in LO, followed by HO. Allowed region in MM solutions is higher than other choices of $\sin^2\theta_{23}$. This will be further explained in Chapter~\ref{sec:ch7}.

\section{Summary }
\label{work2:conc}

The ubiquitous phenomena of neutrino oscillation, has been the torch-bearer of physics beyond the SM. It has undergone innumerable radical changes, paving its way to accurate and precise measurement of the fundamental oscillation parameters. In the previous chapter we observed that a detailed understanding of Earth's Matter effect is inevitable to correctly analyze the data from the upcoming high-precision long-baseline experiments to resolve the remaining fundamental unknowns such as neutrino mass ordering, leptonic CP violation and precision measurements of the oscillation parameters. 

DUNE with 1300 km baseline has significant matter effect and can measure $\delta_{\mathrm{CP}}$ and $\theta_{23}$ with reasonable precision exploiting the information on oscillation pattern at several $L/E$ values. On the other hand, with a relatively shorter baseline and high statistics JD offers an unmatched sensitivity to the $\delta_{\mathrm{CP}}$ free from matter effect. In this chapter, for the first time, we show how the complementary features between DUNE + T2HK can play a crucial role to establish the Earth's matter effect at more than 5$\sigma$ C.L. for any values of oscillation parameters. The complementary informations coming from DUNE and T2HK setups also play an important role to provide a high-precision measurement of $\rho_{\mathrm{avg}}$ and to reduce the allowed regions in $\left(\rho_{\mathrm{avg}} - \delta_{\mathrm{CP}}\right)$ and $\left(\rho_{\mathrm{avg}} - \theta_{23}\right)$ planes considerably. We find that with combined DUNE + T2HK, the relative 1$\sigma$ precision in the measurements of $\rho_{\mathrm{avg}}$ improves to 11\% in NMO, while increasing further to a 9\% in IMO. We also witness the essential complementarity between DUNE and T2HK for furnishing degeneracy-free measurements in the line-averaged constant Earth matter density.

\chapter{ Atmospheric Parameters with DUNE in Light of Current Neutrino Oscillation Data}
\label{sec:Ch6}

A deeply-relevant and much-awaited result concerning neutrinos in recent times is the {\it hint} for violation of the CP symmetry in the leptonic sector. The T2K collaboration~\cite{T2K:2011qtm} in their 2019 results have shown that their neutrino and antineutrino appearance data point towards CP being near-maximally violated $i.e.$ $\vert\sin\delta_{\rm CP}\vert$ is close to 1. They obtain a best-fit value of $\delta_{\rm CP}$ at $252^\circ$ while the CP-conserving values of $\delta_{\rm CP} = 0^\circ\,, \,\pm\, 180^\circ$ are ruled out at $95\%$ confidence level (C.L.). They also report a preference for the normal mass ordering (NMO) over the inverted mass ordering (IMO) at nearly $68\%$ confidence level. NMO and $\delta_{\rm CP} = 270^\circ$ is in fact one of the most favorable parameter combinations for which early hints regarding mass ordering and CP violation can be expected from the currently running long-baseline accelerator experiments~\cite{Prakash:2012az, Agarwalla:2012bv}. The same set of measurements are also being carried out by the NO$\nu$A experiment~\cite{NOvA:2007rmc,NOvA:2004blv} which operates at a longer baseline with more energetic neutrinos. The recent results from NO$\nu$A~\cite{NOvA:2021nfi} also show a preference for NMO, but their  best-fit to $\delta_{\rm CP}$ is not in conjunction with T2K. NO$\nu$A's $\delta_{\rm CP}$ best-fit value of $148^\circ$ is $2.5\sigma$ away from T2K's best-fit. However, the two experiments agree on $\delta_{\rm CP}$ measurements when they assume IMO to be true -- each reporting a best-fit value of around $270^\circ$. The tension between these two data sets is not yet at a statistically significant level and we need to wait for further data from T2K and NO$\nu$A to see if this tension persists. In any case, a $5\sigma$ {\it discovery} of any of the current unknowns in neutrino oscillation physics does not seem to be within the reach of either of these experiments~\cite{Agarwalla:2012bv}. Nonetheless, these results are quite important and play an essential role in the global fit studies.

Fig.~\ref{fig:current-allowed-osc-params} summarizes our current understanding of the six neutrino oscillation parameters in the standard three-neutrino framework. It confirms that we have already attained a remarkable precision on solar oscillation parameters ($\Delta m^2_{21}$ and $\sin^2\theta_{12}$), atmospheric mass-splitting ($\Delta m^2_{31}$), and reactor mixing angle ($\theta_{13}$). In this figure, we compare the $1\sigma$ and $3\sigma$ allowed regions of the oscillation parameters that have been calculated by doing a combined analyses of the existing global oscillation data~\cite{deSalas:2020pgw,Capozzi:2021fjo,Esteban:2020cvm}. All the three studies find that the earlier tension between Solar and KamLAND data has been reduced considerably after incorporating the recent results from Super-K Phase IV 2970 days of solar data (energy spectra and day-night asymmetry)~\cite{Super-Kamiokande:2016yck}. Additionally, both Esteban $et~ al.$~\cite{Esteban:2020cvm, NuFIT} and Capozzi $et ~al.$~\cite{Capozzi:2021fjo} also consider the recent Super-K Phase IV atmospheric data~\cite{yasuhiro_nakajima_2020_3959640}. Note that, while Esteban $et~ al.$ and de Salas $et~ al.$ quote the values of atmospheric mass-splitting in terms of $\Delta m^{2}_{31}$, Capozzi $et~ al.$ express it in terms of $\Delta m^{2} = m^{2}_{3}- (m^{2}_{1} + m_{2}^{2})/2$, where $\Delta m^{2}_{31}$ = $\Delta m^{2} + \Delta m^{2}_{21}/2$ for both NMO and IMO.

A novel aspect of Fig.~\ref{fig:current-allowed-osc-params} is that all three global fits now rule out $\delta_{\rm CP} \in \left[ 0, \sim135^\circ \right]$ at $3\sigma$ confidence level and $\delta_{\rm CP} \in \left[ 0, \sim180^\circ \right]$ at $1\sigma$ confidence level, while predicting the best-fit value to lie somewhere in the range $\left[ 200^\circ,~230^\circ \right]$. The constraint in $\delta_{\rm CP}$ is  essentially due to the data from T2K and NO$\nu$A as discussed earlier. As far as the neutrino mass ordering is concerned, all three global fits show preference for NMO, ruling out IMO at close to $2.5\sigma$~\cite{Esteban:2020cvm,NuFIT,Capozzi:2021fjo,deSalas:2020pgw}. Therefore, for the sake of simplicity, in this work, we show our results assuming NMO both in data and fit. We observe that the results do not change much for IMO.

In Ref.~\cite{Capozzi:2021fjo}, the authors find a preference at 1.6$\sigma$ for $\theta_{23}$ in the LO with respect to the secondary best-fit in HO. They obtain a best-fit value of $\sin^{2}\theta_{23}$ = 0.455 in the LO assuming NMO and disfavor maximal $\theta_{23}$ mixing at $\sim$1.8$\sigma$. However, there is a slight disagreement between the three global fit studies as far as the measurement of $\theta_{23}$ is concerned (see top right panel in Fig.~\ref{fig:current-allowed-osc-params}). In Ref. \cite{deSalas:2020pgw}, de Salas $et~ al.$ find a best-fit in the HO around $\sin^2\theta_{23} \sim 0.57$ assuming NMO, while Capozzi $et~ al.$~\cite{Capozzi:2021fjo} and Esteban $et~ al.$~\cite{Esteban:2020cvm} obtain the best-fit around $\sin^{2}\theta_{23} \sim 0.45$ in the LO. This difference in the best-fit value of $\sin^2\theta_{23}$ is probably due to the recent Super-K Phase I-IV 364.8 kt$\cdot$yrs of atmospheric data~\cite{yasuhiro_nakajima_2020_3959640} that only Capozzi $et~ al.$ and Esteban $et~ al.$ consider in their latest analyses.  

The issue of resolution of octant (if $\sin^{2}2\theta_{23} \neq 1$) have far reaching consequences as far as the models explaining neutrino masses and mixings are concerned. Some examples of such models are quark-lepton complementarity~\cite{Raidal:2004iw,Minakata:2004xt,Ferrandis:2004vp,Antusch:2005ca}, $A_{4}$ flavor symmetry~\cite{Ma:2002ge,Ma:2001dn,Babu:2002dz,Grimus:2005mu}, and $\mu$-$\tau$ permutation symmetry~\cite{Fukuyama:1997ky,Mohapatra:1998ka,Lam:2001fb,Harrison:2002et,Ghosal:2003mq}. The $\mu$-$\tau$ permutation symmetry is of particular interest since the current oscillation data strongly indicates that this symmetry is not exact in Nature. A high-precision measurement of 2-3 mixing angle and the measurement of its octant are inevitable to disclose the pattern of deviations from the above mentioned symmetries, which in turn will help us to explain tiny neutrino masses and one small and two large mixing angles in the lepton sector~\cite{Xing:2014zka,Xing:2015fdg}. It has also been shown that without an accurate measurement of $\theta_{23}$, a precise measurement of $\delta_{\rm CP}$ will not be possible~\cite{Minakata:2013eoa,Singh:2023bky}.

There are several studies in the literature addressing the issues related to the 2-3 mixing angle in the context of various neutrino oscillation experiments.  In this chapter, we analyze in detail the sensitivities of the next generation, high-precision long-baseline neutrino oscillation experiment DUNE~\cite{DUNE:2021cuw} to resolve its octant at high confidence level in light of the current neutrino oscillation data. We also study how much improvement DUNE can offer in the precision measurements of $\sin^{2}\theta_{23}$ and $\Delta m^2_{31}$ as compared to their current precision. While estimating the achievable precision on these parameters in DUNE, we also quantify the contribution from individual appearance and disappearance channels and demonstrate the importance of having both neutrino and antineutrino data.

The layout of this chapter is as follows. In Sec.~\ref{probability}, we discuss the potential of DUNE's baseline and energy in establishing deviation from maximal $\theta_{23}$ at the level of probabilities. In Sec.~\ref{sec:work3:octant}, we quantify the performance of DUNE to settle the correct octant of $\theta_{23}$, and to precisely measure the values of atmospheric oscillation parameters $-$ $\sin^{2}\theta_{23}$ and $\Delta m^{2}_{31}$. We also address several issues which are relevant to achieve the above mentioned goals. In Sec.~\ref{conclusion}, we summarize our findings and make concluding remarks.

\section{Discussion at the level of probabilities}
\label{probability}

\begin{table}[htb!]
		\label{tableprecision}

		\centering
		\resizebox{\columnwidth}{!}{%
			\begin{tabular}{|c|c|c|c|c|c|}
				\hline \hline
				\multirow{2}{*}{\textbf{Parameter}} & \multirow{2}{*}{\textbf{Best-fit}} & \multirow{2}{*}{\textbf{1$\sigma$ range}} & \multirow{2}{*}{\textbf{2$\sigma$ range}} & \multirow{2}{*}{\textbf{3$\sigma$ range}} & \textbf{Relative 1$\sigma$}\\
				& & & & &\textbf{Precision (\%)}\\
				\hline \hline
				$\Delta m^2_{21}/10^{-5}$ $\mathrm{eV^{2}}$ & 7.36 & 7.21 - 7.52 & 7.06 - 7.71 & 6.93 - 7.93 & 2.3\\
				\hline
				$\sin^{2}\theta_{12}/10^{-1}$ & 3.03 & 2.90 - 3.16 & 2.77 - 3.30 & 2.63 - 3.45 & 4.5\\
				\hline
				$\sin^{2}\theta_{13}/10^{-2}$ & 2.23 & 2.17 - 2.30 & 2.11 - 2.37 & 2.04 - 2.44 &  3.0\\
				\hline
				$\sin^2\theta_{23}/10^{-1}$ & 4.55 & 4.40 - 4.73 & 4.27 - 5.81 & 4.16 - 5.99 & 6.7 \\
				\hline
				$\Delta m^2_{31}/10^{-3}$ $\mathrm{eV^2}$ & 2.522 & 2.490 - 2.545 & 2.462 - 2.575 & 2.436 - 2.605 & 1.1\\
				\hline
				$\delta_{\text{CP}}$/$^\circ$ & 223  & 200 - 256 & 169 - 313 & 139 - 355 & 16\\
				\hline \hline
			\end{tabular}
			}
			\caption{Current benchmark values of the oscilation parameters and their corresponding ranges that we consider in our study assuming NMO following Ref.~\cite{Capozzi:2021fjo}.}
			\label{work3:table:one}
	\end{table}

In the three-neutrino framework, the flavor eigenstates $\vert \nu_{\alpha}\rangle~\left(\alpha = e, \mu, \tau \right)$ and the mass eigenstates $\vert \nu_{i} \rangle ~\left(i=1,2,3\right)$ are connected by the $3\times3$ unitary Pontecorvo-Maki-Nakagawa-Sakata (PMNS) matrix $U$:
\begin{equation}
\vert \nu_{\alpha}\rangle  = \sum_{i} U^\ast_{\alpha i}\vert\nu_{i} \rangle ~~~ {\rm and} ~~~
\vert \bar{\nu}_{\alpha} \rangle = \sum_{i} U_{\alpha i}\vert \bar{\nu}_{i} \rangle\, .
\end{equation}
Following the standard Particle Data Group convention~\cite{ParticleDataGroup:2022pth}, the vacuum PMNS matrix $U$ is parametrized in terms of the three mixing angles ($\theta_{23}$, $\theta_{13}$, $\theta_{12}$) and one Dirac-type CP phase ($\delta_{\mathrm{CP}}$). The probability that a neutrino, with flavor $\alpha$ and energy $E$, after travelling a distance $L$, can be detected as a neutrino with flavor $\beta$ is given by
\begin{equation}
P_{\alpha\beta} = \delta_{\alpha\beta} - 4 \sum_{j>i} \mathcal{R} \left( U^\ast_{\alpha j}U_{\beta j} U_{\alpha i} U^\ast_{\beta i}\right)\sin^2\frac{\Delta m^2_{ji} L}{4E} + 2 \sum_{j>i} \mathcal{I} \left( U^\ast_{\alpha j}U_{\beta j} U_{\alpha i} U^\ast_{\beta i}\right)\sin\frac{\Delta m^2_{ji} L}{2E}\, ,
\end{equation}
where, $\Delta m^2_{ji} = m^2_{j} - m^2_{i}$.
Approximate analytical expressions for oscillation probabilities including matter effect have been derived in Ref.~\cite{Akhmedov:2004ny}, retaining terms only up to second order in the small parameters $\sin^2\theta_{13}$ and $\alpha \left(\equiv \Delta m^2_{21}/\Delta m^2_{31}\right)$. The analytical expression for muon neutrino survival probability ($P_{\mu \mu}$) under the constant matter density approximation is given in Eq. 33 of Ref.~\cite{Akhmedov:2004ny}. Considering the current best-fit values of oscillation parameters (see second column in Table~\ref{work3:table:one}), we have $\sin^2\theta_{13}\approx 0.02$, $\alpha \approx 0.03$, $\alpha\sin\theta_{13}\approx 0.004$, and $\alpha^2\approx 0.0008$.

The main sensitivity to settle the octant of $\theta_{23}$ stems from $\nu_{\mu} \rightarrow \nu_{e}$ appearance channel ($P_{\mu e}$), which when expressed up to first order in $\alpha\sin\theta_{13}$ is given by (ignoring the term $\propto$ $\alpha^{2}$ and $\cos\theta_{13}\approx 1$)
\beqa \label{eqpmue}
P_{\mu e} \approx N \sin^2\theta_{23} + O \sin2\theta_{23}\cos\left(\Delta + \delta_{\mathrm {CP}}\right)\, ,
\eeqa
where,
\beqa \label{eqN}
N = 4\sin^2\theta_{13}\frac{\sin^2(\hat A - 1) \Delta}{(\hat A-1)^2}\, ,
\eeqa
\beqa \label{eqO}
O = 2\alpha\sin\theta_{13}\sin2\theta_{12}\frac{\sin\hat A \Delta }{\hat A}\frac{\sin(\hat A -1)\Delta }{\hat A-1}\, .
\eeqa
%
Note that the first term in Eq.~\ref{eqpmue} is sensitive to octant of $\theta_{23}$, while the second term is sensitive to CP phase $\delta_{\mathrm {CP}}$. This leads to an octant\,-\,$\delta_{\mathrm {CP}}$ degeneracy in the measurements made via appearance channel. However, this degeneracy can be resolved with the help of balanced neutrino and antineutrino data in appearance mode as discussed for the the first time in Ref.~\cite{Agarwalla:2013ju}. Since, both the terms in $P_{\mu e}$ contain information on $\theta_{23}$ (see Eq.~\ref{eqpmue}), they contribute towards establishing deviation from maximal $\theta_{23}$ (see discussion around Fig.~\ref{work3:fig:4}) and to precisely measure the value of $\sin^{2}\theta_{23}$ (see discussion in Sec.~\ref{precisionapp} and Fig.~\ref{work3:fig:3}). 

\section{Octant of $\theta_{23}$}
\label{sec:work3:octant} 
In this subsection, we study the potential of DUNE to resolve the octant of 2-3 mixing angle. We define $\Delta\chi^2_{\rm octant}$ in the following fashion
\begin{equation}
\Delta \chi^2_{\text{\rm octant}}(\zeta) = \underset{\boldsymbol{\theta}=(\delta_{\mathrm{CP}},~\Delta m^{2}_{31})}{\mathrm{min}}\left\{ \chi^2\left(\zeta^{\mathrm{test}}\right) - \chi^2\left(\zeta^{\mathrm{true}} \right)\right\}. 
\label{work3:eq:chi-octant} 
\end{equation}
\begin{figure}[htb!]
	\centering
	\includegraphics[width=0.49\linewidth]{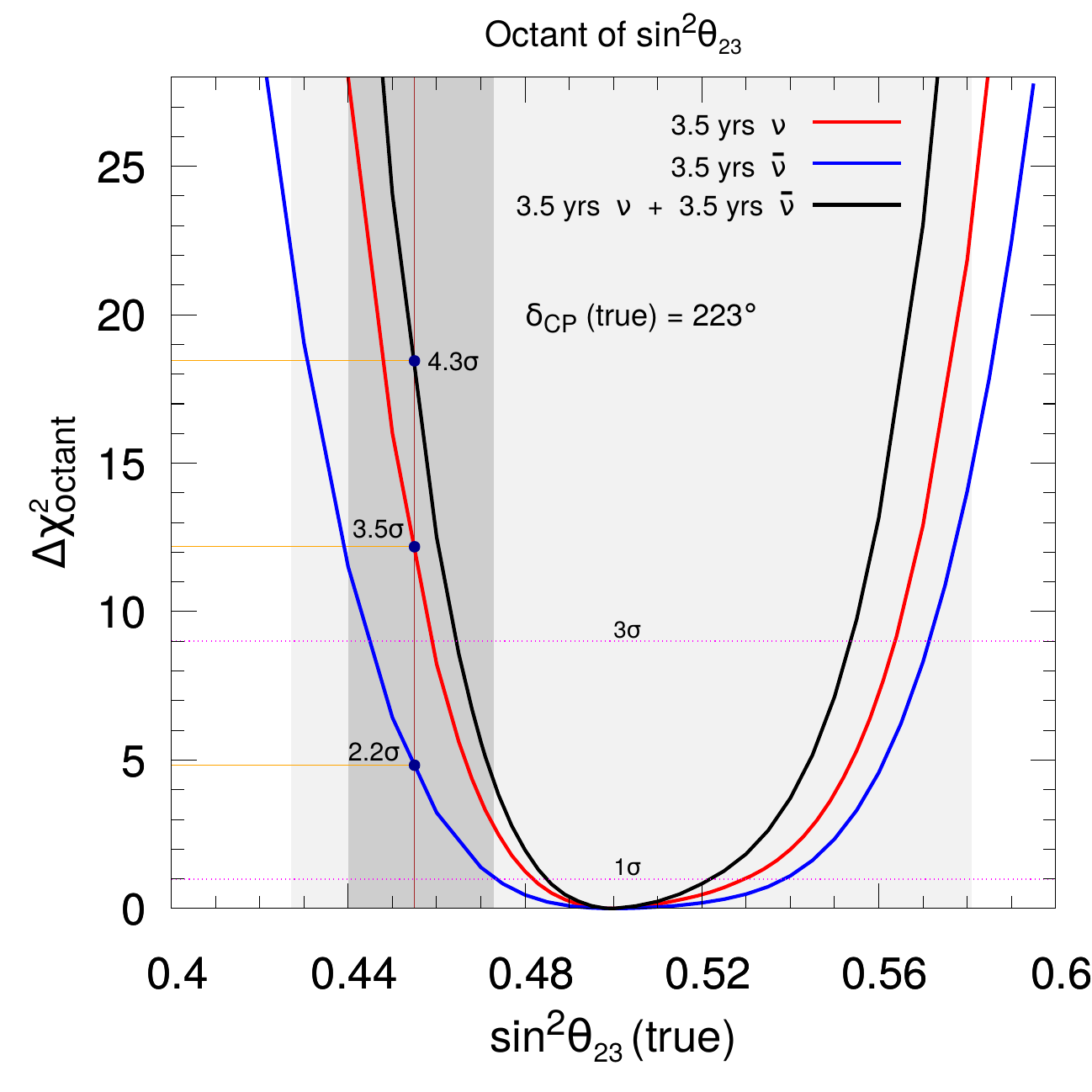}
	\includegraphics[width=0.49\linewidth]{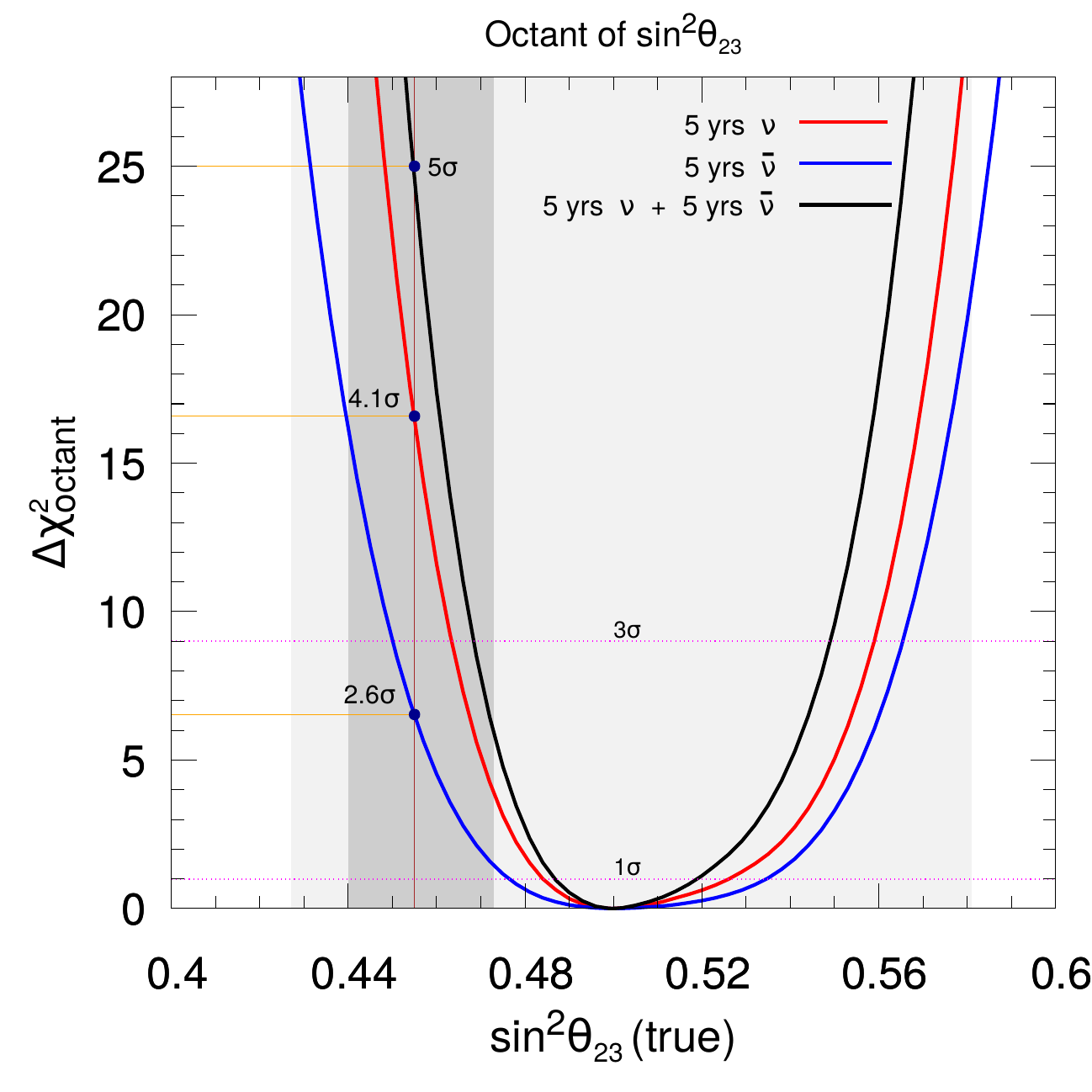}
	\caption{Octant discovery potential of DUNE as a function of true $\sin^2\theta_{23}$. In the fit, we marginalize over $\Delta m^{2}_{31}$ and $\delta_{\rm CP}$, keeping others fixed at their benchmark values (refer to Table~\ref{work3:table:one}), following Eq.~\ref{work3:eq:chi-octant}. For illustrative purposes, the line drawn at the best-fit value of $\sin^{2}\theta_{23} = 0.455$, projects statistical confidence for each curve.\textbf{While both $\nu$ and $\bar{\nu}$ statistics is essential in excluding the wrong octant solutions, $\nu$ dominates.}}
	\label{work2:fig:2}
\end{figure}

Here, $\zeta^{\text{true}}$ is the true value of $\sin^2\theta_{23}$ in lower or upper octant  and $\zeta^{\text{test}}$ is the test value of $\sin^2\theta_{23}$ in opposite octant including the test value of $\sin^2\theta_{23}= 0.5$. $\delta_{\mathrm{CP}}$ and $\Delta m^{2}_{31}$ are the oscillation parameters over which $\Delta\chi^2_{\mathrm{octant}}$ has been minimized in the fit.

In Fig.~\ref{work2:fig:2}, we show $\Delta\chi^2_{\rm octant}$ as a function of true $\sin^2\theta_{23}$, where, for each true value of $\sin^{2}\theta_{23}$, we consider test values of $\sin^2\theta_{23}$ in its present 3$\sigma$ range in the opposite octant including $\sin^2\theta_{23}\, \mathrm{(test)}\,= 0.5$ in the fit and pick up the minimum value of $\Delta \chi^{2}_{\mathrm{octant}}$. In the fit, we minimize over the present 3$\sigma$ range of $\Delta m^{2}_{31}$ and $\delta_{\rm CP}$, while keeping rest of the oscillation parameters fixed at their present best-fit values as shown in Table~\ref{work3:table:one}. The dark (light)-shaded grey area shows the currently allowed $1\sigma$ $(2\sigma)$ region in $\sin^{2}\theta_{23}$ as obtained in the global fit study~\cite{Capozzi:2021fjo} assuming NMO with the best-fit value of $\sin^{2}\theta_{23} = 0.455$ as shown by vertical brown line. The horizontal orange lines show the sensitivity (experessed in $\sigma = \sqrt{\Delta\chi^2_{\rm octant}}$) due to individual runs for the current best-fit value of $\sin^2\theta_{23}$.

It is evident from Fig.~\ref{work2:fig:2} that the combined data from neutrino and antineutrino modes significantly improve the result by breaking the octant - $\delta_{\rm CP}$ degeneracy as discussed before in~\cite{Agarwalla:2013ju}. Assuming the current best-fit values in Table~\ref{work3:table:one} as their true choices and with true NMO, the octant of $\theta_{23}$ can be settled in DUNE at 4.3$\sigma$ (5$\sigma$) using 336 (480) kt$\cdot$MW$\cdot$years of exposure which corresponds to 7 years (10 years) of data taking with equal sharing in neutrino and antineutrino modes. A 3$\sigma$ (5$\sigma$) resolution of $\theta_{23}$ octant is possible in DUNE with an exposure of 336 kt$\cdot$MW$\cdot$years if the true value of $\sin^{2}\theta_{23}$ $\lesssim~0.462~(0.450)$ or $\sin^{2}\theta_{23}$ $\gtrsim~0.553~(0.569)$ assuming true NMO and $\delta_{\rm CP}~(\rm true)~ = 223^{\circ}$. The same is possible with 480 kt$\cdot$MW$\cdot$years of exposure if the true value of $\sin^{2}\theta_{23}$ $\lesssim~0.466~(0.454)$ or $\sin^{2}\theta_{23}$ $\gtrsim~0.548~(0.565)$.

\subsection{Precision measurements of $\sin^2\theta_{23}$ and $\Delta m^2_{31}$}
\label{precisionapp}

\begin{figure}[htb!]
\includegraphics[width=0.49\linewidth]{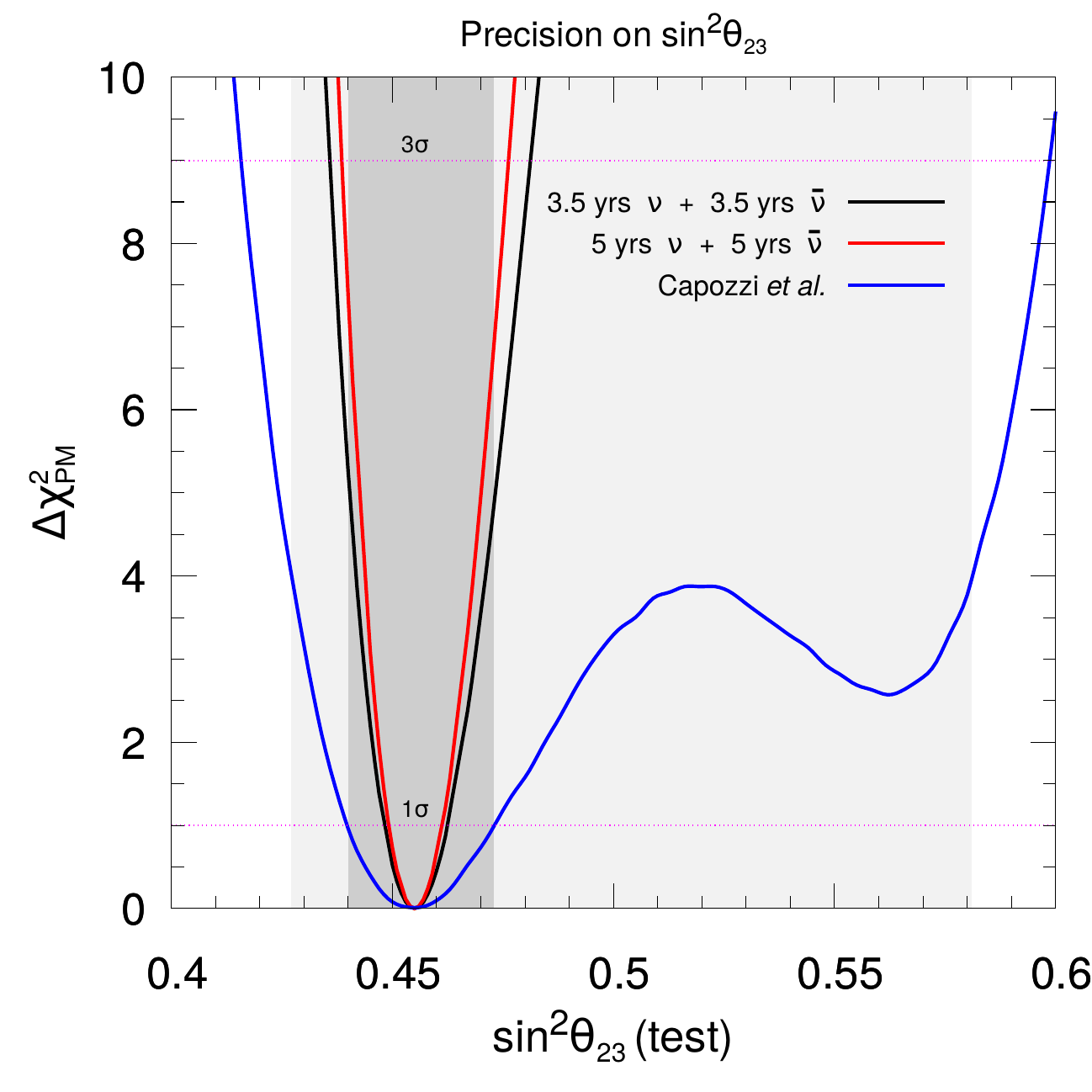}
\includegraphics[width=0.49\linewidth]{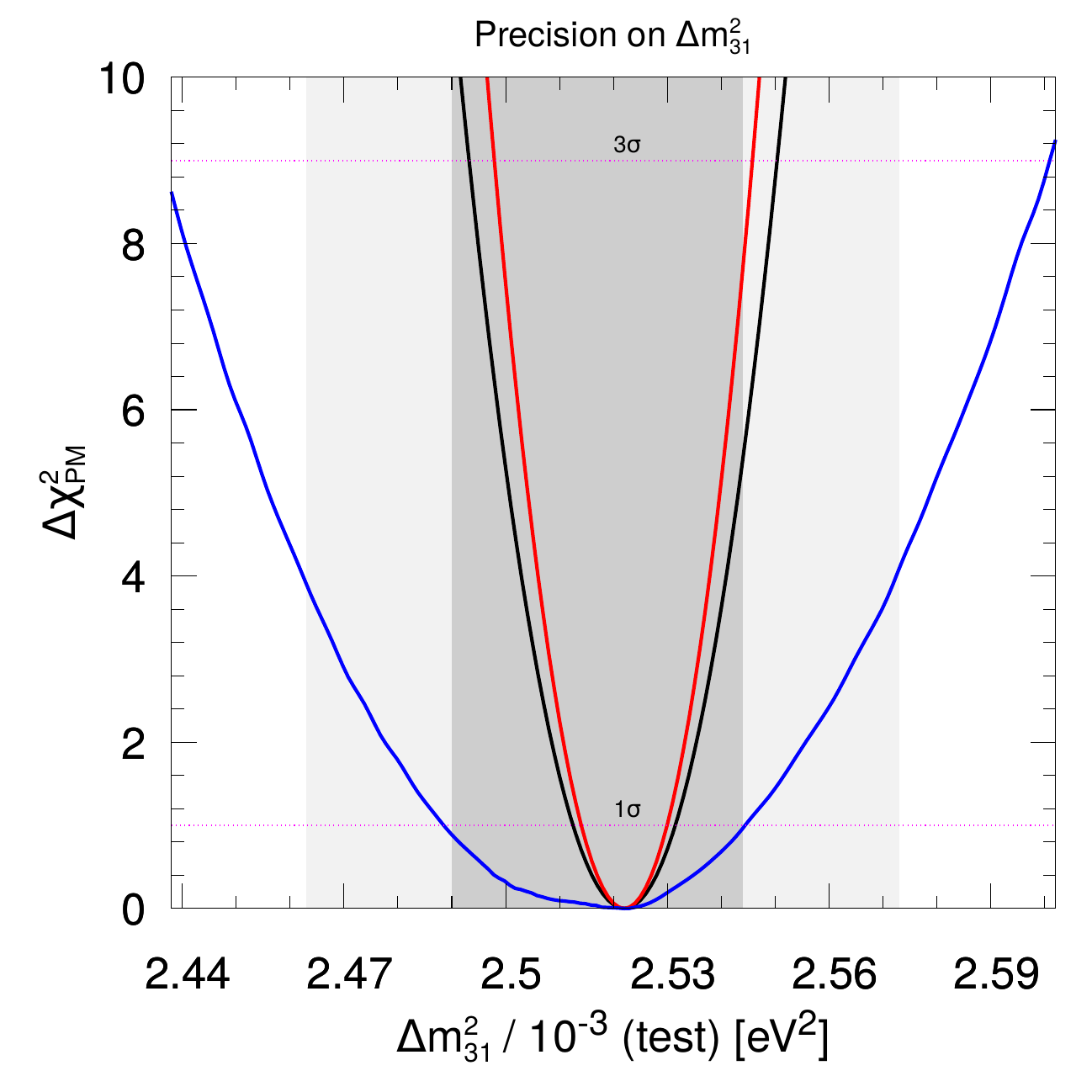}
\caption{Achievable precision in $\sin^2\theta_{23}$ and $\Delta m^2_{31}$ around the respective benchmark values (refer Table~\ref{work3:table:one}) for two proposed run time in DUNE and global-fit study from Ref.~\cite{Capozzi:2021fjo}. In the fit, we minimiize over allowed region in $\delta_{\rm CP}$ and $\Delta m^{2}_{31}$ while determining precision in $\sin^2\theta_{23}$. Similarly, we perform minimization over $\delta_{\rm CP}$ and $\sin^2\theta_{23}$ while determining precision measurements in $\Delta m^{2}_{31}$. Refer Table~\ref{work3:table:3} for quantifying the precision in terms of relative 1$\sigma$, computed following Eq.~\ref{precision1}. \textbf{Change in statistics influences precision in $\Delta m^2_{31}$ more than precision in $\sin^2\theta_{23}$.}}
\label{work3:fig:3}
\end{figure}

In this subsection, we estimate the sensitivity of the DUNE experiment to constrain the oscillation parameters $\Delta m^2_{31}$ and $\sin^2\theta_{23}$. We calculate the relative $1\sigma$-precision with which DUNE can measure these oscillation parameters and compare them with the existing constraints. 
In Fig.~\ref{work3:fig:3}, we show $\Delta\chi^2_{\rm PM}$ as a function of the test oscillation parameters $\sin^2\theta_{23}$ (left panel) and $\Delta m^2_{31}$ (right panel). $\Delta\chi^2_{\rm PM}$ in this chapter is computed as follows:

\begin{equation}
	\Delta \chi^2_{\text{PM}}\,(\zeta^{\rm test}) = \underset{\boldsymbol{\theta}}{\mathrm{min}}\left\{ \chi^2\left(\zeta^{\mathrm{test}}\right) - \chi^2\left(\zeta^{\mathrm{true}} \right)\right\}.
	\label{precision-chisq}
	\end{equation}
	
Here, $\zeta^{\text{true}}$ is the best-fit value of the oscillation parameter under consideration and $\zeta^{\text{test}}$ represents a test value of the same oscillation parameter in its currently allowed 3$\sigma$ range~\cite{Capozzi:2021fjo}. $\boldsymbol{\theta}$ denotes the set of oscillation parameters over which we perform minimization in the fit for a given analysis. In the fit, we minimize $\chi^2\left(\zeta^{\rm test}\right)$ over the systematic uncertainties to obtain $\Delta \chi^2_{\text{PM}}(\zeta^{\rm test})$. We show the results in Table~\ref{work3:table:3}. The relative 1$\sigma$ precision in the measurement of oscillation parameters $\zeta$ is estimated as follows:

\begin{equation}
p(\zeta)\, =\, \frac{\zeta^{\rm max} - \zeta^{\rm min}}{6.0 \,\times \, \zeta^{\rm true}}\, \times \, 100\%\, .
\label{precision1}
\end{equation} 

Here, $\zeta^{\rm max}$ and $\zeta^{\rm min}$ represent the allowed $3\sigma$ upper and lower bounds, respectively. In the fourth column of Table~\ref{work3:table:3}, we mention the current relative 1$\sigma$ precision on $\sin^2\theta_{23}$ and $\Delta m^2_{31}$ from the recent global fit study~\cite{Capozzi:2021fjo}. For a comparative study, we also show the achievable relative 1$\sigma$ precision\footnote{The achievable precision on atmospheric oscillation parameters using the full exposure of T2K and NO$\nu$A is discussed in Ref.~\cite{Agarwalla:2013qfa}.} on $\Delta m^2_{31}$ from the upcoming medium-baseline reactor experiment JUNO~\cite{JUNO:2015zny}. We observe that DUNE can improve the current relative 1$\sigma$ precision on $\sin^2\theta_{23}$ ($\Delta m^2_{31}$) by a factor of 4.4 (2.8) using 3.5 years of neutrino and 3.5 years of antineutrino runs. The total exposure of 10 years equally shared in neutrino and antineutrino modes further improves the precision on these parameters. Also, since precision in $\Delta m^2_{31}$ is mostly dominated by disappearance channel, increasing the statistics improves measurement. Further, precision studies in $\sin^2\theta_{23}$ is also disappearance dominated, but also has appearance influence in excluding the wrong octant solutions, therefore a mere increase in statistics does not help much in precision studies.

\begin{table}[htb!]
\centering
\resizebox{\columnwidth}{!}{%
\begin{tabular}{|c||c|c|c|c|}
\hline
\hline
\multirow{3}{*}{Parameter} & \multicolumn{4}{c|}{Relative 1$\sigma$ precision (\%)}\\
\cline{2-5}
& DUNE & DUNE &  \multirow{2}{*}{Capozzi $et~al.$ \cite{Capozzi:2021fjo}} & \multirow{2}{*}{JUNO \cite{NavasNicolas:2023fza,JUNO:2015zny}} \\
& ($3.5~\nu + 3.5~\bar{\nu}$) yrs  & ($5~\nu + 5~\bar{\nu}$) yrs & &  \\
\hline
$\sin^{2}\theta_{23}$ & 1.53 & 1.31 & 6.72 & ---\\
\hline
$\Delta m^{2}_{31}$ & 0.39 & 0.31 & 1.09 & 0.20\\
\hline
\hline
\end{tabular}
}
\caption{\footnotesize{Relative 1$\sigma$ precision on $\sin^2\theta_{23}$ and $\Delta m^2_{31}$ around the true choices $\sin^{2}\theta_{23} = 0.455$ and $\Delta m^2_{31} = 2.522\times 10^{-3}~\rm eV^2$. Existing precision from Capozzi $et~al.$ follows from Ref.~\cite{Capozzi:2021fjo} and expected relative 1$\sigma$ precision in JUNO, expected with 6 years of run~\cite{NavasNicolas:2023fza}.}}
\label{work3:table:3}
\end{table}

\begin{figure}[htb!]
\centering
\includegraphics[width=0.7\linewidth]{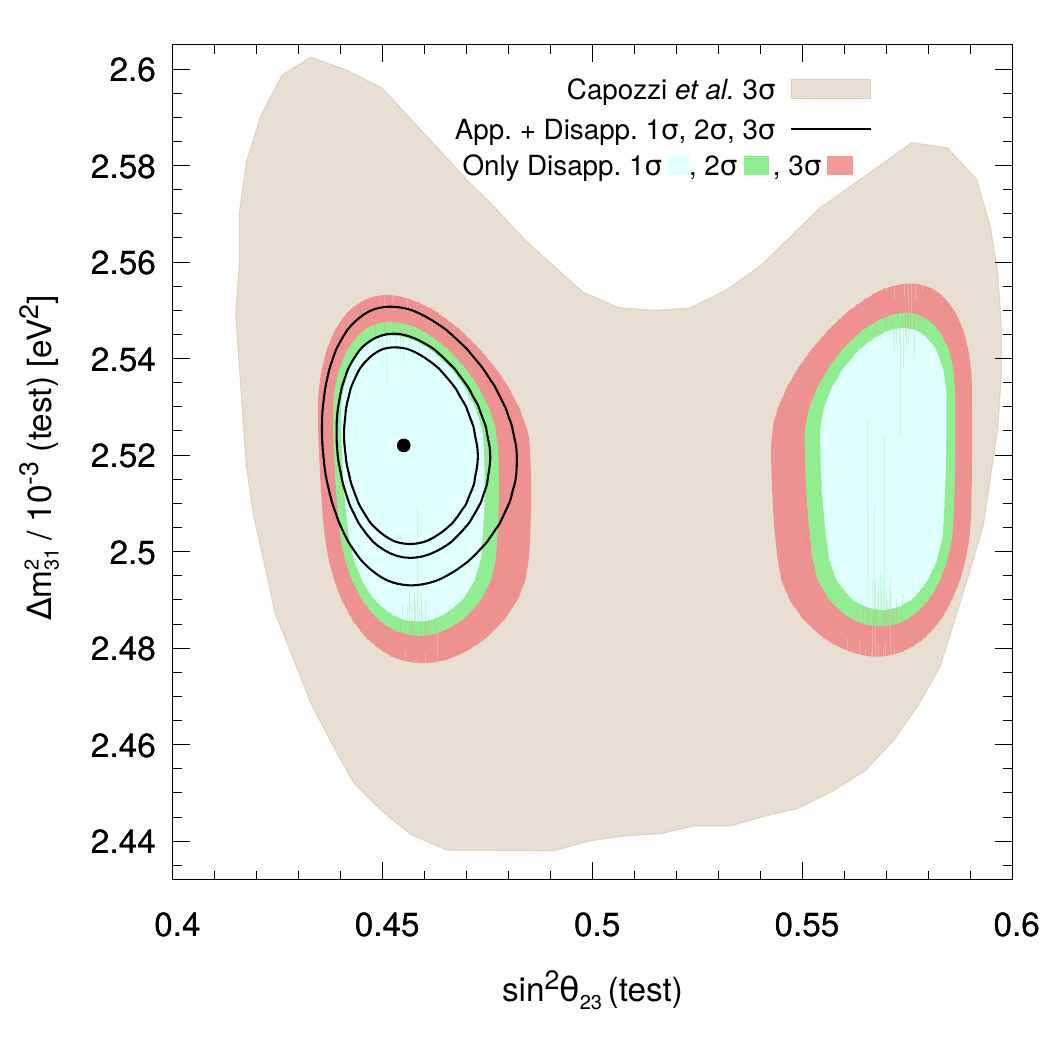}
\caption{Allowed regions in the test-statistic of atmospheric mixing angle, $\sin^2\theta_{23}$ and mass-squared difference $\Delta m^2_{31}$) in DUNE. The true values correspond to the benchmark values following Table~\ref{work3:table:one}. The test-statistic is scanned over $\sin^{2}\theta_{23}$  and $\Delta m^2_{31}$ (refer to Eq.~\ref{precision-chisq}). \textbf{Both appearance and disappearance statistics are essential in excluding the wrong octant solutions in DUNE.}}
\label{work3:fig:4}
\end{figure}

In Fig.~\ref{work3:fig:4}, we show the allowed regions in the test $\sin^2\theta_{23}$ - $\Delta m^2_{31}$ at 1$\sigma$, 2$\sigma$, and 3$\sigma$ for 1 degree of freedom. The shaded light-grey region shows the currently allowed $3\sigma$ values due to the existing oscillation data~\cite{Capozzi:2021fjo}. As can be observed, the allowed region is quite large especially in the parameter $\sin^2\theta_{23}$. Further, both maximal mixing and wrong octant solutions are allowed at the $3\sigma$ C.L. The current best-fit to the global data is the lower octant solution of $\sin^2\theta_{23}=0.455$, shown by the black dot. We give the results for DUNE with 7 years of equally shared exposure in neutrino and antineutrino modes, with only disappearance data (blue, green, and red contours) and combined appearance and disappearance data (solid black contours), considering the true value of $\sin^2\theta_{23}$ to be $0.455$. It can be seen that the disappearance data significantly constrains the allowed range of $\sin^2\theta_{23}$. However, it is still unable to rule out the wrong octant solution even at $1\sigma$. On the other hand, though the appearance data only marginally improves the $\sin^2\theta_{23}$ precision in the right octant, it plays the main role in completely ruling out the wrong octant solution. We also find that the combined appearance and disappearance data improves the precision in measurement of both $\sin^{2}\theta_{23}$ in correct octant and $\Delta m^2_{31}$.

\begin{figure}[htb!]
	\centering
	\includegraphics[width=\linewidth]{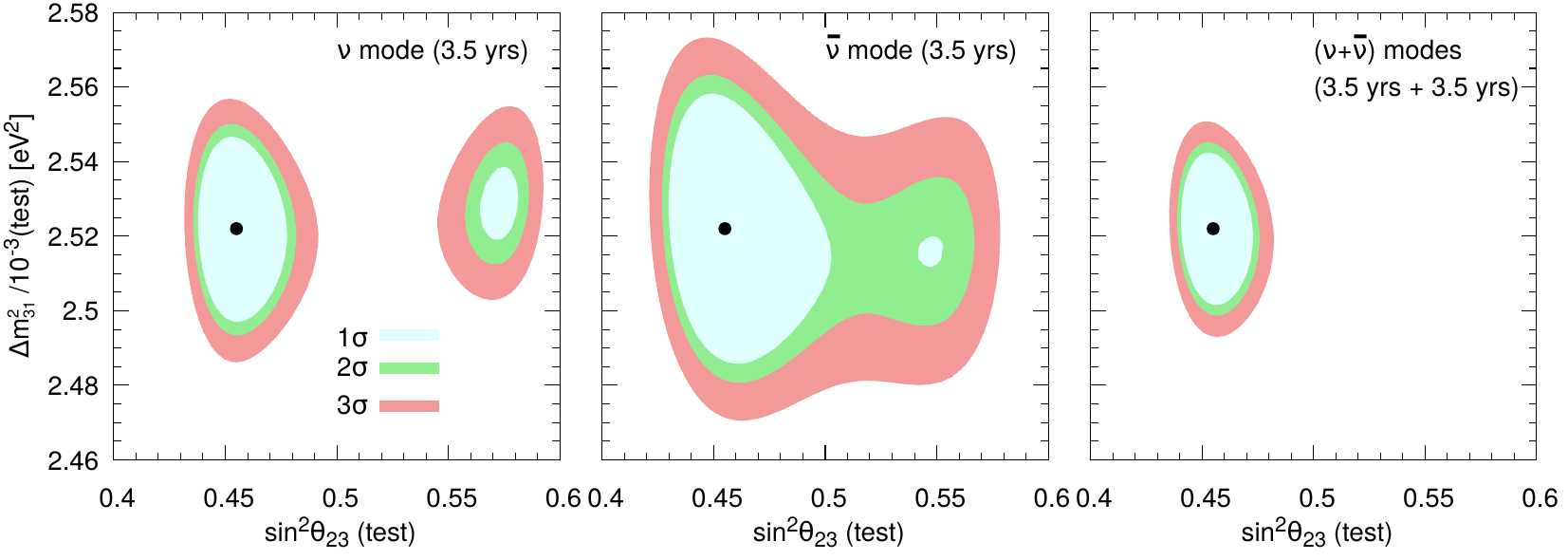}
	\caption{ Allowed regions in the test-statistic of atmospheric mixing angle, $\sin^2\theta_{23}$ and mass-squared difference $\Delta m^2_{31}$) in DUNE. The true values correspond to the benchmark values following Table~\ref{work3:table:one}. The test-statistic is scanned over $\sin^{2}\theta_{23}$  and $\Delta m^2_{31}$ (refer to Eq.~\ref{precision-chisq}). \textbf{Both neutrino and antineutrino statistics are essential for establishing stringent precision in correct octant of $\sin^2\theta_{23}$.} 
}
	\label{work3:fig:5}
\end{figure}

In Fig.~\ref{work3:fig:5}, we display the benefit of having data from both neutrino and antineutrino modes, the merit of which was discussed elaborately in the context of T2K and NO$\nu$A in Ref.~\cite{Agarwalla:2013ju}. The left panel explores the allowed region in the test ($\sin^2\theta_{23}-\Delta m^2_{31}$) plane considering 3.5 years of neutrino run and having contributions from both disappearance and appearance channels. The middle panel depicts the same for 3.5 years of antineutrino run. In the right panel, we demonstrate how the allowed region in the test ($\sin^2\theta_{23}-\Delta m^2_{31}$) plane gets shrinked when we combine the data from 3.5 years of neutrino and 3.5 years of antineutrino runs. From the left and middle panels, we observe that the prospective data from only neutrino or only antineutrino run cannot rule out the wrong octant solution even at $1\sigma$ confidence level, while with only antineutrino run, even maximal mixing solution of $\theta_{23}$ is allowed at $2\sigma$. However, from the right panel of Fig.~\ref{work3:fig:5}, it is evident that the data from both neutrino and antineutrino runs are quite effective in ruling out the wrong octant solution as well as the maximal mixing at $3\sigma$ confidence level. This happens because the combined neutrino and antineutrino data can resolve the octant\,-\,$\delta_{\mathrm{CP}}$ degeneracy~\cite{Agarwalla:2013ju} that exists in the standalone neutrino or antineutrino data. We also notice that the allowed regions for $\sin^2\theta_{23}\, \mathrm{and}\,\Delta m^2_{31}$ around the correct octant get reduced when we combine the data from both neutrino and antineutrino modes (see right panel). The increase in statistics due to both neutrino and antineutrino runs and the possible complementarity between them lead to these improvements in the sensitivity.

\section{Summary}
\label{conclusion}

We have achieved remarkable precision on the solar ($\theta_{12}, \, \Delta m^{2}_{21}$) and atmospheric ($\theta_{23}, \, \Delta m^{2}_{31}$) oscillation parameters over the last few years. According to Ref.~\cite{Capozzi:2021fjo}, the current relative 1$\sigma$ errors on $\sin^{2}\theta_{12}, \, \Delta m^{2}_{21}, \, \sin^{2}\theta_{23}$, and $\Delta m^{2}_{31}$ are 4.5\%, 2.3\%, 6.7\%, and 1.1\%, respectively. The recent hints for normal mass ordering (at $\sim$ 2.5$\sigma$), as well as for lower octant $\theta_{23}$ ($\theta_{23} <$ 45$^{\circ}$) and for $\delta_{\mathrm{CP}}$ in the lower half-plane ($\sin \delta_{\mathrm{CP}} < 0$) signify major developments in the three-flavor neutrino oscillation paradigm. The high-precision measurement of $\theta_{13}$ from the Daya Bay reactor experiment and the possible complementarities between the recent Super-K Phase I-IV atmospheric data and the appearance and disappearance data from the ongoing long-baseline oscillation experiments - NO$\nu$A and T2K, play an important role in providing these crucial hints. An accurate measurement of $\theta_{23}$ and resolution of its octant (if $\theta_{23}$ turns out to be non-maximal) are crucial to transform these preliminary hints into 5$\sigma$ discoveries. A discovery of non-maximal $\theta_{23}$ at high confidence level will serve as a crucial input to the theories of neutrino masses and mixings and it will certainly be a major breakthrough in addressing the the age-old flavor problem. In this chapter, we explore in detail the sensitivities of the upcoming high-precision long-baseline experiment DUNE to establish the possible deviation from maximal $\theta_{23}$ and to resolve its octant at high confidence level in light of the recent neutrino oscillation data. 

We start the chapter by studying the possible correlations and degeneracies among the oscillation parameters $\sin^{2}\theta_{23}, \, \Delta
m^{2}_{31}$, and $\delta_{\mathrm{CP}}$ in the context of $\nu_{\mu} \rightarrow \nu_{e}$ appearance channel at the probability and event levels. We find that both neutrino and antineutrino data are needed to reduce the impact of octant - $\delta_{\mathrm{CP}}$ degeneracy, which in turn allows us to resolve the $\theta_{23}$ octant at a high confidence level. DUNE can settle the issue of $\theta_{23}$ octant at 4.3$\sigma$ (5$\sigma)$ using 336 (480) kt$\cdot$MW$\cdot$years of exposure assuming $\sin^{2}\theta_{23}$ (true) = 0.455, $\delta_{\mathrm{CP}}$ (true) = 223$^{\circ}$, and true NMO. On the other hand, the octant ambiguity of $\theta_{23}$ can be resolved at 3$\sigma$ (5$\sigma$) in DUNE with an exposure of 336 kt$\cdot$MW$\cdot$years if the true value of $\sin^{2}\theta_{23}$ $\lesssim 0.462\, (0.450)$ or $\sin^{2}\theta_{23}$ $\gtrsim 0.553\, (0.569)$ assuming $\delta_{\mathrm{CP}}$ (true) = 223$^{\circ}$ and true NMO. If we increase the exposure to 480 kt$\cdot$MW$\cdot$years (corresponding to 10 years of run), the wrong octant solution can be excluded if $\sin^{2}\theta_{23}$ (true) $\lesssim 0.466 \,(0.454)$ or $\sin^{2}\theta_{23}$ (true) $\gtrsim 0.548 \,(0.565)$ keeping the assumptions on other oscillation parameters same. 

Finally, we quote how accurately DUNE can measure the atmospheric oscillation parameters. We observe that DUNE can improve the current relative 1$\sigma$ precision on $\sin^2\theta_{23}$ ($\Delta m^2_{31}$) by a factor of 4.4 (2.8) using 336 kt$\cdot$MW$\cdot$years of exposure. We analyze how much contribution we obtain from individual appearance and disappearance oscillation channels and also study the importance of having data from both neutrino and antineutrino modes while measuring these parameters.

\chapter{Enhancing Sensitivity to Leptonic CP Violation using Complementarity among DUNE, T2HK, and T2HKK}
\label{sec:ch7}

\section{Introduction and motivation}
\label{sec:1}

One of the fundamental properties of particles is their behavior under the 
CP (charge-parity) transformation and a violation of the CP symmetry may
have an important connection to the observed baryon asymmetry 
in the Universe~\cite{Sakharov:1967dj}. So far, in the quark 
sector of the SM, we have two known sources of  
CP-invariance violation~\cite{ParticleDataGroup:2022pth}. One of them is the 
CP-odd phase in the Cabibbo-Kobayashi-Maskawa (CKM) matrix,
which is known to be large and governs all the CP-violating phenomena
observed so far. The other one is the so-called strong CP-phase 
$\theta_{\rm QCD}$, which is known to be vanishingly small. In the 
lepton sector, we achieved an important breakthrough in 2012
in establishing the standard three-flavor oscillation picture of neutrinos
through the pioneering discovery of the non-zero value of the smallest neutral lepton mixing
angle $\theta_{13}$ by the Daya Bay reactor antineutrino 
experiment~\cite{DayaBay:2012fng}. This landmark finding 
opened the door for a completely new and independent source of 
CP invariance violation in neutrino oscillation experiments. 
The so-called Dirac CP-odd phase $\dcp$ in the $3 \times 3$ 
unitary Pontecorvo-Maki-Nakagawa-Sakata (PMNS) matrix 
is the source of CP-invariance violation in the neutral lepton sector,
which can be probed via neutrino oscillation probabilities. 

After the discovery of non-zero $\theta_{13}$, remarkable precision has 
been achieved on neutrino mass-mixing parameters over the past decade,
which has enabled us to come up with a simple, robust, three-flavor neutrino
oscillation paradigm, which is capable of accommodating most of the oscillation
data~\cite{deSalas:2020pgw,Esteban:2020cvm,Capozzi:2021fjo}.
In 3-$\nu$ oscillation picture, if the value of $\dcp$ turns out to be different 
from both $0^{\circ}$ and $180^{\circ}$ in Nature, then it would cause 
a difference between neutrino and antineutrino transition probabilities -- 
providing a smoking gun signature of CP violation (CPV) in neutrino 
oscillation experiments. In the intensity frontier, the currently running
and upcoming high-precision long-baseline (LBL) neutrino oscillation 
experiments are the most promising avenues to unravel the novel
signatures of CPV. One of the prime scientific goals of these LBL 
experiments is to provide an explicit demonstration of leptonic CPV 
by precisely measuring the differences between the $\theta_{13}$-driven 
oscillations of muon-type neutrinos and antineutrinos into electron-type 
neutrinos and antineutrinos, respectively.

By observing these differences, the currently running LBL experiments 
T2K~\cite{T2K:2021xwb} and NOvA~\cite{NOvA:2021nfi} 
have already started probing the parameter space of $\dcp$ and 
provided hints toward non-zero CPV. Now, it bestows upon the 
next-generation LBL experiments in the neutrino roadmap to convert 
these crucial hints of leptonic CPV into discoveries at high confidence 
level. Such a path-breaking discovery would certainly pave the way 
to elucidate the age-old flavor puzzle and the prevalence of matter 
over antimatter in the Universe~\cite{Sakharov:1967dj}.

In this chapter, after having an insightful discussion on the critical role of intrinsic, 
extrinsic and total CP asymmetries in the appearance channel and extrinsic 
CP asymmetries in the disappearance channel, we study in detail the capabilities 
of the next-generation long-baseline neutrino oscillation experiments DUNE
~\cite{DUNE:2021cuw}
and T2HK~\cite{Hyper-Kamiokande:2018ofw} in isolation and combination to establish the leptonic CPV ($\dcp$ $\neq$ $0^{\circ}$ and $180^{\circ}$) 
at 3$\sigma$ confidence level for at least 75\% choices\footnote{We often mention this performance 
indicator of a given experiment as ``CP coverage", which denotes the values of true $\dcp$ (in \%) in its entire range of $[-180^{\circ} , 180^{\circ}]$,
for which leptonic CPV can be established at $\ge$ 3$\sigma$ confidence level.}
of true $\dcp$ in its entire range of $-180^{\circ}$ to $180^{\circ}$, considering the 
state-of-the-art simulation details of these facilities. We extend our analysis to the 
proposed T2HKK \footnote{This setup consists of two water Cherenkov 
far detectors (187 kt each): the first detector in Japan (JD) and the second detector
in Korea (KD) at a distance of 295 km and 1100 km from the J-PARC
facility, respectively. The JD (KD) will receive the same off-axis ($2.5^{\circ}$),
narrow-band beam from the J-PARC facility covering the first (second) oscillation maximum.
We often denote this setup as JD+KD in this chapter.} setup~\cite{Hyper-Kamiokande:2016srs} 
and explore its CP coverage in standalone mode and also in combination with DUNE.
The main thrust of this chapter is to investigate in detail the possible complementarity 
among these high-precision experiments~\cite{Fukasawa:2016yue,Ballett:2016daj,Liao:2016orc,Ghosh:2017ged,Choubey:2017cba,Agarwalla:2021owd}
to fully exploit the three-flavor interference effects suppressing the parameter degeneracies, 
which in turn enhances the CP coverage. The key point of our analysis is to 
demonstrate how these experiments bring complementary information 
on the CP phase $\dcp$ for different octant choices of the 2-3 mixing angle 
($\theta_{23}$) at different $L/E$ values by means of appearance and disappearance 
channels in neutrino and antineutrino modes. Such a combination is also crucial 
to tackle the underlying degeneracies among $\theta_{23}$ and $\dcp$, 
which reduce the CP coverage in $\dcp$~\cite{Nath:2015kjg,Machado:2015vwa} 
while establishing the leptonic CPV.

The upcoming DUNE is planning to use a 40 kt liquid argon time projection chamber
(LArTPC) as a far detector which will be exposed to an on-axis, high-intensity, wide-band 
neutrino beam covering both the first and second oscillation maxima with a baseline of 1300 km.
The DUNE far detector is expected to have an unmatched kinematic reconstruction capability 
for all the observed particles in the final state, which plays an important role to reject a large 
fraction of the neutral current background. The presence of an efficient near detector will 
significantly minimize the impact of flux and cross-section related systematic uncertainties
at the DUNE far detector. DUNE will experience a significant amount of Earth's matter effect
because of its large baseline of 1300 km and having access to a wide-band
beam whose flux extends up to $\sim$ 6 GeV with a peak at around 2.5 GeV. All these features 
help DUNE to settle the issue of
neutrino mass ordering (refer to Chapter~\ref{sec:intro}) at a very high confidence level irrespective of the choices of other 
oscillation parameters and to measure the values of $\dcp$ and $\theta_{23}$ with satisfactory 
precision utilizing the information on oscillation pattern at several $L/E$ values~\cite{Arafune:1997hd,DUNE:2020jqi}.
On the other hand, in Japan, the proposed gigantic 187 kt Hyper-Kamiokande (HK) 
water Cherenkov detector will serve as the far detector for the T2HK (JD) setup at a distance of
295 km from the J-PARC facility and receive an off-axis ($2.5^{\circ}$), upgraded, 
narrow-band beam with a flux peaking around the first oscillation maximum of $\sim$ 0.6 GeV.
The shorter baseline and high statistics help T2HK (JD) to achieve an unparalleled
precision in the measurement of $\dcp$ and $\theta_{23}$ free from Earth's matter 
effect~\cite{Hyper-KamiokandeProto-:2015xww,Hyper-Kamiokande:2018ofw}. 
The Korean detector (KD) with a baseline of 1100 km and an off-axis beam from 
J-PARC having a peak around the second oscillation maximum is slightly sensitive 
to Earth's matter effect and provides complementary information on $\dcp$ as 
compared to JD~\cite{Hyper-Kamiokande:2016srs}.

In this chapter, we study in detail how the coverage in $\dcp$ for $\ge$ 3$\sigma$ 
leptonic CPV varies with the choice of $\theta_{23}$, exposure, optimal runtime 
in neutrino and antineutrino modes, and systematic uncertainties in these experiments 
in isolation and combination. We find that neither DUNE nor T2HK individually can achieve 
the milestone of 75\% CP coverage for which at least 3$\sigma$ leptonic CPV
can be ensured with their nominal exposures and systematic uncertainties.
In DUNE, we observe that the main bottleneck is $\theta_{23}-\dcp$ 
degeneracy which appears in the picture when $\theta_{23}$ lies in the 
range of 42$^{\circ}$ to 48$^{\circ}$. We notice that this degeneracy cannot
be resolved in DUNE even by doubling the exposure or reducing the current 
systematic uncertainties by a factor of two. While in T2HK, although such 
degeneracy does not play any significant role because of the negligible 
matter effects, the current systematic uncertainties are an obstacle in achieving 
the above-mentioned sensitivity. One of the important conclusions of our work 
is that the complementarity between DUNE and T2HK is essential to obtain the
desired CP coverage irrespective of the value of $\theta_{23}$ in Nature.
Our study shows that for the combination of DUNE and T2HK, only half of their 
nominal exposures are sufficient to establish 3$\sigma$ leptonic CPV for at least 
75\% choices of $\dcp$ for almost all values of $\theta_{23}$, with their nominal 
systematic uncertainties. This becomes possible due to the less systematic 
uncertainties in DUNE as compared to T2HK and high matter-independent 
disappearance statistics in T2HK, that helps in constraining $\theta_{23}$ 
in a narrow range and thus to resolve the $\theta_{23}-\dcp$ degeneracy.
We find that with an improved systematic uncertainty of 2.7\% in appearance 
mode, the standalone T2HK (JD) setup can provide a CP coverage of around 75\% 
for almost all values of $\theta_{23}$ with nominal exposure. We observe that 
with nominal exposure and systematic uncertainties, T2HKK (JD+KD) 
can also achieve the 75\% CP coverage for all values of $\theta_{23}$, 
but its CP coverage is always less than that of DUNE+JD.
At the same time, with only half of their exposures and nominal systematic 
uncertainties, T2HKK+DUNE can achieve a CP coverage of more than 
80\% for almost all values of $\theta_{23}$.

We organize the chapter as follows. We initiate our discussion with a detailed analytical 
understanding of CP asymmetry in Sec.~\ref{sec:2}, which will come in handy while 
analyzing our findings. Sec.~\ref{sec:4} summarizes our results 
and findings wherein we discuss our milestone of achieving 3$\sigma$ leptonic CPV
for at least 75\% choices of $\dcp$ in these experiments in isolation and combination 
as a function of (a) true $\sin^{2}\theta_{23}$ in Sec.~\ref{sec:4a}, 
(b) varying exposure in Sec.~\ref{sec:4b}, 
(c) optimal runtime in neutrino and antineutrino modes in Sec.~\ref{sec:4c}, 
and (d) systematic uncertainties in Sec.~\ref{sec:4d}.
Also, in Sec.~\ref{sec:4e}, we estimate the enhanced CP coverage of these
experiments when we assume the values of true $\dcp$ only in its current 
3$\sigma$ allowed range of $[-175^{\circ} , 41^{\circ}]$, instead of its entire
range of $[-180^{\circ} , 180^{\circ}]$.
Finally, we summarize with our concluding remarks in Sec.~\ref{sec:5}.

\section{Discussion at the oscillation probability level }
\label{sec:2}
%
The mixing matrix in the standard three-neutrino (3$\nu$) framework is written in terms of the three mixing angles ($\theta_{23}$, $\theta_{13}$, and $\theta_{12}$) 
and one complex phase ($\delta_{\mathrm{CP}}$)~\cite{ParticleDataGroup:2022pth}. Following the usual PMNS parameterization, we have :
\begin{eqnarray}
   \nonumber  U_{\mathrm{PMNS}}&=
     \begin{pmatrix}
    c_{12}c_{13} & s_{12}c_{13} & s_{13}e^{-i\delta_{\mathrm{CP}}} \\
     -s_{12}c_{23}-c_{12}s_{23}s_{13}e^{i\delta_{\mathrm{CP}}} & c_{12}c_{23}-s_{12}s_{23}s_{13}e^{i\delta_{\mathrm{CP}}} & s_{23}c_{13} \\
     s_{12}s_{23}-c_{12}c_{23}s_{13}e^{i\delta_{\mathrm{CP}}} & -c_{12}s_{23}-s_{12}c_{23}s_{13}e^{i\delta_{\mathrm{CP}}} & c_{23}c_{13}
     \end{pmatrix}\,,
 \end{eqnarray}
 where we notice that the Dirac phase $\delta_{\mathrm{CP}}$ is always coupled to the mixing angles. This explains that sensitivity in the CP phase strongly depends on the knowledge of other mixing parameters. 
At LBL experiments, we mostly probe the $\nu_\mu (\bar{\nu}_{\mu})\to\nu_{\mu} (\bar{\nu}_{\mu})$ (disappearance) and the $\nu_{\mu} (\bar{\nu}_{\mu})\to\nu_e (\bar{\nu}_{e})$ (appearance) channels. Following the approach in Ref.~\cite{Agarwalla:2021bzs}, we can further simplify the appearance probability expression in Ref.~\cite{Akhmedov:2004ny} dropping the small $\alpha^2$ terms, where $\alpha=\Delta m_{21}^2/\Delta m_{31}^2$\,, as follows: 
\begin{eqnarray}
  P_{\mu e}\approx N\sin^2\theta_{23}+O\sin2\theta_{23}\cos(\Delta+\delta_{\mathrm{CP}})\,,
  \label{eq:1}
\end{eqnarray}
where,
\begin{eqnarray}
  N&=&4\sin^2\theta_{13}\frac{\sin^2[(\hat A-1)\Delta]}{(\hat A-1)^2}\, ,
  \label{eq:2}
  \end{eqnarray}
  \begin{eqnarray}
  O&=&2\alpha\sin\theta_{13}\sin2\theta_{12}\frac{\sin\hat A\Delta}{\hat A}\frac{\sin[(\hat A-1)\Delta]}{\hat A-1}\,.
  \label{eq:3}
\end{eqnarray}
This grouping of terms helps to visualize the dependence on the atmospheric mixing angle ($\theta_{23}$). In the above set of equations, $\Delta=\Delta m_{31}^2 L/4E$, and $\hat A=A/\Delta m_{31}^2$, wherein the Wolfenstein matter term, $A=2\sqrt{2}G_F N_e E \approx 2\times7.6\times Y_{e} \times 10^{-5} \times \rho_{\mathrm{avg}} \,$ (g/cm$^3$) $\times E $\, (GeV). Here $\rho_{\mathrm{avg}}$ is the line-averaged constant Earth matter density which we consider as 2.848 g/cm$^3$, 2.7 g/cm$^3$, and 2.8 g/cm$^3$ in DUNE~\cite{Roe:2017zdw}, JD~\cite{Hyper-Kamiokande:2018ofw}, and KD~\cite{Hyper-Kamiokande:2016srs}, respectively. Also, assuming that Earth's matter is electrically neutral and isoscalar, we obtain $N_{e} = N_{p} = N_{n}$ where  $N_{p}$ and $N_{n}$ are proton and neutron number densities in Earth, respectively. Thus, the relative number density given by: $Y_{e} \equiv N_{e}/(N_{p}+ N_{n})$ is estimated as 0.5. Further, from Eq.~\ref{eq:1}, we observe that the CP-violating term contains $\sin2\theta_{23}$ and thus is insensitive to the octant of the atmospheric angle~\cite{Fogli:1996pv, Barger:2001yr, Minakata:2002qi, Minakata:2004pg,Hiraide:2006vh,Das:2017fcz}. Moreover, changing from neutrino to antineutrino mode notably changes signs of $\hat A$, thus leading to matter-induced or fake (extrinsic) CPV~\cite{Tanimoto:1998sn,Barger:1980tf,Arafune:1997hd}. This will have a dominant contribution as it is present in the coefficient of the leading term: N (refer to Eq.~\ref{eq:2}). While the presence of the Dirac CP phase in the sub-leading term gives rise to the genuine (intrinsic) CPV (refer to Eq.~\ref{eq:3}).

\begin{table}[htb!]
\resizebox{\columnwidth}{!}{%
  \centering
  \begin{tabular}{|c|c|c|c|c|c|c|}
    \hline \hline 
    \multirow{2}{*}{$\sin^2 \theta_{12}$} & \multirow{2}{*}{$\sin^2\theta_{23}$} & \multirow{2}{*}{$\sin^2 \theta_{13}$} &
    $\Delta m^2_{31}$ (eV$^2$) & $\Delta m^2_{21}$ (eV$^2$) & $\delta_{\rm CP}$
    & Mass \\
    & & & $\times10^{-3}$ & $\times10^{-5}$ & ($^{\circ}$) & Ordering\\
    \hline \hline
    0.303 & [0.4 , 0.6] & 0.0223 & 2.522 (-2.418) & 7.36
    & [- 180 , 180] & NMO (IMO) \\
    \hline \hline 
  \end{tabular}}
  \caption{The benchmark values of six oscillation parameters used in our analysis assuming NMO (IMO) following Ref.~\cite{Capozzi:2021fjo}.}
  \label{work4:table:one}
\end{table}
%

To probe CPV in oscillation experiments, one needs to note the difference between neutrino and antineutrino oscillation probabilities. The quantity which is strongly correlated to the sensitivity in $\delta_{\mathrm{CP}}$ is the CP asymmetry~\cite{Bernabeu:2019npc,Bernabeu:2018use,Ohlsson:2014cha,Nunokawa:2007qh}. In the following subsections, we will discuss how CP asymmetries affect CPV at the probability level. Throughout our simulation, we use the constant values of $\theta_{12}, \theta_{13}, \Delta m^{2}_{21},$ and $\Delta m^{2}_{31}$ as our benchmark values (refer to Table~\ref{work4:table:one}). Moreover, we will only consider here the NMO case, discussing the IMO case in Sec. \ref{sec:5}. We often vary $\theta_{23}$ in some cases as will be mentioned wherever necessary, while $\dcp$ is always varied in its entire range, except for Sec~\ref{sec:4e} where we make use of the current 3$\sigma$ constraints.

\subsection{Extrinsic, intrinsic, and total CP asymmetries in appearance channel}
\label{sec:2a}
\begin{figure}[htb!]
    \centering
    \includegraphics[scale = 0.97]{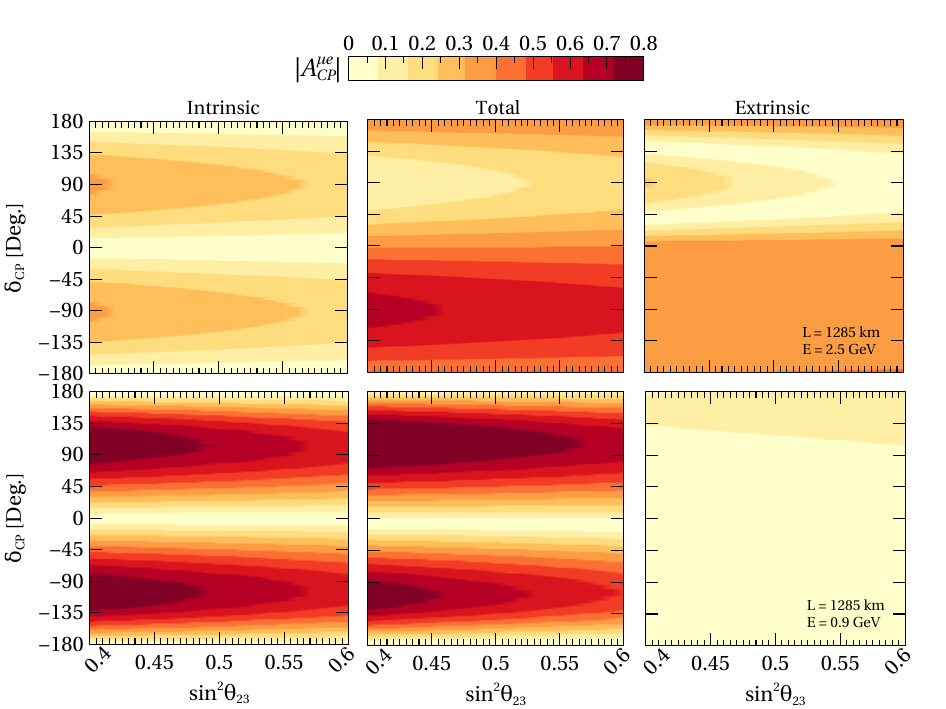}
    \caption {Absolute CP asymmetry ($|\Amue|$) as a function of $\delta_{\mathrm{CP}}$ and $\sin^{2}\theta_{23}$ for first oscillation maximum ($L$ = 1285 km, $E$ = 2.5 GeV) and second oscillation maximum ($L$ = 1285 km, $E$ = 0.9 GeV) in DUNE, assuming NMO. Values of other oscillation parameters are taken from Table~\ref{work4:table:one}. \textbf{Intrinsic CP asymmetry is more distinctly visible in second oscillation maxima than the first.}}
    \label{work4:fig:1}
\end{figure}
\begin{figure}[htb!]
    \centering
    \includegraphics[scale = 0.97]{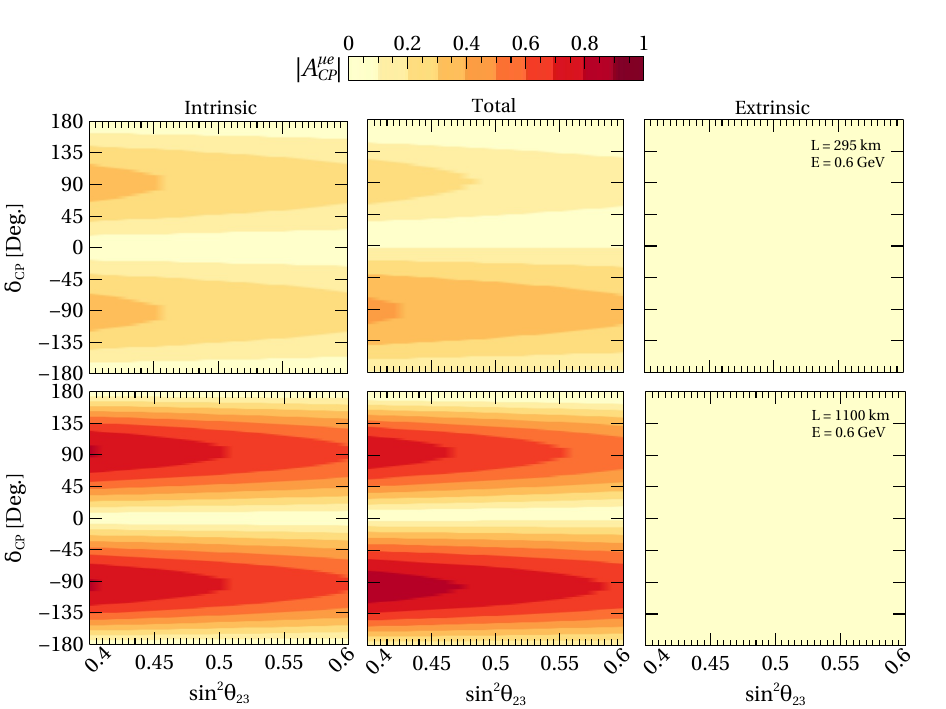}
    \caption{$|\mathcal{A}^{\mu e}_{\mathrm{CP}}|$ as a function of $\delta_{\mathrm{CP}}$ and $\sin^{2}\theta_{23}$ for first oscillation maximum in T2HK (JD) ($L$ = 295 km, $E$ = 0.6 GeV) and second oscillation maximum in T2HKK ($L$ = 1100 km, $E$ = 0.6 GeV) assuming NMO. Values of other oscillation parameters are taken from Table~\ref{work4:table:one}.}
    \label{work4:fig:2}
\end{figure}

The CP asymmetry in the appearance channel is defined as
\begin{eqnarray}
 \Amue=\dfrac{P_{\mu e}-\bar P_{\mu e}}{P_{\mu e}+\bar P_{\mu e}}\,.
\end{eqnarray}
Different expansions have been done to understand the behavior of such asymmetry in terms of mixing angles \cite{Giarnetti:2021wur}. However, to realize the role of the atmospheric mixing angle in the $\delta_{\mathrm{CP}}$ sensitivity, we fix the remaining mixing angles at their benchmark values ($\sin\theta_{13}\sim 1/7$ and $\sin\theta_{12}\sim 1/\sqrt{3}$). Also, since the value of matter parameter in the considered LBL experiments is not large, we can expand in $\hat A$ up to the first order. The resulting asymmetry is written as follows: 
\begin{eqnarray}
  \Amue=\Amuevac+\hat A \Amuemat +\mathcal{O}(\hat A^2) \,,
\end{eqnarray}
where
\begin{equation}
  \Amuevac = \frac{-28 \alpha\Delta\cos\theta_{23}\sin\delta_{\mathrm{CP}}\sin\Delta }{3\sqrt{2}\sin\theta_{23}\sin\Delta+28\alpha\Delta \cos\theta_{23}\cos\delta_{\mathrm{CP}}\cos\Delta}\label{eq:Avac}
\end{equation} 
\begin{equation}  
\resizebox{.9\hsize}{!}{ $\Amuemat = -\sin^2\theta_{23}(\Delta\cos\Delta-\sin\Delta)\frac{126\alpha\Delta\cos\theta_{23}\cos\delta_{\mathrm{CP}}\cos\Delta+18\sin^2\theta_{23}\sin\Delta}{(3\sin^2\theta_{23}\sin\Delta+7\sqrt{2}\alpha\cos\delta_{\mathrm{CP}}\cos\Delta\sin^2(2\theta_{23}))^2}$}
\label{Amat} 
\end{equation} 
It is clear that when the value of $\theta_{23}$ increases, the denominator of both the contributing term in Eq.~\ref{eq:Avac} and Eq.~\ref{Amat} increases. For this reason, the absolute value of the asymmetry becomes smaller, and we expect less CPV sensitivity. So, at the first oscillation maximum, ($\Delta=\pi/2$)\footnote{To be maximally sensitive to the oscillation probability, we must have $\Delta = (2n + 1) \frac{\pi}{2}$, where $n = 0, 1, 2,...$} the asymmetry reduces to:
\begin{eqnarray}
  \Amue&\approx &-\frac{7}{3}\alpha\sqrt{2}\pi\cot\theta_{23}\sin\delta_{\mathrm{CP}}+2\hat A
  \label{FOMmue}\,,
\end{eqnarray}
whose modulus decreases with an increase in $\theta_{23}$. The genuine or intrinsic CP has a sign opposite to the extrinsic or fake matter-induced contribution. Thus, there exists some combination of $\theta_{23}$ and $\dcp \in [0^{\circ},180^{\circ}]$, such that this asymmetry vanishes. It is interesting to note that the vacuum contribution becomes three times larger when considering the second oscillation maximum ($\Delta=3\pi/2$). Therefore, observing the CPV at such $L/E$ combination can give much more sensitivity to $\delta_{\mathrm{CP}}$ \cite{Rout:2020emr,Blennow:2019bvl,Tang:2019wsv}. The exact numerical behavior of the CP asymmetry in the appearance channel ($|\Amue|$) is shown in Fig.~\ref{work4:fig:1} for ($L$ = 1285 km, $E$ = 2.5 GeV), ($L$ = 1285 km, $E$ = 0.9 GeV)  which corresponds to the first and second oscillation maxima in DUNE. For each $L/E$ combination, we show three panels: in the left column, we show the vacuum or $\delta_{\mathrm{CP}}$-induced contribution (intrinsic). In the center, we illustrate the total asymmetry, and in the right column, we display the contribution due to the matter effects (extrinsic). In all the panels, we only plot the absolute value of the asymmetries since, in this work, the most important aspect is to stress on the difference between the asymmetries in the CP-violating and the CP-conserving cases. From the top left panel in Fig.~\ref{work4:fig:1}, we observe that the intrinsic contribution is the same in both the maximal CP-violating values of $\delta_{\mathrm{CP}}$ (90 and -90$^\circ$). Moreover, keeping the CP phase fixed to any value, the asymmetry reduces when we increase the value of $\theta_{23}$ from lower octant\footnote{Existence of non-maximal $\theta_{23}$ gives rise to two degenerate solutions: $\theta_{23} < 45^{\circ}$ (LO) and $\theta_{23} > 45^{\circ}$ (HO).} (LO) to higher octant (HO) (as expected from Eq.~\ref{eq:Avac} and Eq.~\ref{Amat}). Contrastingly, the extrinsic CP asymmetry (top right panel), which occurs solely due to the matter effect, is asymmetric, being larger for favorable choices of $\delta_{\mathrm{CP}}$ in NMO (negative half plane \ie~ $\dcp \in [180^{\circ}, 360^{\circ}$) and smaller for unfavorable choices of $\delta_{\mathrm{CP}}$ in NMO (positive half plane \ie~ $\dcp \in [0^{\circ},180^{\circ}$). Therefore, the total asymmetry (middle panel) is no longer the same for the maximal CP-violating values of $\delta_{\mathrm{CP}}$. Further, due to contribution from $\hat A$, the intrinsic $|\Amue|$, which is zero at CP-conserving values ($\delta_{\mathrm{CP}} = 0^\circ\, \text{and}\,180^\circ$), now has a finite value. Hence, from the top middle panel, it is clear that the asymmetries in CP-violating cases tend to shift closer to the CP-conserving value when $\theta_{23}$ increases. For the bottom row, where we plot the CP asymmetry at the second oscillation maxima ($E$ = 0.9 GeV), the matter effect becomes less, and the intrinsic component completely dominates the total asymmetry. Moreover, the asymmetry values are amplified in the bottom panel, which we expect as here the $L$ and $E$ are similar to the second oscillation maximum in DUNE.
Similarly, in Fig.~\ref{work4:fig:2}, we plot the CP asymmetry for the T2HK (JD) setup with $L=295$ km and $E=0.6$ GeV at the first oscillation maxima (top row) and T2HKK with $L=1100$ km and $E=0.6$ GeV at the second oscillation maxima (bottom row).
The top panel does not observe any significant contribution in the extrinsic panel due to less matter effect ($L$ = 295 km). Thus, we expect the J-PARC based experiments to provide a cleaner environment for the measurements in $\delta_{\mathrm{CP}}$, even though the values reached by the asymmetries in these cases are not as high as the DUNE. On the other hand, the bottom panels of Fig.~\ref{work4:fig:2}, which correspond to the second oscillation maximum $L/E$ choice for T2HKK, behave just like the bottom panels in Fig.~\ref{work4:fig:1}.

\subsection{Extrinsic CP asymmetry in disappearance channel}
\label{sec:2b}

\begin{figure}[htb!]
     \centering
     \includegraphics[scale=1]{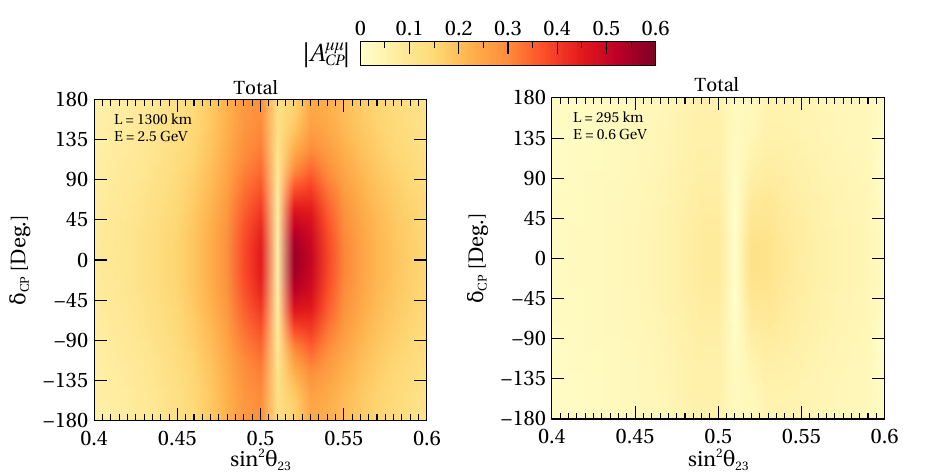}
     \caption{ $|\mathcal{A}^{\mu \mu}_{\mathrm{CP}}|$ as a function of $\delta_{\mathrm{CP}}$ and $\sin^{2}\theta_{23}$ assuming NMO for first oscillation maximum in DUNE ($L$ = 1285 km, $E$ = 2.5 GeV) and T2HK ($L$ = 295 km, $E$ = 0.6 GeV). Values of other oscillation parameters are taken from Table~\ref{work4:table:one}. \textbf{Extrinsic (fake) CP asymmetry is maximum around maximal mixing solutions of $\sin^{2}\theta_{23}$\,, decreasing abruptly on either side.}} 
     \label{work4:fig:3}
 \end{figure}
The $\nu_{\mu}$ disappearance channel, being T invariant, is directly CP-conserving in a vacuum-like scenario. However, the Earth matter potential interacts differently with neutrinos and antineutrinos, which can further induce a fake or extrinsic CPV in this channel \cite{Giarnetti:2021wur,Ohlsson:2014cha}. Thus, although the disappearance channel does not directly affect the measurements in CP phase, it is important to realize its ability to generate fake or extrinsic CPV. This is crucial in our study as we will discuss later its effect in our results (Sec.~\ref{sec:4a}).
Following the same convention, as discussed in Ref.~\cite{Agarwalla:2021bzs}, we write the disappearance probability as
\begin{eqnarray}
  P_{\mu\mu}\approx 1-M\sin^2(2\theta_{23})-N\sin^2\theta_{23}-R\sin2\theta_{23}+T\sin4\theta_{23}\,,
\end{eqnarray}
where:
\begin{eqnarray}
 \nonumber M&=&\sin^2\Delta-\alpha\cos^2\theta_{12}\Delta\sin2\Delta+ \\
  & &+\frac{2}{\hat A-1}\sin^2\theta_{13}\left(\sin\Delta\cos(\hat A\Delta)\frac{\sin[(\hat A-1)\Delta]}{\hat A-1}-\frac{\hat A}{2}\Delta \sin2\Delta\right)\,,\label{eq:disapp_M} \\
  R&=&2\alpha\sin\theta_{13}\sin2\theta_{12}\cos\delta_{\mathrm{CP}}\cos\Delta\frac{\sin\hat A\Delta}{\hat A}\frac{\sin[(\hat A-1)\Delta]}{\hat A-1} \,, \label{eq:disapp_R}\\
  T&=&\frac{1}{\hat A-1}\alpha \sin\theta_{13}\sin2\theta_{12}\cos\delta_{\mathrm{CP}}\sin\Delta\left(\hat A \sin\Delta -\frac{\sin\hat A \Delta}{\hat A}\cos[(\hat A-1)\Delta]\right)\,.\label{eq:disapp_T}\nonumber \\ && \,
\end{eqnarray}
and N has already been defined in Eq. \ref{eq:2}. The detailed analytical discussion of fake CP asymmetry in the disappearance channel results in a cumbersome expression. However, for first oscillation minimum ($\Delta=\pi/2$) and the approximated numerical values of the solar and the reactor mixing angles ($\sin\theta_{12}=1/\sqrt{3}$ and $\sin\theta_{13}=1/7$), one can calculate the CP asymmetry in the $\nu_\mu\rightarrow\nu_\mu$ disappearance channel which is defined as
\begin{equation}
\mathcal{A}^{\mu\mu}_{\mathrm{CP}}=\dfrac{P_{\mu \mu}-\bar P_{\mu \mu}}{P_{\mu \mu}+\bar P_{\mu \mu}}\,
\end{equation}
Substituting the discussed approximation in the above expression and neglecting the higher order terms, we get
\begin{eqnarray}
 \mathcal{A}^{\mu\mu}_{\mathrm{CP}}\approx\hat{A}\frac{24\sin^2\theta_{23}+7\sqrt{2}(\pi^2-4)\alpha\cos\delta_{\mathrm{CP}}\sin2\theta_{23}}{6+141\cos2\theta_{23}}\,.
 \label{eq:4}
\end{eqnarray}
This asymmetry increases with the increase in $\theta_{23}$ until the expansion breaks at $\cos2\theta_{23}= - 6/141$. This occurs for $\sin^{2}\theta_{23} > 0.5$ (HO). While after this value, the magnitude of asymmetry starts decreasing with the increase in $\theta_{23}$. In Fig.~\ref{work4:fig:3}, we exhibit the absolute value of the disappearance asymmetry ($|\mathcal{A}^{\mu\mu}_{\mathrm{CP}}|$) for ($L$ = 1285 km, $E$ = 2.5 GeV) and ($L$ = 295 km, $E$ = 0.6 GeV) that also corresponds to DUNE and JD at their respective first oscillation maxima energy. We do not show the plots corresponding to the second oscillation maxima since the fake CP asymmetry in disappearance is solely due to the interaction with the Earth matter potential, whose effect becomes minimal in such conditions. Thus, we do not expect $\mathcal{A}^{\mu\mu}_{\mathrm{CP}}$ in DUNE and KD working at their second oscillation maxima to have any significant contribution to our analysis. We notice that JD, bearing a relatively smaller baseline ($L$ = 295 km), has very small matter effects and thus exhibits very minute fake asymmetry even at first oscillation maximum, as shown in the right panel of Fig.~\ref{work4:fig:3}.  On the other hand, DUNE has a larger baseline ($L$ = 1285 km), thus exhibiting consequential $|\mathcal{A}^{\mu\mu}_{\mathrm{CP}}|$ that reaches as high as $\approx 0.6$ (see the left panel in Fig.~\ref{work4:fig:3}), which is almost comparable to the total $|\mathcal{A}^{\mu e}_{\mathrm{CP}}|$ ($\approx$ 0.8 ) (see the top middle panel in Fig.~\ref{work4:fig:1}). We observe that $|\mathcal{A}^{\mu\mu}_{\mathrm{CP}}|$ is minimal at the two extremes of octant of $\theta_{23}$ for any value of $\delta_{\mathrm{CP}}$. The asymmetry gradually increases while proceeding towards the maximal mixing (MM) corresponding to $\sin^{2}\theta_{23}=0.5$, from either side for almost all $\delta_{\mathrm{CP}}$. However, $|\mathcal{A}^{\mu\mu}_{\mathrm{CP}}|$ manifests two maxima around $\delta_{\mathrm{CP}} = 0^{\circ}$, one each in the two octants: LO ($\sin^2\theta_{23} \approx 0.49$) and HO ($\sin^2\theta_{23} \approx 0.52$). As discussed previously in the analysis of Eq.~\ref{eq:4}, we observe a critical point in HO in the figure as well, around which the nature of $\mathcal{A}^{\mu\mu}_{\mathrm{CP}}$ changes. Correspondingly, around this point, our analytical expansion also breaks. However, in our expression, this completely vanishes ($\mathcal{A}^{\mu\mu}_{\mathrm{CP}} \approx 0$) as we have neglected the higher order terms. This nature of fake $\mathcal{A}^{\mu\mu}_{\mathrm{CP}}$ is crucial in our result, as we will elaborate on this further in our results (Sec.~\ref{sec:4a}).
%
\subsection{$\theta_{23}-\delta_{\mathrm{CP}}$ degeneracy}
\label{sec:2c}
%
From the above discussion, we observe that the value of the atmospheric mixing angle can influence CPV sensitivity. We can further expect this sensitivity to be affected by the octant of $\theta_{23}-\delta_{\mathrm{CP}}$ degeneracy in appearance channel~\cite{Agarwalla:2013ju,Minakata:2013eoa,Coloma:2012wq}. The persisting issue of octant of $\theta_{23}$~\cite{Fogli:1996pv} makes it highly probable for some $\bar\theta_{23}$ and $\bar\delta_{\mathrm{CP}}$ to exist such that for a given $\theta_{23}$, $\delta_{\mathrm{CP}}$,
\begin{eqnarray}
 P_{\mu e}(\theta_{23},\delta_{\mathrm{CP}})&=&P_{\mu e}(\bar\theta_{23},\bar\delta_\mathrm{{CP}}) \,,
 \label{systema}\\
 \bar P_{\mu e}(\theta_{23},\delta_{\mathrm{CP}})&=&\bar P_{\mu e}(\bar\theta_{23},\bar\delta_{\mathrm{CP}})\,,
 \label{systemb}
\end{eqnarray}
holds true.
 For instance, fixing $\theta_{23}=45^\circ$, $\sin\theta_{13}=1/7$, and $\sin\theta_{12}=1/\sqrt{3}$ in presence of matter effect, $\nu_{\mu}\to \nu_{e}$ oscillation probability is given as:
\begin{eqnarray}
 P_{\mu e}(45^{\circ}+x,\bar\delta_{\mathrm{CP}})=P_{\mu e}(45^{\circ},\bar\delta_{\mathrm{CP}})+\frac{4x \sin[(\hat{A}-1)\Delta]^2}{49(\hat{A}-1)^2}\,.
\end{eqnarray}
For the above scenario, the system of equations in Eq.~\ref{systema} and Eq.~\ref{systemb}
reduces to the following two equations-
\begin{subequations}
\begin{equation}
 \frac{\sqrt{2}\alpha\sin(\hat{A}\Delta)}{3\hat{A}}[\cos(\Delta-\delta_{\mathrm{CP}})-\cos(\Delta-\bar\delta_{\mathrm{CP}})] = x\frac{\sin[(\hat{A}-1)\Delta]}{7(\hat{A}-1)}\, \mathrm{and}
  \label{rhsa}
 \end{equation}
  \begin{equation}
 \frac{\cos(\Delta+\delta_{\mathrm{CP}})-\cos(\Delta+\bar\delta_{\mathrm{CP}})}{\cos(\Delta-\delta_{\mathrm{CP}})-\cos(\Delta-\bar\delta_{\mathrm{CP}})} =  \frac{\sin[(\hat{A}-1)\Delta]}{\sin[(\hat{A}+1)\Delta]}\frac{1+\hat{A}}{\hat{A}-1}\,.
 \label{rhsb}
\end{equation}
\end{subequations}
So, for each value of $\Delta$, the above two equations will have different solutions. Thus, in principle, spectral analysis can reduce the above-mentioned degeneracy. Then, it is clear that in the vacuum-like scenario ($A\to0)$, Eq.~\ref{rhsa} gives $x=0$, while in Eq.~\ref{rhsb} $\bar\delta_{\mathrm{CP}}=\delta_{\mathrm{CP}}$. Therefore in an experiment with negligible matter effect, the role of this degeneracy will not be crucial in determining the sensitivity of $\delta_{\mathrm{\mathrm{CP}}}$. While in the presence of substantial matter effect and taking as an example $\delta_{\mathrm{CP}}=90^\circ$, the above equations will always have a solution for $\bar\delta_{\mathrm{CP}}$ except when $\Delta= \pi/2$. Moreover, we have checked that given a certain matter potential, there is always a value of $\Delta < \pi/2$,   for which we obtain $\bar\delta_{\mathrm{CP}}=0$. This infers that, in the presence of matter effect, we can always have identical solutions for maximal CP-violating and a CP-conserving case. The corresponding value of $x$ in such cases is always positive, implying that the degenerate solution lies in the higher octant. While on the other hand, when $\delta_{\mathrm{CP}}=-90^\circ$, we obtain degenerate solutions for $\bar\delta_{\mathrm{CP}}$, but to get the CP-conserving degenerate solution, we need to find a specific $L/E$ ratio that exceeds the value which we obtain at the atmospheric peak. Moreover, in this case, the corresponding deviation of the mixing angle is negative; hence the degenerate solution lies in the lower octant. 

Next, we discuss the $\theta_{23}-\delta_{\mathrm{CP}}$ degeneracy when considering both disappearance and appearance channels. For the degeneracy to occur with all other mixing angles and mass-splittings kept fixed, there should exist $\bar\theta_{23}$ and $\bar\delta_{\mathrm{CP}}$ such that
\begin{eqnarray}
 P_{\mu \mu}(\theta_{23},\delta_{\mathrm{CP}})&=&P_{\mu \mu}(\bar\theta_{23},\bar\delta_{\mathrm{CP}}) \, \\
 \bar P_{\mu \mu}(\theta_{23},\delta_{\mathrm{CP}})&=&\bar P_{\mu \mu}(\bar\theta_{23},\bar\delta_{\mathrm{CP}}).
\end{eqnarray}
However, the disappearance probability only has a mild dependence on $\delta_{\mathrm{CP}}$, which plays an important role when there is a substantial matter effect and $\theta_{23}$ around maximal mixing, as elaborated previously in the discussion of Fig.~\ref{work4:fig:3}. Otherwise, the disappearance channel precisely measures the atmospheric mixing angle in $\delta_{\mathrm{CP}}$ independent way. Thus disappearance channel helps in constraining $\theta_{23}$ when it is not around maximal mixing, which reduces the effect of $\theta_{23}-\delta_{\mathrm{CP}}$ degeneracy for these $\theta_{23}$ in the appearance channel as well. Thus a combined analysis between the appearance and disappearance channels is expected to break the $\theta_{23}-\delta_{\mathrm{CP}}$ degeneracy only if (a) the matter effects are negligible or (b) true values of the atmospheric mixing angle are far away from the maximal mixing scenario, which for DUNE is true if $\sin^2\theta_{23}\notin[0.48 , 0.55]$ (refer to the discussion around Fig.~\ref{work4:fig:3}). Therefore, in this region we obtain a  $\theta_{23}$ independent $\delta_{\mathrm{CP}}$ measurement.  
\subsection{Discussion at event-level}
\begin{table}
\resizebox{\columnwidth}{!}{%
\centering
\begin{tabular}{ |c|c|c|c|c|}
\hline
\multirow{2}{*}{Parameter} &\multirow{2}{*}{$\theta_{23}$}&$\delta_{\mathrm{CP}}$= 0$^{\circ}$ & $\delta_{\mathrm{CP}}=90^{\circ}$ & $\delta_{\mathrm{CP}}=-90^{\circ}$ \\
&&($\nu_e$, $\bar{\nu}_e$, $\mathcal{N}^{\mu e}_{\mathrm{CP}}$)&($\nu_e$, $\bar{\nu}_e$, $\mathcal{N}^{\mu e}_{\mathrm{CP}}$)&($\nu_e$, $\bar{\nu}_e$, $\mathcal{N}^{\mu e}_{\mathrm{CP}}$)\\
\hline
\multirow{3}{4em}{DUNE} &
$40^{\circ}$ & 1965, 812, \it{0.41} &1657, 857, \it{0.31} &2303, 728, \it{0.52}\\
& $45^{\circ}$ & 2215, 875, \it{0.43} &1902, 920, \it{0.34} & 2558, 790, \it{0.53}\\
& $50^{\circ}$ & 2470, 938, \it{0.45} &2161, 982, \it{0.37} & 2807, 854, \it{0.53} \\
\hline
\multirow{3}{4em}{ T2HK} & $40^{\circ}$ & 1644, 1420, \it{0.074} & 1277, 1687, \it{-0.14}  & 2024, 1113, \it{0.29}\\
& $45^{\circ}$ & 1890,1594, \it{0.085}  & 1517, 1868, \it{-0.1}  & 2276, 1286, \it{0.28} \\
& $50^{\circ}$ &  2137, 1770, \it{0.093} & 1770, 2041, \it{-0.07}  & 2518, 1476, \it{0.26}\\
\hline
\end{tabular}}
\caption{
Illustrative total appearance event rates (signal + background) in neutrino, antineutrino modes, and 
$\mathcal{N}^{\mu e}_{\mathrm{CP}}$ for DUNE (T2HK). \textbf{$\mathcal{N}^{\mu e}_{\mathrm{CP}}$ in the CP-violating case tends to come closer to its corresponding value in the CP-conserving case as $\theta_{23}$ increases in both the experiments.}
}
\label{table:three}

\end{table}

In Table~\ref{table:three}\,, we summarize the number of neutrino and antineutrino events for DUNE and JD for three different choices of $\delta_{\mathrm{CP}}$ ($0^\circ,\,90^\circ,\,-90^\circ)$ and three choices of $\theta_{23}$: LO ($40^\circ$), MM ($45^\circ$), and in the HO ($50^\circ$). For the sake of comparison, we also compute the values of the integrated asymmetries, defined as
\begin{eqnarray}
 \mathcal{N}^{\mu e}_{\mathrm{CP}}=\frac{N_{\mu e}-\bar{N}_{\mu e}}{N_{\mu e}+\bar{N}_{\mu e}}\,,
\end{eqnarray}
where $N_{\mu e}$ ($\bar N_{\mu e}$) is the number of events in the neutrino (antineutrino) mode. JD statistics are higher than DUNE due to the higher exposure: 2431 kt$\cdot$MW$\cdot$yrs in the J-PARC based and 480 kt$\cdot$MW$\cdot$yrs in the Fermilab based experiment. Moreover, as expected, for the favorable choice of parameters, i.e., NMO, $\delta_{\mathrm{CP}}=-90^\circ$ ($\delta_{\mathrm{CP}}= 90^\circ$), and HO, we observe the highest neutrino (antineutrino) events. The integrated asymmetries follow the same nature as we observe at the probability level (refer to Fig.~\ref{work4:fig:1}). Here also $\mathcal{N}^{\mu e}_{\mathrm{CP}}$ in the CP-violating case tends to come closer to its value in the CP-conserving case as $\theta_{23}$ increases in both the experiments.
%
\section{Our Results}
\label{sec:4}

In this section, we discuss the abilities of DUNE, T2HK (JD), and their combination in achieving the landmark of 75\% CP coverage in true $\dcp$ for leptonic CPV at 3$\sigma$ C.L. for all the values of $\sin^{2}\theta_{23}$ in Nature. We also explore the capability of the T2HKK (JD + KD) setup. Further, we also inspect the effect of changing overall exposure in each setup while determining the CP coverage. Next, we survey the optimal runtime in neutrino and antineutrino modes to effectively increase CP coverage in different experimental setups. Finally, we address the importance of systematic uncertainties and how their variation can significantly affect our results.

In our analysis, we define the CP coverage as the percentage of true $\delta_{\mathrm{CP}}$ for which an experiment establishes at least a 3$\sigma$ CPV sensitivity. For this, we generate our prospective data assuming the entire range of true $\delta_{\mathrm{CP}} \in [-180^{\circ} , 180^{\circ}]$, while in the fit, we minimize over the test $\delta_{\mathrm{CP}} = 0^{\circ}$ and $180^{\circ}$ and choose the minimum. We use the Poissonian $\chi^2$~\cite{Baker:1983tu} and estimate the median sensitivity~\cite{Cowan:2010js} of a given LBL experiment in the frequentist approach~\cite{Blennow:2013oma}. To evaluate the sensitivity towards leptonic CPV, we use the following definition of $\Delta\chi^2$:
\begin{equation}
	\Delta \chi^2 = \underset{(\delta^{\mathrm{test}}_{\mathrm{CP}}, \,\sin^2\theta_{23})}{\mathrm{min}} \,\bigg[\chi^2(\delta^{\mathrm{true}}_{\mathrm{CP}})-\chi^2(\delta^{\mathrm{test}}_{\mathrm{CP}}=0^{\circ}\, \text{and}\, 180^{\circ})\bigg]\,.
	\label{work4:eq:chi2-cp-coverage}
\end{equation}
The fit is performed by minimizing over $\theta_{23}$ in its current 3$\sigma$ range of [0.4 , 0.6] while keeping
all other parameters fixed at the benchmark values as mentioned in Table~\ref{work4:table:one}. The present global neutrino oscillation data~\cite{Capozzi:2021fjo,deSalas:2020pgw,Esteban:2020cvm} measures atmospheric mass splitting with a very high precision of 1.1\%, which will further improve to 0.5\% with six years of data taking by JUNO~\cite{JUNO:2022mxj}. Therefore, we do not minimize over the present uncertainty in $\Delta m^{2}_{31}$. Also, we do not minimize over the wrong mass ordering in any of the results as there are hints towards NMO from the global oscillation data~\cite{Capozzi:2021fjo,deSalas:2020pgw,Esteban:2020cvm}. Moreover, the currently running LBL experiments: NO$\nu$A and T2K, and the atmospheric experiments: Super-K and DeepCore will further strengthen the mass ordering measurements in the near future. Also, by the time DUNE accumulates data for CPV searches, it is expected to already fix the mass ordering. For all the simulations, we use the GLoBES software \cite{Huber:2004ka,Huber:2007ji}, using the definition of $\chi^2$ in Eq.~\ref{eq:chi2-in-all}. 
%
\subsection{Impact of $\theta_{23}$ on CP coverage}
\label{sec:4a}
 \begin{figure}[htb!]
     \centering
     \includegraphics[width=\linewidth]{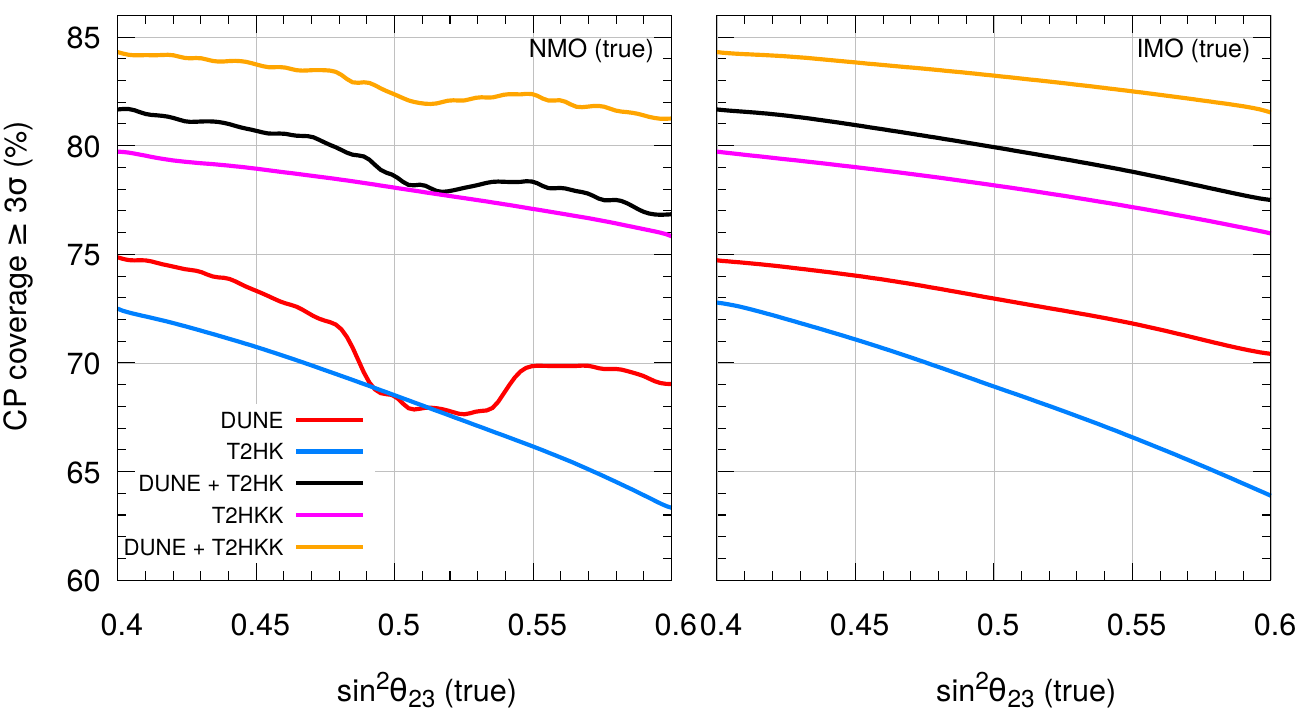}
     \caption{Coverage in true $\dcp$ for achieving $\geq 3 \sigma$ leptonic CPV as a function of true $\sin^{2}\theta_{23}$. We minimize over the current 3$\sigma$ uncertain range of $\sin^{2}\theta_{23}$ (refer to Table~\ref{work4:table:one}) in the theory following Eq.~\ref{work4:eq:chi2-cp-coverage} and assuming benchmark exposure, and the nominal runtime as mentioned in Table~\ref{table:contrasting-features-dune-t2hk} in both data and theory. \textbf{Complementarity due to combination is the only solution to achieve 75\% CP coverage for $3\sigma$ CPV, irrespective of the mass ordering and $\sin^{2}\theta_{23}$ in Nature.}}
 \label{work4:fig:4}
 \end{figure}
 \begin{figure}[htb!]
     \centering
     \includegraphics[width=\linewidth]{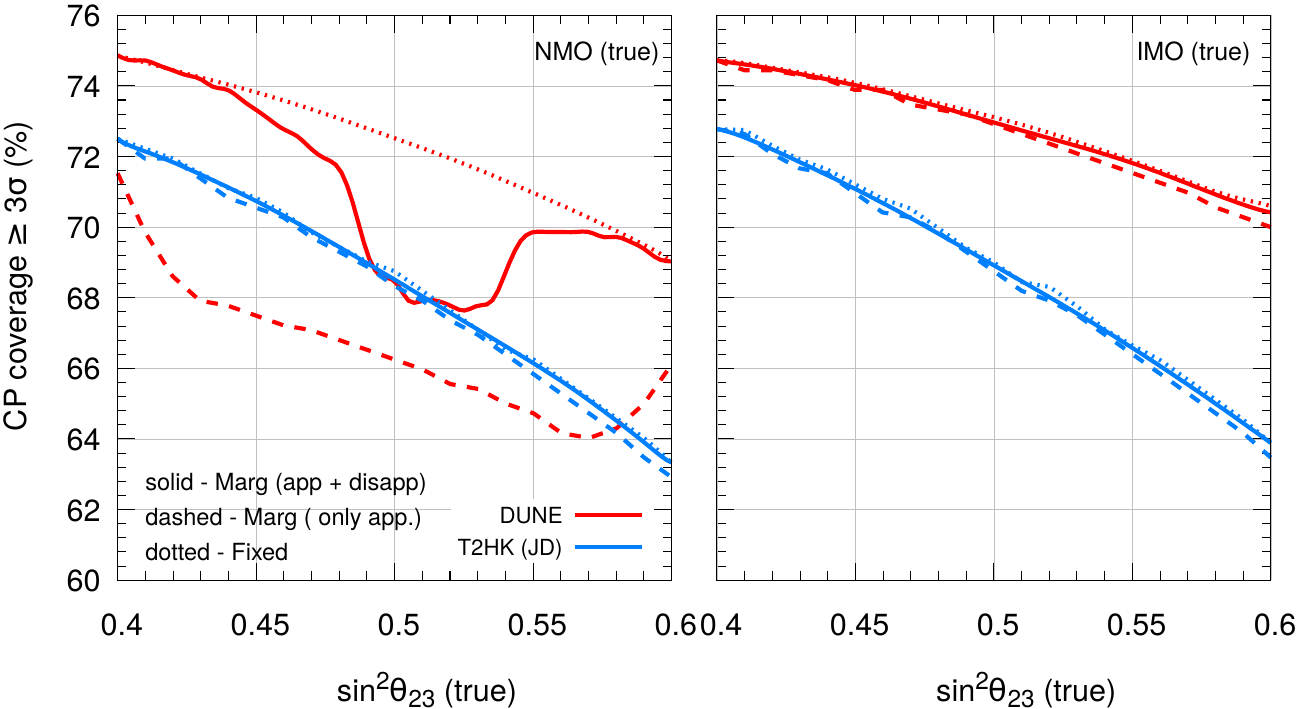}
     \caption{Supplement to Fig.~\ref{work4:fig:4}, separate contribution of detection channels in establishing CP coverage as a function of true $\sin^{2}\theta_{23}$. We assume NMO in the left and IMO in the right panel. See Sec.~\ref{sec:4a} for details. \textbf{Letting $\sin^{2}\theta_{23}$ free in the fit leads to $\sin^{2}\theta_{23} - \dcp$ degeneracy in the appearance statistics of DUNE as it has considerable matter effect.} }
     \label{work4:fig:5}
\end{figure}

 \begin{figure}[htb!]
     \centering
     \includegraphics[width=\linewidth]{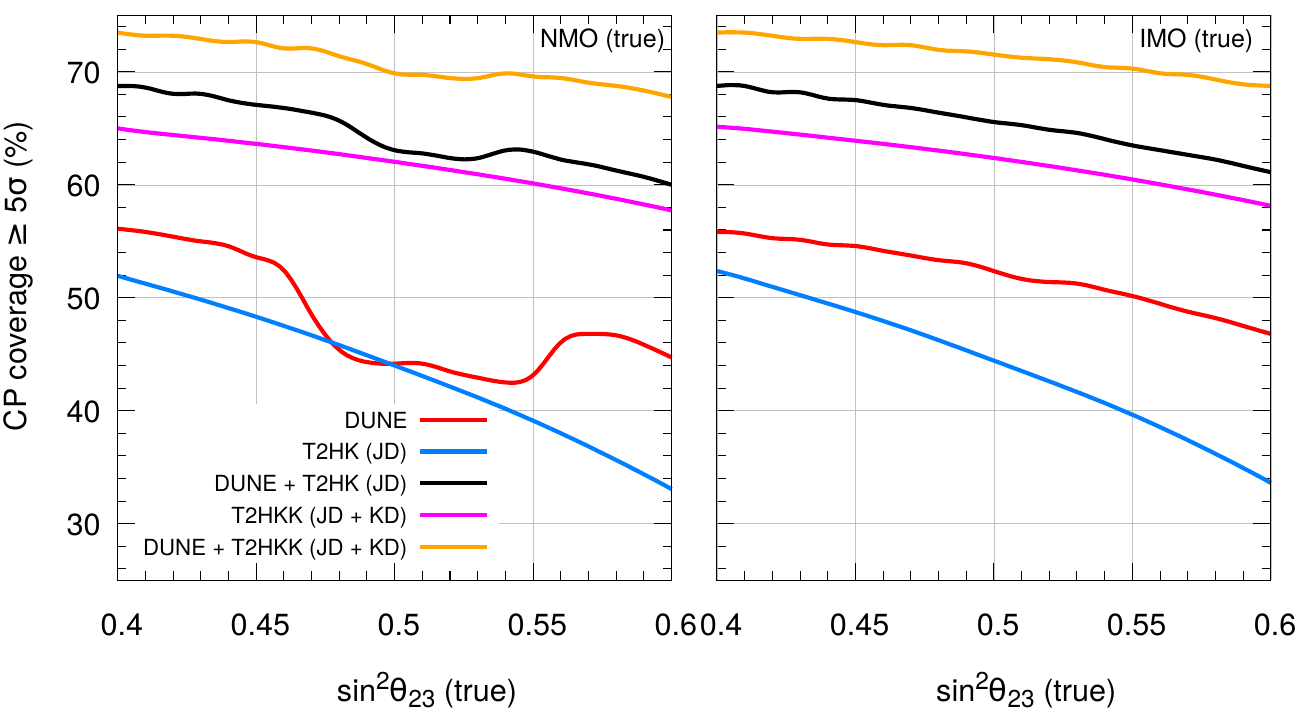}
     \caption{Coverage in true $\dcp$ for achieving $\geq 5 \sigma$ leptonic CPV as a function of true $\sin^{2}\theta_{23}$. We minimize over the current 3$\sigma$ uncertain range of $\sin^{2}\theta_{23}$ (refer to Table~\ref{work4:table:one}) in the theory following Eq.~\ref{work4:eq:chi2-cp-coverage} and assuming benchmark exposure, and the nominal runtime as mentioned in Table~\ref{table:contrasting-features-dune-t2hk} in both data and theory. \textbf{Combined DUNE + T2HK is able to project a 5$\sigma$ discovery of CP-violation for $\sim 60\%$ of leptonic CP phase, irrespective of the mass ordering and $\sin^{2}\theta_{23}$ in Nature.}}
 \label{work4:fig:6}
 \end{figure}

As previously discussed in Sec.~\ref{sec:2}, the atmospheric mixing angle can play an important role in determining the CPV sensitivity. In Fig.~\ref{work4:fig:4}, we depict how the ability of an experimental setup to measure CP coverage changes with the true value of $\sin^{2}\theta_{23}$. We generate our data for each $\sin^{2}\theta_{23}$ (true) by varying $\sin^{2}\theta_{23}$ in our theory throughout the uncertain range of [0.4 , 0.6]. Each colored curve corresponds to the CP coverage of a particular setup as a function of the true value of $\sin^2\tzm$. We observe that none of the individual experiments: DUNE (red curve) or JD (blue curve) can achieve the milestone of 75\%. However, their combination makes CP coverage for the entire canvas of $\sin^{2}\theta_{23}$ above 75\%. This points out that the complementarity between DUNE and JD or JD and KD can help attain a better CP coverage irrespective of the $\sin^{2}\theta_{23}$ value in Nature. Thus, if we are concerned about only CP coverage, there may be no remarkable advantages in adding a second detector to T2HK as DUNE + JD attains a better CP coverage than JD + KD combined. In all the setups, there is a general notion of CP coverage decreasing as we increase $\sin^{2}\theta_{23}$ in the data; the reason for that can be found in the behavior of the appearance asymmetry, as previously discussed in text around figures~\ref{work4:fig:1} and~\ref{work4:fig:2}. In DUNE, the $3\sigma$ CP coverage decreases from 75\% to 68\% when $\theta_{23}$ increases from $40^\circ$ ($\sin^2\theta_{23}=0.4$) to $60^\circ$ ($\sin^2\theta_{23}=0.6$), while in JD the coverage decreases from 73\% to 63\% of true $\dcp$ for the same. The performance of DUNE is observed to be better than JD, but not around maximal mixing. DUNE and combined setups with DUNE (DUNE + JD and DUNE + JD + KD) exhibit an additional worsening around maximal mixing choices, which is absent in JD and JD + KD. 

While on going from NMO to IMO, both $\alpha$ and $\hat A$ have a negative sign. Therefore, apart from an overall minus sign and tiny modifications due to the benchmark values of the parameters under the IMO hypothesis (refer to Table~\ref{work4:table:one}), the behavior of the asymmetry in the electron appearance channel is exactly the same as discussed previously in Sec.~\ref{sec:4a}. In the disappearance channel, the change in sign of the mass splittings ratio, $\alpha$ modifies the absolute value of the asymmetry of the disappearance channel shown in Eq. \ref{eq:4}. This leads to a smaller matter-induced asymmetry in the $\nu_\mu\to\nu_\mu$ channel, which in turn slightly improves the capability of the disappearance channel to measure the atmospheric mixing angle, independent of the $\delta_\mathrm{CP}$ value, in the IMO scenario \cite{Fukasawa:2016yue}. Thus, by simply looking at the asymmetries, one would expect the results on the $\dcp$ coverage to be very similar for both mass orderings. Nonetheless, at the probability level, it is possible to notice that in IMO, the $\theta_{23}-\delta_\mathrm{CP}$ degeneracy is relatively milder. For instance, we find that if $\Delta m_{31}^2<0$ and $\dcp$ is maximal, Eq. \ref{rhsb} has no solutions for $\bar{\delta}_{\mathrm{CP}}$. This is also valid for other values of $\delta_{\mathrm{CP}}$, even when matter effects are not negligible. This exemplifies that, in IMO, the appearance channel in itself measures $\dcp$ with good precision even when the atmospheric mixing angle is around the maximal mixing in DUNE.

To further explore this, we try to understand this nature in individual setups: DUNE and JD. In Fig.~\ref{work4:fig:5}, we represent CP coverage in three scenarios for both JD and DUNE. In the first (dotted blue and red curves) we fix the same value of atmospheric mixing angle in both theory and data; the second (solid blue and red curves) is the result from previously discussed Fig.~\ref{work4:fig:4}, wherein we minimize in theory over $\sin^{2}\theta_{23}$ in the current 3$\sigma$ uncertain range of [0.4 , 0.6]; in the third (dashed blue and red curves) we show the contribution from only appearance channel minimized over the allowed $3\sigma$ range in $\sin^2\theta_{23}$. In the fixed parameter case, there is no role of $\theta_{23}-\delta_{\mathrm{CP}}$ degeneracy since both are fixed to their true value in the fit.  Thus, we see monotonically decreasing CP coverage with increasing $\theta_{23}$. This follows the nature of CP asymmetry that we discussed at both probability and event levels. However, once we consider the freedom of uncertainty of $\theta_{23}$ in theory, the CP coverage drastically decreases around the maximal true value of $\tzm$ in the case of DUNE. This signifies that there are CP phases in DUNE, which, when considered in a fixed parameter scenario, gives a 3$\sigma$ or larger sensitivity towards CPV, but they fail to attain the same when the minimization is performed. So, these act as unfavorable CP phases in DUNE, which are absent in JD. For instance, in DUNE, if we fix both data and theory at $\delta_{\mathrm{CP}}$ = -148$^{\circ}$ and  $\sin^{2}\theta_{23}$ = 0.5, we obtain $\Delta \chi^{2} = 9.82$, which drops down to 6.44 when we vary $\sin^{2}\theta_{23}$ in theory; however for the same set of parameters in JD, $\Delta \chi^{2} = 10.6$ changes only to $\Delta \chi^{2} = 10.4$ when minimized over $\sin^{2}\theta_{23}$. For this reason, in JD, minimization has a negligible effect on the CP coverage.  Further, there are instances where minimized $\Delta \chi^{2}$ chooses the opposite octant in the fit in the case of DUNE. For instance, when we generate the data with $\delta_{\mathrm{CP}}$ = -145$^{\circ}$ and  $\sin^{2}\theta_{23}$ = 0.48 (LO), we obtain a minimized $\Delta \chi^{2}$ for $\sin^{2}\theta_{23}$ = 0.531 (HO) in DUNE leading to $\theta_{23}-\delta_{\mathrm{CP}}$ degeneracy, while this is not the case in JD. Therefore this degeneracy becomes crucial when the matter potential is considerable (like in DUNE), while the same degeneracy almost vanishes when we are in a vacuum-like scenario.

Now, the dashed curves represent the contribution from only appearance, which is quite less in DUNE when compared with the CP coverage from both appearance and disappearance (solid red). However, solid and dashed blue-colored curves are almost overlapping. This signifies that the effect of disappearance events is much more crucial in DUNE than in JD. This is because, in JD, the absence of any significant matter effect drastically reduces the impact of the $\theta_{23}-\delta_{\mathrm{CP}}$ degeneracy. Since the leading term in the disappearance channel is dependent on $\sin^{2}2\theta_{23}$, it strongly constrains the $\theta_{23}$ parameter in a $\delta_{\mathrm{CP}}$ independent manner for regions far from maximal mixing. For these values of $\sin^{2}2\theta_{23}$, the appearance channel suffers less from the $\theta_{23}-\delta_{\mathrm{CP}}$ degeneracy in DUNE just like JD. However, for $\sin^{2}\theta_{23}$ around MM, as discussed in Fig~\ref{work4:fig:3}, extrinsic or matter-induced fake CP asymmetry also plays an essential role because of the substantial matter effect in DUNE. This worsens the CP coverage since the effect of $\theta_{23}-\delta_{\mathrm{CP}}$ degeneracy is much more dominant. On the other hand, JD remains almost independent of this fake CP asymmetry as the matter effect is negligible here. Therefore, we must notice that despite the bigger systematic uncertainties in T2HK, it can achieve a better CP coverage of true $\dcp$ in leptonic CPV than DUNE around the MM of $\sin^{2}\theta_{23}$. 

For sake of completeness, we also show the $5\sigma$ coverage in function of $\sin^2\theta_{23}$. The nature of curves in Fig.~\ref{work4:fig:6} follows the tendency we observed in Fig.~\ref{work4:fig:4} assuming NMO and IMO, respectively. In Fig.~\ref{work4:fig:6}, the maximum CP coverage reached by DUNE (T2HK) is 56\% (52\%), while the minimum CP coverage is 45\% (33\%). While a projected 5$\sigma$ discovery of CP-violation is achievable for $\sim 60\%$ CP phase irrespective of the mass ordering and $\sin^{2}\theta_{23}$ in Nature with the combined DUNE + T2HK setup. Further, the projected discovery potential of DUNE + T2HK is better than T2HKK, hence reducing the impact of the second detector in T2HK. However, DUNE + T2HKK in the best case scenario (LO) could raise the 5$\sigma$ coverage up to an astonishing 74\%. Since IMO does not observe any deterioration due to $\theta_{23}-\dcp$ degeneracy, we will focus more on NMO in the following sections.
%
 \subsection{CP coverage as a function of exposure}
\label{sec:4b}
%
 \begin{figure}[htb!]
     \centering
     \includegraphics[width=1.0\linewidth]{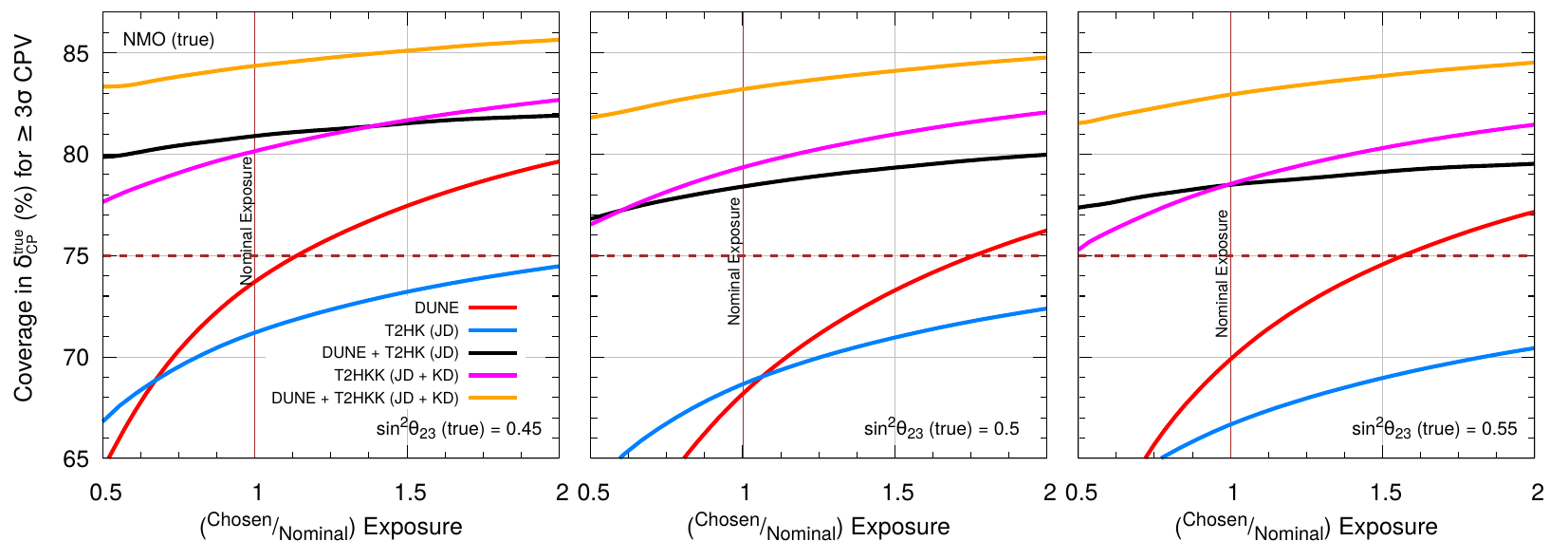}
     \caption{Coverage in true $\dcp$ for $\geq 3\sigma$ leptonic CPV as a function of scaled exposure assuming true NMO. We obtain these curves by generating the data with true $\sin^{2}\theta_{23}$ = 0.45 (LO), 0.5 (MM), and 0.55 (HO) and minimizing over $\theta_{23}$ in the fit (refer to Table~\ref{work4:table:one}) following Eq.~\ref{work4:eq:chi2-cp-coverage}. See Table~\ref{table:contrasting-features-dune-t2hk} for benchmark exposure. \textbf{DUNE + T2HK can achieve $\sim 76\%$ CP coverage independent of the exposure, while standalone experiments need full exposure for the same. } } 
     \label{work4:fig:7}
 \end{figure}
In this section, we discuss the CP coverage of various experimental setups when the total exposure in the experiment is varied. Recently, the DUNE collaboration had an extensive study on how they expect to achieve desired exposure in a staged manner~\cite{DUNE:2021mtg}. However, to study its effect in our analysis, we take a more simplistic approach and vary the full exposure by reducing it to half of its nominal value and increasing it to twice. In Fig. \ref{work4:fig:7}, we show how the CP coverage is influenced by the change in a total exposure of the experiments for $\sin^{2}\theta_{23}=0.45$, 0.5, and 0.55. The curves are shown for DUNE (red), JD (blue), DUNE + JD (black), JD + KD (magenta), and DUNE + JD + KD (orange). We obtain these by minimizing the atmospheric angle. We observe that by keeping the true value for $\sin^{2}\theta_{23}$ fixed in LO and doubling the exposure from the nominal value of 2431 (480) kt$\cdot$MW$\cdot$yrs in JD (DUNE), the coverage increases from 74\% to 79\% in DUNE and from 71\% to 75\% in JD. On the other hand, if the exposure is reduced up to half the nominal value, the CP coverage drastically reduces for both experiments. Also, DUNE outperforms JD when we compare both experiments at their 70\% of nominal exposures. Even though going from LO to MM, we observe a similar trend of increasing CP coverage with exposure, the maximum reachable coverage is now reduced (76\% for DUNE, 72\% for JD). Moreover, at MM and nominal exposures, JD outperforms DUNE. This happens due to strong $\tzm-\dcp$ degeneracy in DUNE near MM, which results in the reduction of CP coverage as compared to JD.
In the HO case, coverage worsens further for JD (maximum coverage 70\%), but not for DUNE, which now suffers less from the $\theta_{23}-\delta_{\mathrm{CP}}$ degeneracy under the non-maximal case. Following the previous discussion on fixing the value of $\sin^{2}\theta_{23}$, DUNE's performance improves considerably in MM, but only a little in the HO scenario.
 
It is interesting to note that the complementarity between DUNE and T2HK (JD) can achieve more than 75\% coverage even if we consider only half of their individual nominal exposures in the three panels. As discussed earlier, JD + KD further increases the CP coverage significantly for all the three values of true $\sin^{2}\theta_{23}$. Moreover, under the nominal exposure, the combination of DUNE + JD + KD establishes CP coverage always above $\sim 80\%$. 
%
\subsection{Optimizing runtime for maximal CP coverage}
\label{sec:4c}
%
 \begin{figure}[htb!]
     \centering
     \includegraphics[width=1.0\linewidth]{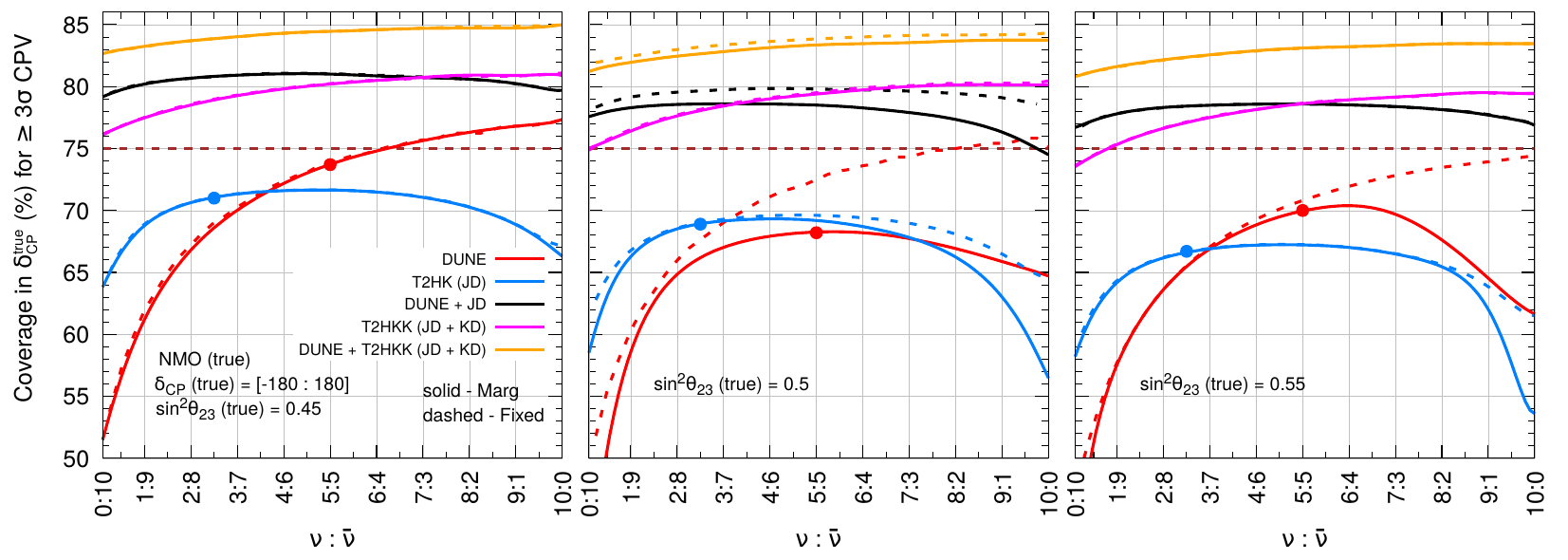}
     \caption{Coverage in true $\dcp$ for $\geq 3\sigma$ leptonic CPV as a function of the ratio of the runtime in neutrino and antineutrino ($\nu : \bar{\nu}$) modes, assuming true NMO. The dashed lines are obtained by considering identical $\sin^{2}\theta_{23}$ in both data and theory, while the solid lines follow Eq.~\ref{work4:eq:chi2-cp-coverage} and the benchmark parameters in Table~\ref{work4:table:one}. Nominal runtime in DUNE and T2HK is labelled (refer to Table~\ref{table:contrasting-features-dune-t2hk}). \textbf{Sensitvity to establish CPV at 3$\sigma$ is independent of the choice of distribution of runtime in $\nu$ and $\bar{\nu}$ modes for DUNE + T2HK, irrespective of $\sin^{2}\theta_{23}$ in Nature. } }
     \label{work4:fig:8}
 \end{figure}
%
While discussing the total exposure, it is also important to determine the optimal runtime in neutrino and in antineutrino modes for higher CP coverage. The two collaborations have proposed different approaches. 
  With the intent of having a similar number of neutrino and antineutrino events (see in Table~\ref{table:contrasting-features-dune-t2hk}), the T2HK (JD) plans to split the total exposure of 10 years into 2.5 years in neutrino and 7.5 years in antineutrino mode (refer to blue filled circles in Fig.~\ref{work4:fig:8}). This choice ensures very small integrated asymmetries in the CP-conserving cases, which help in highlighting easily the differences from the intrinsic asymmetries. Moreover, this choice has already been proven to be very useful for resolving degeneracies~\cite{Agarwalla:2013ju}. Contrastingly in DUNE, they propose a balanced ratio of runtime in neutrino and antineutrino modes of [5 $\nu$ yrs + 5 $\bar{\nu}$ yrs] (refer to red filled circles in Fig.~\ref{work4:fig:8}). However, this will reduce the number of events in antineutrino mode considerably but simultaneously improve the potential for the increased number of neutrino events. To visualize the discussion in Fig.~\ref{work4:fig:8}, we represent how the ratio between neutrino and antineutrino runtime affects the coverage for all the considered setups and the usual three choices of $\sin^{2}\theta_{23}$ (true) (LO, MM, and HO). We distinctly show two possible scenarios: first by fixing the same set of oscillation parameters in both data and theory (dashed curves) and second by minimizing $\sin^{2}\theta_{23}$ (solid curves) in the theory and fixing all other parameters to the benchmark choices from Table~\ref{work4:table:one}. In the LO case, while the nominal choice for JD [2.5 $\nu$ yrs + 7.5 $\bar{\nu}$ yrs] turns out to be the best, DUNE has no advantage of running in antineutrino mode. Instead, we observe that the best coverage (77\%) for DUNE is acquired when only neutrino mode is employed for the full 10 years of runtime. This is because of the $\delta_{\mathrm{CP}}$ independent measurement of $\sin^{2}\theta_{23}$ by the disappearance channel in LO. Once the atmospheric angle is constrained by disappearance, the appearance channel benefits more from the increment in statistics by running only in neutrino mode for 10 years instead of a balanced number of neutrino and antineutrino events because of the small appearance systematic uncertainties in DUNE (2\%).
 In the maximal mixing case, the JD remains almost the same, in contrast to DUNE which establishes the best coverage for the balanced runtime scenario. This is because here, the subdued abilities of the disappearance channel in the minimized $\tzm$ scenario are overcome by the balanced runtime of [5 $\nu$ yrs + 5 $\bar{\nu}$ yrs], thus achieving the best coverage of 68\%. While the HO case is intermediate: the best coverage in DUNE is neither obtained by the balanced [5 $\nu$ yrs + 5 $\bar{\nu}$ yrs] nor with the highest number of events by fully running in only neutrino mode, rather [6.5 $\nu$ yrs + 3.5 $\bar{\nu}$ yrs] scenario is best. This can be understood from previous discussions. We observe that $\sin^{2}\theta_{23}$ (true) = 0.55 is still in the dip region (refer to solid red in Fig.~\ref{work4:fig:5}) but not completely, thus disappearance is not able to constrain $\sin^{2}\theta_{23}$ in the fit but does not deteriorate the significance as much it did in the MM scenario. So, we still feel the effect of the $\theta_{23} - \delta_{\mathrm{CP}}$ degeneracy that requires contribution from both neutrino and antineutrino modes.
 
 However, the complementarity between the two setups plays a crucial role which is essentially independent of $\sin^{2}\theta_{23}$ in Nature. It is quite interesting to observe that DUNE + JD makes the choices of runtime almost irrelevant in establishing coverage in true $\dcp$ for CPV with $\geq$ 3$\sigma$ C.L. around 75\% in all the three panels, given they both run in their full exposure. DUNE + T2HKK can further improve this coverage to about 80\% for the three choices of $\theta_{23}$.
 
 %
  \subsection{Impact of systematic uncertainties on CP coverage}
\label{sec:4d}

\begin{figure}[htb!]
     \centering
     \includegraphics[width=1.0\linewidth]{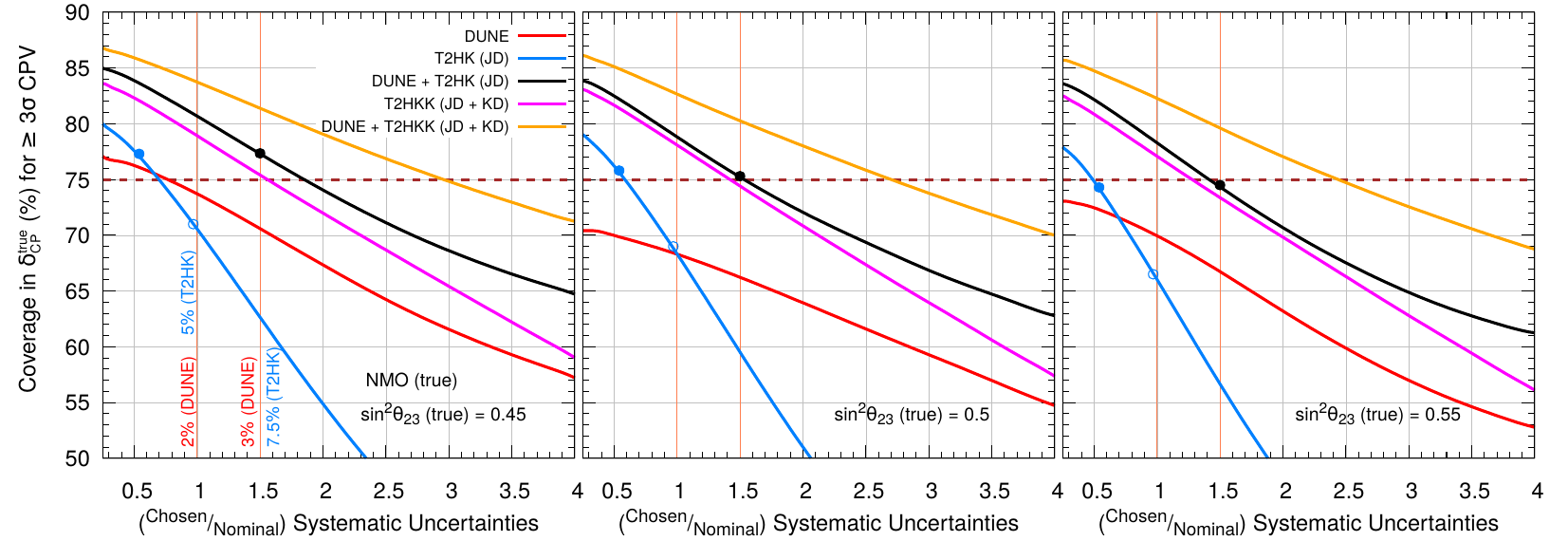}   \caption{Coverage in true $\dcp$ for $\geq 3\sigma$ leptonic CPV as a function of scaled appearance systematic uncertainties, assuming NMO. Refer to Table~\ref{table:contrasting-features-dune-t2hk} for benchmark systematic uncertainties. In the left, middle, and right panels, we obtain the results by considering true $\sin^{2}\theta_{23}$ = 0.45 (LO), 0.5 (MM), and 0.55 (HO), and minimizing over the allowed range of $\sin^{2}\theta_{23}$ in the fit. The blue-colored filled and empty circles in the figure depict CP coverage corresponding to 2.7\% and 4.9\% systematic uncertainties in T2HK, respectively. \textbf{Combined DUNE + T2HK can achieve 75\% CP coverage with just 1.5 times their standalone nominal systematic uncertainties, irrespective of $\sin^{2}\theta_{23}$ in Nature.} } 
     \label{work4:fig:9}
 \end{figure}
%
In Fig. \ref{work4:fig:9}, we illustrate the effect of appearance systematic uncertainties in the coverage in true $\dcp$ for determining CPV with at least 3$\sigma$ C.L. when minimized over $\sin^{2}\theta_{23}$ in the fit. It is clear that the JD curve (refer to blue colored curve) has a steeper slope than  DUNE (refer to red colored curve); however, one must note that the nominal appearance systematic uncertainties in JD (5\%) is more than twice than that of DUNE (2\%). Recently the T2K collaboration~\cite{T2K:2019bcf} has been considering the conservative uncertainties in the appearance systematics of about 4.9\%, which they further expect to improve to about 2.7\% by the time T2HK starts taking data~\cite{Munteanu:2022zla}. Thus we also discuss these two possibilities in Fig. \ref{work4:fig:9} (refer to blue empty and filled circles). Comparing the CP coverage at the expected T2HK systematics of 2.7\% (filled blue circles) with the nominal in DUNE (2\%), we observe that T2HK outperforms DUNE in the three possible choices of $\theta_{23}$. We also confirm that in DUNE, the impact of the minimization becomes negligible  when systematics are higher than 5\%, so the coverage in true $\dcp$ becomes completely systematics dominated.

Contrastingly, in Nature, if the real appearance systematic uncertainty turns out to be about 1.5 times higher than its nominal value in both DUNE and T2HK setups, then the complementarity between them is the only solution to achieve 75\% of coverage in true $\dcp$ for the three possible choices of $\theta_{23}$: 0.45, 0.5, and 0.55 (refer to the coordinates of black filled circles in each panel). Also, when we include the second proposed detector: KD, in the analysis along with DUNE (refer to the orange curve), we achieve the milestone of 75\% coverage, even if appearance systematics is increased by a factor of 2.5 for  three possible choices of true $\theta_{23}$.

\subsection{Effect of current 3$\sigma$ allowed range in $\delta_{\mathrm{CP}}$ on CP coverage}
\label{sec:4e}

 \begin{figure}[t!]
    \centering
    \includegraphics[width = 0.8\linewidth]{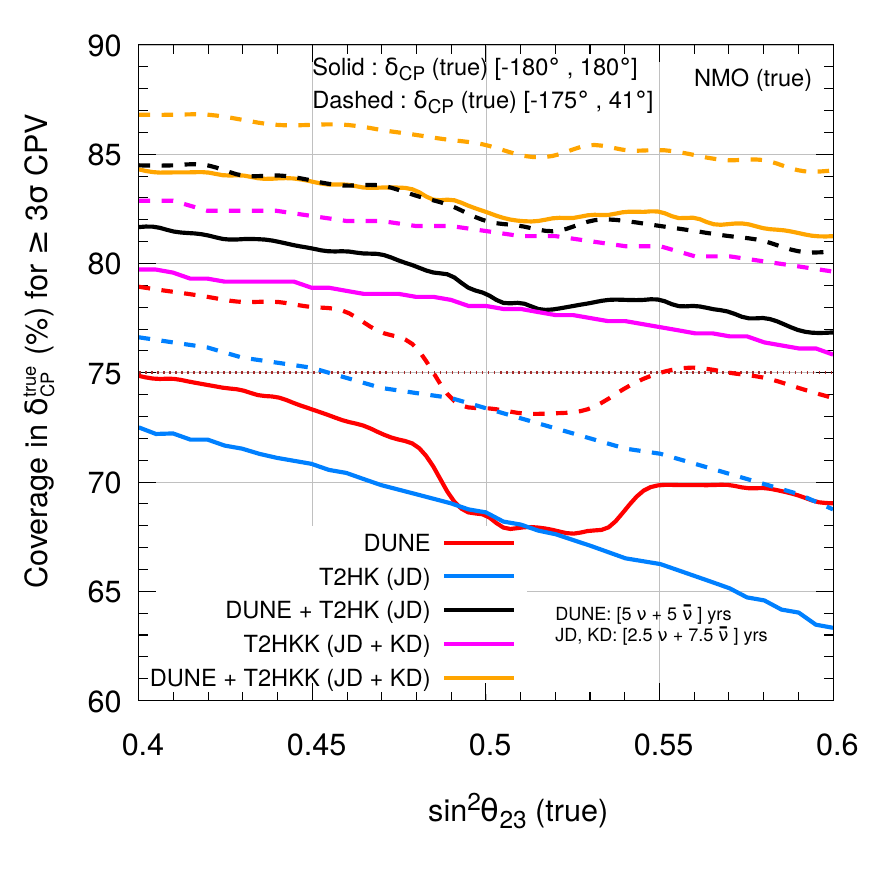}
    \caption{Coverage in true $\dcp$ for $\geq$ 3$\sigma$ leptonic CPV as a function of $\sin^{2}\theta_{23}$ (true) while 
    minimizing over $\theta_{23}$ in the fit. The solid curves represent the results wherein we generate data assuming 
    the entire range of $\delta_{\mathrm{CP}}$ (true) $\in [-180^{\circ} , 180^{\circ}]$ and exclude test $\delta_{\mathrm{CP}}$ 
    = $0^{\circ}$ and $180^{\circ}$ in the fit. On the other hand, the dashed curves show the results when we generate data 
    considering only the present 3$\sigma$ allowed range of $\delta_{\mathrm{CP}}$ (true) 
    $\in [-175^{\circ} , 41^{\circ}]$~\cite{Capozzi:2021fjo} and exclude test $\delta_{\mathrm{CP}}$ = $0^{\circ}$ in the fit.  We assume NMO in both data and theory. \textbf{} }
    \label{fig:9}
\end{figure}

In all the previous results, we consider the entire allowed range of $\dcp \in$ [$-180^{\circ} , 180^{\circ}$] for generating 
the prospective data. However, the presently running LBL accelerator experiments: T2K~\cite{T2K:2019bcf} and 
NOvA~\cite{NOvA:2021nfi} along with the high-precision measurement of $\theta_{13}$ from the Daya Bay reactor 
antineutrino experiment~\cite{kam_biu_luk_2022_6683712} have played an important role to constrain the allowed
parameter space of $\dcp$, which is apparent from the global fit analyses of the world neutrino oscillation 
data~\cite{NuFIT,Capozzi:2021fjo,deSalas:2020pgw,Esteban:2020cvm}. Therefore, it becomes imperative to
estimate the CP coverage of the next-generation LBL experiments like DUNE and T2HK considering the currently 
allowed range of $\dcp$ assuming that the present hints on the allowed values of $\dcp$ will be converted into 
a discovery as more data will become available in the coming years.

In this subsection, we repeat some of our analysis using the present 3$\sigma$ allowed range 
of $\delta_{\mathrm{CP}} \in [-175^{\circ} , 41^{\circ}]$  with a relative 1$\sigma$ uncertainty of 
16\% as obtained in Ref.~\cite{Capozzi:2021fjo}. Since one of the CP-conserving cases, $\dcp=180^\circ$ 
is ruled out in the present $3\sigma$ bound, we now use only $\dcp = 0^\circ$ in the fit and define 
the Poissonian $\Delta \chi^{2}$~\cite{Baker:1983tu} following the frequentist 
approach~\cite{Blennow:2013oma} as follows
\begin{equation}
	\Delta \chi^2 = \underset{(\sin^2\theta_{23})}{\mathrm{min}} \,\bigg[\chi^{2}(\delta^{\mathrm{true}}_{\mathrm{CP}} \in [-175^{\circ} , 41^{\circ}])-\chi^2(\delta^{\mathrm{test}}_{\mathrm{CP}}= 0^{\circ})\bigg]\,.
	\label{eq:3sig_chi2}
\end{equation}
 We calculate CP coverage for different experimental setups using Eq.~\ref{eq:3sig_chi2}. In Fig.~\ref{fig:9}, we show coverage in true $\dcp$ which can establish CPV with at least 3$\sigma$ C.L. as a function of $\sin^2\tzm$. The dashed colored curves represent the CP coverage of a given setup calculated using the $3\sigma$ bounds on $\dcp$.
We observe that with better constraints on $\dcp$, we improve the coverage in $\dcp$ consistently for each $\sin^{2}\theta_{23}$ by almost (4 - 5)\% in both DUNE and JD. Previously in the unconstrained scenario (red solid curve), DUNE does not attain the benchmark of 75\% coverage for any value of $\sin^{2}\theta_{23}$, however, with the new definition (dashed red curve) DUNE can attain 75\% of CP coverage for about 58\% of $\sin^{2}\theta_{23}$ in Nature. Therefore, if in Nature $\sin^{2}\theta_{23}$ turns out to be in any value in LO, then DUNE can easily achieve the milestone of 75\% coverage with the nominal appearance systematics and exposure. Similarly, JD which previously with the entire range of $\dcp$, could achieve only 72\% of CP coverage in the most favorable zone (solid blue curve) improves further to 75\% (dashed blue curve) of coverage if $\sin^{2}\theta_{23} \in [0.4 , 0.45]$ in Nature. 
As expected the combined setups can achieve enhanced coverage as well. This increment in CP coverage is quite expected as here we generate data with a more constrained bound on $\dcp$, and also in the fit, we consider only one CP-conserving ($\dcp = 0^{\circ}$) value for studying CP violation.

%
\section{Summary }
\label{sec:5}

The current knowledge of the active three-neutrino mixing angles and two independent 
mass-squared differences has reached unprecedented precision. One of the remaining 
goals is to establish CPV in the leptonic sector. Even though hints of non-vanishing 
$\delta_{\mathrm{CP}}$ are emerging from the current neutrino data, it is worth testing 
the capability of the future long-baseline experiments such as DUNE, T2HK, and T2HKK 
to establish leptonic CPV at $\ge$ 3$\sigma$ C.L. for a large choices of true $\dcp$ 
in its entire range of $[-180^{\circ}, 180^{\circ}]$. Here, we summarize the main findings
of our paper and mention some crucial points that bring out the novelty of this paper.

\begin{itemize}

\item 
In this paper, we extensively discuss the abilities of next-generation high-precision
LBL experiments DUNE, T2HK, and T2HKK in establishing leptonic CPV at $\ge$ 
3$\sigma$ C.L. in isolation and combination, considering their latest state-of-the-art 
configuration details. We emphasize on the fact that DUNE + T2HK is not a mere 
combination of two LBL experiments, but a necessity to achieve the desired 
milestone by reducing their inherent parameter degeneracies that exist in isolation.

\item 
In Sec.~\ref{sec:2a} and Sec.~\ref{sec:2b}, we discuss in detail the intrinsic and 
extrinsic CP asymmetries in appearance and disappearance channels both 
analytically and numerically, explaining their connections to the measurement
of $\delta_{\mathrm{CP}}$. We observe a non-trivial behavior around MM in 
extrinsic CP asymmetry in DUNE, where the matter effect is more important 
and the experiment is less sensitive in measuring $\theta_{23}$ independently 
of the value of $\delta_{\mathrm{CP}}$.

\item 
We observe that assuming NMO, benchmark values of oscillation parameters, 
nominal systematics and exposures, neither DUNE nor T2HK in isolation 
can achieve 75\% CP coverage in true $\dcp$ irrespective of the choices 
of $\theta_{23}$. While the complementarity between DUNE + T2HK can 
enable us to achieve more than 77\% CP coverage irrespective of the 
values of $\theta_{23}$ in its entire 3$\sigma$ allowed range of 
$\sin^2\theta_{23}\in[0.4-0.6]$ (see discussions in Sec.~\ref{sec:4a}).

\item 
We notice that the capabilities of DUNE in establishing leptonic CP violation 
are significantly worse around the maximal mixing value of $\theta_{23}$. 
This is due to the fact that when $\theta_{23}$ is close to $45^{\circ}$, 
the disappearance channel fails to provide a robust measurement of the 
atmospheric mixing angle $\theta_{23}$ independent of the value of 
$\dcp$ (see Fig.~\ref{work4:fig:5}). The underlying physics reason behind 
this is the presence of substantial extrinsic CP asymmetry in the 
disappearance channel around $\sin^{2}\theta_{23}$ $\approx$ 0.5 
in DUNE, which is clearly visible in Fig. 3 and can be easily understood 
from the analytical expression given in Eq.~\ref{eq:4}.

\item
Another interesting observation that we make is that assuming true NMO 
and the benchmark values of oscillation parameters, the CP coverage that 
DUNE + T2HK can achieve with just half of their individual exposures, 
cannot be attained by these experiments in isolation even with twice of 
their nominal exposures for three different choices of true $\theta_{23}$ 
in LO, MM, and HO (see Fig.~\ref{work4:fig:6}). Also, considering our general 
CP coverage estimates as a function of $\theta_{23}$ in Fig.~\ref{work4:fig:4}, 
the results shown in Fig.~\ref{work4:fig:6} as a function of exposure are valid 
for any values of $\theta_{23}$ in its entire 3$\sigma$ allowed range.

\item 
We observe that the different ratios of runtime in neutrino and antineutrino 
modes are needed in DUNE and T2HK to achieve the better CP coverage 
depending upon the values of true $\sin^{2}\theta_{23}$. For instance, 
T2HK always prefers a ratio of $\nu$ and $\bar\nu$ runtimes for which 
the number of appearance events in neutrino and antineutrino modes 
are almost similar irrespective of the choice of true $\sin^{2}\theta_{23}$. 
This is not the case in DUNE. If true $\sin^{2}\theta_{23}$ lies in the 
lower octant, then the expected number of appearance events in DUNE 
becomes very less and therefore, it prefers to have more run in the 
neutrino mode (see discussions in Sec.~\ref{sec:4c}).

\item 
We notice the pronounced effect of larger nominal systematic uncertainties 
in T2HK (5\%) in comparison with DUNE (2\%) in achieving the desired
CP coverage in true $\dcp$ for $\ge$ 3$\sigma$ leptonic CPV. Further, 
we also observe that in a pessimistic scenario in which the systematic 
uncertainties in DUNE and T2HK turn out to be around 1.5 times larger than 
their nominal ones, the combination of the datasets from DUNE and T2HK
is the only solution to achieve the milestone of 75\% CP coverage 
(see discussions in Sec.~\ref{sec:4d}).

\item 
While studying the CP coverage of the T2HKK setup, we discern that the 
combination of the data from both the Japanese and Korean detectors 
(T2HKK/JD+KD) maybe enough to achieve more than 75\% CP coverage 
in true $\dcp$ for $\ge$ 3$\sigma$ leptonic CPV for all the currently allowed 
values of $\theta_{23}$. At the same time, the combination of DUNE and 
T2HKK can attain an unprecedented CP coverage of around 80\% to 85\%
depending upon the value of true $\sin^2\theta_{23}$ (see Fig.~\ref{work4:fig:4}).

\item 
We also study the effect of the currently allowed 3$\sigma$ range in $\dcp$ and 
observe an improvement of about 5\% in CP coverage in both DUNE and T2HK.

\item 
Finally, we also analyze the CP coverage as a function of $\sin^{2}\theta_{23}$ assuming true IMO. We notice that in this case, the $\theta_{23}-\dcp$ degeneracy is milder in the appearance channel. Therefore, we do not observe any decrease in the coverage around maximal mixing in $\sin^{2}\theta_{23}$ by DUNE. Apart from this feature, the projected coverage attainable by the experiments assuming IMO is similar to NMO.

\end{itemize}

\begin{table}[h!]
    \centering
    \begin{tabular}{|c|c|c|c|}
    \hline \hline
        \multirow{1.25}{*}{$\sin^{2}\theta_{23}$} & \multicolumn{3}{c|}{Coverage in true $\dcp$ for $\geq$ 3$\sigma$ CPV (\%)}\\
        \cline{2-4}
       (true) & DUNE & T2HK   & DUNE + T2HK \\
        \hline
        0.45 & 73 & 71 (77) & 81  (83)\\
        0.5 & 68 & 69 (76)&  78 (79)\\
        0.55 & 70 & 66 (74) & 79 (80) \\
      \hline \hline
    \end{tabular}
    \caption{Coverage in true $\dcp$ for $\ge 3\sigma$ leptonic CPV 
    with nominal exposures and appearance systematic uncertainties 
    in DUNE, T2HK, and DUNE + T2HK for three different true choices 
    of $\sin^{2}\theta_{23}:$ 0.45, 0.5, and 0.55. We follow Eq.~\ref{work4:eq:chi2-cp-coverage} and Fig.~\ref{work4:fig:5} for generating this table.
    The values in the parenthesis have been obtained by considering 
    an improved appearance systematic uncertainty of 2.7\% 
    in Eq.~\ref{work4:eq:chi2-cp-coverage} for T2HK instead of the nominal value of 5\%.}
    \label{work4:table:four}
\end{table}

In Table~\ref{work4:table:four}, we mention the CP coverage in true 
$\delta_{\mathrm{CP}}$ for $\ge$ 3$\sigma$ leptonic CPV 
achievable by DUNE, T2HK, and DUNE + T2HK considering 
their nominal exposures and a systematic uncertainty of 2\% 
(5\%) in $\nu_e/\bar\nu_e$ appearance channel in DUNE (T2HK) in the NMO case.
The bracketed values give the same when an improved systematic 
uncertainty of 2.7\% in $\nu_e/\bar\nu_e$ appearance channel
is considered for T2HK as recently suggested by the 
collaboration~\cite{Munteanu:2022zla}. The complementarity 
between DUNE and T2HK increases the CP coverage achievable 
by DUNE (T2HK) in isolation by about 10\% (13\%) when we 
combine the data from these two experiments. When improved 
systematics are taken into account, we observe a significant 
enhancement in CP coverage by T2HK for the three benchmark 
choices of true $\sin^2\theta_{23}$, outperforming DUNE's performance 
in each case. The combined DUNE + T2HK setup, on the other hand, 
does not exhibit much improvement since the DUNE's contribution 
remains limited by the so-called ($\theta_{23} - \dcp$) degeneracy.
\chapter{Flavor-dependent Long-range Neutrino Interactions in DUNE \& T2HK: Alone they Constrain, Together they Discover}
\label{sec:ch8}

\begin{figure}[t!]
\centering
 \includegraphics[width=0.8\linewidth]{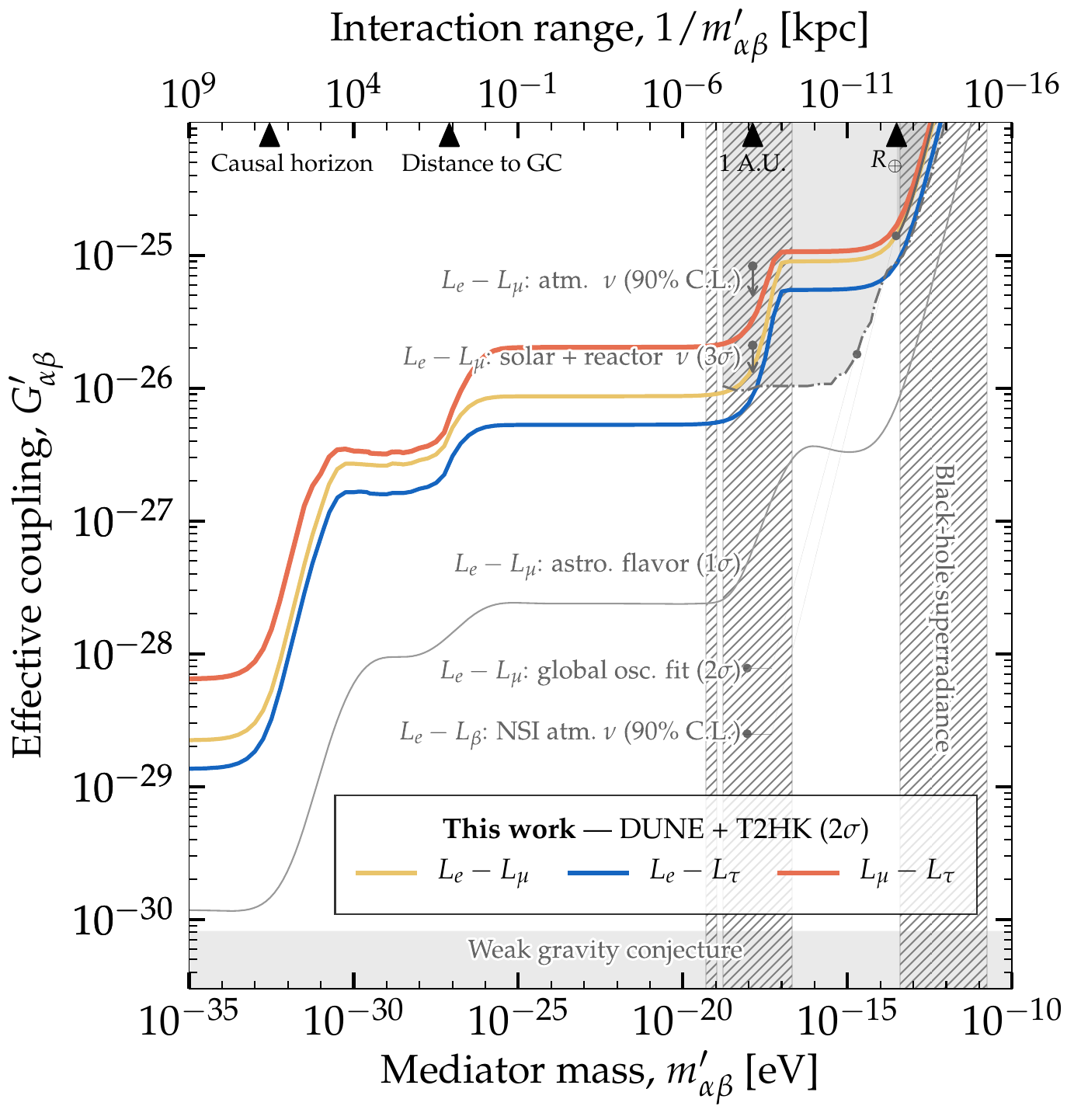}
 \caption{\textbf{\textit{Projected upper limits on the effective coupling, $G_{\alpha\beta}^\prime$ (Eq.~(\ref{equ:Gab})), of the new boson, $Z_{\alpha\beta}^\prime$, with mass $m_{\alpha\beta}^\prime$, that mediates flavor-dependent long-range neutrino interactions, using DUNE, T2HK, and their combination.}}  DUNE runs for 5 years in $\nu$ mode and 5 years in $\bar{\nu}$ mode.  T2HK runs for 2.5 years in $\nu$ mode and 7.5 years in $\bar{\nu}$ mode.  For this plot, we assume that the neutrino mass ordering is normal. Existing limits are from a recent global oscillation fit~\cite{Coloma:2020gfv} (2$\sigma$), atmospheric neutrinos~\cite{Joshipura:2003jh} (90\% C.L.), solar and reactor neutrinos~\cite{Bandyopadhyay:2006uh} (3$\sigma$), and non-standard interactions~\cite{Super-Kamiokande:2011dam, Ohlsson:2012kf, Gonzalez-Garcia:2013usa} (90\% C.L.).  We show the projected sensitivity ($1\sigma$) expected from flavor-composition measurements of high-energy astrophysical neutrinos in IceCube-Gen2~\cite{Bustamante:2018mzu}.  These limits are for the $L_e - L_\mu$ symmetry; see \figu{constraints_g_vs_m} for others.  Indirect limits~\cite{Wise:2018rnb} are from black-hole superradiance (90\% C.L.)~\cite{Baryakhtar:2017ngi}, and the weak gravity conjecture~\cite{Arkani-Hamed:2006emk}, assuming a lightest neutrino mass of $0.01$~eV.  {\it Our projected limits may improve on existing ones, especially for ultra-light mediators of masses below about $10^{-18}$~eV.}  See Section~\ref{sec:results_constraints} for details,  \figu{constraints_g_vs_m} for constraints using DUNE or T2HK separately, and Figs.~\ref{fig:g_vs_m_discovery}--\ref{fig:ranges_th23_V} for discovery plots. 
 }
 \label{fig:moneyplot}
\end{figure}

Discovering a new fundamental interaction would be striking evidence of physics beyond the Standard Model.  Yet, because new interactions are likely feeble, they are difficult to detect.  And because they may manifest in a variety of ways, they are difficult to search for comprehensively.  So far, there is no evidence for them, despite a long history of searches, though there are stringent limits on their strength~\cite{Lee:1955vk, Okun:1995dn, Williams:1995nq, Dolgov:1999gk}.

Starting in the 2030s, the next-generation long-baseline neutrino experiments, DUNE~\cite{DUNE:2021cuw} and T2HK~\cite{Hyper-Kamiokande:2016srs, Hyper-Kamiokande:2018ofw}, presently under construction, will bring about an opportunity to search for new physics, via neutrinos, more incisively than ever before.  DUNE and T2HK target the potential of neutrino to probe beyond three-flavor via rich physics programs, both within the standard neutrino paradigm and beyond it, that stem from their high expected event rates and well-characterized neutrino beams.  We focus on their capability to look for new neutrino-matter interactions: because, in the Standard Model, neutrinos interact only weakly, the presence of an additional neutrino interaction may be more easily spotted, even if it is feeble.  

We consider flavor-dependent neutrino-matter interactions, originally introduced in \Refes~\cite{He:1990pn,Foot:1990uf, Foot:1990mn, He:1991qd, Foot:1994vd}, and explored and constrained in earlier literature.  Two reasons motivate our choice.  First, if these interactions are long-range, \ie, if they act across long distances, then large collections of nearby and distant matter may source a sizable matter potential that affects neutrino flavor oscillations appreciably.  Thus, we concentrate on interactions mediated by new, ultra-light mediators, with masses below $10^{-10}$~eV, that subtend ultra-long interaction ranges.  Second, the flavor-dependent interactions we consider are born from gauging, anomaly-free, global symmetries of the Standard Model~\cite{Pontecorvo:1957qd, Pontecorvo:1967fh, Gribov:1968kq, Hisano:1998fj, Cirigliano:2005ck, Altarelli:2010gt}: $L_e-L_\mu$, $L_e-L_\tau$, and $L_\mu-L_\tau$, where $L_e$, $L_\mu$, and $L_\tau$ are the electron, muon, and tau lepton numbers.  This makes them arguably natural and economical extensions of the Standard Model.  Gauging each one introduces a single new neutral vector boson that mediates new neutrino interactions with electrons or neutrons (interactions with other particles are suppressed, as we elaborate on later).

Previous works have explored the sensitivity of existing and future long-baseline neutrino experiments to flavor-dependent long-range interactions.  However, they either fixed the interaction range, typically to be equal to the Sun-Earth distance (see, \eg, \Refe~\cite{Chatterjee:2015gta}), or considered mediator masses only as small as about $10^{-18}$~eV (see, \eg, \Refe~\cite{Coloma:2020gfv}).  We abandon both limitations and explore mediator masses down to $10^{-35}$~eV.  Doing so opens up a largely unexplored regime of ultra-long-range interactions.  As pointed out in \Refe~\cite{Bustamante:2018mzu}, a mediator this light allows for electrons and neutrons in the Earth, Moon, Sun, Milky Way, and the cosmological distribution of matter to affect neutrino oscillations.  To make our forecasts realistic, we base them on detailed simulations of DUNE and T2HK, including their different detection channels, efficiency, backgrounds, and run times.  

Figure~\ref{fig:moneyplot} conveys the novel perspectives revealed by our work.  It shows the first half of our main results, concerning constraints: \textbf{\textit{separately or, as in \figu{moneyplot}, together, DUNE and T2HK may place the strongest constraints on long-range interactions, especially for mediators lighter than $10^{-18}$~eV}}.  (Future sensitivity from flavor measurements of high-energy astrophysical neutrinos in the IceCube-Gen2 neutrino telescope might be comparable~\cite{Bustamante:2018mzu}, but, for now, they are subject to large uncertainties in the neutrino flux, not pictured in \figu{moneyplot}, unlike the constraints from DUNE and T2HK.)  The other half of our main results, not contained in \figu{moneyplot}, concerns discovery.  We find that, separately, DUNE and T2HK will likely be unable to discover subdominant flavor-dependent long-range neutrino interactions, due to degeneracies between their effect on neutrino oscillations and that of the standard mixing parameters.  Yet, \textbf{\textit{together, their complementary capabilities may lift degeneracies and enable the discovery of the new interactions; see \figu{g_vs_m_discovery}.}}  Below, we elaborate on these perspectives.

This chapter is organized as follows.  Section~\ref{sec:lri} introduces lepton-number gauge symmetries, long-range interactions, and their effect on neutrino oscillations.  Section~\ref{sec:experiments} shows oscillation probabilities and event rates in DUNE and T2HK.  Section~\ref{sec:results} shows projected constraints and discovery prospects.  Section~\ref{sec:conclusions} summarizes and concludes. 
%
\section{Flavor-dependent long-range neutrino  interactions}
\label{sec:lri}
\subsection{Gauged lepton-number symmetries}
\label{sec:lri_symmetries}
\begin{figure}
\includegraphics[width=\linewidth]{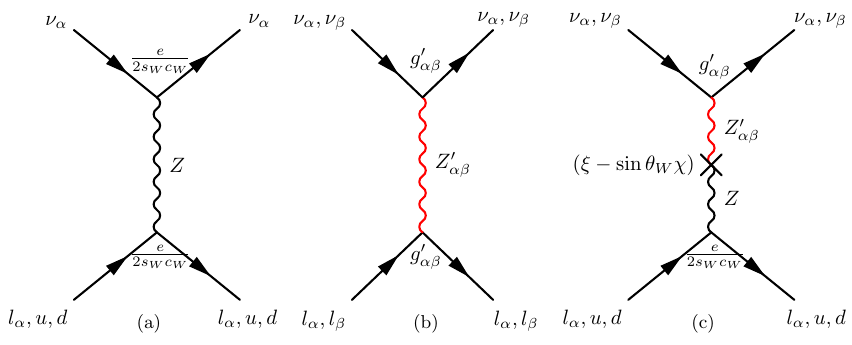}
\caption{\textbf{\textit{Feynman diagrams for neutrino-matter interactions.}}  Each diagram corresponds to a term in the Lagrangian, \equ{lagrangian_total}: (a) SM contribution mediated by the $Z$ boson, (b) new contribution from the gauge symmetry $U(1)_{L_\alpha-L_\beta}$, mediated by the new boson $Z_{\alpha\beta}^\prime$, and (c) mixing between $Z$ and $Z_{\alpha\beta}^\prime$.  In our analysis, (a) is significant only for neutrinos inside the Earth. For $L_{e} - L_\beta$ symmetries, (b) is the only additional contribution sourced by electrons.  For $L_\mu - L_\tau$, (c) is instead the only additional contribution sourced by neutrons.  See Section~\ref{sec:lri_symmetries} for details.}
\label{fig:feynman}
\end{figure}
%
In the Standard Model (SM), the baryon number and the lepton numbers, $L_e$, $L_\mu$, and $L_\tau$, are accidental $U(1)$ global symmetries.  Linear combinations of the lepton-number symmetries can be gauged anomaly-free, \ie, without introducing a new fermion or right-handed neutrino (although not simultaneously)~\cite{Foot:1990uf}.  To showcase the capabilities of DUNE and T2HK, we explore three such new $U(1)$ gauge symmetries, generated by $L_e-L_\mu$, $L_e-L_\tau$, and $L_\mu-L_\tau$, that introduce new flavor-dependent neutrino-matter interactions; later, we show how they affect neutrino oscillations.  (Other combinations of baryon and lepton numbers can also be gauged anomaly-free; see \Refe~\cite{Coloma:2020gfv}.) 

Figure~\ref{fig:feynman} shows the Feynman diagrams for neutrino-matter interaction that we consider.  For a particular lepton-number symmetry, the corresponding effective Lagrangian is
\begin{equation}
 \label{equ:lagrangian_total} 
 \mathscr{L}_{\rm eff}
 =
 \mathscr{L}_{\rm SM}
 +
 \mathscr{L}_{Z^\prime}
 +
 \mathscr{L}_{\text{mix}} \;.
\end{equation}
The first term describes the SM contribution, mediated by the $Z$ boson, \ie,
\begin{equation}
 \label{equ:lagrangian_sm}
 \mathscr{L}_{\text{SM}}
 =
 \frac{e}{\sin \theta_{W} \cos \theta_{W}}Z_\mu 
 \left[-\frac{1}{2}\bar{l}_{\alpha}\gamma^\mu P_{L} l_{\alpha}+\frac{1}{2}\bar{\nu}_{\alpha}\gamma^\mu P_{L} \nu_{\alpha}+\frac{1}{2}\bar{u}\gamma^\mu P_{L} u-\frac{1}{2}\bar{d}\gamma^\mu P_{L} d
 \right] \;,
\end{equation}
where $e/(\sin\theta_W\cos\theta_W) = 0.723$, $e$ is the unit charge, $\theta_W$ is the Weinberg angle, $\nu_\alpha$ and $l_\alpha$ are a neutrino and charged lepton of flavor $\alpha = e, \mu, \tau$, $P_{L}$ is the left-handed projection operator, and $u$ and $d$ are up and down quarks.  Because the $Z$ boson is heavy, the interaction that it mediates is short-range; in our work, it matters only inside the Earth.  (Equation~(\ref{equ:lagrangian_sm}), and also \equ{lag_mix} below, assumes that matter is electrically neutral, \ie, that it has equal abundance of electrons and protons~\cite{Heeck:2010pg}, which is also what we assume later when computing the new matter potential; see Section~\ref{sec:lri_potential}.)

The second term in \equ{lagrangian_total} describes the interaction between $\nu_\alpha$ and $l_\alpha$ mediated by the new $Z_{\alpha\beta}^\prime$ boson~\cite{He:1990pn, He:1991qd, Heeck:2010pg}, \ie, for the $L_\alpha-L_\beta$ symmetry,
\begin{equation}
 \label{equ:lagrangian_zprime}
 \mathscr{L}_{Z^\prime}
 =
 g_{\alpha \beta}^\prime Z^\prime_{\sigma}(\bar{l}_{\alpha}\gamma^{\sigma}l_{\alpha}-\bar{l}_{\beta}\gamma^{\sigma}l_{\beta}+ \bar{\nu}_{\alpha}\gamma^{\sigma}P_{L}\nu_{\alpha}-\bar{\nu}_{\beta}\gamma^{\sigma}P_{L}\nu_{\beta}) \;,
\end{equation}
where $g_{\alpha\beta}^\prime$ is a dimensionless coupling constant.  Due to the dearth of naturally occurring muons and tauons with which neutrinos can interact, we neglect this contribution under $L_\mu-L_\tau$ and consider it only under $L_e-L_\mu$ and $L_e-L_\tau$, for which the interaction is sourced by comparatively abundant electrons.

The final term in \equ{lagrangian_total} describes the mixing between $Z$ and $Z_{\alpha\beta}^\prime$~\cite{Babu:1997st, Heeck:2010pg, Joshipura:2019qxz}, which can arise directly or by radiative mixing~\cite{Holdom:1985ag,Tomalak:2020zfh}.  In the physical basis, this term is~\cite{Babu:1997st} $\mathscr{L}_{ZZ^\prime} \supset
 (\xi-\sin\theta_W\chi) Z'_{\mu}Z^{\mu}$,
where $\chi$ is the kinetic mixing angle between the two bosons and $\xi$ is the rotation angle between gauge eigenstates and physical states.  This introduces a four-fermion interaction between neutrinos and charged leptons, protons, and neutrons via $Z$--$Z_{\alpha\beta}^\prime$ mixing, \ie,
\begin{equation}
 \label{equ:lag_mix}
 \mathcal{L}_{\rm mix}
 =
 -g_{\alpha\beta}^\prime
 (\xi-\sin\theta_W\chi)\frac{e}{\sin\theta_W \cos\theta_W}
 J'_\rho J_3^\rho \;,
\end{equation}
where $J^\prime_\rho = \bar{\nu}_\alpha \gamma_\rho P_L\nu_\alpha-\bar{\nu}_\beta \gamma_\rho P_L\nu_\beta$ and $J_3^\rho = -\frac{1}{2}\bar{e}\gamma^\rho P_L e+\frac{1}{2}\bar{u}\gamma^\rho P_L u-\frac{1}{2}\bar{d}\gamma^\rho P_L d$.
However, the contribution of electrons is nullified by that of protons, leaving only neutrons to source the new interaction via mixing.  The term $(\xi - \sin\theta_{W}\chi)$ effectively describes the strength of the $Z$--$Z_{\alpha\beta}^\prime$ mixing.  Its value is unknown, but there are upper limits on it~\cite{Schlamminger:2007ht, Adelberger:2009zz, Heeck:2010pg}. We do not consider its value independently, but together with $g_{\alpha\beta}^{\prime}$\,, as an effective coupling (more on this below).  In order to showcase the effect of mixing, we include $\mathscr{L}_{\rm mix}$ only under $L_\mu-L_\tau$.

In summary, under $L_e - L_\mu$ and $L_e - L_\tau$, the new interactions are described by $\mathscr{L}_{Z^\prime}$, and are sourced by electrons only, whereas under $L_\mu-L_\tau$, the new interactions are described by $\mathscr{L}_{\rm mix}$, and are sourced by neutrons only.  In all cases, in addition, standard neutrino-electron interactions, described by $\mathscr{L}_{\rm SM}$, are active only inside the Earth.
\subsection{Long-range matter potential}
\label{sec:lri_potential}
The above interactions induce flavor-dependent Yukawa potentials, sourced by electrons and neutrons, that affect the mixing of neutrinos.  Under $L_e-L_\beta$ ($\beta = \mu, \tau$), a neutrino located at a distance $d$ from a collection of $N_e$ electrons experiences a potential
\begin{equation}
 V_{e\beta}
 =
 G_{\alpha\beta}^{\prime 2}
 \frac{N_e}{4\pi d}
 e^{-m'_{e\beta}d} \;,
 \label{equ:potential_e_beta}
\end{equation}
where $m_{e\beta}^\prime$ is the mass of the mediating $Z_{e\beta}^\prime$ boson.  Under $L_\mu-L_\tau$\,, a neutrino located at a distance $d$ from a collection of $N_n$ neutrons experiences a potential
\begin{equation}
 V_{\mu\tau}
 =
 G_{\alpha\beta}^{\prime 2}
  \frac{e}{\sin\theta_W\cos\theta_W}
 \frac{N_n}{4 \pi d}
 e^{-m'_{\mu\tau}d} \;,
 \label{equ:potential_mu_tau}
\end{equation}
where $m_{\mu\tau}^\prime$ is the mass of the mediating $Z_{\mu\tau}^\prime$ boson. In Eqs.~(\ref{equ:potential_e_beta}) and~(\ref{equ:potential_mu_tau}), the effective coupling strength is
\begin{equation}
 G_{\alpha \beta}^\prime
 =
 \left\{
  \begin{array}{lll}
   g^{\prime }_{e \mu} & , & ~{\rm for}~\alpha, \beta = e, \mu \\
   g^{\prime }_{e \tau} & , & ~{\rm for}~\alpha, \beta = e, \tau \\
   \sqrt{g^{\prime}_{\mu \tau} (\xi-\sin \theta_W \chi)} & , & ~{\rm for}~\alpha, \beta = \mu, \tau \\
  \end{array}
 \right. \;.
 \label{equ:Gab}
\end{equation}
At distances longer than the interaction range of $1/m_{\alpha\beta}^\prime$, the potential is suppressed due to the mediator mass.  Like \Refe~\cite{Bustamante:2018mzu}, we explore light mediators with $m_{\alpha\beta}^\prime = 10^{-35}$--$10^{-10}$~eV, corresponding to interaction ranges from $10^3$~Gpc --- larger than the observable Universe --- to hundreds of meters; see \figu{moneyplot}.

We adopt the methods introduced in \Refe~\cite{Bustamante:2018mzu} to compute the total potential sourced by nearby and faraway electrons and neutrons in the Earth ($\oplus$), Moon ($\leftmoon$), Sun ($\astrosun$), Milky Way (MW), and by the cosmological distribution of matter (cos) in the local Universe, \ie,
\begin{equation}
 \label{equ:pot_total}
 V_{\alpha \beta}
 =
 V_{\alpha \beta}^\oplus + V_{\alpha \beta}^{\leftmoon} + V_{\alpha \beta}^{\astrosun} + V_{\alpha \beta}^{\rm MW} +  V_{\alpha \beta}^{\rm cos} \;.
\end{equation} 
The specific value of $m_{\alpha\beta}^\prime$ determines the relative size of the contributions of the above sources to the total potential.  We do not compute the changing potential along the underground trajectories of the neutrinos from source to detector inside the Earth; see \Refe~\cite{Coloma:2020gfv} for such treatment. Instead, like \Refe~\cite{Bustamante:2018mzu}, we compute the average potential experienced by the neutrinos at their point of detection.  This approximation is especially valid for mediators lighter than about $10^{-14}$~eV, for which the interaction range is longer than the radius of the Earth (see \figu{moneyplot}), and so all of the electrons and neutrons on Earth contribute to the potential experienced by a neutrino regardless of its position along its trajectory.  Below $10^{-14}$~eV is also where we place novel projected limits.

We assume that the matter that sources the potential is electrically neutral, so that the number of electrons and protons is the same, and isoscalar, so that the number of electrons and neutrons is the same, except for the Sun~\cite{Heeck:2010pg} and for the cosmological distribution of matter~\cite{Hogg:1999ad, Steigman:2007xt, Planck:2015fie}.  We treat the Moon ($N_{e,\leftmoon} = N_{n,\leftmoon}\sim 5 \cdot 10^{49}$) and the Sun ($N_{e,\astrosun} \sim 10^{57}$, $N_{n,\astrosun} = N_{e,\astrosun}/4$) as point sources of electrons and neutrons, and the Earth ($N_{e, \oplus} \approx N_{n, \oplus} \sim 4 \times 10^{51}$), the Milky Way ($N_{e, {\rm MW}} \approx N_{n, {\rm MW}} \sim 10^{67}$), and the cosmological matter ($N_{e,\mathrm{cos}} \sim 10^{79}$, $N_{n,\mathrm{cos}} \sim 10^{78}$) as continuous distributions. We defer to \Refe~\cite{Bustamante:2018mzu} for a detailed calculation of \equ{pot_total}, but adopt two differences introduced by \Refe~\cite{Agarwalla:2023sng}.  First, unlike \Refe~\cite{Bustamante:2018mzu}, which studied extragalactic neutrinos and so averaged the contribution of cosmological matter over redshift, here we consider $V_{\alpha\beta}^{\rm cos}$ to be only the contribution from the local Universe, \ie, we evaluate Eq.~(A8) in \Refe~\cite{Bustamante:2018mzu} at redshift $z = 0$.  Second, unlike \Refe~\cite{Bustamante:2018mzu}, which only computed the potential sourced by electrons under $L_e-L_\mu$ and $L_e-L_\tau$, here we compute also the potential sourced by neutrons under $L_\mu-L_\tau$.
%
\subsection{Neutrino oscillation probabilities under long-range interactions}
\label{sec:lri_osc_prob}
We consider mixing between the three active neutrinos, $\nu_e$, $\nu_\mu$, and $\nu_\tau$.  Under the $L_\alpha-L_\beta$ symmetry, the Hamiltonian that drives neutrino propagation, in the flavor basis, is
\begin{equation}
 \label{equ:hamiltonian_tot}
 \mathbf{H}
 =
 \mathbf{H}_{\rm vac}
 +
 \mathbf{V}_{\rm mat}
 +
 \mathbf{V}_{\alpha\beta} \;.
\end{equation}
The first two terms on the right-hand side induce standard oscillations, including SM matter effects; the third one, oscillations due to the new interactions.  

In vacuum, oscillations are driven by
\begin{equation}
 \mathbf{H}_{\rm vac}
 =
 \frac{1}{2 E}
 \mathbf{U}~
 {\rm diag}(0, \Delta m^2_{21}, \Delta m^2_{31})
 ~\mathbf{U}^{\dagger} \;,
\end{equation}
where $E$ is the neutrino energy, $\Delta m^2_{ij} \equiv m^2_i-m^2_j$ are the mass-squared splittings between neutrino mass eigenstates, and $\mathbf{U}$ is the Pontecorvo-Maki-Nakagawa-Sakata (PMNS) mixing matrix, parametrized~\cite{ParticleDataGroup:2022pth} in terms of the  mixing angles $\theta_{12}$, $\theta_{23}$, and $\theta_{13}$, and the CP-violating phase, $\dcp$.  In the main text, we show results assuming normal neutrino mass ordering (NMO), where $m_1 < m_2 < m_3$; Table~\ref{tab:mix_param_benchmark} shows the values of the mixing parameters that we use, taken from \Refe~\cite{Capozzi:2021fjo}.  Section~\ref{app:imo} contains results obtained under the inverted mass ordering (IMO), where $m_3 < m_2 < m_1$.

In \equ{hamiltonian_tot}, the contribution of SM coherent forward scattering on electrons, mediated by the $W$ boson, is
\begin{equation}
 \mathbf{V}_{\rm mat}
 =
 {\rm diag}(V_{\rm CC}, 0, 0) \;,
\end{equation}
where $V_{\rm CC} = \sqrt{2} G_F n_e \simeq 7.6 Y_e [ \rho / (10^{14}~{\rm g}~{\rm cm}^{-3})]$~eV is the charged-current neutrino-electron interaction potential, $G_F$ is the Fermi constant, $n_e$ is the electron number density, $Y_e \equiv n_e / (n_p + n_n)$ is the electron fraction, \ie, its abundance relative to that of protons and neutrons, $n_p$ and $n_n$, and $\rho$ is the matter density.  In our work, this contribution is relevant only inside Earth, where matter densities are high.  We take $\rho$ to be the average density of underground matter along the trajectory from source to detector, calculated using the Preliminary Reference Earth Model~\cite{Dziewonski:1981xy}: $2.848$~g~cm$^{-3}$ for DUNE and 2.8~g~cm$^{-3}$ for T2HK.  The potential above is for neutrinos; for antineutrinos, it flips sign, \ie, $\mathbf{V}_{\rm mat} \to -\mathbf{V}_{\rm mat}$.

Finally, in \equ{hamiltonian_tot} the contribution from the new matter interaction is
\begin{equation}
 \label{equ:pot_lri_matrix}
 \mathbf{V}_{\alpha\beta}
 =
 \left\{
  \begin{array}{ll}
   {\rm diag}(V_{e\mu}, -V_{e\mu}, 0), & {\rm for}~ \alpha, \beta = e, \mu \\
   {\rm diag}(V_{e\tau}, 0, -V_{e\tau}), & {\rm for}~ \alpha, \beta = e, \tau \\
   {\rm diag}(0, V_{\mu\tau}, -V_{\mu\tau}), & {\rm for}~ \alpha, \beta = \mu, \tau \\   
  \end{array}
 \right. \;,
\end{equation}
where the potential, $V_{\alpha\beta}$, \equ{pot_total}, depends on the mediator mass, $m_{\alpha\beta}^\prime$, and effective coupling, $G_{\alpha\beta}^\prime$.  The potential above is for neutrinos; for antineutrinos, it flips sign, \ie, $\mathbf{V}_{\alpha\beta} \to -\mathbf{V}_{\alpha\beta}$.

The $\nu_\alpha \to \nu_\beta$ transition probability associated to the Hamiltonian, \equ{hamiltonian_tot}, is
\begin{equation}
 \label{equ:osc_prob}
 P_{\nu_{\alpha}\rightarrow \nu_{\beta}}
 =
 \Bigg| 
 \sum_{i=1}^{3} 
 U^\prime_{\alpha i}
 \exp \left( \dfrac{\Delta \tilde{m}^2_{i1}L}{2E} \right)
 U^{\prime \ast}_{\beta i}
 \Bigg|^{2} \;,
\end{equation}
where $L$ is the distance traveled by the neutrino from production to detection, $\Delta \tilde{m}^2_{ij} \equiv \tilde{m}_i^2 - \tilde{m}_j^2$, with $\tilde{m}_i^2/2E$ the eigenvalues of the Hamiltonian, modified from those of $\mathbf{H}_{\rm vac}$ by matter effects, and $\mathbf{U}^\prime$ is the unitary matrix that diagonalizes the Hamiltonian.   We parametrize $\mathbf{U}^\prime$ with the same shape as the PMNS matrix, but evaluated at mixing parameters $\theta_{12}^m$, $\theta_{23}^m$, $\theta_{13}^m$, and $\dcp^m$ modified by matter effects.  In our work, we compute the oscillation probability, \equ{osc_prob}, exactly and numerically to arbitrary precision.

For the new matter interactions to affect the oscillation probability, the new matter potential must be at least comparable to the standard contributions in \equ{hamiltonian_tot}, \ie, in vacuum, $V_{\alpha \beta} \gtrsim (\Delta m_{31}^2/2E)$ [inside the Earth, this is instead $V_{\alpha \beta} \gtrsim \max \left( \Delta m_{31}^2/2E, V_{\rm CC} \right)$].  In DUNE and T2HK, where the first oscillation maxima occur at 2.6~GeV and 0.6~GeV, respectively, this implies that they become important for $V_{\alpha\beta} \gtrsim 10^{-13}$~eV.  This sets the scale of the potential to which our analysis is sensitive.  Later, in Section~\ref{sec:experiments_probabilities}, we show how the new interactions affect the probabilities in DUNE and T2HK.
%
\subsection{Existing limits}
\label{sec:prev_limits}
%
Figure~\ref{fig:moneyplot} (also \figu{constraints_g_vs_m}) shows existing limits on flavor-dependent long-range neutrino interactions.  Below, we summarize them.  We focus on light mediators; the complementary case for heavy mediators was first studied in \Refes~\cite{Foot:1990mn, He:1990pn, He:1991qd, Foot:1994vd, Dutta:1994dx}.

Pioneering studies in \Refes~\cite{Joshipura:2003jh} and \cite{Bandyopadhyay:2006uh} identified the potential of neutrino oscillations to test new long-range interactions, possibly more stringently than gravitational probes.  They focused on interactions with a range equal to the Earth-Sun distance [though ignoring the Yukawa suppression in \equ{potential_e_beta}] and sourced by solar electrons.  Reference~\cite{Joshipura:2003jh} used Super-Kamiokande atmospheric neutrino data to find $g_{e\mu}^\prime < 8.32 \times 10^{-26}$ and $g_{e\tau}^\prime < 8.97 \times 10^{-26}$, at 90\%~confidence level (C.L.)  Reference~\cite{Bandyopadhyay:2006uh} used solar and reactor neutrino data from KamLAND to find $g_{e\mu}^\prime < 2.06 \times 10^{-26}$ and $g_{e\tau}^\prime < 1.77 \times 10^{-26}$, at 3$\sigma$, assuming $\theta_{13} = 0^{\circ}$.

Reference~\cite{Heeck:2010pg} studied the effect of the $L_\mu-L_\tau$ symmetry via kinetic mixing (see Section~\ref{sec:lri_symmetries}) on $\nu_\mu$ in the long-baseline experiment MINOS.  By comparing the potential $V_{\mu\tau}$ sourced by a neutron in the Sun to the fifth-force gravitational potential sourced by it, and applying upper limits on the strength of the latter from torsion-balance experiments~\cite{Schlamminger:2007ht, Adelberger:2009zz}, \Refe~\cite{Heeck:2010pg} set an upper limit on the mixing strength of $(\xi - \sin\theta_W\chi) < 5 \times 10^{-24}$ at 95\%~C.L.~for a long-range interaction with range equal to the Earth-Sun distance.  (This is the limit that we saturate  when computing the $V_{\mu\tau}$ potential, \equ{potential_mu_tau}; see Section~\ref{sec:lri_symmetries}.)  This translates into an upper limit of $g_{\mu\tau}^\prime \leq 2.51 \times 10^{-26}$.  For an interaction with a range of the size of the Earth, the upper limit degrades to $g_{\mu\tau}^\prime \leq 10^{-24}$.  

Reference~\cite{Wise:2018rnb} showed that upper limits on the coefficients that parametrize the strength of non-standard neutrino interactions (NSI) can be translated into upper limits on the coupling strength of flavor-dependent long-range interactions.  Figure~\ref{fig:moneyplot} shows the resulting limits, based on the NSI limits from \Refes~\cite{Super-Kamiokande:2011dam, Ohlsson:2012kf, Gonzalez-Garcia:2013usa}.

Recently, \Refe~\cite{Coloma:2020gfv} performed a global oscillation analysis of new $U(1)$ symmetries, including $L_e-L_\mu$ and $L_e-L_\tau$, by using the same experimental data sets used in NuFIT~5.0~\cite{Esteban:2020cvm, NuFIT}. Unlike our analysis, \Refe~\cite{Coloma:2020gfv} computed the changing long-range matter potential due to underground matter in the Earth along the trajectory of the neutrinos.  Their procedure is more detailed than ours for mediators lighter than $10^{-14}$~eV.  However, they explore masses only as low as $10^{-18}$~eV, \ie, an interaction range of 1~A.U.

Reference~\cite{Bustamante:2018mzu} first showed that the flavor composition of TeV--PeV astrophysical neutrinos, \ie, the relative number of $\nu_e$, $\nu_\mu$, and $\nu_\tau$, can be used to probe long-range interactions under $L_e-L_\mu$ and $L_e-L_\tau$ sourced by the same collections of nearby and distant electrons that we consider here.  Reference~\cite{Agarwalla:2023sng} refined the statistical methods and included also $L_\mu-L_\tau$.  The main effect is that, if the potential sourced by electrons or neutrons were to be dominant, oscillations would turn off, and the flavor composition emitted by the astrophysical sources and received at Earth would be the same; see also \Refe~\cite{Farzan:2018pnk}.  Figure~\ref{fig:moneyplot} shows the projected upper limits obtained in \Refe~\cite{Bustamante:2018mzu} based on estimates of flavor-composition measurements in the envisioned IceCube-Gen2 neutrino telescope~\cite{IceCube-Gen2:2020qha}.
  
Finally, following \Refe~\cite{Wise:2018rnb}, \figu{moneyplot} includes two indirect limits.  First, \Refe~\cite{Baryakhtar:2017ngi} excluded three mediator mass windows (``Black-hole superradiance'') by considering the superradiant growth rate of a gravitationally bound accumulation of light vector bosons around selected stellar-mass and supermassive black holes.  Second, \Refe~\cite{Arkani-Hamed:2006emk} placed a tentative lower limit on the coupling (``Weak gravity conjecture'') by studying low-energy effective theories that contain gravity and $U(1)$ gauge fields where at least one particle charged under $U(1)$ is essential for gravity to be the weakest force.  

In \figu{moneyplot}, we show existing limits as they were published in their original references.  Hence, they do not extend to mediators lighter than $10^{-14}$--$10^{-20}$~eV, depending on the limit (except for the proof-of-principle sensitivity based on projected IceCube-Gen2 measurements of the flavor composition~\cite{Bustamante:2018mzu}).  These limits could be recomputed and extended to span lighter mediators, using the same long-range matter potential that we have used, \equ{pot_total}, though doing so lies beyond the scope of this work.


\section{Long-range interactions in DUNE and T2HK}
\label{sec:experiments}
\begin{table}[htb!]
 \centering
 \begin{tabular}{|c|c|c|c|c|c|c|}
  \hline 
  \multirow{2}{*}{} & \multicolumn{6}{c|}{Standard mixing parameters (NMO)} \\
   & $\sin^2 \theta_{12}$ & $\sin^2\theta_{23}$ & $\sin^2 \theta_{13}$ &
  $\frac{\Delta m^2_{31}}{10^{-3}\,\text{eV}^2}$  & $\frac{\Delta m^2_{21}}{10^{-5}\,\text{eV}^2}$  & $\delta_{\rm CP}\, (^\circ)$\\[0.8ex]
  \hline
  Benchmark & 0.303 & 0.455 & 0.0223 & 2.522 & 7.36 & 223  \\
  Status in fits & Fixed & Minimized & Fixed & Minimized & Fixed & Minimized \\
  Range & -- & [0.4, 0.6] & -- & [2.438, 2.602] & -- & [139, 355]\\
  \hline 
 \end{tabular}
 \caption{\textbf{\textit{Values of the standard mixing parameters used in our analysis.}}  We assume normal neutrino mass ordering (NMO) in the main text.  The benchmark values are the best-fit values from \Refe~\cite{Capozzi:2021fjo}.  For each parameter over which we minimize our test statistic (see Section~\ref{sec:results_stat_methods}), the minimum is searched for within the range shown, which is the $3\sigma$ allowed range from \Refe~\cite{Capozzi:2021fjo}.  We assume no correlation between the parameters. Table~\ref{tab:mix_param_benchmark_imo} shows the parameter ranges that we use in Sec.~\ref{app:imo} to obtain results under the inverted mass ordering (IMO) instead.}
 \label{tab:mix_param_benchmark}
\end{table}
%

\begin{figure}[htb!]
 \centering
 \includegraphics[width=\linewidth]{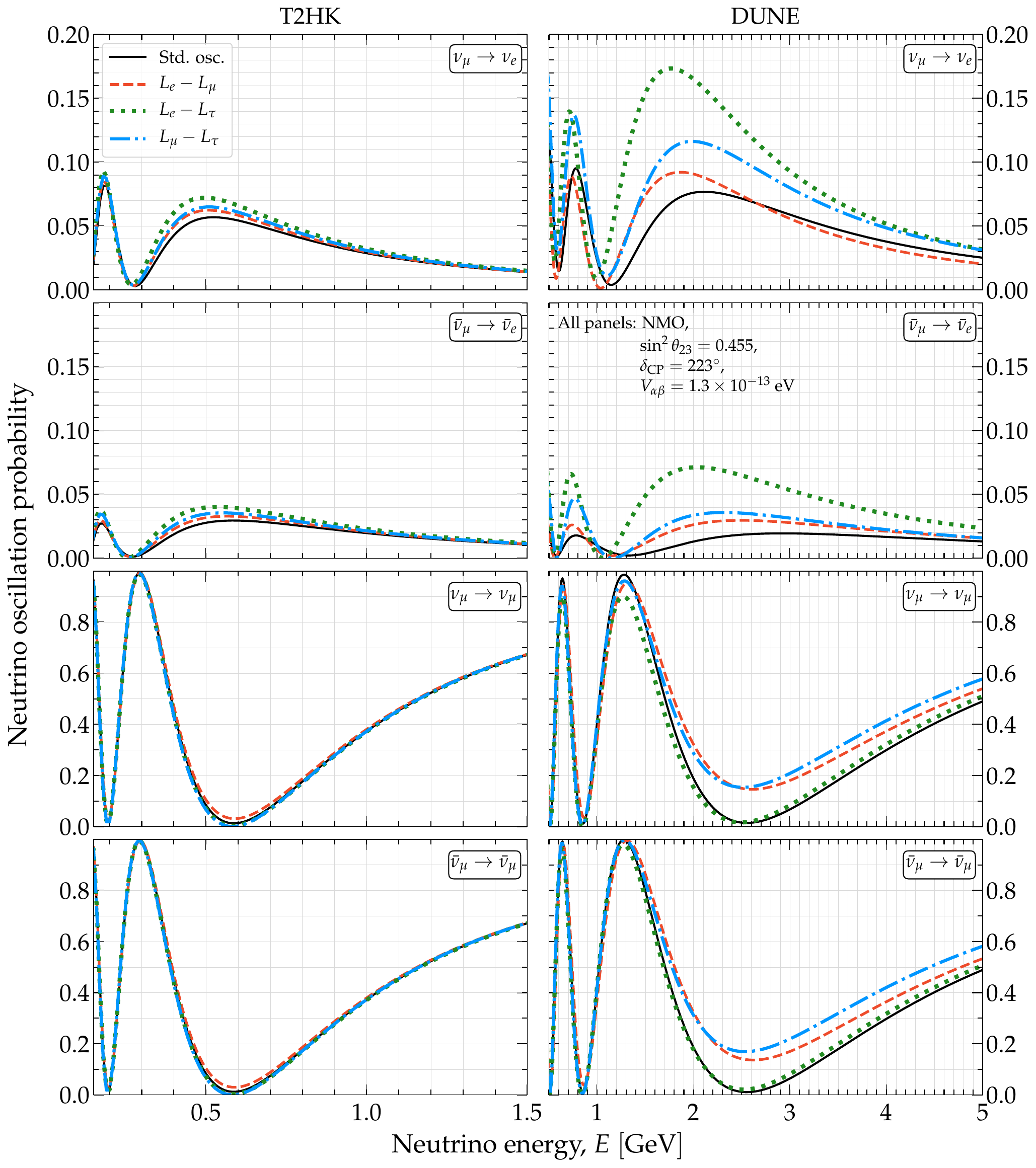}
 \caption{\textbf{\textit{Neutrino oscillation probabilities for T2HK (left column) and DUNE (right column) under flavor-dependent long-range neutrino interactions.}}  The interactions are induced by the lepton-number symmetry $L_e-L_\mu$, $L_e-L_\tau$, or $L_\mu-L_\tau$.  For this figure, we fix the long-range potential to $V_{\alpha\beta} = 1.3 \times 10^{-13}$~eV as illustration, and the standard mixing parameters to their benchmark values from Table~\ref{tab:mix_param_benchmark}.  See Section~\ref{sec:experiments_probabilities} for details.}
 \label{fig:probabilities}
\end{figure}
\subsection{Oscillation probabilities}
\label{sec:experiments_probabilities}
\section{Effects of long-range interactions on neutrino oscillation parameters}
\label{app:evol_mix_param} 

\begin{figure}[t!]
 \includegraphics[width=\linewidth]{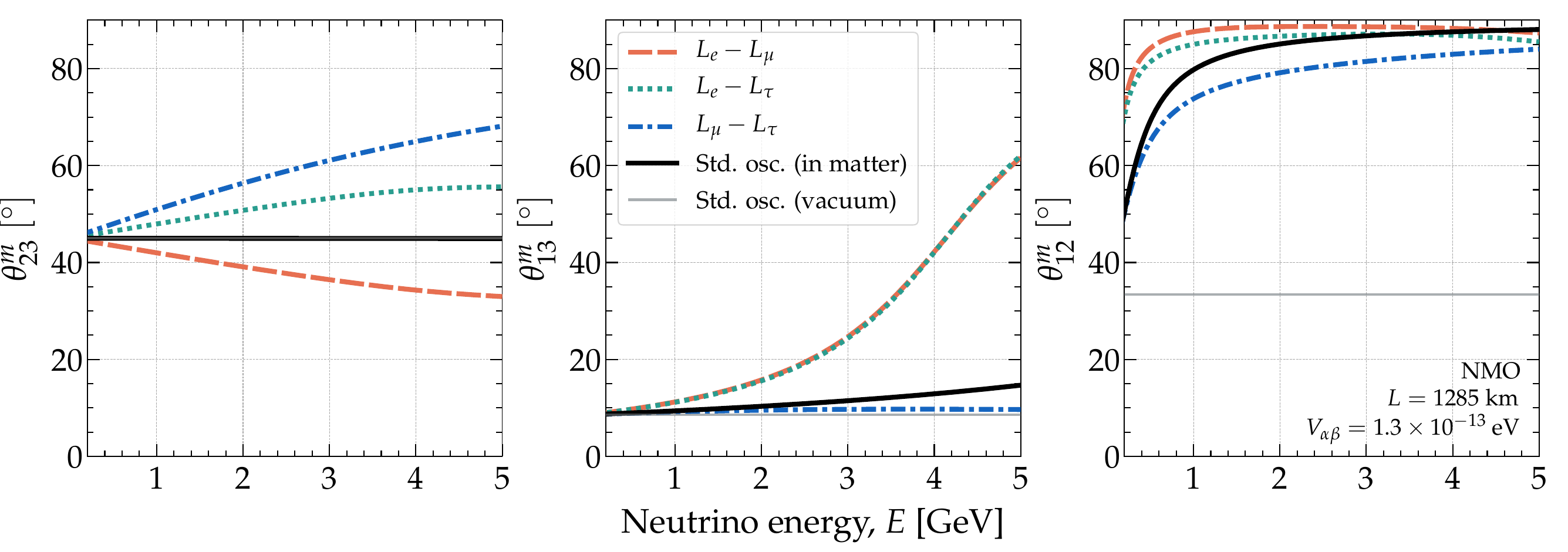}
 \caption{\textbf{\textit{Variation of the effective neutrino mixing angles with neutrino energy, for the three lepton-number symmetries, $L_\alpha-L_\beta$.}}  For this plot, we adopt the baseline, average matter potential, and approximate energy range of DUNE; see Section~\ref{sec:experiments-overview}.
 For all the symmetries, we adopt an illustrative value of the new matter potential of $V_{\alpha\beta} = 1.3 \times 10^{-13}$~eV.  For comparison, we include results using only standard matter effects and in a vacuum.  The values of the mixing parameters in vacuum are from Table~\ref{tab:mix_param_benchmark}, except with $\sin^{2}\theta_{23} = 0.5$. See Sec.~\ref{app:evol_mix_param} for details.}
 \label{fig:evol_mix_param}
\end{figure}

\begin{figure}[t!]
 \includegraphics[width=\linewidth]{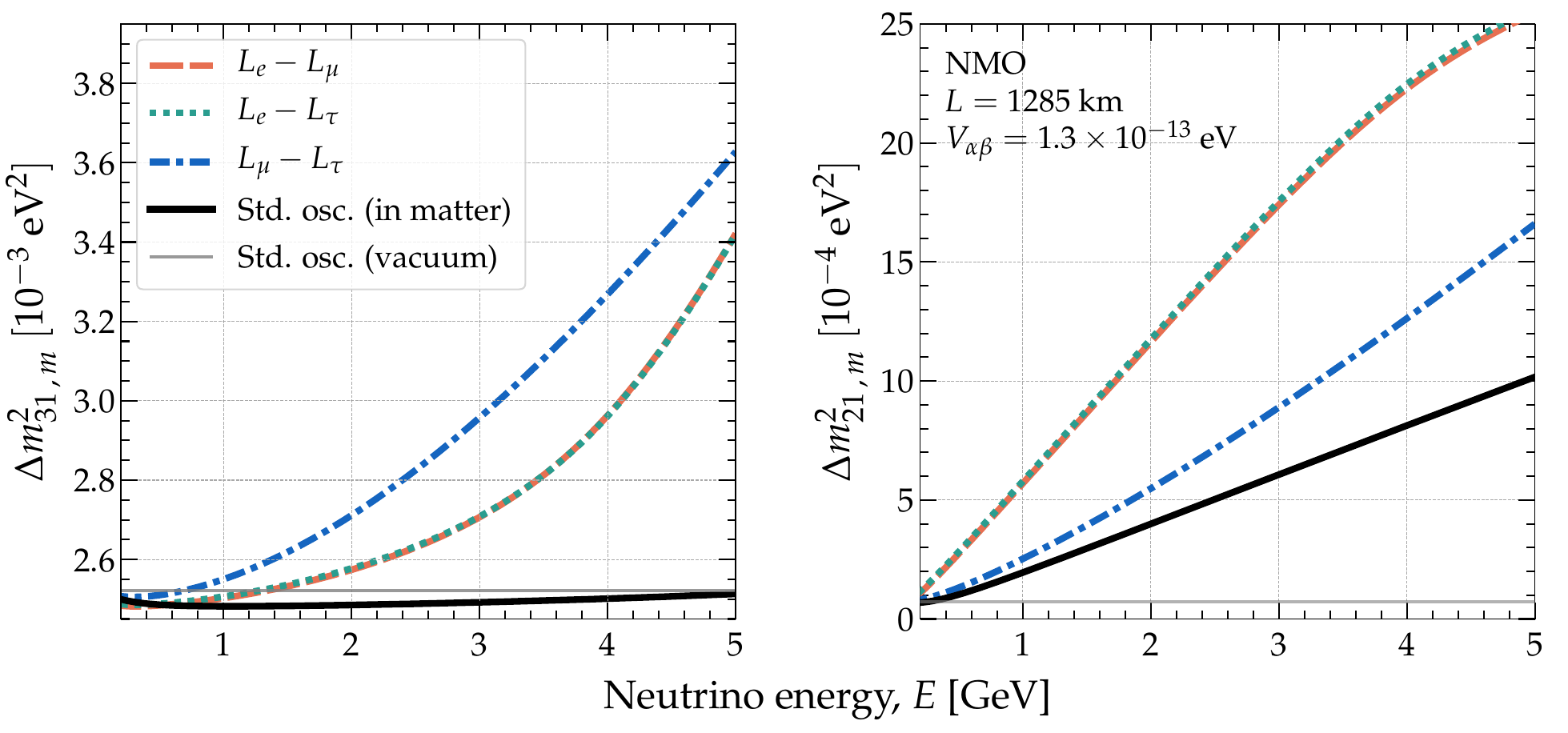}
 \caption{
 \textbf{\textit{Variation of the effective neutrino mass splittings with neutrino energy, for the three lepton-number symmetries, $L_\alpha-L_\beta$.}}  Same as \figu{evol_mix_param}, but for the mass splittings.  See Sec.~\ref{app:evol_mix_param} for details.}
 \label{fig:evol_mass_square}
\end{figure}

Figures~\ref{fig:evol_mix_param} and \ref{fig:evol_mass_square} show the variation with neutrino energy of, respectively, the mixing angles and mass splittings under the new matter interactions for the three symmetries, and geared at DUNE for illustration.  Their behavior when geared at T2HK is similar.  

Regarding the mixing angles, their behavior in \figu{evol_mix_param} backs the explanation of the behavior of the oscillation probabilities in Section~\ref{sec:experiments_probabilities}, in agreement with \Refes~\cite{Chatterjee:2015gta, Khatun:2018lzs, Agarwalla:2021zfr}.

Regarding the mass splittings, the energy at which the first oscillation maximum occurs in the probabilities (\figu{probabilities}) is determined mostly by $\Delta m^2_{31,m}$.  Figure~\ref{fig:evol_mass_square} shows that, under $L_e-L_\mu$ and $L_e-L_\tau$, $\Delta m^2_{31,m}$ evolves similarly with energy; this explains why the oscillation maxima for these two symmetries occur at approximately the same energy.  The change with energy of $\Delta m^2_{21,m}$ affects the $\nu_\mu \to \nu_\mu$ and $\nu_\mu \to \nu_\tau$ probabilities, but only for baselines of around 10000~km, as shown in \Refe~\cite{Agarwalla:2015cta}.
%
Figure \figu{evol_mix_param}, shows in detail the effects of long-range interactions on the modified mixing parameters $\theta_{12}^m$, $\theta_{23}^m$, and $\theta_{13}^m$; here, we summarize them.  Differences in their behavior under the different symmetries stem from differences in the flavor structure of the new matter potential, $\mathbf{V}_{\alpha\beta}$ in \equ{hamiltonian_tot}.

The solar angle in matter rapidly approximates its maximum value of $\theta_{12}^m = 90^\circ$ already at a few GeV, for all symmetries. (We use this later, in Section~\ref{sec:results_stat_methods}, to justify why we neglect the effect on our forecasts of the uncertainty in its value in vacuum, $\theta_{12}$.)  For DUNE and T2HK, the mixing angles that drive the probabilities are the atmospheric angle, $\theta_{23}^m$, and the reactor angle, $\theta_{13}^m$.  Assuming $\theta_{12}^m = 90^\circ$, \Refe~\cite{Khatun:2018lzs} showed that the transition probabilities for $\nu_\mu \to \nu_e$ and $\bar{\nu}_\mu \to \bar{\nu}_e$ are $\propto \sin^2 \theta_{23}^m \sin^2 \theta_{13}^m$ and the survival probabilities for $\nu_\mu \to \nu_\mu$ and $\bar{\nu}_\mu \to 
\bar{\nu}_\mu$ are $\propto \sin^2 2\theta_{23}^m$, with a more nuanced dependence on $\theta_{13}^m$.  The deviation of $\theta_{23}^m$ relative to $\theta_{23}$ grows with energy, though there are differences depending on which symmetry is active: $\theta_{23}^m$ grows under $L_e-L_\tau$ and $L_\mu-L_\tau$, and shrinks under $L_e-L_\mu$.  The reactor angle in matter, $\theta_{13}^m$, grows appreciably with energy under $L_e-L_\mu$ and $L_\mu-L_\tau$, and falls to about $0^\circ$ under $L_\mu-L_\tau$.

Figure~\ref{fig:probabilities} shows the oscillation probabilities, \equ{osc_prob}, computed under the three symmetries in each of the four detection channels listed in Section~\ref{sec:experiments-overview}, for DUNE and T2HK.  To illustrate the effects of long-range interactions, we pick a relatively high value of the potential, $V_{\alpha\beta} = 1.3 \times 10^{-13}$~eV; later, when producing our results, we vary this value.  Via \equ{pot_total}, multiple combinations of $m_{\alpha\beta}^\prime$ and $g_{\alpha\beta}^\prime$ can yield this value of the potential, or any other.  Because the baseline for DUNE is longer than for T2HK, the effects of long-range interactions with underground matter on the probabilities in the former are more prominent than in the latter~\cite{Agarwalla:2021zfr}.  The effects are more clearly visible in the transition probabilities: the oscillation maxima shift to lower energies, due to a change in the effective mass-splitting $\Delta m_{31,m}^2$, in agreement with \Refe~\cite{Chatterjee:2015gta}, and the oscillation amplitudes grow, especially after the first maximum.  The effects are more prominent under $L_e-L_\tau$ because $\theta_{23}^m$ and $\theta_{13}^m$ are enhanced, whereas under $L_e-L_\mu$ and $L_\mu-L_\tau$ only one of them is; see Fig.~\ref{app:evol_mix_param} for details.  Naturally, for weaker potentials, the above effects are lessened. 
%
\subsection{Event rates}
\label{sec:experiments_event_rates}
\begin{table}[t!]
 \centering
 \begin{tabular}{ | c  *{6}{>{\centering\arraybackslash}p{2.2cm} |}}
 \hline
 \multicolumn{2}{|c|}{\multirow{3}{*}{Detector}}
 & \multicolumn{4}{c|}{Mean number of events (standard oscillations, NMO)} \\
          & & \multicolumn{2}{c|}{Appearance} 
          & \multicolumn{2}{c|}{Disappearance} \\
          & & $\nu$ mode & $\bar{\nu}$ mode & $\nu$ mode & $\bar{\nu}$ mode \\
 \hline
 \multirow{2}{*}{DUNE} & Signal &    1390 & 387 & 15574 & 8975 \\
                       & Bkg.   & 690  & 457  & 347   & 210   \\
 \hline
 \multirow{2}{*}{T2HK} & Signal & 1374 & 1166 & 10083 & 13905 \\
                       & Bkg.   & 802  & 991  & 1686  & 1769  \\
 \hline
 \end{tabular}
 \caption{\textbf{\textit{Mean number of signal and background events, summed over all background channels, expected in DUNE and T2HK after their full run times.}}  For this table, whose aim is illustrative only, we assume standard oscillations and normal mass ordering (NMO).  DUNE runs for 5 years in $\nu$ mode and 5 years in $\bar{\nu}$ mode.  T2HK runs for 2.5 years in $\nu$ mode and 7.5 years in $\bar{\nu}$ mode.  To compute the rates in this table, we fix the values of the standard mixing parameters to their benchmark values from \Refe~\cite{Capozzi:2021fjo}; see Table~\ref{tab:mix_param_benchmark}.  In the main text, to produce results, we also compute event rates in the presence of the new long-range neutrino interactions (not shown in this table).  In those cases, the relative sizes of the event rates in the different detection channels are roughly as in this table.  See Section~\ref{sec:experiments_event_rates} for details.
 }
 \label{tab:event}
\end{table}
%
\begin{figure}[t!]
 \centering
 \includegraphics[width=\linewidth]{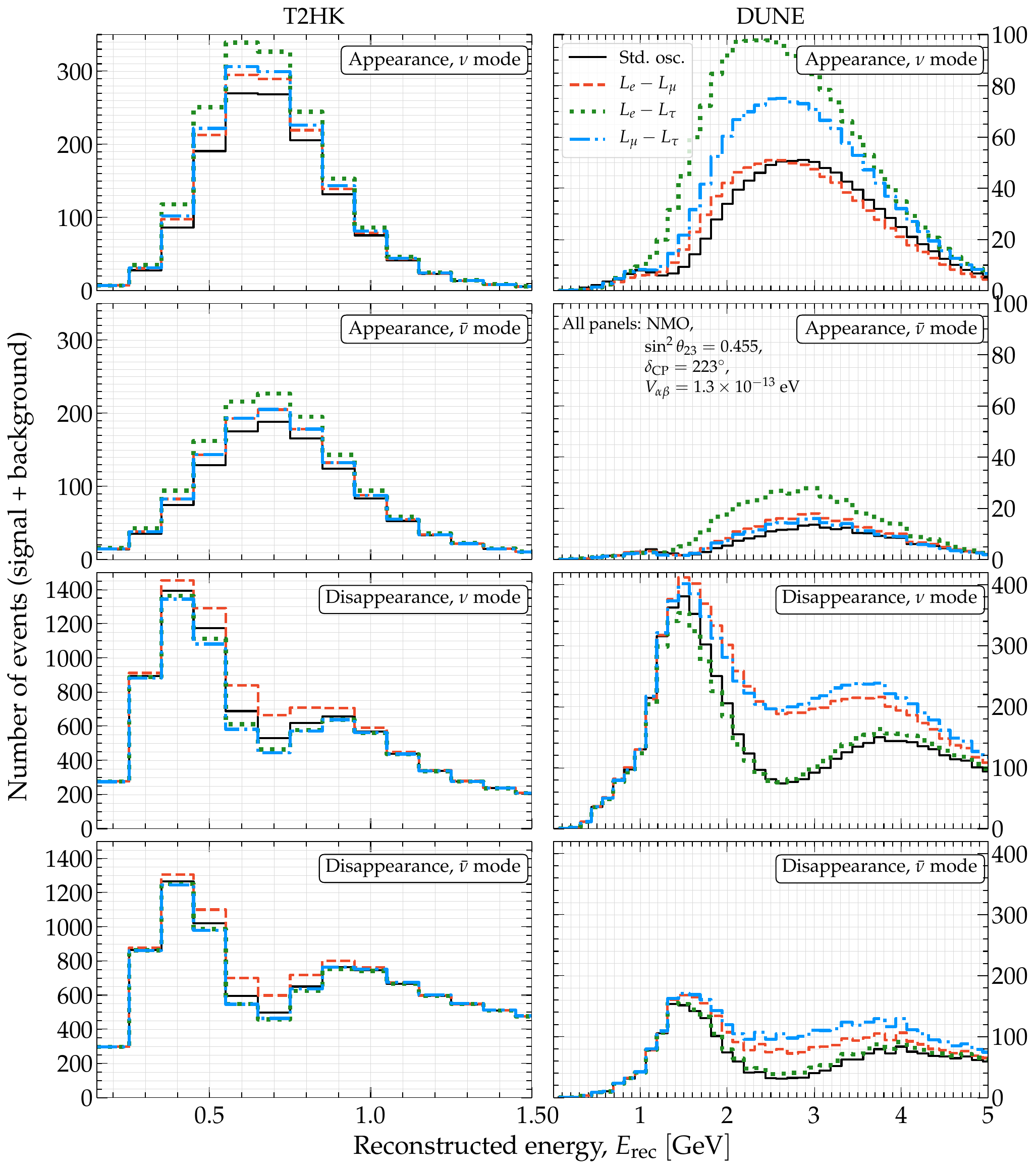}
 \caption{\textbf{\textit{Expected mean number of detected events in T2HK (left column) and DUNE (right column) under flavor-dependent long-range neutrino interactions.}}  The interactions are induced by the symmetry $L_e-L_\mu$, $L_e-L_\tau$, or $L_\mu-L_\tau$.  For T2HK, we use 2.5 years in $\nu$ mode and 7.5 years in $\bar{\nu}$ mode.  For DUNE, we use 5 years in $\nu$ mode and 5 years in $\bar{\nu}$ mode.  For this figure, we fix the long-range potential to $V_{\alpha\beta} = 1.3 \times 10^{-13}$~eV as illustration, and the standard mixing parameters to their benchmark values from Table~\ref{tab:mix_param_benchmark}.  See Section~\ref{sec:experiments_event_rates} for details.}
 \label{fig:event_rates}
\end{figure}
%
We compute event rates in DUNE and T2HK using {\sc GLoBES}~\cite{Huber:2004ka, Huber:2007ji},  extended with the {\sc snu} matrix-diagonalization library~\cite{Kopp:2006wp, Kopp:2007ne},  by modeling their technical design specifications~\cite{Hyper-Kamiokande:2018ofw, DUNE:2021cuw} of efficiency, operation times, and backgrounds.  Because we are interested in assessing the mean sensitivity of the experiments (Section~\ref{sec:results_stat_methods}), we compute only mean event rates and do not generate event spectra that include fluctuations from the mean rates.  We bin event rates in reconstructed energy, $E_{\rm rec}$, built from the detected secondaries born in neutrino interactions.  In both experiments, because the far detectors cannot distinguish between neutrinos and antineutrinos, there is irreducible contamination from ``wrong-sign'' events; we add it to the signal.

\medskip

Figure~\ref{fig:event_rates} shows the mean event-rate spectra under long-range interactions for each detection channel in DUNE and T2HK, including all the above backgrounds, and computed using the same illustrative value of the long-range potential $V_{\alpha\beta}$ as in \figu{probabilities}.  The event rates in T2HK are higher than in DUNE due to its larger size.  The shapes of the event spectra in \figu{event_rates} reflect those of the oscillation probabilities in \figu{probabilities}.  Long-range interactions affect each detection channel differently, but there are common features among them.  Broadly stated, in the appearance channels, they enhance the event rates relative to the standard-oscillations rates (with the exception of $L_e-L_\mu$ in neutrino mode for DUNE).  For DUNE, additionally, they slightly shift the event rates to lower energies, reflecting the shift in the oscillation maxima.  In the disappearance channels, the effect of long-range interactions is more nuanced; the event rate is enhanced or reduced depending on the symmetry and the energy.  The above features in the event spectra hold for other values of the potential, though, naturally, their prominence varies depending on the value.

Table~\ref{tab:event} shows the mean expected number of signal and background events for each detection channel, assuming standard oscillations.  In all channels, the signal is dominant.  In T2HK, unlike DUNE, neutrino and antineutrino event rates are comparable, due to the 1:3 ratio between run times in neutrino and antineutrino modes that compensates for the smaller antineutrino cross sections.  In DUNE, neutrino event rates are higher than T2HK due to its longer run time in neutrino mode.  These general features of the event rates hold also in the presence of long-range interactions.  

Below, we show how the above features grant DUNE and T2HK the capability to probe long-range interactions, and how they organically complement each other.
%
\section{Projected constraints and discovery potential}
\label{sec:results}
\subsection{Statistical methods}
\label{sec:results_stat_methods}
We forecast the capability of DUNE and T2HK to probe long-range interactions that stems from the modification of the oscillation probabilities (Section~\ref{sec:experiments_probabilities}), based on the detailed computation of event rates outlined above (Section~\ref{sec:experiments_event_rates}).  Our forecasts are two-fold: we forecast {\it constraints} on long-range interactions --- on the long-range matter potential and ultimately on the mediator mass and coupling --- assuming that no evidence for them is found, and we forecast prospects of {\it discovering} them and measuring their parameter values.

We study each symmetry, $L_e-L_\mu$, $L_e-L_\tau$, and $L_\mu-L_\tau$, separately.  For a given symmetry, we generate two sets of event spectra, including signal plus backgrounds, for each of the four detection channels of T2HK and DUNE (Section~\ref{sec:experiments_event_rates}): a ``true'' spectrum, which we take to be the observed spectrum, and a set of ``test'' spectra, generated for test values of the parameters, that we compare against it.  When forecasting constraints, in Section~\ref{sec:results_constraints}, we compute the true spectrum fixing the true value of the potential to be $V_{\alpha\beta}^{\rm true} = 0$, which corresponds to standard oscillations.  When forecasting discovery prospects, in Section~\ref{sec:results_discovery}, we compute the true spectrum fixing $V_{\alpha\beta}^{\rm true}$ to a specific nonzero choice.  We expand on this below. 

To compare true and test event spectra, we follow \Refes~\cite{Baker:1983tu, Cowan:2010js, Blennow:2013oma} and adopt a Poissonian $\chi^2$ function as defined in Sec.~\ref{sec:stats}. 

In Table~\ref{tab:normalization_err}, we show the assumed
normalization errors on the signal ($\pi_{e,c}^s$) and background ($\pi_{e,c,k}^b$) rates, which lie between 2\% and 10\%; see Table~\ref{tab:normalization_err} in Table~\ref{tab:normalization_err} for their  values, taken from \Refes~\cite{Hyper-Kamiokande:2018ofw, DUNE:2021cuw}.  The background rates do not vary significantly upon changing the mass ordering.

\begin{table}[htb!]
 \resizebox{\columnwidth}{!}{%
  \centering
  \begin{tabular}{|c|c|c|c|c|c|c|c|c|}
   \hline
   \multirow{3}{*}{Experiment}  & \multicolumn{8}{c|}{Normalization errors~[\%]}  \\
   & \multicolumn{4}{c|}{Signal, $\pi_{e,c}^s$} & \multicolumn{4}{c|}{Background, $\pi_{e,c,k}^b$}\\
   &  App.~$\nu$ & App.~$\bar{\nu}$ & Disapp.~$\nu$ & Disapp.~$\bar{\nu}$ & $\nu_{e}$, $\bar{\nu}_{e}$ CC & $\nu_{\mu}$, $\bar{\nu}_{\mu}$ CC & $\nu_{\tau}$, $\bar{\nu}_{\tau}$ CC & NC \\ 
   \hline
   DUNE & 2 & 2 & 5 & 5 & 5 & 5 & 20 & 10\\
   T2HK & 5 & 5 & 3.5 & 3.5 & 10 & 10 & -- & 10\\
   \hline
  \end{tabular}}
 \caption{\textbf{\textit{Normalization errors of the event rates associated to the signal and background detection channels in DUNE and T2HK.}}  They are shown separately for the neutrino ($\nu$) and antineutrino ($\bar{\nu}$) modes, and for the appearance (``App.'') and disappearance (``Disapp.'') channels.  Normalization errors are used to compute test event spectra, \equ{num_test}.  The values are taken from \Refe~\cite{Hyper-Kamiokande:2016srs, DUNE:2021cuw}.  See Section~\ref{sec:results_stat_methods} for details.}
 \label{tab:normalization_err}
\end{table}

For T2HK or DUNE, separately or together, we compute the total $\chi^2$ by adding the contributions of all the detection channels, \ie,
\begin{eqnarray}
 \label{equ:chi2_dune}
 \chi_{\rm DUNE}^{2}
 (V_{\alpha\beta}, \boldsymbol{\theta}, o)
 &=&
 \sum_c \chi_{{\rm DUNE}, c}^{2}
 (V_{\alpha\beta}, \boldsymbol{\theta}, o)
 \;, \\
 \label{equ:chi2_t2hk}
 \chi_{\rm T2HK}^{2}
 (V_{\alpha\beta}, \boldsymbol{\theta}, o)
 &=&
 \sum_c \chi_{{\rm T2HK}, c}^{2}
 (V_{\alpha\beta}, \boldsymbol{\theta}, o) 
 \;, \\
 \label{equ:chi2_dune_t2hk}
 \chi^2_{{\rm DUNE}+{\rm T2HK}}
 (V_{\alpha\beta}, \boldsymbol{\theta}, o)
 &=&
 \chi^2_{\rm DUNE}
 (V_{\alpha\beta}, \boldsymbol{\theta}, o)
 +
 \chi^2_{\rm T2HK}
 (V_{\alpha\beta}, \boldsymbol{\theta}, o)
 \;, 
\end{eqnarray}
and we take the contributions of different channels to be uncorrelated.  

We report sensitivity by comparing the minimum value of the $\chi^2$ function, $\chi_{e, \rm min}^2$, reached when it is evaluated at the true values of the parameters, $V_{\alpha\beta}^{\rm true}$, $\boldsymbol{\theta}^{\rm true}$, and $o^{\rm true}$, against the value of $\chi^2$ evaluated at test values of the parameters.  We treat $\boldsymbol{\theta}$ and $o$ as nuance parameters and profile over them (more on this below).  For instance, for DUNE,
\begin{equation}
 \label{equ:delta_chi2_dune}
 \Delta\chi^2_{\rm DUNE}(V_{\alpha\beta})
 =
 \underset
 {\{ \boldsymbol{\theta}, o\}}{\mathrm{min}} 
 \left[
 \chi_{\rm DUNE}^{2}
 (V_{\alpha\beta}, \boldsymbol{\theta}, o) 
 -
 \chi_{\rm DUNE, min}^{2}
 \right]
 \;,
\end{equation}
and similarly for T2HK and DUNE+T2HK.  In the main text, we fix $\boldsymbol{\theta}^{\rm true}$ to its present-day best-fit value under NMO, from \Refe~\cite{Capozzi:2021fjo} (Table~\ref{tab:mix_param_benchmark}) and $o^{\rm true}$ also to NMO; we fix them to inverted mass ordering in Sec.~\ref{app:imo}.  When reporting {\it constraints} on long-range interactions, we set $V_{\alpha\beta}^{\rm true} = 0$ and extract from $\Delta \chi_{\rm DUNE}^2$, $\Delta \chi_{\rm T2HK}^2$, and $\Delta \chi_{\rm DUNE+T2HK}^2$ the upper limits on the inferred value of $V_{\alpha\beta}$, for 1~degree of freedom (d.o.f).  When reporting {\it discovery potential}, we fix $V_{\alpha\beta}^{\rm true}$ to a nonzero illustrative value and report the inferred range of values of $V_{\alpha\beta}$, again for 1~d.o.f.  When reporting discovery, we also study the correlation between $V_{\alpha\beta}$ and $\dcp$ or $\sin^2\theta_{23}$.  In those cases, we use instead, respectively,
\begin{eqnarray}
 \label{equ:delta_chi2_2dof_dune_vs_dcp}
 \Delta\chi^2_{\rm DUNE}
 (V_{\alpha\beta}, \dcp)
 &=&
 \underset
 {\{ \sin^2 \theta_{23}, \left\vert \Delta m_{31}^2 \right\vert, o\}}{\mathrm{min}} 
 \left[
 \chi_{\rm DUNE}^{2}
 (V_{\alpha\beta}, \boldsymbol{\theta}, o) 
 -
 \chi_{\rm DUNE, min}^{2}
 \right]
 \;, \\
 \label{equ:delta_chi2_2dof_dune_vs_th23}
 \Delta\chi^2_{\rm DUNE}
 (V_{\alpha\beta}, \sin^2 \theta_{23})
 &=&
 \underset
 {\{ \dcp, \left\vert \Delta m_{31}^2 \right\vert, o\}}{\mathrm{min}} 
 \left[
 \chi_{\rm DUNE}^{2}
 (V_{\alpha\beta}, \boldsymbol{\theta}, o) 
 -
 \chi_{\rm DUNE, min}^{2}
 \right]
 \;,
\end{eqnarray}
and similarly for T2HK and DUNE + T2HK, and we show allowed regions for 2~d.o.f.  After placing bounds on $V_{\alpha\beta}$, we translate them into bounds on $g_{\alpha\beta}^\prime$ for varying $m_{\alpha\beta}^\prime$ by means of the definition of the long-range matter potential, \equ{pot_total}. 

When profiling, we minimize the $\Delta\chi^2$ functions above with respect to $\sin^{2}\theta_{23}$, $\delta_{\mathrm{CP}}$, and $\vert \Delta m^{2}_{31} \vert$ by varying them within their $3\sigma$ allowed ranges from \Refe~\cite{Capozzi:2021fjo}; see Table~\ref{tab:mix_param_benchmark}.  We do not include correlations between them, since these are expected to weaken in coming years (see, \eg, \Refe~\cite{Song:2020nfh}), nor do we include pull terms on the mixing parameters in the test-statistic, in order to be conservative.  We keep $\theta_{13}$ and $\theta_{12}$ fixed at their present-day best-fit values~\cite{Capozzi:2021fjo}.  For $\theta_{13}$, the precision that Daya Bay has achieved, of 2.8\%~\cite{DayaBay:2022orm}, is not expected to be improved upon by upcoming experiments.  For $\theta_{12}$, whose present-day uncertainty is of 4.5\%, we expect only weak sensitivity in the oscillation probabilities (see Section~\ref{sec:experiments_probabilities}), so fixing its value is a justified approximation.  For the neutrino mass ordering, we adopt a simplified approach where switching from NMO to IMO amounts only to flipping the sign of $\Delta m_{31}^2$ to make it negative.  This approach is motivated by the fact that the present-day 3$\sigma$ allowed ranges of the mixing parameters are similar in the NMO and IMO~\cite{Capozzi:2021fjo, Esteban:2020cvm, NuFIT}, except for $\dcp$ and $\Delta m_{31}^2$.  (Further, in the next decade, DUNE is expected to determine the mass ordering, though, admittedly, non-standard oscillations like those induced by long-range interactions may confound this~\cite{Chatterjee:2015gta}.) 

Below, we compute the test-statistics by varying the long-range matter potential in the range $10^{-15} \leq V_{\alpha\beta}/{\rm eV} \leq 3 \times 10^{-13}$, where its effects are potentially visible in DUNE and T2HK.  This range is wide enough to comfortably place constraints or make measurements with significant statistical confidence.
\subsection{Projected constraints on long-range interactions}
\label{sec:results_constraints}
%
\begin{figure}[t!]
 \centering
 \includegraphics[width=\linewidth]{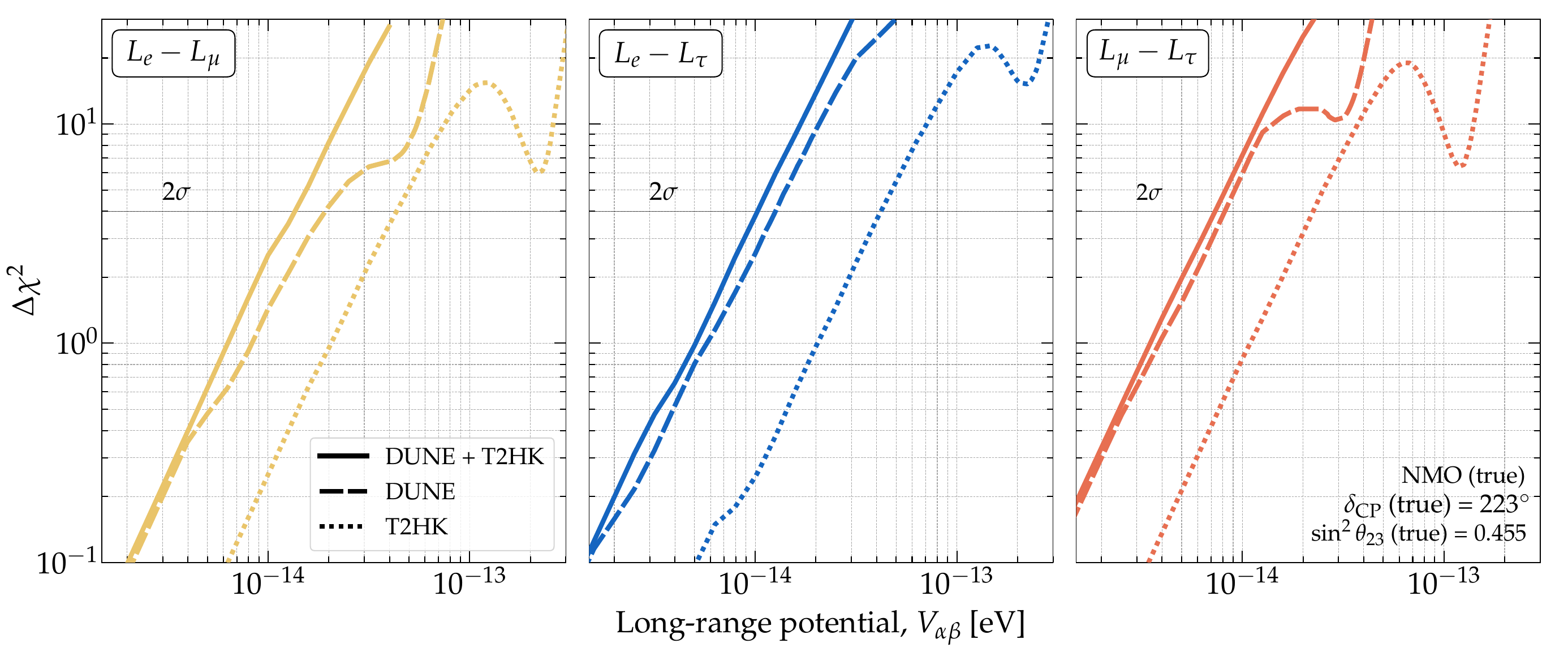}
 \caption{\textbf{\textit{Projected test-statistic used to constrain the long-range matter potentials $V_{e\mu}$, $V_{e\tau}$, and $V_{\mu\tau}$, using DUNE, T2HK, and their combination.}}  The $\Delta \chi^2$ function is \equ{delta_chi2_dune} and similar ones, assuming $V_{\alpha\beta}^{\rm true} = 0$ for the true value of the potentials.  We profile over the values of the most relevant standard mixing parameters and over the neutrino mass ordering; see Table~\ref{tab:mix_param_benchmark}.  See Table~\ref{tab:upper_limits_potential} for the resulting upper limits on the potentials and \figu{constraints_g_vs_m} for the corresponding constraints on the mass and coupling of the new mediator.  \textit{Combining DUNE and T2HK not only provides sensitivity to lower values of the potential, but also removes degeneracies in the test-statistics that would otherwise weaken the sensitivity.} See Sections~\ref{sec:results_stat_methods} and \ref{sec:results_constraints} for details.}
 \label{fig:test_statistics}
\end{figure}
%
Figure~\ref{fig:test_statistics} shows how the test-statistics for constraints, \eg, \equ{delta_chi2_dune}, vary with the potential, for the three symmetries and for DUNE, T2HK, and their combination.  As expected, they are smallest closer to the true value of the potential, $V_{\alpha\beta}^{\rm true} = 0$, and grow as they move away from it.  At high values of $V_{\alpha\beta}$, the test-statistics for DUNE and T2HK dip, reflecting a loss of sensitivity due to $V_{\alpha\beta}$ being degenerate with $\theta_{23}$ and $\dcp$.  Combining DUNE and T2HK removes the dips: T2HK lifts the degeneracies due to $\theta_{23}$ and $\dcp$, while DUNE fixes the mass ordering, \ie, the sign of $\Delta m_{31}^2$.  Thus, our results reveal novel insight: \textbf{\textit{the interplay of DUNE and T2HK facilitates degeneracy-free constraints on flavor-dependent long-range neutrino interactions.}}

\begin{figure}[htb!]
 \centering
 \includegraphics[width=\linewidth]{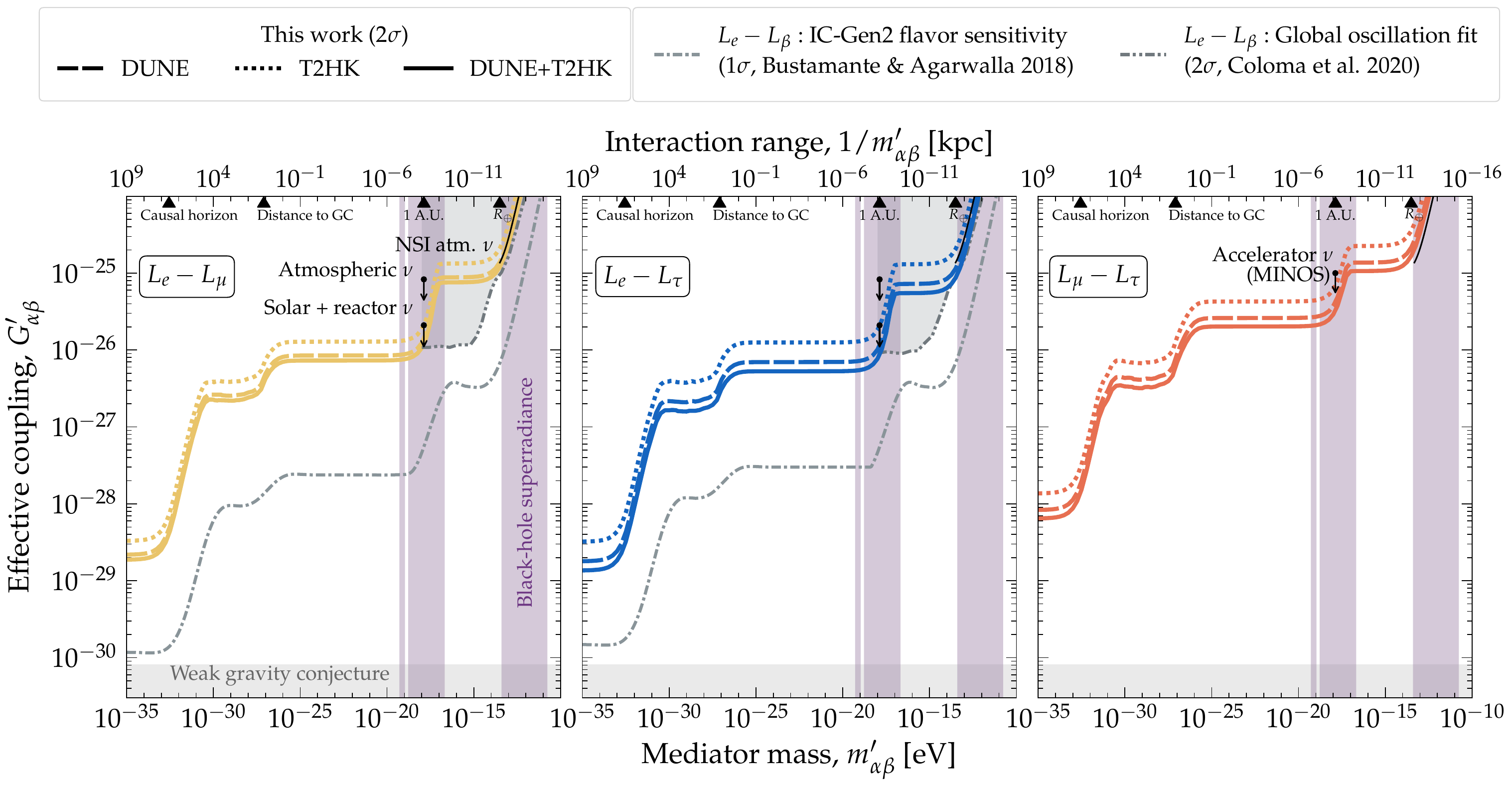}
 \caption{\textbf{\textit{Projected upper limits on the effective coupling, $G_{\alpha\beta}^\prime$ (refer to Eq.~\ref{equ:Gab}), of the new boson, $Z_{\alpha\beta}^\prime$, with mass $m_{\alpha\beta}^\prime$, that mediates flavor-dependent long-range neutrino interactions, using DUNE, T2HK, and their combination.}}  Same as \figu{moneyplot}, but now showing also limits using DUNE and T2HK separately.  {\it Left:} For neutrino-electron interactions under the $L_e-L_\mu$ symmetry.  {\it Center:} For neutrino-electron interactions under $L_e-L_\tau$.  {\it Right:} For neutrino-neutron interactions under $L_\mu-L_\tau$; existing limits from accelerator neutrinos in MINOS (95\% C.L.) are from \Refe~\cite{Heeck:2010pg}, and for non-standard interactions (NSI) we follow Appendix C in Ref.~\cite{Agarwalla:2023sng}. The span in the values of the coupling for $L_\mu-L_\tau$ is different because $V_{\mu\tau}$ scales $\propto g^\prime_{\mu\tau}$, whereas $V_{e\mu}$ and $V_{e\tau}$ scales $\propto g^{\prime 2}_{e\mu}$ and $\propto g^{\prime 2}_{e\tau}$; see Section~\ref{sec:lri}.  See Section~\ref{sec:results_constraints} for details.}
 \label{fig:constraints_g_vs_m}
\end{figure}

Table~\ref{tab:upper_limits_potential} shows the resulting upper limits on the potential.  They are strongest for $V_{\mu\tau}$, followed by $V_{e\tau}$ and then $V_{e\mu}$.  For DUNE, the limits are driven predominantly by the runs in neutrino mode, which contribute most of the total event rate; see Table~\ref{tab:event} and \figu{event_rates}.  For T2HK, the runs in neutrino and antineutrino modes contribute comparably.  For $L_\mu-L_\tau$, the limits on $V_{\mu\tau}$ are strongest because long-range interactions affect mainly the disappearance probabilities, $\nu_\mu \to \nu_\mu$ and $\bar{\nu}_\mu \to \bar{\nu}_\mu$ (see \figu{probabilities}), whose associated  disappearance detection channels have high event rates (see \figu{event_rates}), making deviations from standard oscillations easier to spot.  For $L_e-L_\tau$, long-range interactions enhance instead the appearance probabilities, $\nu_\mu \to \nu_e$ and $\bar{\nu}_\mu \to \bar{\nu}_e$, but the appearance detection channels have lower rates, so the limits on $V_{e\tau}$ are weaker.  For $L_e-L_\mu$, long-range interactions affect both the appearance probabilities --- though less so than under the other two symmetries --- and the disappearance probabilities --- though less so than under the $L_\mu-L_\tau$ symmetry; as a result, the limits on $V_{e\mu}$ are the weakest.  
%
\begin{table}[b!]
\centering
 \begin{tabular}{ | c | *{4}{>{\centering\arraybackslash}p{2.25cm} |}}
 \hline
 \multirow{2}{*}{Detector} &
 \multicolumn{3}{c|}{Upper limits ($2\sigma$) on  potential [$10^{-14}$~eV]} \\
  & $V_{e\mu}$ & $V_{e\tau}$ & $V_{\mu\tau}$ \\
 \hline
 DUNE        & $1.9$ & $1.3$ & $0.82$ \\
 T2HK        & $4.4$ & $4.2$ & $2.2$  \\
 DUNE + T2HK & $1.4$ & $1.0$ & $0.73$ \\
 \hline
 \end{tabular}
 \caption{\textbf{\textit{Projected upper limits (2$\sigma$) on the long-range matter potentials $V_{e\mu}$, $V_{e\tau}$, and $V_{\mu\tau}$, using DUNE, T2HK, and their combination.}}  See \figu{test_statistics} for the test-statistics from whence they originate and \figu{constraints_g_vs_m} for constraints on the mass and coupling of the associated mediator. See Sections~\ref{sec:results_stat_methods} and \ref{sec:results_constraints} for details.}
 \label{tab:upper_limits_potential}
\end{table}
Figure~\ref{fig:constraints_g_vs_m} (also \figu{moneyplot}) shows the corresponding upper limits on $G_{\alpha\beta}^\prime$ (refer to Eq.~\ref{equ:Gab}) for varying $m_{\alpha\beta}^\prime$, translated from the upper limits on $V_{\alpha\beta}$ in Table~\ref{tab:upper_limits_potential} via the definition of the potential, \equ{pot_total}.  Each curve in \figu{constraints_g_vs_m} is an isocontour of potential that saturates each of the upper limits in Table~\ref{tab:upper_limits_potential}.  The curves show step-like transitions at various values of $m_{\alpha\beta}^\prime$: as explained in \Refe~\cite{Bustamante:2018mzu} (see also Section~\ref{sec:lri_potential}), each transition reflects the interaction range becoming long enough for a new source of electrons or neutrons to contribute to the total potential, \equ{pot_total}.  For $m_{\alpha\beta}^\prime \sim 10^{-18}$--$10^{-10}$~eV, the Earth and the Moon dominate the upper limits; for $m_{\alpha\beta}^\prime \lesssim 10^{-18}$~eV, the Sun dominates; for $m_{\alpha\beta}^\prime \lesssim 10^{-27}$~eV, the Milky Way dominates; and, for $m_{\alpha\beta}^\prime \lesssim 10^{-33}$~eV, cosmological electrons and neutrinos dominate.  

Down to $m_{\alpha\beta}^\prime \sim 10^{-18}$~eV, where direct limits on flavor-dependent long-range neutrino interactions exist,  our projected limits improve on existing ones that use atmospheric neutrinos~\cite{Joshipura:2003jh}, are comparable to limits that use solar, reactor~\cite{Bandyopadhyay:2006uh}, and accelerator neutrinos~\cite{Heeck:2010pg}, and to limits culled from non-standard interactions~\cite{Super-Kamiokande:2011dam, Ohlsson:2012kf, Gonzalez-Garcia:2013usa}, but are weaker than limits from a global fit to oscillation data~\cite{Coloma:2020gfv}.  
%

Below $m_{\alpha\beta}^\prime \sim 10^{-18}$~eV, our projected limits tread into a largely unexplored range.  To our knowledge, the only constraints that exist there, other than the indirect, tentative one from the weak gravity conjecture~\cite{Arkani-Hamed:2006emk}, are from measurements of the flavor composition of high-energy astrophysical neutrinos in the IceCube neutrino telescope, from \Refe~\cite{Bustamante:2018mzu}, which, however, were only produced at the $1\sigma$ level as a proof of principle of the sensitivity.  A recent recalculation~\cite{Agarwalla:2023sng} found comparable results at higher statistical significance, using more solid statistical methods.  Ideally, the sensitivity that could be reaped from high-energy astrophysical neutrinos is unmatched due to them having energies in the TeV--PeV range --- which enhances the possible contribution of long-range interactions relative to standard oscillations --- and to the fact that neutrinos of all flavors are detected.  However, it is presently downplayed by large astrophysical uncertainties, limited event rates, and the difficulty in measuring the flavor composition in neutrino telescopes.  These issues will likely be surmounted in the future~\cite{Ackermann:2019cxh, Arguelles:2019rbn, Ackermann:2022rqc}.  For now, \figu{constraints_g_vs_m} shows the projected proof-of-principle sensitivity of the envisioned IceCube-Gen2 upgrade~\cite{IceCube-Gen2:2020qha}, from \Refe~\cite{Bustamante:2018mzu}.  Our limits from DUNE and T2HK improve on it significantly due to high event rates and well-characterized neutrino beams.

\subsection{Discovering subdominant long-range interactions}
\label{sec:results_discovery}
\begin{figure}[htb!]
 \centering
 \includegraphics[width=\linewidth]{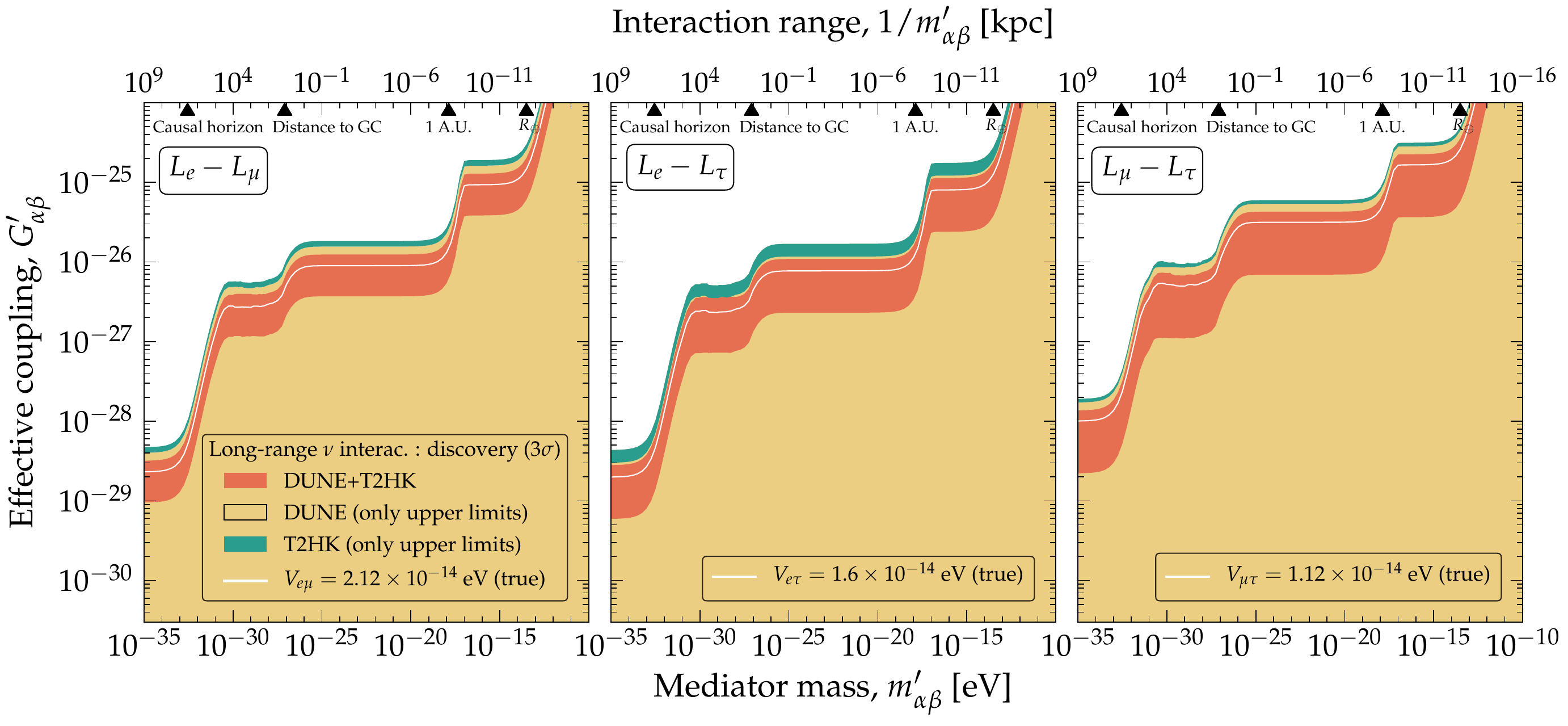}
 \caption{\textbf{\textit{Projected discovery potential of flavor-dependent long-range neutrino interactions.}}  We show allowed ranges ($3\sigma$) of the effective coupling, $G_{\alpha\beta}^\prime$ (refer to Eq.~\ref{equ:Gab}), of the new boson, $Z_{\alpha\beta}^\prime$, with mass $m_{\alpha\beta}^\prime$, that mediates the interactions, using DUNE, T2HK, and their combination.  The $\Delta \chi^2$ function is \equ{delta_chi2_dune} and similar ones, fixing the true value of the long-range potential, $V_{\alpha\beta}^{\rm true}$, at test values chosen to make the long-range interactions subdominant.  Like in \figu{constraints_g_vs_m}, we either fix or profile over the standard mixing parameters and the neutrino mass ordering; see Table~\ref{tab:mix_param_benchmark}.  See related Figs.~\ref{fig:ranges_dcp_V} and \ref{fig:ranges_th23_V}.  \textit{DUNE or T2HK may not be able to discover long-range interactions separately, but their combination may.} See Sections~\ref{sec:results_stat_methods} and \ref{sec:results_discovery} for details.}
 \label{fig:g_vs_m_discovery}
\end{figure}
%
\begin{figure}[htb!]
 \includegraphics[width=\linewidth]{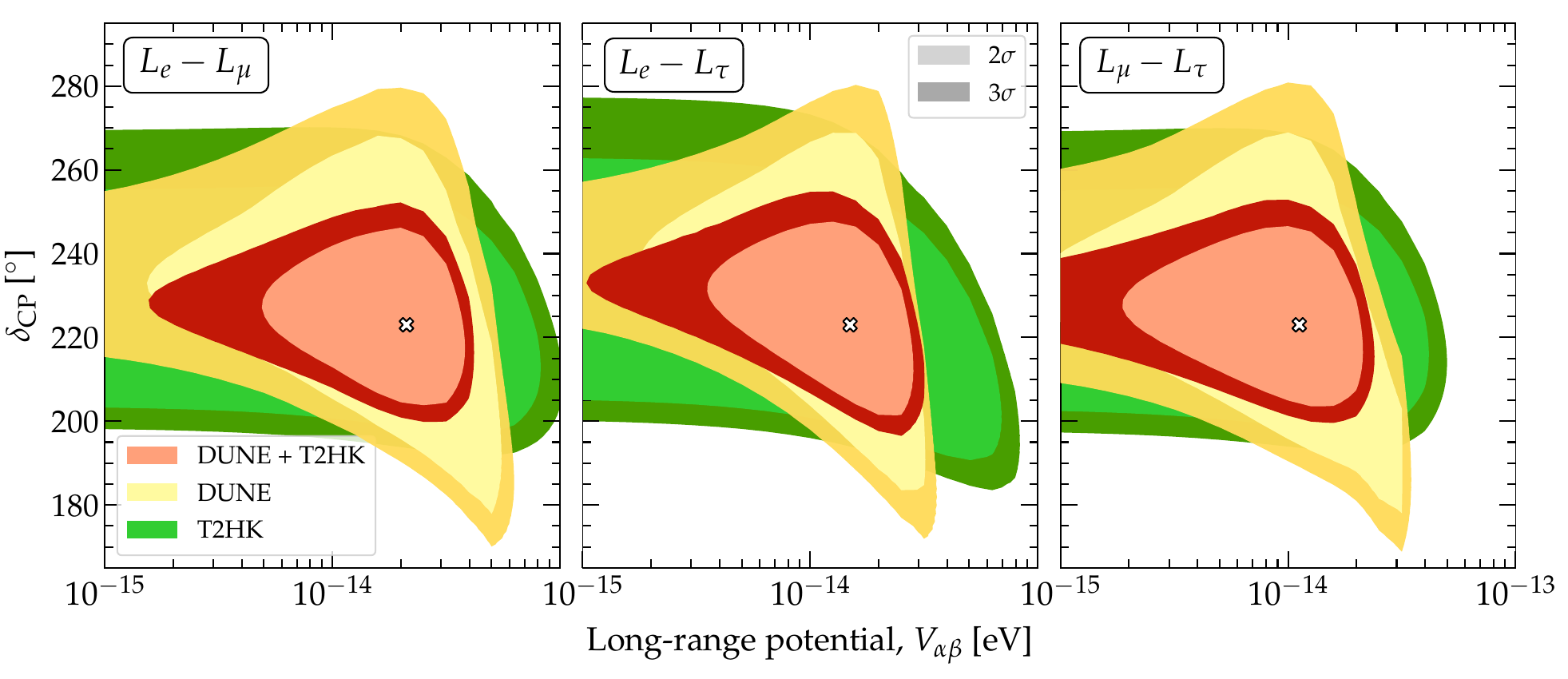}
 \caption{\textbf{\textit{Allowed regions of the long-range potential, $V_{\alpha\beta}$, and the CP-violating phase, $\dcp$.}}  The true values of the potentials are the same as in \figu{g_vs_m_discovery}.  The test-statistic is profiled over $\sin^2\theta_{23}$, $\lvert \Delta m^2_{31} \rvert$, and the mass ordering; see, \eg, \equ{delta_chi2_2dof_dune_vs_dcp}.  See Sections~\ref{sec:results_stat_methods} and \ref{sec:results_discovery} for details.}
 \label{fig:ranges_dcp_V}
\end{figure}
%
Figure~\ref{fig:g_vs_m_discovery} shows the inferred allowed ranges of the coupling, for varying mediator mass, for three illustrative choices of the true value of the long-range potential, one for each symmetry: $V_{e\mu}^{\rm true}$ = $2.12 \times 10^{-14}$~eV, $V_{e\tau}^{\rm true}$ = $1.6 \times 10^{-14}$~eV, and $V_{\mu\tau}^{\rm true}$ = $1.12 \times 10^{-14}$~eV.  They represent long-range interactions that are subdominant to standard oscillations; see Section~\ref{sec:lri_osc_prob}.   To generate \figu{g_vs_m_discovery}, first we compute the allowed ranges of $V_{\alpha\beta}$ using the discovery test-statistics, \eg, \equ{delta_chi2_dune} with $V_{\alpha\beta}^{\rm true}$ fixed to the above illustrative choices, and then we use \equ{pot_total} to translate those into allowed ranges of $g_{\alpha\beta}^\prime$ for different values of $m_{\alpha\beta}^\prime$.  We present results for a modest discovery significance of $3\sigma$.

Figure~\ref{fig:g_vs_m_discovery} shows that DUNE and T2HK, by themselves, can only place upper limits on $g_{\alpha\beta}^\prime$.  Thus, our discovery forecasts also reveal novel insight: \textbf{\textit{DUNE and T2HK, by themselves, may be unable to discover subdominant long-range interactions, but their combined action may.}}  For the illustrative choices of the potentials, the allowed $3\sigma$ ranges combining DUNE and T2HK are $V_{e\mu} \in [3.62 \times 10^{-15}, 4.04 \times 10^{-14}]$~eV, $V_{e\tau} \in [1.41 \times 10^{-15}, 3.13 \times 10^{-14}]$~eV, and $V_{\mu\tau} \in [5.87 \times 10^{-16}, 2.24 \times 10^{-14}]$~eV, implying a relative measurement uncertainty of 90\%--100\%.  For larger values of the true potential, discovery claims should be stronger and the uncertainty in its measurement should shrink.

\begin{figure}[htb!]
 \includegraphics[width=\linewidth]{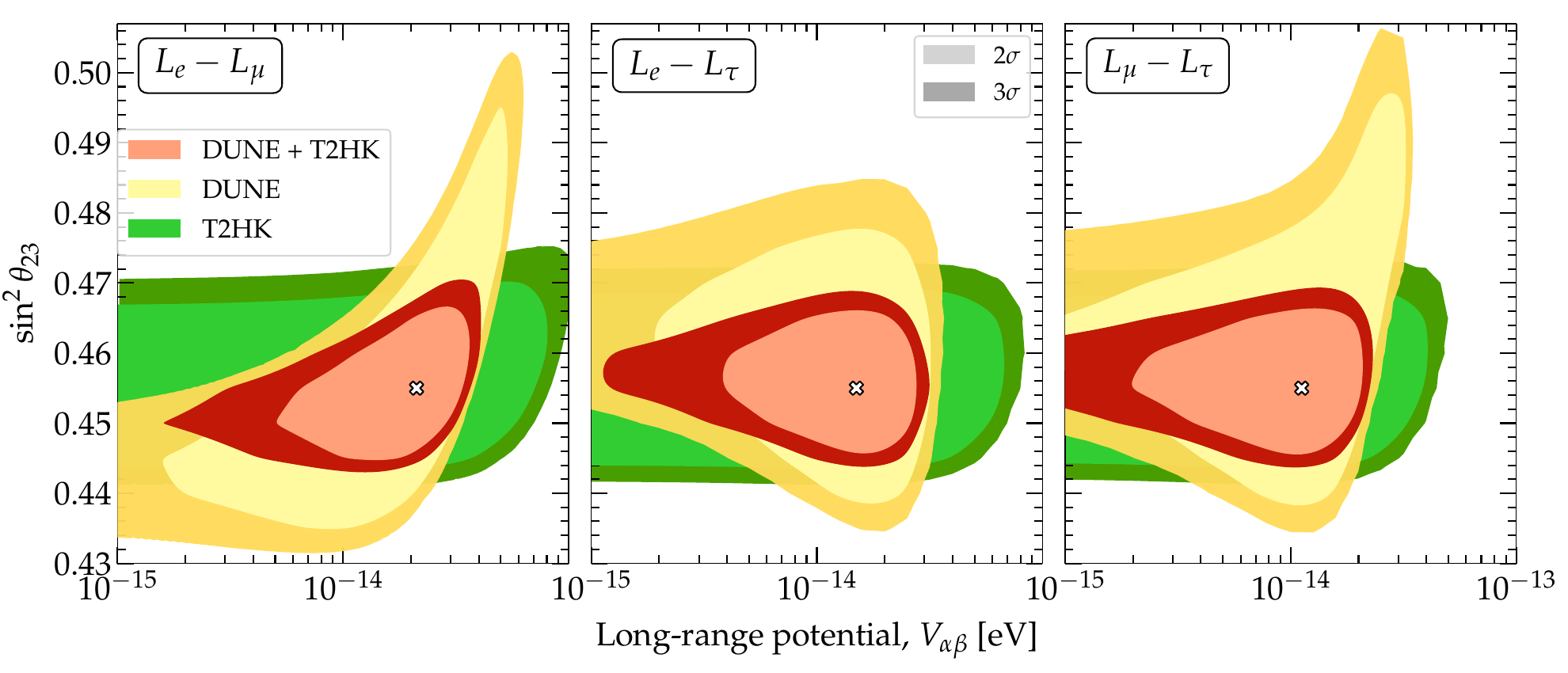}
 \caption{\textbf{\textit{Allowed regions of the long-range potential, $V_{\alpha\beta}$, and the atmospheric mixing angle, $\sin^2\theta_{23}$.}}  Same as \figu{ranges_dcp_V}, but profiling the test-statistic instead over $\dcp$, $\lvert\Delta m_{31}^2\rvert$, and the mass ordering; see, \eg, \equ{delta_chi2_2dof_dune_vs_th23}.  See Sections~\ref{sec:results_stat_methods} and \ref{sec:results_discovery} for details.}
 \label{fig:ranges_th23_V}
\end{figure}

Figures~\ref{fig:ranges_dcp_V} and \ref{fig:ranges_th23_V} reveal that the reason behind the difficulty of T2HK and DUNE to discover subdominant long-range interactions by themselves are the uncertainties in $\dcp$, $\theta_{23}$, and the neutrino mass ordering.  On the one hand, in T2HK, the shorter baseline provides less contamination from fake CP violation induced by SM matter effects and, therefore, higher precision in measuring $\dcp$~\cite{Ballett:2016daj, Bernabeu:2018twl, Bernabeu:2018use, King:2020ydu, Agarwalla:2022xdo}, while the high event rates in the disappearance channels provide high precision on $\theta_{23}$.  However, at the same time, the shorter baseline reduces the sensitivity to the mass ordering and provides a shorter neutrino travel time during which long-range interactions may act.  On the other hand, in DUNE, the longer baseline helps to pin down the mass ordering, but introduces more contamination from fake CP violation, which degrades the sensitivity to $\dcp$ compared to T2HK.  Also, in the presence of long-range interactions, DUNE can measure $\theta_{23}$ significantly less precisely than T2HK.  
 
Thus, combining DUNE and T2HK improves the sensitivity with which $\dcp$, $\theta_{23}$, and the mass ordering can be measured, weakens the degeneracies between them and the long-range potential, and allows for its measurement at a high statistical significance.
%
%
\section{Constraints assuming inverted mass ordering }
\label{app:imo}

\begin{table}[b!]
 \centering
 \begin{tabular}{|c|c|c|c|c|c|c|}
  \hline 
  \multirow{2}{*}{} & \multicolumn{6}{c|}{Standard mixing parameters (IMO)} \\
   & $\sin^2 \theta_{12}$ & $\sin^2\theta_{23}$ & $\sin^2 \theta_{13}$ &
  $\frac{\Delta m^2_{31}}{10^{-3}\,\text{eV}^2}$  & $\frac{\Delta m^2_{21}}{10^{-5}\,\text{eV}^2}$  & $\delta_{\rm CP}\, (^\circ)$\\[0.8ex]
  \hline
  Benchmark & 0.303 & 0.569 & 0.0223 & 2.418 & 7.36 & 274  \\
  Status in fits & Fixed & Minimized & Fixed & Minimized & Fixed & Minimized \\
  Range & -- & [0.4, 0.6] & -- & [2.341, 2.501] & -- & [193, 342]\\
  \hline 
 \end{tabular}
 \caption{\textbf{\textit{Values of the standard mixing parameters used in our analysis, assuming that the true neutrino mass ordering is inverted.}}  Same as Table~\ref{tab:mix_param_benchmark}, made assuming that the true mass ordering is normal, but instead assuming that it is inverted (IMO).  The benchmark values are the best-fit values from \Refe~\cite{Capozzi:2021fjo}.}
 \label{tab:mix_param_benchmark_imo}
\end{table} 

\begin{figure}[t!]
 \centering
 \includegraphics[width=\linewidth]{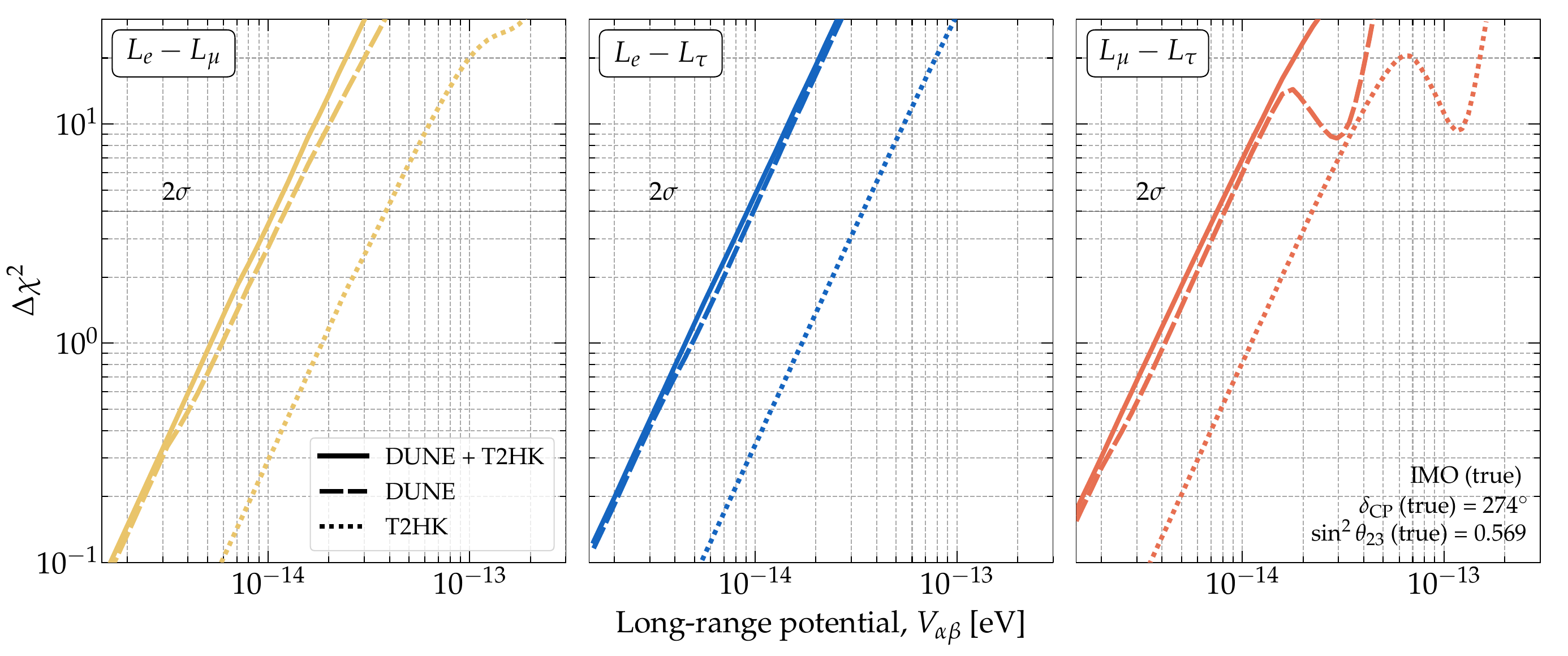}
 \caption{\textbf{\textit{Projected test-statistic used to constrain the long-range matter potentials $V_{e\mu}$, $V_{e\tau}$, and $V_{\mu\tau}$, using DUNE, T2HK, and their combination, assuming that the true neutrino mass ordering is inverted.}}  Same as \figu{test_statistics}, made assuming that the true mass ordering is normal, but instead assuming that it is inverted.  See Table~\ref{tab:upper_limits_potential_imo} for the resulting upper limits on the potentials.  See Sections~\ref{sec:results_stat_methods} and \ref{sec:results_constraints} in the main text, and Sec.~\ref{app:imo} for details.}
 \label{fig:test_statistics_imo}
\end{figure}

Previously, we showed results obtained assuming that the true neutrino mass ordering is normal.  Here we show constraints  assuming instead that the true ordering is inverted.  They are broadly similar to the ones obtained under normal ordering (Section~\ref{sec:results_constraints}).  

Table~\ref{tab:mix_param_benchmark_imo} shows the values and allowed ranges of the mixing parameters that we use when assuming that the inverted mass ordering is true, taken from the global oscillation fit of \Refe~\cite{Capozzi:2021fjo}.  The values are similar as for normal ordering (Table~\ref{tab:upper_limits_potential}), with the important difference that for inverted ordering the benchmark value of $\theta_{23}$ lies in the higher octant, which has consequences for the sensitivity, as we point out below.

Figure~\ref{fig:test_statistics} shows how the test-statistics for constraints, \eg, \equ{delta_chi2_dune}, vary with the long-range potential, for the three symmetries and for DUNE, T2HK, and their combination, assuming the true mass ordering is inverted.  Their behavior is broadly similar to that in \figu{constraints_g_vs_m} for normal ordering.  However, in \figu{test_statistics_imo} the weakening of the sensitivity due to parameter degeneracies is milder than in \figu{test_statistics} because, assuming inverted mass ordering, $\theta_{23}$ lies in the higher octant, which lessens the influence of $\dcp$~\cite{Ballett:2016daj,Agarwalla:2022xdo}.

Table~\ref{tab:upper_limits_potential_imo} shows the resulting upper limits on the long-range potentials.  Like for normal ordering (Table~\ref{tab:upper_limits_potential}), they are strongest for $V_{\mu\tau}$.  Compared to normal ordering, the limits on $V_{e\mu}$ are stronger by a factor of 2 or more, and the limits on $V_{e\tau}$ are similarly weaker.

\begin{table}[t!]
\centering
 \begin{tabular}{ | c | *{4}{>{\centering\arraybackslash}p{2.25cm} |}}
 \hline
 \multirow{2}{*}{Detector} &
 \multicolumn{3}{c|}{Upper limit ($2\sigma$) on potential [$10^{-14}$~eV]} \\
  & $V_{e\mu}$ & $V_{e\tau}$ & $V_{\mu\tau}$ \\
 \hline
 DUNE        & $0.99$ & $3.4$ & $0.92$ \\
 T2HK        & $1.2$ & $3.9$ & $1.1$  \\
 DUNE + T2HK & $0.82$ & $2.2$ & $0.75$ \\
 \hline
 \end{tabular}
 \caption{\textbf{\textit{Projected upper limits (2$\sigma$) on the long-range matter potentials $V_{e\mu}$, $V_{e\tau}$, and $V_{\mu\tau}$, using DUNE, T2HK, and their combination, assuming that the true neutrino mass ordering is inverted.}}  Same as Table~\ref{tab:upper_limits_potential}, made assuming that the true mass ordering is normal, but instead assuming that it is inverted.  See \figu{test_statistics_imo} for the test-statistics from whence they originate and \figu{constraints_g_vs_m}. See Sections~\ref{sec:results_stat_methods} and \ref{sec:results_constraints}, and Sec.~\ref{app:imo} for details.}
 \label{tab:upper_limits_potential_imo}
\end{table}

\section{Summary }
\label{sec:conclusions}

Neutrinos are powerful probes of new physics.  Extant uncertainties in their properties leave room for the conceivable possibility that they experience interactions with matter beyond weak ones.  Discovering them would not only further our view of neutrino physics, but also represent striking evidence of physics beyond the Standard Model.  In the 2030s, the next-generation long-baseline neutrino experiments DUNE and T2HK may provide us with an opportunity to look for new neutrino interactions more incisively than ever before, thanks to high event rates and well-characterized neutrino beams.  We have forecast their reach.  Because we use detailed simulations of the detectors, including efficiencies, run times, and backgrounds, our predictions are realistic.  

Our forecasts are geared at new flavor-dependent neutrino interactions that are introduced by gauging three different accidental global lepton-number symmetries of the Standard Model, generated by $L_e-L_\mu$, $L_e-L_\tau$, and $L_\mu-L_\tau$, that have received prior attention in other experimental settings~\cite{Joshipura:2003jh, Bandyopadhyay:2006uh,  Heeck:2010pg, Chatterjee:2015gta, Khatun:2018lzs, Wise:2018rnb, Bustamante:2018mzu, Coloma:2020gfv,Singh:2023nek}.  We focus on them because they can be gauged anomaly-free, so the only new particle introduced is a neutral vector boson that mediates the interaction.  Its mass and coupling strength are a priori undetermined.  Gauging $L_e-L_\mu$ and $L_e-L_\tau$ introduces new neutrino-electron interactions.  Gauging $L_\mu-L_\tau$ introduces new neutrino-neutron interactions.  

Under these interactions, electrons and neutrons source a flavor-dependent potential that may affect neutrino oscillations.  We concentrate on ultra-light mediators, with masses below $10^{-10}$~eV.  They induce interactions whose range is ultra-long --- ranging from hundreds of meters to Gpc, depending on the mass --- so that neutrinos may experience the potential sourced by a large number of nearby and distant electrons and neutrons in the Earth, the Moon, the Sun, the Milky Way, and the cosmological matter distribution~\cite{Bustamante:2018mzu}.  Yet, because the coupling strength may be tiny, their effects on the oscillation probabilities may be subtle and, therefore, testable with future experiments, like DUNE and T2HK.  

Our forecasts reveal two novel, promising perspectives.  First, while DUNE and T2HK, individually, should be able to improve on present-day upper limits on the coupling strength of the new interaction, their individual sensitivities are hampered by degeneracies due to uncertainties in the mixing angle $\theta_{23}$, the CP-violating phase, $\dcp$, and the neutrino mass ordering.  Yet, DUNE and T2HK have complementary capabilities: while T2HK is especially well-suited to measure $\theta_{23}$ and $\dcp$, DUNE is especially well-suited to measure the neutrino mass ordering.  Thus, \textbf{\textit{combining DUNE and T2HK removes parameter degeneracies, which tightens the upper limits on long-range neutrino interactions.}}  Second, and more importantly, \textbf{\textit{neither DUNE nor T2HK, by itself, may discover subdominant long-range interactions, owing to parameter degeneracies, but their combination may}}.  Thus, our forecasts stress the need for combining measurements in DUNE and T2HK to probe long-range interactions.

More broadly, our results illustrate the known need for complementarity in the future long-baseline neutrino program, not only to measure the standard mixing parameters, but to search for new physics. For flavor-dependent long-range neutrino interactions, a future global fit to oscillation data is poised to deliver substantially improved limits or transformative discovery.
\chapter{Summary and Future Perspective}
\label{sec9:conclusion}

Neutrino oscillation, a remarkable phenomenon in particle physics, unveils the intriguing nature of neutrinos, the elusive and ghost-like subatomic particles. Unlike their electrically charged counterparts, neutrinos carry no charge and interact only weakly with matter, making them exceptionally difficult to detect. First hypothesized by physicist Wolfgang Pauli in 1930 to account for the conservation of energy and momentum in beta decay, these enigmatic particles became the subject of intense scrutiny and experimental investigations in the latter half of the 20th century.

The concept of neutrino oscillation stems from a surprising revelation that neutrinos can transform into different types or flavors as they propagate through space. These flavors are associated with the three known types of neutrinos: electron neutrinos, muon neutrinos, and tau neutrinos. Initially thought to be distinct and immutable entities, experimental evidence began to challenge this view in the late 20th century.

In the quest to unravel the mysteries of neutrino oscillation, physicists designed ingenious experiments utilizing neutrino beams produced in particle accelerators or generated from natural cosmic sources. Through these experiments, they observed a fascinating phenomenon: neutrinos transitioning between different flavors during their journey across vast distances. This groundbreaking discovery not only provided definitive evidence for neutrino mass, which had long been an open question in the Standard Model of particle physics, but it also hinted at the possibility of physics beyond the known realm.

It is now a renowned fact that the phenomenon of neutrino oscillation gets affected in presence of matter. Therefore, a detailed understanding of Earth's Matter effect is inevitable to correctly analyze the data from the upcoming high-precision long-baseline experiments to resolve the remaining fundamental unknowns such as neutrino mass ordering, leptonic CP violation and precision measurements of the oscillation parameters. In this thesis, for the first time, we have explored in detail the capability of Deep Underground Neutrino Experiment (DUNE) to establish the matter oscillation as a function of $\delta_{\mathrm{CP}}$ and $\theta_{23}$ by excluding the vacuum oscillation. We find that DUNE is sensitive to Earth's matter effect at more than 2$\sigma$ C.L. irrespective of the choice of the oscillation parameters. The relative 1$\sigma$ precision in the measurement of line-averaged constant Earth matter density ($\rho_{\mathrm{avg}}$ ) for maximal CP-violating choices of $\delta_{\mathrm{CP}}$ is around 10\% to 15\% depending on the choice of neutrino mass ordering. If $\delta_{\mathrm{CP}}$ turns out to be around -90$^{\circ}$ or 90$^{\circ}$ , the precision in measuring $\rho_{\mathrm{avg}}$ is better in DUNE as compared to what are achievable from the Super-K atmospheric data, combined data from Solar and KamLand, and full exposure of T2K and NO$\nu$A. We also observe new interesting degeneracies among $\rho_{\mathrm{avg}}$ , $\delta_{\mathrm{CP}}$ , and $\theta_{23}$ and notice that the present uncertainty in $\delta_{\mathrm{CP}}$ dilutes more the measurement of $\rho_{\mathrm{avg}}$ compared to $\theta_{23}$ . To lift these degeneracies, we have incorporated the prospective data from the upcoming Tokai to Hyper-Kamiokande (T2HK). With a relatively shorter baseline and high statistics at first oscillation maximum, T2HK offers unprecedented sensitivity to establish genuine CP violation and to measure $\delta_{\mathrm{CP}}$. We have explored interesting complementarities among these possible setups and we find that the combined data from DUNE and T2HK can establish Earth's matter effect at more than 5$\sigma$ C.L. irrespective of the choices of mass ordering, $\delta_{\mathrm{CP}}$, and $\theta_{23}$.

Present global analyses of the existing neutrino oscillation data points to near-percent-level-relative 1$\sigma$ precision in oscillation parameters such as $|\Delta m^2_{31}|$ (1.1\%), $\Delta m^2_{21}$ (2.3\%), $\sin^{2}\theta_{13}$ (3.0\%), and $\sin^2\theta_{12}$ (4.5\%). All these analyses show a preference for the normal mass ordering, thus favoring the inverted mass ordering at nearly $\sim 2.5\sigma$, while the best-fit for $\theta_{23}$ in the higher octant at $\sin^2\theta_{23} \sim 0.57$. All three of them allow the solution in the other octant at 2$\sigma$ or less. The primary goal of DUNE is to conclusively find out the sign of $\Delta m^2_{31}$, the value of $\sin^2\theta_{23}$, and the value of CP phase. From this thesis work, we find that disappearance data from DUNE can improve the current precision in the measurements of $|\Delta m^2_{31}|$ and $\sin^2\theta_{23}$ by a factor of three while the appearance data can very effectively eliminate the wrong-octant solution. Our results show that DUNE can resolve the octant of $\sin^2\theta_{23}$ at $4.2\sigma\, (5\sigma)$ using 7 (10) years of run, assuming $\sin^2\theta_{23}=0.455\,, \, \dcp = 223^\circ$, and NMO. DUNE can improve the current relative $1\sigma$ precision on $\sin^2\theta_{23} (\Delta m^2_{31})$ by a factor of 4.4 (2.8) using 7 years of run.

After the landmark discovery of non-zero $\theta_{13}$ by the modern reactor experiments, unprecedented precision on neutrino mass-mixing parameters has been achieved over the past decade. This has set the stage for the discovery of leptonic CP violation (LCPV) at high confidence level in the next-generation long-baseline neutrino oscillation experiments. In this thesis work, we have focused on complementarity among the on-axis DUNE and off-axis T2HK experiments to enhance the sensitivity to LCPV suppressing the $\theta_{23}-\delta_{\mathrm{CP}}$ degeneracy. We find that none of these
experiments individually can achieve the milestone of 3$\sigma$ LCPV for at least 75\% choices of $\delta_{\mathrm{CP}}$ in its entire range of $[-180^{\circ} , 180^{\circ}]$, with their nominal exposures and systematic uncertainties. However, their combination can attain the same for all values of $\theta_{23}$
with only half of their nominal exposures. We also observe for the first time that the proposed T2HKK setup in combination with DUNE can further increase the CP coverage to more than 80\% with only half of their nominal
exposures. We have studied in detail how the coverage in $\delta_{\mathrm{CP}}$ for $\ge$ 3$\sigma$ LCPV depends on the choice of $\theta_{23}$, exposure, optimal runtime in neutrino and antineutrino modes, and systematic uncertainties in these experiments in isolation and combination. We find that with an improved systematic uncertainty of 2.7\% in appearance mode, the standalone T2HK
setup can provide a CP coverage of around 75\% for all values of $\theta_{23}$. We have also discussed the pivotal role of intrinsic, extrinsic, and total CP asymmetries in the appearance channel and
extrinsic CP asymmetries in the disappearance channel while analyzing our results.

Once DUNE and T2HK establishes unprecedented precision on oscillation parameters, their next goal would be to look for beyond the realm of standard 3-$\nu$ interaction. Discovering new neutrino interactions would represent evidence of physics beyond the Standard Model. In this thesis, we have focused on new flavor-dependent long-range neutrino interactions mediated by ultra-light mediators, with masses below $10^{-10}$~eV, introduced by new lepton-number gauge symmetries $L_e-L_\mu$, $L_e-L_\tau$, and $L_\mu-L_\tau$. Because the interaction range is ultra-long, nearby and distant matter --- primarily electrons and neutrons --- in the Earth, Moon, Sun, Milky Way, and the local Universe, may source a large matter potential that modifies neutrino oscillation probabilities. The upcoming DUNE and T2HK long-baseline neutrino experiments will provide an opportunity to search for these interactions, because of their high event rates and well-characterized neutrino beams.  We forecast their probing power.  Our results reveal novel perspectives. Alone, DUNE and T2HK may strongly constrain long-range interactions, setting new limits on their coupling strength for mediators lighter than $10^{-18}$~eV.  However, if the new interactions are subdominant, then both DUNE and T2HK, together, will be needed to discover them, since their combination lifts parameter degeneracies that weaken their individual sensitivity.  DUNE and T2HK, especially when combined, provide a valuable opportunity to explore physics beyond the Standard Model.

This thesis work has placed a great deal of emphasis on the complementarity between the two upcoming long-baseline experiments: DUNE and T2HK. Both experiments will be able to probe very different aspects of neutrino physics. By combining the data from them, we can gain deeper insight into the unknowns of the neutrino sector. The future holds great promise for leveraging the collaborative synergy between DUNE and T2HK, extending its potential impact from standard three-neutrino interactions to broader scientific horizons. We are confident that the combined data from these experiments will significantly advance our comprehension of neutrino properties, acting as a catalyst for a more profound understanding of the universe. Such revelations could prove instrumental in unraveling the enigmas surrounding dark matter, dark energy, and the origins of matter itself. The collaborative efforts of DUNE and T2HK may indeed mark the initiation of a transformative era in neutrino physics, paving the way for groundbreaking discoveries and advancements in our understanding of the cosmos.

\bibliography{bib}
\bibliographystyle{ieeetr}

\end{document}